\documentclass{report}

\usepackage{graphicx}
\usepackage{xcolor}
\usepackage{physics}
\usepackage{dcolumn}
\usepackage{multicol}
\usepackage{bm}
\usepackage{amsmath}
\usepackage{hyperref}
\usepackage{amssymb}
\usepackage[export]{adjustbox}[2011/08/13] 
\usepackage{float} 
\usepackage[style=trad-unsrt,citestyle=numeric-comp]{biblatex}
\usepackage{verbatim}
\usepackage{bbold}
\usepackage{mathtools}

\DeclarePairedDelimiter\floor{\lfloor}{\rfloor}

\newcommand\sbullet[1][.5]{\mathbin{\vcenter{\hbox{\scalebox{#1}{$\bullet$}}}}}

\numberwithin{equation}{chapter}
\begin{document}

\begin{titlepage}
\begin{center}
    \includegraphics[width=5.1cm]{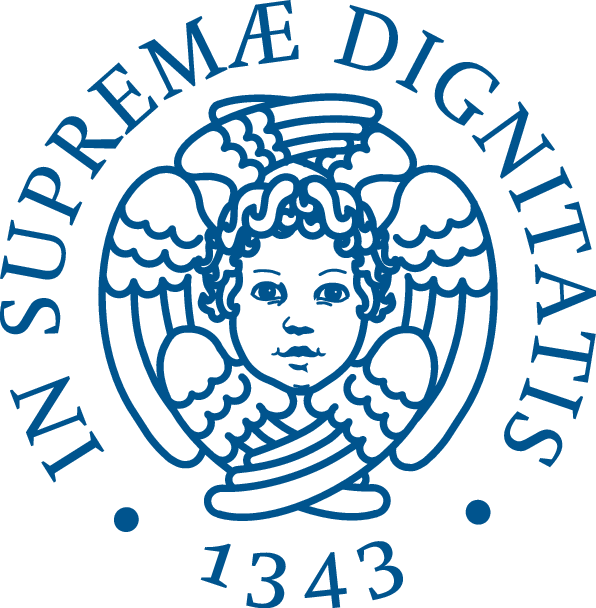} \\
    \vspace{1 cm}
    \LARGE
    \textsc{Università di Pisa} \\
    \vspace{0.25cm}
    \Large
    \textsc{Dipartimento di Fisica ``Enrico Fermi''} \\
    \vspace{1 cm}
    \LARGE
    Corso di Laurea Magistrale in Fisica \\
    \vspace{2.1 cm}
    \textbf{The DSSYK Model: \\ Charge and Holography} \\
\end{center}

\vspace{2.2 cm}
    \Large
    \noindent
    \textit{Relatore:} \hfill \textit{Candidato:} \\
    \textbf{Dr. Guilherme L. Pimentel} \hfill \textbf{Mattia Arundine} \\
    \phantom{a} \\
    \textit{Tutor:} \\ \textbf{Prof. Alessandro Vichi} \\
\vspace{2 cm}

\hrule
\begin{center}
    \Large ANNO ACCADEMICO 2023/2024
\end{center}
\end{titlepage}

\newpage

\begin{abstract}
	The Anti-de Sitter/Conformal Field Theory correspondence (AdS/CFT) is one of the most significant findings in theoretical physics and forms the basis of this thesis. Although highly powerful, the main limitation of AdS/CFT is that AdS does not appear in the real world outside of very specific limits. This limitation justifies the attempt to generalize the holographic principle to other spacetimes. In this thesis, we will pursue this direction and seek spacetimes that have at least some connection to de Sitter (dS), whose cosmological interest is evident. dS is, in fact, not only the geometry that best represents our Universe on large scales in the present, but also during the inflationary epoch that followed the Big Bang, where our current description of Nature fails completely.
	
	First, we will present some basic facts about the AdS/CFT correspondence. Then, the gravitational path integral will be introduced. After presenting the Sachdev-Ye-Kitaev (SYK) model and dilaton-gravity models, a holographic link between the two will be established. Next, we will discuss the double-scaled limit of SYK, known as DSSYK. We will then consider a ``charged'' variation of SYK with Dirac fermions, for which we will determine the fermionic two-point function. After studying the thermodynamic properties of the gravitational dual of DSSYK and the quasinormal modes of massive real scalars propagating in this geometry, we will conjecture how to modify the duality when considering the Dirac version of the model, showing that several bounds constrain the space of possible dual theories. Finally, we will summarize our findings and present an outlook on possible future developments based on the results described here.
\end{abstract}

\newpage
\pagestyle{plain}
\tableofcontents
\newpage

\chapter{Introduction}
The Anti-de Sitter/Conformal Field Theory correspondence (AdS/CFT) is one of the most significant findings in theoretical physics and forms the basis of this thesis. It is a vast and active field of research, with several introductory reviews \cite{AHARONY2000183,natsuume2016adscft,ZaffaroniAdSCFT,SISSAAdSCFT}. We will assume that the reader has a basic knowledge of CFTs, so we will not present their properties here: some references in this case are \cite{DiFrancesco:1997nk,Simmons-Duffin:2016gjk,Poland_2019}.

The AdS/CFT duality states that every theory of quantum gravity living in a $(d+1)$-dimensional (asymptotically) Anti-de Sitter spacetime (AdS$_d$ is reviewed in Section \ref{sec:NearHorizon} and Appendix \ref{app:AdS}) can be described in terms of a dual non-gravitational $d$-dimensional CFT and vice versa. In particular, the CFT can be seen as ``living'' on the timelike boundary of (asymptotic) AdS. We will give a very brief overview of the minimum amount of knowledge required on this topic in Chapter 2.

One of the interesting facts about AdS/CFT is that it is a duality between a strongly-coupled regime on one side and a weakly-coupled regime on the other: therefore, it is a way to compute quantum effects in a strongly-coupled theory by performing computations in a classical gravitational setup (and vice versa). Although very powerful on its own, the main limit of AdS/CFT is that AdS$_d$ does not appear in the real world outside of very specific limits, one of which will be the focus of the first part of this thesis and will link the infrared (IR) sector of the Sachdev-Ye-Kitaev (SYK) model to (near-)extremal black holes.

This limitation is the main reason why a lot of effort is being made towards generalizing the holographic principle to other spacetimes. In this thesis, we will follow this direction and look for spacetimes that have at least some connection to de Sitter, whose cosmological interest is evident as it is able to describe both the Universe right now and during the inflationary epoch, which took place immediately after the Big Bang. Making contact with the origin of our Universe is an invaluable possibility, since we currently do not possess any theory that is capable of describing what appears to be a singularity at the beginning of reality itself.

The outline of the thesis will be the following:
\begin{itemize}
    \item In Chapter 2, as already stated, we will present some basic facts about the AdS/CFT correspondence, which will shed light on the way we will concretely act in the final part of this work.
    
    \item In Chapter 3, the gravitational path integral will be introduced. Although we only know how to compute its semiclassical approximation and loop corrections by considering the classical solutions that come from the action, this will be enough to determine several quantum effects that could not be deduced through General Relativity alone, such as the Hawking temperature of black holes and the Bekenstein-Hawking entropy formula. This is the main mathematical instrument that we will employ to study the properties of the gravitational theories dual to the modifications of the SYK model that we will consider, thus gaining insight into their structure.
    
    \item In Chapter 4, we will present the SYK model along with several of its properties. We will mostly focus on its IR limit, as the action that emerges for the soft reparametrization modes will be linked to quantum theories of gravity in $(1+1)$ dimensions.
    
    \item In Chapter 5, we will introduce the relevant class of these gravitational theories: the dilaton-gravity models, and Jackiw-Teitelboim (JT) gravity in particular. These models (and SYK as a consequence) are able to describe the near-horizon region of (near-)extremal black holes, so their appeal goes beyond their mere mathematical structure.
    
    \item In Chapter 6, we will discuss an extremely successful example of modification to the SYK model: the double-scaled limit (DSSYK). In this case, the existence of an interesting dual Hilbert space allows us to obtain many more explicit results. We will review the construction of the dual space, then compute the partition function and two point functions of operators as an application.
    
    \item In Chapter 7, we will consider another possible modification to the SYK model, one in which Dirac fermions appear instead of the usual Majorana ones. This will allow us to introduce a chemical potential that couples to the conserved $U(1)$ charge. Again, we will focus on the double-scaled limit of this new model. Through our computations, we will be able to determine the fermionic two point function of the theory and document its behavior.
    
    \item In Chapter 8, we will study the thermodynamic properties of the gravitational dual of the DSSYK model (determined in \cite{Blommaert_2024,Blommaert:2023wad,Blommaert:2024ydx}) at the semiclassical level. We will then investigate the quasinormal modes of a massive real scalar field that propagates in this geometry and in an effective one. Finally, we will conjecture how to modify the duality when considering the Dirac version of the model, showing that several bounds constrain the space of possible dual theories.
    
    \item In the Conclusions, we will summarize our findings and we will present an outlook on possible future developments of the results described here.
\end{itemize}

We underline the fact that the first six chapters are to be considered review material, while the seventh and the eighth chapter contain the original work of this thesis and its novel results. More precisely, our original contributions are the following:
\begin{itemize}
	\item We have determined which differential equations are solved by the first correction $g(\tau)$ to the free fermionic two point function of the ``charged'' DSSYK model (cDSSYK) at finite inverse temperature $\beta$ and with a chemical potential $\mu$. These are derived from the saddle point of the partition function of cDSSYK, which we have simplified to an integral over $g(\tau)$ alone. We have then solved the resulting equations, focusing in particular on the $\beta\abs{\mu} \ll 1$ and $\beta\abs{\mu} \gg 1$ limits, and we have compared our approximated solutions to the numerical ones. Interesting phenomena take place depending on the values of $\beta$ and $\mu$, which we have documented and discussed.
	\item We have studied how scalar fields propagate in the sine dilaton geometry dual to the DSSYK model in the case of minimal and non-minimal coupling to gravity. The different behavior that emerges in the two scenarios is a very interesting result, because it allows us to distinguish these two probes, of which only one is able to feel the presence of de Sitter regions in the bulk theory.
	\item We have conjectured a duality between the cDSSYK model and a gravitational theory that slightly modifies the one determined in \cite{Blommaert_2024,Blommaert:2023wad,Blommaert:2024ydx}, up to three unknown functions. We have then performed a semiclassical match of the partition functions of the two theories, through which we have shown some bounds that these three functions have to satisfy.
\end{itemize}

Even within the first six chapters, though, we will present several extra arguments and explanations to better understand the relevant physics, while also correcting numerous results and computations in the literature. In the Appendix, similarly, we have explicitly proven several essential results of which only claims were found in the literature.

\newpage
\chapter{A (Very Brief) Introduction to AdS/CFT}
\section{The GKPW Dictionary}
The relation between a gravitational theory in AdS$_{d+1}$ and a CFT$_d$ can be stated mathematically in several ways. The first statement of the duality is:
\begin{gather}
    \mathcal{Z}_{\rm CFT} = \mathcal{Z}_{\rm AdS}.
\end{gather}

Here, we are saying that the partition functions of the two theories are equivalent. We will always assume the Euclidean signature for both. This is the result we will be looking for, and the main way we will tackle the problem of searching for dualities: we will start from non-gravitational $d$-dimensional theories and we will match their partition function to that of $d+1$ gravitational theories, thus identifying the first ones as the holographic duals of the second ones. In particular, our objective is to find gravitational theories that do not necessarily live in AdS by studying different versions of the SYK model.

The relation between the two theories can be further characterized through the GKPW dictionary \cite{HartmanLectures,witten1998anti}:
\begin{gather}
    \mathcal{Z}_{\rm AdS}[\phi_0^i(x);\partial M] = \left\langle \exp \left( -\sum_i \int d^dx \: \phi_0^i(x)\mathcal{O}(x) \right) \right \rangle_{\mathrm{CFT \, on} \: \partial M} \equiv \mathcal{Z}_{\rm CFT}[\phi_0^i(x)]. \label{eq:AdSCFT}
\end{gather}

Here, the left-hand side (l.h.s.) is the gravitational partition function with the following boundary conditions for the scalar fields $\phi^i$ living in the spacetime $M$ in terms of the (asymptotic) Poincaré coordinates $(z,\vec x), \, z > 0, \, \vec x \in \mathbb{R}^d$:
\begin{gather}
    \phi^i(z \to 0,x) = z^{d-\Delta}\phi_0^i(x) + z^\Delta \tilde \phi_0^i(x). \label{eq:sourcescale}
\end{gather}

$\phi_0^i(x)$ is the ``source'' term and can be set freely, while the subleading $\tilde \phi_0^i(x)$ is fixed by requiring regularity of the solution for $z \to +\infty$. The right-hand side (r.h.s.), on the other hand, is an expectation value to be computed in the CFT. The operator $\mathcal{O}$ is a primary of the CFT with scaling dimension $\Delta$, which is connected to the mass of the bulk scalar by:
\begin{gather}
    m^2\ell_{\rm AdS}^2 = \Delta(\Delta-d), \quad \Delta = \frac{d}{2} + \sqrt{\frac{d^2}{4}+m^2\ell_{\rm AdS}^2}. \label{eq:massdeltarelation}
\end{gather}

Indeed, $\phi_0^i(x)$ has to be the source term and not $\tilde \phi_0^i(x)$. We can check this by observing that we want to treat the source as a spurion of the CFT, namely the perturbation $\int d^dx \: \phi_0^i(x) \mathcal{O}(x)$ has to be invariant under a rescaling of the coordinates, $z \to \lambda z, \: \vec x \to \lambda \vec x$. This means that $\phi_0^i(x) \to \lambda^{\Delta-d}$ necessarily, and $\phi^i$ is invariant under this diffeomorphism of AdS$_{d+1}$, as expected of a scalar field.

Let us elaborate on the relation between $\phi$ and $\mathcal{O}$. First of all, it is known that the Euclidean conformal algebra of a CFT$_{d}$ can be mapped to $so(d+1,1)$, which in turn can be seen as acting on $\mathbb{R}^{d+1,1}$. The way this is done is through the embedding formalism and the Poincaré section of the projective null hyperboloid. By ``projective'', we mean that we consider as distinct elements of our space the equivalence classes induced by identifying $P^A \sim \lambda P^A$ for every $P^A P_A = 0$ in $\mathbb{R}^{d+1,1}$. In this space, the Poincaré section is then defined as the equivalence classes whose representatives are of the kind:
\begin{equation}
\begin{gathered}
    P^A = (P^+, P^-, P^\mu) = (1,x^2,x^\mu), \quad x^\mu \in \mathbb{R}^d, \\
    P^\pm = P^{d+2} \pm P^{d+1}, \quad P_A P'^B = P_\mu P'^\mu - \frac{P^+ P'^- + P^- P'^+}{2}.
\end{gathered}
\end{equation}

This way, conformal transformations acting on $x^\mu$ have been mapped into $SO(d+1,1)$ transformations acting on the projective hyperboloid. Similarly, operators living in $\mathbb{R}^d$ can be embedded into the hyperboloid, and conformal transformations acting on them can be seen as the aforementioned $SO(d+1,1)$ transformations. At the same time, though, $SO(d+1.1)$ is the group of isometries of Euclidean AdS$_{d+1}$, which can also be embedded in $\mathbb{R}^{d+1,1}$. Since the projective hyperboloid is made up of rays, we can blow the Poincaré section to infinity and identify it as the boundary of AdS$_{d+1}$.

At this point, we can interpret the equation of motion of a minimally coupled bulk scalar:
\begin{gather}
    \nabla^2 \phi = m^2 \phi, \quad \nabla^2 = \nabla_\mu \nabla^\mu,
\end{gather}

as looking for eigenvectors (with eigenvalue $m^2\ell_{\rm AdS}^2$) of the quadratic Casimir of $SO(d+1,1)$, which for a given irreducible representation of the conformal algebra (that is, $so(d+1,1)$) is proportional to the identity with constant $\Delta(\Delta-d) + \ell(\ell+d-2)$ ($\ell$ being the integer spin of the primary of the representation). Indeed, calling $J^{AB}$ the generators of $SO(d+1,1)$ and taking their representation as differential operators acting on $\phi$, it can be shown that the quadratic Casimir $C_2 = \frac{1}{2} J^{AB} J_{AB}$ satisfies:
\begin{gather}
    C_2 \phi = \ell_{\rm AdS}^2 \nabla^2 \phi.
\end{gather}

Basically, solving the classical equation for a field is equivalent to looking for eigenvalues of the Casimir, but we already know their structure thanks to what one learns from CFTs; also, the Casimir is the same for both AdS and the CFT because of the common symmetry group. We find the dual primary $\mathcal{O}$ in the CFT by finally pushing the quantized field $\phi$ on the boundary in the following way:
\begin{gather}
    \mathcal{O}(P^A) = \frac{1}{\sqrt{c_\Delta}} \lim_{\lambda \to +\infty} \lambda^\Delta \, \phi(X = \lambda P^A + \mathrm{subleading}), \quad c_\Delta = \frac{\Gamma(\Delta)}{2\pi^{d/2}\Gamma(\Delta-\frac{d}{2})},
\end{gather}

where $P^A$ is in the Poincaré section and $X$ is a point in AdS whose distance from the null hyperboloid goes to 0 as $\lambda$ goes to infinity.

The GKPW dictionary can also be extended to all the other bulk fields and the metric $g_{\mu\nu}$ in particular, though in this case the boundary conditions also involve the choice of a topology (hence the dependence on $\partial M$ of $\mathcal{Z}_{\rm AdS}$). It is now clear what we mean by gravitational theories living in AdS: we are fixing $\partial M$ to be that of Anti-de Sitter, so that the asymptotic geometry is under control. We clearly ask this of the saddle point spacetime too, that is, the semiclassical approximation of the theory. For all bulk fields, (part of) their boundary configuration acts as an external source in the CFT that couples to some primary operator with the same spin and a mass - scaling dimension relation, so that correlation functions of the latter can be obtained through functional derivatives of the source. In the case of a scalar field:
\begin{gather}
    \langle \mathcal{O}_1(x_1) \dots \mathcal{O}_n(x_n) \rangle_{\rm CFT} = \frac{(-1)^n}{\mathcal{Z}_{\rm CFT}[0]} \frac{\delta^n \mathcal{Z}_{\rm CFT}[\phi_0^i(x)]}{\delta \phi_0^1(x_1) \dots \delta \phi_0^n(x_n)} \Bigg|_{\phi_0^i = 0}. 
\end{gather}

The operator in the CFT that is dual to the metric is the stress-energy tensor $T_{\mu\nu}$ which, we recall, is a quasiprimary in a CFT$_2$ (it only transforms ``well'' under \textit{global} conformal transformations). It is a conserved current of the theory with fixed scaling dimension $\Delta = d$. Analogously, bulk vector fields $A_\mu$ are dual to spin-1 operators $J_\mu$ with $\Delta \geq d-1$ (this is a unitarity bound), where the equality holds if and only if $J_\mu$ is conserved and if and only if the bulk field is massless: in fact, the relation between the vector field's mass $m$ and $J_\mu$'s conformal weight $\Delta$ is:
\begin{gather}
    m^2\ell_{\rm AdS}^2 = (\Delta-1)(\Delta-d+1) \stackrel{m = 0}{\implies} \Delta = d-1. \label{eq:DeltaMassEq}
\end{gather}

$\Delta = 1$ is excluded by unitarity bounds for $d > 2$ and coincides with $d-1$ for $d = 2$. So: $m=0$ implies $\Delta = d-1$, and vice versa. This is a general, important feature of AdS/CFT: global symmetries on the boundary mutually imply gauge symmetries in the bulk. Here is our argument for this statement. For a conserved current, we have the following in the CFT:
\begin{gather}
    \int d^dx \: (A_a(x) + \partial_a \alpha(x)) J^a(x) = \int d^dx \: (A_a(x) J^a(x) - \alpha(x) \partial_a J^a(x)) = \int d^dx \: A_a(x) J^a(x).
\end{gather}

Note that $A_a = P^\mu_a A_\mu$ is the projection of $A_\mu$ on the boundary: in the usual Poincaré coordinates, $A_a$ would not have the $z$ component. On one hand, a gauge symmetry in the bulk implies a global symmetry in the CFT because a gauge transformation $A_\mu(x,z) + \partial_\mu \alpha(x,z)$ reduces to the above on the boundary, so that necessarily $\partial_a J^a = 0$. On the other hand, if we have a global symmetry and \eqref{eq:AdSCFT} holds, the gravitational path integral that results from a $A_\mu(x,z) \to A_\mu(x,z) + \partial_\mu \alpha(x,z)$ transformation will be unmodified because the one in the CFT does not change, hence the theory possesses a gauge symmetry. For $d = 1$, then, we can only exclude $\Delta = 1$ because $m = 0$ implies gauge symmetry in the bulk, which in turn implies a coupling to a conserved current with $\Delta = 0$ in the CFT.

Finally, we observe that CFTs are UV-complete (UV stands for ultraviolet), therefore \eqref{eq:AdSCFT} is a non-perturbative formulation of a UV-complete theory of quantum gravity that miracolously makes use of QFTs without gravity.

\section{Holographic Renormalization Group}
In this section, we present the holographic renormalization group \cite{Boer_2000,de_Boer_2001}. The cited papers prove what we are going to state in the case of AdS$_5 \times S^5$ supergravity / $\mathcal{N} = 4$ SYM, but we can take this lesson to be a general one. This notion will be particularly useful when discussing the properties of JT gravity.

What the articles show is that the classical evolution of bulk fields $\phi^I$ in Euclidean AdS along the radial direction (we can think of $z = e^{-r/\ell_{\rm AdS}}$ in Poincaré coordinates, where $r$ is the coordinate used in the papers) is described by:
\begin{equation}
\begin{gathered}
    ds^2 = dr^2 + a^2(x,r) \hat{g}_{\mu\nu} dx^\mu dx^\nu, \\
    \frac{\partial a}{\partial r} \propto a, \quad a \frac{\partial \phi^I}{\partial a} = \beta^I(\phi).
\end{gathered}
\end{equation}

Since these fields are coupled to gravity, the backreaction of their stress-energy tensor alters the AdS geometry. Moreover, their rescaled boundary values couple to operators in the CFT and pictorially act as (position-dependent) ``couplings'' of the theory. If one computes the on-shell action of supergravity in a cut off AdS with maximum radius $r_0$, there are several terms that diverge as $r_0 \to +\infty$ that one needs to regulate through local counterterms (we will see that this also happens in the computations of the next chapter). The resulting regularized action is the quantum effective action of the CFT that one can use to obtain correlation functions. The final result is that one obtains a Callan-Symanzik equation for the correlators on the boundary, where $a$ acts as the energy scale of the theory and the (renormalized) $\beta^I$ are the beta functions of the couplings.

Consider now the case of AdS$_{d+1}$, where in Poincaré coordinates $a = 1/z$. We deduce from this discussion that the radial coordinate can be interpreted as an energy scale in the CFT, $1/z = \mu \sim E$, with the boundary being the UV and the deep interior of AdS being the IR. Empty AdS is then an RG flow that stays at the CFT fixed point at any energy scale since it corresponds to $\phi^I = 0$. The estabilished link between states at a certain energy in the CFT and the dual bulk geometry tells us, for example, that finite energy excitations are dual to asymptotic AdS geometries with a modified interior: the reason is that these states look like the vacuum from a UV point of view and are only distinguishable from it in the IR.

The presence of non-null sources (bulk fields) is nothing more than perturbing the CFT with the dual operators $\mathcal{O}$:
\begin{gather}
    S_{\rm CFT} \to S_{\rm CFT} + \int J \mathcal{O}.
\end{gather}

We remember that, in AdS/CFT, a bulk field near the boundary behaves as $\phi(z\to 0, x) \sim z^{d-\Delta} \phi_0(x)$, where $\phi_0(x)$ is a regular source for a CFT operator of dimension $\Delta$. There are therefore three scenarios, that we interpret in the following way.
\begin{itemize}
    \item If the operator $\mathcal{O}$ has dimension $\Delta < d$ (relevant), it is associated to a bulk field that vanishes on the boundary. As a consequence, the backreaction that its stress tensor induces is negligible and the asymptotic geometry is still AdS, although we expect the IR region to be heavily modified.
    \item If the operator $\mathcal{O}$ has dimension $\Delta = d$ (marginal), instead, the bulk field has a finite limit which could potentially disrupt the associated geometry. Moreover, such a deformation is expected to have influence over all energy scales, therefore over the entirety of the bulk.
    \item If the operator $\mathcal{O}$ has dimension $\Delta > d$ (irrelevant), finally, the IR region is preserved, but not the UV one. In particular, if we follow the RG flow starting from the interior towards the boundary, the trajectory is deflected from the original CFT fixed point as it follows an irrelevant direction towards a different asymptotic geometry and a different CFT' fixed point. When we turn an irrelevant deformation on, in fact, we correspondingly find a bulk field in the gravitational theory with a blowing up, non-normalizable mode near the boundary, which heavily backreacts on the AdS geometry and destroys it. This blow-up is exactly the deflection from the fixed point towards the UV: what one does in this case is to introduce a radial IR cutoff in AdS, namely a UV cutoff in the field theory (just like what is usually done in EFTs). Any state below the cutoff is on the IR part of the RG flow and is thus described by gravity in AdS with blowing up boundary conditions. We observe that any $m > 0$ scalar is associated to a $\Delta > d$ operator on the boundary, and correspondingly the source term blows up as we reach the boundary of the geometry: when considering the example of a scalar field in a fixed AdS background, though, we always ignore its backreaction, which is absent for a null source anyway.
\end{itemize}

We will encounter the last scenario when studying a varying dilaton field in AdS$_2$ in Section \ref{sec:Dilatons}: it will diverge near the boundary and will therefore require an IR cutoff. \\

In the next chapter, we are going to introduce the notion of gravitational path integrals and showcase several of their uses by performing saddle point computations.

\newpage
\chapter{Gravitational Path Integral}
In a quantum description of gravitational theories, we have to require that the metric $g_{\mu\nu}$ (that is, spacetime itself) is a quantum field, whose fluctuations we have to integrate over in the partition function. The previous chapter suggested to us that, at least for manifolds with a boundary, a reasonable choice of boundary conditions are the topology of $\partial M$ and the value of the metric there (these are the so-called ``Dirichlet conditions''). Ideally, we have a situation of this kind:
\begin{gather}
    \mathcal{Z} = \int \mathcal{D}g_{\mu\nu} \: e^{-S_E[g]},
\end{gather}

with $S_E$ such that we have a UV-complete theory. There is also the matter content, which we do not report explicitly. Note that the Euclidean signature will be crucial to our discussion. Unfortunately, our current situation is hardly of this kind: the Einstein-Hilbert action in four dimensions:
\begin{gather}
    S_{E}[g] = -\frac{1}{16\pi G_N} \int_M d^4x \, \sqrt{g} \: (R-2\Lambda)
\end{gather}

is non-renormalizable \cite{Goroff:1985sz,Goroff:1985th} (although it is one-loop finite when not coupled with matter \cite{tHooft:1974toh}) and, in principle, we should worry about gauge-fixing, as is usual in gauge theories. The situation is different in two dimensions, where the coupling $G_N^{-1}$ is dimensionless, and in three dimensions, where there are no propagating gravitons and the theory is finite to all orders \cite{Witten:1988hc} as long as it is not coupled with matter \cite{Anselmi_2004}.
Even worse, the action is not positive definite, so we expect the partition function to diverge even after factoring the infinite volume of the gauge group (all diffeomorphisms of the coordinates) out. Indeed, it suffices to consider the dynamics of the conformal mode of the metric $\tilde g_{\mu\nu} = \Omega^2 g_{\mu\nu}$ in order to obtain \cite{GIBBONS1978141}:
\begin{gather}
    S_E[\tilde g] = -\frac{1}{16\pi G_N} \int_M d^4x \, \sqrt{g} \: [\Omega^2(R-2\Lambda) + 6 \partial_\mu \Omega \, \partial^\mu \Omega],
\end{gather}

which can be made arbitrarily negative by choosing a rapidly varying $\Omega(x)$.

Fortunately, we can ignore all these issues when considering the semiclassical computation of the path integral. In a formal $G_N \to 0$ limit (Newton's gravitational constant), the path integral is dominated by its saddle point, which is simply the solution to Einstein's equations $\bar g_{\mu\nu}$, and is therefore equal to:
\begin{gather}
    \mathcal{Z} \approx e^{-S_E[\bar g]}.
\end{gather}

One important thing to note before proceeding is that the Einstein-Hilbert action is actually incomplete when considering spacetimes with a boundary, and one should also add the Gibbons-Hawking-York term \cite{PhysRevD.15.2752}. If we vary the Lorentzian action with respect to the inverse metric $g^{\mu\nu}$, in fact, we obtain \cite{Wald:1984rg}:
\begin{equation}
\begin{gathered}
    \delta S_{\rm EH} = \frac{1}{16\pi G_N} \left[ \int_M d^dx \, \sqrt{-g} \: \left( R_{\mu\nu} - \frac{1}{2} R g_{\mu\nu} \right) \delta g^{\mu\nu} + \int_M d^dx \, \partial_\mu (\sqrt{-g} \, v^\mu) \right], \\
    v_\mu = \nabla^\nu (\delta g_{\mu\nu}) - g^{\nu\rho} \nabla_\mu (\delta g_{\nu\rho}). \label{eq:MetricVariation}
\end{gathered}
\end{equation}

We have to be careful when lowering and raising indices in the expression for $v_\mu$. The second term in $\delta S_{\rm EH}$ is a boundary term that is zero only if we vary $g^{\mu\nu}$ while keeping both its boundary value \textit{and} its first derivative orthogonal to the boundary surface constant. Usually, though, only its value \textit{or} its derivative is kept fixed: these are the Dirichlet and the Von Neumann boundary conditions, respectively.

Assuming Dirichlet conditions, we can use the following consequence of Stokes' theorem:
\begin{equation}
\begin{gathered}
    \int_M d^dx \, \partial_\mu (\sqrt{-g} \, v^\mu) = \int_{\partial M} d^{d-1}y \, \sqrt{|h|} \, \varepsilon n_\mu v^\mu, \\
    h_{ab} = \frac{\partial x^\mu}{\partial y^a} \frac{\partial x^\nu}{\partial y^b} g_{\mu\nu} \equiv \partial_a x^\mu \partial_b x^\nu g_{\mu\nu}, \quad a,b = 1,\dots,d-1. \label{eq:Stokes}
\end{gathered}
\end{equation}

$h_{ij}$ is the metric induced by $g_{\mu\nu}$ on $\partial M$, that we assume is parametrized by $\mathbf{y} \in \mathbb{R}^{d-1}$, while $n_\mu(\mathbf{y})$ is the versor field orthogonal to the boundary and $\varepsilon = n_\mu n^\mu = \pm 1$ (+ if $\partial M$ is timelike, - if it is spacelike). A thorough derivation of \eqref{eq:Stokes} is presented in Appendix \ref{app:Stokes}. We have to evaluate:
\begin{equation}
\begin{gathered}
    \int_{\partial M} d^{d-1}y \, \sqrt{|h|} \, \varepsilon n^\mu (\nabla^\nu (\delta g_{\mu\nu}) - g^{\nu\rho} \nabla_\mu (\delta g_{\nu\rho})), \\
    \delta g_{\mu\nu}|_{\partial M} = 0 \implies \partial_a x^\rho \partial_\rho (\delta g_{\mu\nu}) = 0  \quad (\mathrm{derivative \: along \: \partial}M)\\
    \implies n^\mu (\nabla^\nu (\delta g_{\mu\nu}) - g^{\nu\rho} \nabla_\mu (\delta g_{\nu\rho})) = n^\mu g^{\nu\rho} (\nabla_\rho (\delta g_{\mu \nu}) - \nabla_\mu (\delta g_{\nu \rho})) = n^\mu g^{\nu \rho} (\partial_\rho (\delta g_{\mu \nu}) - \partial_\mu (\delta g_{\nu \rho})).
\end{gathered}
\end{equation}

In the third line, the $\nu \leftrightarrow \rho$ symmetry of $g^{\nu\rho}$ cancels the Christoffel symbols of the covariant derivatives. Since $g^{\nu\rho}|_{\partial M} = \frac{d}{d-1} \partial_a x^\nu \partial_b x^\rho h^{ab}$\footnote{This follows from $h_{ab} h^{ab} = \partial_a x^\mu \partial_b x^\nu g_{\mu\nu} h^{ab} = d-1 = \frac{d-1}{d} g^{\mu\nu} g_{\mu\nu}$.}, the first term in the last equality cancels out and we are left with $- n^\mu g^{\nu\rho} \partial_\mu \delta g_{\nu\rho}$. We need to modify the Einstein-Hilbert action by adding a term whose variation cancels this boundary term: this is exactly the Gibbons-Hawking-York term:
\begin{gather}
    S_{\rm GHY} = \frac{1}{8\pi G_N} \int_{\partial M} d^{d-1}y \, \sqrt{|h|} \, \varepsilon K.
\end{gather}

$K$ is the trace of the extrinsic curvature of $\partial M$:
\begin{gather}
    K_{\mu\nu} = \nabla_\mu n_\nu, \quad K = K^\mu_\mu = g^\mu_\nu \nabla_\mu n^\nu.
\end{gather}

Let us find how this term varies when we vary the metric. Since $g_{\mu\nu}$ is kept fixed at the boundary, so are the induced metric $h_{ab}$ and the inverse metric $g^{\mu\nu}$, hence we only need to consider:
\begin{gather}
    \delta K = -g^{\mu\nu} (\delta \Gamma^\alpha_{\mu \nu}) n_\alpha = -\frac{1}{2} n_\alpha g^{\mu\nu} g^{\alpha \rho} [\partial_\mu (\delta g_{\nu \rho}) + \partial_\nu (\delta g_{\rho \mu}) - \partial_\rho (\delta g_{\mu \nu})] = \frac{1}{2} n^\rho g^{\mu\nu} \partial_\rho (\delta g_{\mu\nu}).
\end{gather}

The object that is defined naturally is $n_\mu$, so this is the one that does not change when varying the metric, and not $n^\mu$ (although the opposite choice yields the same $\delta K$). The first two terms cancel because they are also evaluating the derivative of the metric along the boundary. The addition of this term cancels the boundary contribution of the Einstein-Hilbert action exactly, leaving us with:
\begin{gather}
    \delta (S_{\rm EH} + S_{\rm GHY}) = \int_M d^dx \, \sqrt{-g} \: \left( R_{\mu\nu} - \frac{1}{2} R g_{\mu\nu} \right) \delta g^{\mu\nu},
\end{gather}

which yields the Einstein's equations we were looking for all along. Unfortunately, as we will see, the story is not completely over yet as we will sometimes need to add an additional term to the action.

\section{Spacetime Thermodynamics}
In this section, we will show how turning to the Euclidean signature lets us determine the thermodynamic properties of spacetimes \cite{natsuume2016adscft}. We will consider, in particular, static metrics of the form:
\begin{gather}
    ds^2 = -f(r) \, dt^2 + \frac{dr^2}{f(r)} + r^2 d\Omega_{d-2}^2. \label{eq:metric}
\end{gather}

We first want to define the concept of ``surface gravity''. For a spherically symmetric metric such as the ones we are considering, we can always assume a particle to be moving on the fixed $\theta = \pi/2$ plane.
By considering its action
\begin{gather}
    S = -m\int ds = -m \int d\lambda \, \sqrt{-g_{\mu\nu} \dot x^\mu \dot x^\nu}, \quad \dot x^\mu \equiv \frac{dx^\mu}{d\lambda},
\end{gather}

we know that an independence of the metric from a coordinate $x^\mu$ implies a conservation of its conjugate momentum (thanks to the Euler-Lagrange equations):
\begin{gather}
    \frac{d}{d\lambda} p_\mu = \frac{d}{d\lambda} \left( m \frac{1}{\sqrt{-g_{\mu\nu} \dot x^\mu \dot x^\nu}} \frac{dx_\mu}{d\lambda} \right) = 0.
\end{gather} 

We will choose $\lambda$ to be the proper time $s$. In this case, we know that $p_0 = -mE$ and $p_\phi = mL$ are conserved, so that the equation $p^2 = -m^2$ reduces for radial motions to:
\begin{equation}
\begin{gathered}
	-\frac{1}{f(r)} (p_0)^2 + \frac{1}{f(r)} (p^r)^2 = -m^2 \\
    \left( \frac{dr}{ds} \right)^2 = E^2 - f(r) \implies \frac{d^2 r}{ds^2} = -\frac{1}{2} f'.
\end{gathered}
\end{equation}

In special relativity, given the covariant acceleration $a^\mu$, one can compute its norm in the comoving inertial frame of the particle, thus finding that it is exactly the acceleration that it perceives. In general relativity, this generalizes to:
\begin{gather}
    a^2 = g_{\mu\nu} a^\mu a^\nu = \frac{f'^2}{4f} \implies a = -\frac{f'}{2f^{1/2}}.
\end{gather}

The sign indicates whether the acceleration points towards increasing or decreasing radii. We have used that $a^0 = \frac{d}{ds}(-m E/f) = 0$ in the frame where the particle is at rest. This quantity diverges when evaluated at an horizon $r_0$ (where $f(r_0) = 0$), if present. The surface gravity is defined as the force per unit mass $a_\infty(r_0)$ that an asymptotic observer needs to exert to hold the particle hovering at the horizon. If the observer pulls the particle positioned at $r$ by a proper distance $\delta L$ through a massless string, there are two works at play:
\begin{equation}
\begin{split}
    W_\infty = a_\infty \delta L \quad & \mathrm{(asymptotic \: infinity)}, \\
    W_r = a \, \delta L \quad & \mathrm{(location} \: r).
\end{split}
\end{equation}

If $W_r$ is entirely converted into radiation, this reaches infinity with a different energy $E_\infty$ due to the gravitational redshift:
\begin{gather}
    E_\infty = \sqrt{\frac{f(r)}{f(\infty)}} W_r = \sqrt{\frac{f(r)}{f(\infty)}} \, a \, \delta L.
\end{gather}

Because of energy conservation, $W_\infty = E_\infty$ has to hold, hence we deduce:
\begin{gather}
    a_\infty = \sqrt{\frac{f(r)}{f(\infty)}} \, a = -\frac{f'(r_0)}{2f(\infty)^{1/2}}, \quad \kappa \equiv |a_\infty|.
\end{gather}

In the case of asymptotically flat spacetimes, $f(\infty) = 1$. If our spacetime possesses a cosmological horizon, it is best to define the surface gravity in an ``opposite'' way. Take, for example, the static patch of de Sitter, that is, the region in causal contact with a comoving observer (who may sit at the north pole without loss of generality) \cite{HartmanLectures}:
\begin{gather}
    ds^2 = -\left( 1-\frac{r^2}{\ell_{\rm dS}^2} \right) dt^2 + \frac{dr^2}{1-\frac{r^2}{\ell_{\rm dS}^2}} \, dr^2 + r^2 d\Omega_{d-2}^2. \label{eq:staticpatch}
\end{gather}

In this case, the observer should be located at the center of the patch ($r = 0$) while holding the particle still at $r = \ell_{\rm dS}$. A reasoning analogous to what we have just described would then yield:
\begin{gather}
    a_O = \frac{f'(r_0)}{2f(0)^{1/2}}, \quad \kappa \equiv |a_O|.
\end{gather}

We will soon show the connection of this concept with the Hawking temperature. Let us now analytically continue \eqref{eq:metric} by taking the Euclidean time $t_E = it$:
\begin{gather}
    ds_E^2 = +f(r) \, dt_E^2 + \frac{dr^2}{f(r)} + r^2 d\Omega_{d-2}^2.
\end{gather}

We will not report the angular part in the following, since it factors out in the near-horizon limit. If there exists a radius $r_0$ such that $f(r_0) = 0$, we can consider the near-horizon limit of the metric $f(r) \approx f'(r_0)(r-r_0)$:
\begin{gather}
    ds_E^2 \approx \frac{dr^2}{f'(r_0)(r-r_0)} + f'(r_0)(r-r_0) \, dt_E^2 = d\rho^2 + \rho^2 d \left( \frac{|f'(r_0)|}{2} t_E \right)^2, \quad \rho \equiv 2 \, \sqrt{\frac{|r-r_0|}{|f'(r_0)|}}.
\end{gather}

The near-horizon region of the Euclidean spacetime is a plane in polar coordinates if $|f'(r_0)|\, t_E/2$ has period $2\pi$, otherwise it presents a conical singularity in $\rho = 0$. The request of regularity\footnote{By this, we mean a smooth saddle geometry. If this were not the case, in fact, the curvature would feature a delta function at the conical defect and the equations of motion would not be satisfied there.} implies that our euclidean time is periodic, so the spacetime is a thermodynamic system with inverse temperature at equilibrium given by:
\begin{gather}
    \beta = \frac{4\pi}{|f'(r_0)|} \implies T = \frac{|f'(r_0)|}{4\pi}.
\end{gather}

\begin{figure}
    \centering
    \includegraphics[width = 0.55 \textwidth]{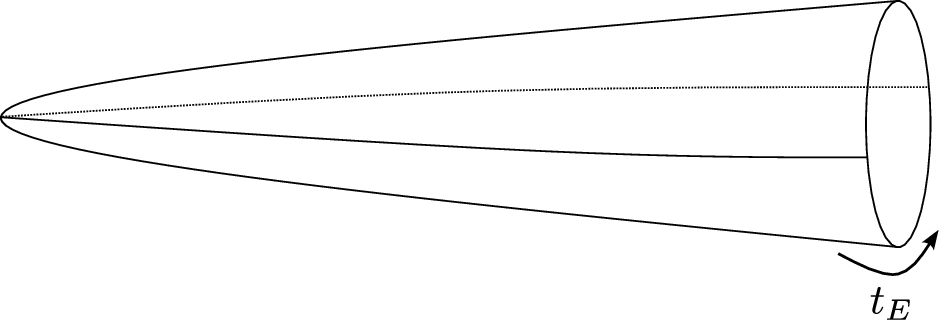}
    \caption{The Euclidean ``cigar'', with the $(d-2)$-sphere suppressed. The manifold caps off at $r_0$, so there is no inner (outer) horizon region and the manifold is regular there. The domain of the radial coordinate is $r \geq r_0$ for event horizons and $r \leq r_0$ for cosmological horizons.}
    \label{fig:Sigaro}
\end{figure}

The thermodynamic stability of the system depends on the sign of its heat capacity. If $f(\infty) \: (f(0)) = 1$, we find that the temperature is $1/2\pi$ times the surface gravity. The Euclidean spacetime is peculiar in the sense that it caps off regularly to a single point for $r = r_0$ (the origin of the polar coordinates), hence there is no event (cosmological) horizon and no inner (outer) horizon region as opposed to the Lorentzian case. A nice visual representation of the Euclidean manifold is shown in Figure \ref{fig:Sigaro}.

If $f'(r_0) = 0$, we actually need to consider $f(r_0) \approx f''(r_0)(r-r_0)^2/2$, so that the absence of conical singularities alone does not fix the temperature uniquely. One such example is the extremal Reissner-Nordström black hole (which we describe in Section \ref{sec:NearHorizon}). In this case, though, its temperature is zero for several reasons: its surface gravity is null, it does not emit radiation and $T = 0$ emerges as a limit of the near-extremal black hole. An interesting discussion about the ``black hole gap'' and its connection to this limit procedure is described in Section \ref{sec:NearHorizon}. We can expect a black hole with null surface gravity to not emit radiation as a consequence of the original computations by Hawking \cite{Hawking:1974rv, Hawking:1975vcx}, which reassure us that $T=0$ in this situation even though this is not imposed by the Euclidean picture.

In the next three sections we will study three examples, one for each possible cosmological constant ($\Lambda = 0,$ $\Lambda > 0$, $\Lambda < 0$) and asymptotic spacetime (flat, de Sitter and Anti-de Sitter respectively).

\section{First Example: the Schwarzschild Black Hole} \label{sec:firstex}
We will now compute the thermodynamic quantities associated to a specific example of \eqref{eq:metric}, the four-dimensional Euclidean Schwarzschild black hole, in a manner similar to \cite{Harlow_2016}:
\begin{gather}
    ds_E^2 = \left( 1 - \frac{r_s}{r} \right) dt_E^2 + \frac{1}{1-\frac{r_s}{r}} \, dr^2 + r^2 d\Omega_2^2, \quad r_s = 2GM.
\end{gather}

The time coordinate has the topology of $S^1$ and periodicity:
\begin{gather}
    \beta = \frac{4\pi}{|f'(r_s)|} = 4 \pi r_s \implies T = \frac{1}{8\pi G M}.
\end{gather}

We need to consider the partition function $\mathcal{Z}(\beta)$ in the semiclassical approximation. What this partition function describes is a canonical ensemble made of a black hole surrounded by radiation at a fine-tuned temperature $T$, which keeps it stationary by exactly feeding back the energy it loses through the evaporation process. In order to solve the Dirichlet problem, we have to fix the topology of the boundary $\partial M$ and the induced metric there. Formally, our black hole spacetime doesn't have a boundary, so we introduce a regularization: we cut off the geometry at a finite two-sphere radius $r_c$, that we will then send to infinity. The resulting Euclidean spacetimes that contribute to the path integral are compact, their boundaries have topology $S^1 \times S^2$ and induced metric:
\begin{gather}
    ds_E^2|_{\partial M} = \left( 1 - \frac{r_s}{r_c} \right) dt^2_E + r_c^2 d\Omega_2^2.
\end{gather}

Note that, in our reference, $g_{tt}$ at the boundary was incorrectly set to $1$. There are only two classical geometries that are compatible with these conditions: not only the cut off Schwarzschild geometry $g_{\rm sch}$ (with the topology of a disk due to the cigar times a two-sphere), but also the cut off flat spacetime $g_{\rm flat}$ (with the topology of a circle times a three-dimensional ball)
\begin{gather}
    ds_E^2 = \left( 1 - \frac{r_s}{r_c} \right) dt^2_E + dr^2 + r^2 d\Omega_2^2.
\end{gather}

In both cases, $t_E$ has the same periodicity $\beta$. Rescaling the time coordinate here trivially recovers the usual Euclidean metric. The requirement of fixing both an induced metric and a periodicity of the time coordinate excludes any other black hole solution from contributing. Here is our proof of this fact. Imagine another metric of the kind:
\begin{equation}
\begin{gathered}
    ds_E^2 = g(r)\, dt_E^2 + \frac{dr^2}{g(r)} + r^2 d\Omega_2^2, \quad g(r) = 1-\frac{\tilde r_s}{r}+\sum_{i\geq 2} \frac{a_i}{r^i}, \\
    \beta' = \frac{4\pi}{g'(r_0)}, \quad g(r_0) = 0.
\end{gathered}
\end{equation}

The metric induced on the hypersurface at that same $r_c$ would be:
\begin{gather}
    ds_E^2|_{\partial M} = g(r_c) \, dt_E^2 + r_c^2 d\Omega_2^2
\end{gather}

This time coordinate has periodicity $\beta'$, so one would need to rescale it to obtain the required periodicity $\beta$ while also matching $g_{tt}$. This implies a system of equations:
\begin{equation}
    \begin{cases}
        \beta = \lambda \beta' \\
        \left( 1 - \frac{r_s}{r_c} \right) \lambda^2 = 1 - \frac{\tilde r_s}{r_c} + \sum_{i \geq 2} \frac{a_i}{r_c^i}
    \end{cases}
\end{equation}

We are particularly concerned with the $r_c \to \infty$ limit, so we need to match the terms in the second equation order by order, thus obtaining:
\begin{gather}
    \lambda = 1, \quad r_s = \tilde r_s, \quad a_{i \geq 2} = 0.
\end{gather}

It would appear that flat spacetime itself is excluded by this reasoning, but in that case there is no fixed formula for $\beta'$ and it can take whatever value is needed. This allows $\lambda$ to be an arbitrary function of $r_c$, so we cannot match the terms in the second equation the way we just did. For any spacetime with a fixed temperature, though, it is clear that we only have $\lambda(r_s,\tilde r_s, a_i)$ and the match can be performed, yielding our result.

It follows that:
\begin{equation}
\begin{gathered}
    \mathcal{Z} = \int \mathcal{D}g_{\mu\nu} \, e^{-S_E[g]} \approx e^{-S_E[g_{\rm sch}]} + e^{-S_E[g_{\rm flat}]}, \\
    S_E[g] = -\frac{1}{16\pi G_N} \int_M d^4x \, \sqrt{g} \: R - \frac{1}{8\pi G_N} \int_{\partial M} d^3y \, \sqrt{h} \, K.
\end{gathered}
\end{equation}

For both topologies, only the boundary term gives a non-null contribution since $R = 0$ everywhere. We now report the detailed computations of the trace of the extrinsic curvature and of the Euclidean actions, which are absent in our reference:
\begin{equation}
\begin{gathered}
    K = \nabla_\mu n^\mu|_{\partial M} = \frac{1}{\sqrt{g}} \partial_\mu (\sqrt{g} \, n^\mu)|_{\partial M}, \quad n^\mu n_\mu = 1, \\
    \sqrt{g} = r^2 \sin\theta, \quad \sqrt{h} = r_c^2 \sin \theta \sqrt{1-\frac{r_s}{r_c}}, \\
    n^\mu = (n^0, n^r, n^\theta, n^\phi) \implies
    \begin{cases}
        n^\mu = (0,1,0,0) \: & (\mathrm{flat}) \\
        n^\mu = \left( 0, \sqrt{1-r_s/r},0,0 \right) \: & (\mathrm{Schwarzschild})
    \end{cases} \\
    \implies K = \frac{2}{r_c} \: (\mathrm{flat}), \quad
        K = \frac{2}{r_c} \sqrt{1-\frac{r_s}{r_c}} + \frac{r_s}{2r_c^2 \sqrt{1-\frac{r_s}{r_c}}} \: (\mathrm{Schwarzschild}).
\end{gathered}
\end{equation}

It is easy to evaluate the GHY integrals:
\begin{equation}
\begin{gathered}
    S_E[g_{\rm flat}] = -\frac{\beta r_c}{G_N} + \frac{\beta r_s}{2 G_N} + \mathcal{O}\left( \frac{1}{r_c} \right), \\
    S_E[g_{\rm sch}] = -\frac{\beta r_c}{G_N} + \frac{\beta r_s}{2 G_N} + \frac{\beta r_s}{4 G_N}.
\end{gathered}
\end{equation}

We find that the dominant contribution to the action always comes from the flat part of the spacetime. This should not surprise us, because we are studying a thermodynamic system at equilibrium. As mentioned earlier, it is known that black holes in asymptotically flat spacetime evaporate \cite{Hawking:1974rv, Hawking:1975vcx}, so for this system to be at equilibrium it is necessary for the black hole to be surrounded by a gas of radiation that counterbalances its evaporation and appears to dominate the thermal ensemble, with a contribution to the partition function that is actually divergent for an infinite volume ($r_c \to \infty$). If we want to only consider the contribution of the black hole to the partition function, we should only keep the difference between the two actions $S_E[g_{\rm sch}] - S_E[g_{\rm flat}]$:
\begin{gather}
    \mathcal{Z}_{\rm BH}(\beta) \approx e^{-\frac{\beta^2}{16 \pi G_N}}.
\end{gather}

We have used that $\beta = 4\pi r_s$ and we have removed the regulator by taking $r_c = \infty$. We should notice that we are ignoring the flat spacetime saddle even though it always has a lower free energy than the black hole solution, which means that it is infinitely more relevant in the $G_N \to 0$ limit we are considering right now: this is because we wanted to isolate the subdominant contribution of the black hole alone. As to this discussion about dominant saddles, we will actually study a surprising, non-trivial situation in the third example.

The Schwarzschild black hole has taught us a general lesson on how to compute the partition function of asymptotically flat spacetimes. We can in fact limit ourselves to the spacetime of interest and ignore the contribution from the flat topology, provided that we first compute the action with a counterterm and a $r_c$ regulator that we send to infinity only at the end (in the Euclidean signature, $\varepsilon = n_\mu n^\mu = 1$ always):
\begin{gather}
    S_E[g] = -\frac{1}{16\pi G_N} \int_M d^dx \, \sqrt{g} \: R - \frac{1}{8\pi G_N} \int_{\partial M} d^{d-1}y \, \sqrt{h} \, K + \frac{1}{8\pi G_N} \int_{\partial M} d^{d-1}y \, \sqrt{h} \, K_0,
\end{gather}

where $K_0$ is the trace of the extrinsic curvature of the boundary of the cut off flat spacetime. The topology of this boundary coincides with the one of the spacetime of interest, so we can also define $K_0$ as the quantity that arises when $\partial M$ is embedded in a flat spacetime. Note that the counterterm does not affect the field equations when varying the action while keeping the metric on the boundary fixed. Its insertion in the action is the choice of a flat spacetime having a null action and thus giving no contribution to the partition function. The lesson is actually even more general than this once we also consider manifolds that are non-compact, but not asymptotically flat (for example asymptotic AdS). We are in fact allowed to add \textit{any} counterterm at the cutoff boundary $\partial M$ that is function of the induced metric $h_{ab}$ only, so that it doesn't affect the equations of motion. We will need this notion in the third example.

We can now use the usual formulas for a canonical ensemble to obtain:
\begin{equation}
\begin{gathered}
    E = -\partial_\beta \log \mathcal{Z}_{\rm BH} = M, \\
    C = -\beta^2 \partial_\beta E = -\frac{1}{8\pi G T^2} = - 8\pi GM^2, \\
    S = (1-\beta \partial_\beta) \log \mathcal{Z}_{\rm BH} = \frac{4 \pi r_s^2}{4G} = \frac{A}{4G}
\end{gathered}
\end{equation}

$A$ is the proper surface of the black hole horizon. We have found two astonishing results:
\begin{itemize}
    \item The black hole is characterized by an entropy, which is proportional to the surface of its horizon. This is the Bekenstein-Hawking formula, and can be explained by noting that, although a black hole is only characterized macroscopically by its mass, electric charge, magnetic charge (if there exists one) and angular momentum, many different initial setups of matter in the spacetime may form the same final black hole. The entropy may then be considered a counting of the different ``microstates'' that contribute to the same black hole. Besides, an external observer cannot access what lies beyond the event horizon, so this lack of information also implies a non-null entropy. This is a counterintuitive result: entropy usually follows a volume law, not an area law. If black holes were to admit an holographic description, though, this scaling would actually be the most natural one, as it would be linked to the boundary degrees of freedom.
    
    \item The heat capacity of a Schwarzschild black hole is negative. This implies that it is unstable: if it is inserted in a thermal bath that is initially hotter, absorbing radiation will make it colder and increase the difference in their temperatures; vice versa, if the bath is initially colder, emitting radiation will make the black hole hotter and increase the difference in their temperatures in this scenario, too. \\
    While evaporating, a black hole's mass decreases and so does its entropy, while its temperature rises. This seems to violate the second law of thermodynamics, but one should also account for the entropy of the emitted radiation. What happens to an evaporating black hole entropy-wise has been known for a long time as the ``black hole information paradox'', but this is beyond the scope of this thesis and we will not talk about it further. We limit ourselves to saying that the AdS/CFT correspondence addresses the issue of unitary evaporation nicely (at least for black holes living in AdS), since every bulk setup is connected to a state (pure or mixed) of the CFT where things evolve with a unitary map, so both the starting point (infalling matter that will form the black hole) and the ending point (radiation only) have to be pure states.
\end{itemize}

An observation is due. The actual temperature that an observer experiences at a certain position in the spacetime is redshifted just like any other form of energy \cite{PhysRevD.33.2092}:
\begin{gather}
    \beta(r) = \beta \sqrt{f(r)} \implies T(r) = \frac{T}{\sqrt{f(r)}}.
\end{gather}

The latter is called the Tolman temperature: at the horizon, the physical Tolman temperature that one perceives is infinite. This means that what we have computed are the thermodynamic quantities from the point of view of an asymptotic observer, since we have taken derivatives with respect to $\beta(\infty) = \beta$. It is possible to perform a similar analysis for higher-dimensional black holes, whose metric has been studied in \cite{Tangherlini:1963bw,Myers:1986un}. What we would learn from the computations is that the Bekenstein-Hawking formula for the entropy still holds:
\begin{gather}
    S = \frac{A}{4G_N} = \frac{1}{4G_N} \frac{2\pi^{\frac{d-1}{2}}}{\Gamma\left( \frac{d-1}{2} \right)} \, r_0^{d-2}.
\end{gather}

Finally, note that any matter content in the theory would appear with an action of order $G_N^0$ as opposed to the $G_N^{-1}$ dependence of the gravitational action, so we can effectively neglect its contribution at the saddle point.

\section{Second Example: Thermal de Sitter}
Let us go back to Equation \eqref{eq:staticpatch}. If we consider the Euclidean manifold, the absence of a conical singularity at $r = \ell_{\rm dS}$ implies:
\begin{gather}
    \beta = \frac{4\pi}{|f'(r_0)|} = 2\pi \ell_{\rm dS}.
\end{gather}

There is a significant difference here with respect to the previous example: the temperature is completely determined by the fixed parameter $\Lambda$ (which determines $\ell_{\rm dS}$), whereas in the case of black holes we have a free parameter $r_s$. In order to have the possibility of thermodynamic laws, then, we need to add matter of some kind to dS$_d$. A way to do this is to consider black hole solutions in dS with tunable mass $M$ \cite{spradlin2001les}, but in this section we will stick to the case of thermal dS.

The cosmological horizon closes to a single point: this means that $M$ is a compact manifold with no boundary. We don't need the GHY term and there is no possibility of adding a counterterm, so we only need to evaluate the volume integral without adding any regulator. We can refer to the formulas in Appendix \ref{app:AdS}, up to a sign:
\begin{gather}
    R = \frac{d(d-1)}{\ell_{\rm dS}^2}, \quad \Lambda = \frac{(d-1)(d-2)}{2\ell_{\rm dS}^2}.
\end{gather}

Let us evaluate the action and the partition function:
\begin{equation}
\begin{gathered}
    \int_M d^d x \, \sqrt{g} \: (R-2\Lambda) = \frac{2(d-1)}{\ell^2_{\rm dS}} \beta \frac{\ell_{\rm dS}^{d-1}}{d-1} \, \frac{2\pi^{\frac{d-1}{2}}}{\Gamma\left( \frac{d-1}{2} \right)}  = 4\pi \frac{2\pi^{\frac{d-1}{2}}}{\Gamma\left( \frac{d-1}{2} \right)} \ell_{\rm dS}^{d-2} \\
    \implies S_E = -\frac{1}{4G_N} \frac{2\pi^{\frac{d-1}{2}}}{\Gamma\left( \frac{d-1}{2} \right)} \ell_{\rm dS}^{d-2}, \quad \mathcal{Z}_{\rm dS} \approx \exp(-S_E).
\end{gathered}
\end{equation}

Since the temperature is a fixed parameter, taking derivatives with respect to $\beta$ makes no sense. We know that energy is a boundary term in General Relativity (Chapter 4 of \cite{Poisson:2009pwt}), so $E = 0$ here. We can determine the entropy $S$ through the free energy $F$ in the following way:
\begin{gather}
    F = -T\log \mathcal{Z}_{\rm dS} = E - TS \implies S = \log \mathcal{Z}_{\rm dS} = \frac{A}{4G_N}.
\end{gather}

We have found the Bekenstein-Hawking formula once again, with $A$ being the proper area of the cosmological horizon. The explanation for a positive entropy is similar to the case of the black hole: it evaluates the lack of information of a comoving observer sitting at the center of the static patch, due to the fact that they cannot access anything beyond the cosmological horizon.

\section{Third Example: Black Holes in \texorpdfstring{AdS$_5$}{AdS5}}
Finally, we study the case of Schwarzschild black holes in Euclidean AdS$_5$. We will follow the presentation contained in Chapter 16 of \cite{HartmanLectures}, but we will perform every computation, whereas the reference only gives the final results. The metric is:
\begin{gather}
    ds^2_E = f(r) \, dt_E^2 + \frac{dr^2}{f(r)} + r^2 d\Omega_3^2, \quad f(r) = 1 - \frac{\mu}{r^2} + \frac{r^2}{\ell^2}. 
\end{gather}

$\ell$ is the usual $\ell_{\rm AdS}$, but we will drop the subscript here. The horizon is the outermost solution $r_+$ of $f(r_+) = 0$:
\begin{equation}
\begin{gathered}
    r_+^2 = \frac{\ell^2}{2} \left( \sqrt{1+\frac{4\mu}{\ell^2}} - 1 \right), \\
    \beta = \frac{4\pi}{|f'(r_+)|} = \frac{2\pi \ell^2 r_+}{2r_+^2 + \ell^2}
    \implies r_+ = \frac{\pi \ell^2}{2\beta} \left[ 1 \pm \sqrt{1-\frac{2\beta^2}{\pi^2 \ell^2}} \right].
\end{gathered}
\end{equation}

We observe that the range of the possible temperatures has a minimum:
\begin{gather}
    T_{\rm min} = \frac{\sqrt{2}}{\ell \pi}.
\end{gather}

\begin{figure}
    \centering
    \includegraphics[width = 0.63 \textwidth]{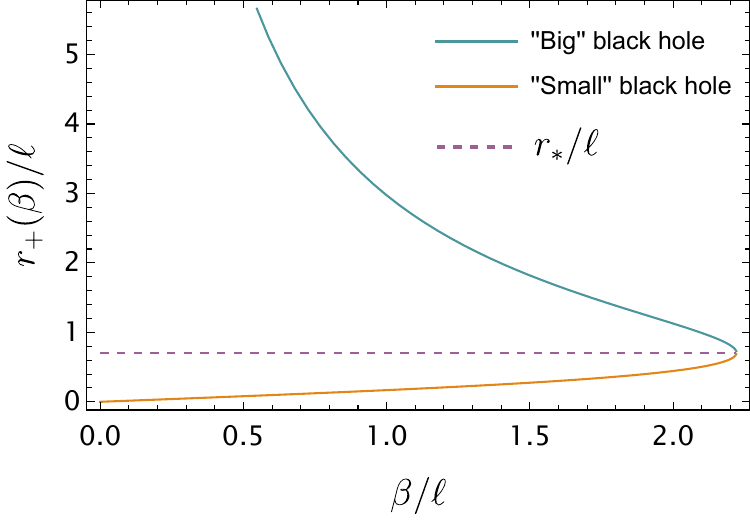}
    \caption{Dependence of the radii of the ``big'' and ``small'' black holes in AdS$_5$ on the inverse temperature $\beta$, using dimensionless units. Big black holes always have a bigger radius than $r_*$, whereas small black holes always have a smaller one.}
    \label{fig:AdSBlack}
\end{figure}

We also notice that at fixed $\beta$ there are two solutions, one whose radius decreases as $\beta$ increases and one whose radius increases as $\beta$ increases. The first black hole is always bigger than the second and they have equal size only at $T_{\rm min}$, where $r_+ = r_* = \ell/\sqrt{2}$. For each $\beta$, then, we have one solution with $r_+ > r_*$ and one with $r_+ < r_*$. This situation is shown in Figure \ref{fig:AdSBlack}.

When evaluating the partition function at fixed $\beta$, both black hole solutions will contribute. Just like for the Schwarzschild black hole, though, we also have to account for the thermal AdS solution. In 5 dimensions:
\begin{gather}
    R = -\frac{20}{\ell^2}, \quad \Lambda = -\frac{6}{\ell^2}.
\end{gather}

Repeating what we did in the first example, we cut off our manifold at $r_c$. We are going to study the Dirichlet problem with boundary topology $S^1 \times S^3$ and induced metric:
\begin{gather}
    ds_E^2 = f(r_c) \, dt_E^2 + r_c^2 d\Omega_3^2.
\end{gather}

$t_E$ has periodicity $\beta$. We start by considering the black holes:
\begin{gather}
    S_{E, \mathrm{BH}}[g] =  -\frac{1}{16\pi G_N} \int_M d^5x \, \sqrt{g} \: \left( -\frac{8}{\ell^2} \right) - \frac{1}{8\pi G_N} \int_{\partial M} d^4y \, \sqrt{h} \: (K-L_{\rm ct}[h]),
\end{gather}

where $L_{\rm ct}$ is a scalar function of the induced metric that cancels any divergent term and that we will determine soon. The sum of the first two terms is:
\begin{equation}
\begin{gathered}
    \frac{1}{2\pi G_N \ell^2} \int d\beta \, dr \, d\Omega_3 \: r^3 - \frac{r_c^3 \sqrt{f(r_c)}}{8 \pi G_N} \int d\beta \, d\Omega_3 \: K, \\
    K = \frac{1}{r^3} \partial_r \left( r^3 \sqrt{f(r)} \right)\Big|_{r = r_c} = \frac{3}{r_c} \sqrt{f(r_c)} + \frac{f'(r_c)}{2\sqrt{f(r_c)}} \\
    \implies S^{(0)}_{E, \rm BH} = \frac{\pi \beta}{4G_N \ell^2}(r_c^4 - r_+^4) + \left( -\frac{\pi \beta r_c^4}{G_N \ell^2} - \frac{3\pi\beta r_c^2}{4G_N} + \frac{\pi \beta \mu}{2 G_N} \right)
\end{gathered}
\end{equation}

The counterterm here is not as obvious as the flat case, but it is not surprising anyway in form:
\begin{equation}
\begin{gathered}
    L_{\rm ct}[h] = \frac{3}{\ell} + \frac{\ell}{4} R[h], \quad R[h] = \frac{6}{r_c^2} \\
    \implies \frac{1}{8\pi G_N} \int_{\partial M} d^4y \, \sqrt{h} \: L_{\rm ct}[h] = \frac{\pi\beta r_c^3}{4G_N} \left( \frac{3}{\ell} + \frac{3\ell}{2r_c^2} \right) \sqrt{f(r_c)} \approx \frac{3\pi\beta r_c^4}{4G_N \ell^2} + \frac{3\pi\beta r_c^2}{4 G_N} + \frac{3\pi\beta}{32 G_N}(\ell^2 - 4\mu).
\end{gathered}
\end{equation}

$R[h]$ is the Ricci scalar of the induced metric. $L_{\rm ct}$ is fixed uniquely by requiring that the $r_c^4$ and $r_c^2$ divergences are canceled. Summing all of these contributions together, we finally find:
\begin{gather}
    S_{E, \mathrm{BH}}[g] = \beta F_{\rm BH} = \frac{\pi\beta}{8G_N \ell^2}\left( r_+^2 \ell^2 - r_+^4 + \frac{3\ell^4}{4} \right).
\end{gather}

We have used that $\mu = r_+^2 + r_+^4/\ell^2$. If we compare the free energies of the two black holes with the same inverse temperature $\beta$, we obtain:
\begin{gather}
    F_{\rm BH, big} - F_{\rm BH, small} = -\frac{\pi^5 \ell^6}{8G_N \beta^4}\left( 1-\frac{\beta^2}{\beta_{\rm max}^2} \right)^{3/2} < 0.
\end{gather}

From this point onwards, then, we can ignore the existence of the small black hole and only consider the big one. We can easily compute the energy, the heat capacity and the entropy of this black hole:
\begin{equation}
\begin{gathered}
    E = \partial_\beta S_{E,\rm BH} = \frac{3\pi}{8G_N} \left( \mu + \frac{\ell^2}{4} \right), \\
    C = -\beta^2 \partial_\beta E = \frac{3\pi^2 r_+^3}{2 G_N} \frac{2r_+^2 + \ell^2}{2r_+^2-\ell^2}, \\
    S = \beta E - S_{E, \rm BH} = \frac{2\pi^2 r_+^3}{4G_N} = \frac{A}{4G_N}.
\end{gathered}
\end{equation}

Once again, the entropy obeys the Bekenstein-Hawking formula. The result for the energy is made up of two contributions: the first term is due to the mass of the black hole, while the second one is a Casimir energy that is present in thermal AdS. As opposed to the Schwarzschild black hole, in this case $C > 0$: we could have expected this from Figure \ref{fig:AdSBlack}, which tells us that the black hole grows in size (and entropy) as its temperature increases. This means that by definition:
\begin{gather}
    C = \frac{dE}{dT} = T \frac{dS}{dT} > 0.    
\end{gather}

We have applied the first law of thermodynamics to black holes. Indeed, all the work we have done in this chapter should convince the reader that they obey the usual laws. Conversely, then, we would have found that a small black hole has a negative heat capacity, just like the one in flat spacetime: indeed, it is the only one that can achieve a size much smaller than $\ell$, at which point it is unable to notice that it lives in AdS rather than Minkowski. We can explain the stability of big black holes heuristically: the radiation they emit bounces at the conformal boundary of AdS and is fed back to it in a short enough time (given the size of the black hole with respect to $\ell$) that the black hole does not evaporate.

It is now the turn of thermal AdS. We remind the reader that its metric is obtained by simply taking $\mu = 0$. Since it has no horizons, its temperature $\beta$ is not fixed by regularity and can assume any value: therefore, it is a solution that one always has to account for in a thermal partition function. Since we are fixing $h_{ab}$ at the boundary, the time coordinate that has the periodicity $\beta$ of the black hole is such that the metric is:
\begin{gather}
    ds_E^2 = \frac{f(r_c)}{g(r_c)} \, g(r) \, dt_E^2 + \frac{dr^2}{g(r)} + r^2 d\Omega_3^2, \quad g(r) = 1 + \frac{r^2}{\ell^2}.
\end{gather}

$f(r)$ is the usual function with $\mu \neq 0$. The sum of the first two terms in the action is now:
\begin{equation}
\begin{split}
    S_{E,\rm AdS}^{(0)} & = \frac{\pi \beta}{4G_N \ell^2} \sqrt{\frac{f(r_c)}{g(r_c)}}\, r_c^4 - \frac{\pi\beta r_c^3}{4G_N} \left( \frac{3}{r_c} \sqrt{f(r_c) \, g(r_c)} + \frac{g'(r_c)}{2} \sqrt{\frac{f(r_c)}{g(r_c)}} \right) \\
    & = -\frac{3\pi \beta}{4G_N}\left( \frac{r_c^4}{\ell^2} + r_c^2 - \frac{\mu}{2} \right) + \mathcal{O}\left( \frac{1}{r_c} \right).
\end{split}
\end{equation}

The divergent part of the action is exactly the same one as before, so the counterterm that we need is the same, too. If we add it to our action, we finally obtain:
\begin{gather}
    S_{E, \rm AdS} = \frac{3\pi\beta \ell^2}{32 G_N}.
\end{gather}

Interestingly enough, this regularized action is what we obtain if we put $r_+ = 0$ in the black hole's one. Its linear dependence on $\beta$ gives us:
\begin{gather}
    E = \frac{3\pi \ell^2}{32 G_N}, \quad C = 0, \quad S = 0.
\end{gather}

Its energy is simply the Casimir energy of the spacetime, regardless of its temperature: this is one of the several features that justify comparing AdS to a box, even though it is a manifold with infinite volume. In this case, in order to determine the actual heat capacity and entropy, we should go beyond the semiclassical approximation and consider $\mathcal{O}(G_N^0)$ contributions to them. As opposed to the flat spacetime case, which of the two saddles (presence or absence of the big black hole) dominates depends non-trivially on the temperature:
\begin{equation}
\begin{gathered}
    F_{\rm BH} - F_{\rm AdS} = \frac{\pi}{8G_N \ell^2}(r_+^2 \ell^2 - r_+^4) \\
    \implies F_{\rm BH} < F_{\rm AdS} \iff r_+ > \ell \iff \beta < \frac{2}{3} \pi \ell.
\end{gathered}
\end{equation}

The existence of a black hole is only favorable if its radius is bigger than the AdS length $\ell$, that is, at high enough temperatures. In the context of AdS/CFT, this is a really interesting result. The connection lies in our thermal spacetimes being dual to a thermal state in the CFT, which has the same periodicity in the Lorentzian time $t_L \sim t_L + i\beta$ when computing expectation values of observables. This is our understanding of the reason behind this proposition. Computing the gravitational path integral at finite temperature with fixed boundary conditions is conceptually the following:
\begin{gather}
    \mathcal{Z}[g,\partial M] = \int_{\substack{g_{\mu\nu}|_{\partial M} \: \mathrm{fixed} \\ g_{\mu\nu}(0)=g_{\mu\nu}(\beta)}} \mathcal{D}g_{\mu\nu} \: e^{-S_E[g]} = \langle g_{\partial M}| e^{-\beta H} | g_{\partial M} \rangle.
\end{gather}

\begin{figure}
    \centering
    \includegraphics[width = 0.45 \textwidth]{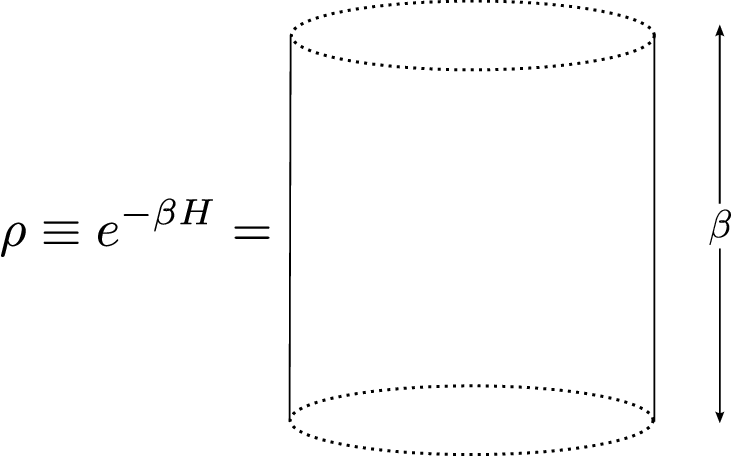}
    \caption{Any state can be defined through its overlap with a basis. A thermal state is characterized by its matrix elements $\langle \phi_1 | \rho | \phi_2 \rangle$ and can therefore be visualized as a ``cylinder'' at whose extremities one inserts any ``in'' and ``out'' state, which are connected through an Euclidean time evolution (that is, a path integral).}
    \label{fig:densitymatrix}
\end{figure}

By assuming the existence of an underlying and unknown quantum theory of gravity, $|g_{\partial M} \rangle$ is the state of the Hilbert space associated to the fixed boundary conditions on $\partial M$. Recall that field states are defined through their configurations on a codimension-1 submanifold that is then ``time''-evolved through an appropriate operator: an example are fields in a radially quantized CFT, where a state is defined by its values at a fixed radius on the plane and is then ``time''-evolved through the dilatation operator. Our path integral takes $|g_{\partial M} \rangle$, evolves it for an Euclidean time $\beta$ and then considers its overlap with itself. The Hamiltonian $H$ that appears is the generator $\partial_t$ of the time translations associated to our choice of the time coordinate. A visual representation of this explanation is shown in Figure \ref{fig:densitymatrix}. In the cases of interest, $\partial M$ also contains the periodic coordinate $t_E$.

The bottom line is that a thermal spacetime is associated to a thermal density matrix in our unknown theory of quantum gravity that we ``prepare'' at Lorentzian time $t_L = 0$ through our Euclidean path integral. Then, thanks to the AdS/CFT correspondence \eqref{eq:AdSCFT}, we conclude that on the CFT side we are also considering expectation values with respect to a dual thermal state.

A CFT that has a low entropy at low temperatures and a high entropy at high temperatures is a theory that undergoes a phase transition between a confined and a deconfined phase. From a microcanonical point of view entropy is, in fact, a quantity that counts the density of states in the energy spectrum:
\begin{gather}
    S(E) = \log \rho(E).
\end{gather}

Here we have a sharp transition from a $\mathcal{O}(G_N^0)$ entropy to a $\mathcal{O}(G_N^{-1})$ one, hence a sharp change in the density of energy eigenstates from low to high as the temperature (energy) rises. We can think of what happens for example in $SU(N)$ QCD: at low temperatures, the theory is confined and our physical states are $\mathcal{O}(1)$ color singlet hadrons, while at high temperatures the theory is deconfined and our physical states are $\mathcal{O}(N^2)$ gluons and quarks. 

We make a final observation on our results. We have obtained $E = 0$ for thermal dS because it has no boundary, but asymptotically flat and AdS spacetimes also have no boundary, technically. The difference lies in the fact that thermal dS is compact and does not need any regularization, since its Euclidean cosmological horizon closes to a point. Asymptotically flat and AdS spacetimes, on the other hand, do have a notion of timelike boundary, sitting at spatial infinity. Additionally, the existence of this ``effective'' boundary is maintained even in their thermal counterparts. This boundary is the one that emerges after a conformal compactification, which yields a new manifold $\tilde M$ through a Weyl rescaling of the metric of the initial manifold $M$: the boundary of interest is $\partial \tilde M$. Note that this process is the ordinary procedure that yields Penrose-Carter diagrams of spacetimes, which has the merit of describing manifolds through coordinates with finite range. In the case of AdS$_{d>2}$, for example, we read from Appendix \ref{app:AdS} that its metric can be written as:
\begin{gather}
    ds^2 = \frac{\ell^2_{\rm AdS}}{\cos^2 \theta} (-dt^2+d\theta^2+\sin^2\theta d\Omega^2_{d-2}),
\end{gather}

with $t \in \mathbb{R}$ and $\theta \in [0,\pi/2)$. The conformal compactification here is the process of performing a Weyl rescaling of the metric that gets rid of the $\ell^2_{\rm AdS}/\cos^2\theta$ prefactor, so that the resulting manifold $\tilde M$ is a cylinder that has a regular behaviour in $\theta = \pi/2$, which is not part of the original AdS$_d$ but is clearly $\partial \tilde M$. The border of the full cylinder is therefore the conformal boundary of AdS$_{d > 2}$, where the dual CFT lives. When taking a regulator and sending it to infinity, we are pushing the cutoff surface until it reaches the conformal boundary of these manifolds, therefore giving us our non-null results. \\

With these examples, we have successfully introduced the first instrument that we will need in our future analysis and shown its utility. We can now temporarily forget about gravity and talk about something else entirely in the next chapter: the SYK model.

\newpage
\chapter{The SYK Model}
In the following, we will describe the Sachdev-Ye-Kitaev (SYK) model and several of its properties as presented mainly in \cite{Sarosi_2018, Trunin_2021, Maldacena_2016}, although we will be more detailed and pedagogical in the presentation.

The SYK model is an ensemble of quantum mechanical models made of $N$ Majorana fermions in $(0+1)$ dimensions. A member of such ensemble is specified by the following Lagrangian and Hamiltonian:
\begin{equation}
\begin{gathered}
    \label{eq:SYKLag} L = \frac{1}{2} \sum_{i=1}^N \psi_i \partial_\tau \psi_i - i^{p/2} \sum_{1 \leq i_1 < \dots < i_p \leq N} J_{i_1 \dots i_p} \psi_{i_1} \dots \psi_{i_p}, \\
    H = \sum_{i=1}^N \frac{\partial L}{\partial_\tau \psi_i} \partial_\tau \psi_i - L = i^{p/2} \sum_{1 \leq i_1 < \dots < i_p \leq N} J_{i_1 \dots i_p} \psi_{i_1} \dots \psi_{i_p}, \\
    \{ \psi_i, \psi_j \} = \delta_{ij}, \quad i,j = 1,\dots,N.
\end{gathered}
\end{equation}

We will always assume $N$ and $p$ to be even integers. The couplings $J_{i_1 \dots i_p}$ are different for each member of the ensemble and are extracted independently from a Gaussian distribution with zero mean and variance:
\begin{gather}
    \label{eq:JvsJ} \langle J^2_{i_1 \dots i_p} \rangle_J = \frac{J^2(p-1)!}{N^{p-1}} = \frac{2^{p-1}}{p} \frac{\mathcal{J}^2(p-1)!}{N^{p-1}}.
\end{gather} 

The $\langle \sbullet \rangle_J$ symbol indicates a mean over the ensemble. The couplings are real since $H$ has to be Hermitian:
\begin{equation}
\begin{gathered}
    H^\dagger = (-i)^{p/2} \sum_{1 \leq i_1 < \dots < i_p \leq N} J^*_{i_1 \dots i_p} \psi_{i_p} \dots \psi_{i_1} = (-i)^{p/2} (-1)^{p(p-1)/2} \sum_{1 \leq i_1 < \dots < i_p \leq N} J^*_{i_1 \dots i_p} \psi_{i_1} \dots \psi_{i_p} \\
    = i^{p/2} (-1)^{p^2/2} \sum_{1 \leq i_1 < \dots < i_p \leq N} J^*_{i_1 \dots i_p} \psi_{i_1} \dots \psi_{i_p} = H \implies J^*_{i_1 \dots i_p} = J_{i_1 \dots i_p}.
\end{gathered}
\end{equation}

First of all, we want to build an Hermitian representation of the Clifford algebra. To this end, we consider:
\begin{equation}
\begin{gathered}
    c_i = \frac{1}{\sqrt{2}}(\psi_{2i}-i\psi_{2i-1}), \quad c_i^\dagger = \frac{1}{\sqrt{2}}(\psi_{2i}+i\psi_{2i-1}), \quad i = 1,\dots,N/2, \label{eq:cicidag} \\
    \{ c_i, c_j \} = \{ c_i^\dagger, c_j^\dagger \} = 0, \quad \{ c_i, c_j^\dagger \} = \delta_{ij}.
\end{gathered}
\end{equation}

These are the usual anticommutation relations for fermions, so we can build a Fock space by choosing a vacuum annihilated by all the $c_i$ and picking the following states as a basis:
\begin{gather}
    (c_1^\dagger)^{n_1} \dots (c_{N/2}^\dagger)^{n_{N/2}} \ket{\Omega}, \quad n_k = 0,1.
\end{gather}

The resulting Hilbert space has dimension $2^{N/2}$. A recursion relation for the representation matrices is:
\begin{equation}
\begin{gathered}
    \psi_i^{(K)} = \psi_i^{(K-1)} \otimes
    \begin{pmatrix}
        -1 & 0 \\ 0 & 1
    \end{pmatrix}, \quad i = 1,\dots,N-2, \\
    \psi_{N-1}^{(K)} = \frac{1}{\sqrt{2}} I_{2^{K-1}} \otimes
    \begin{pmatrix}
        0 & 1 \\ 1 & 0
    \end{pmatrix}, \\
    \psi_{N}^{(K)} = \frac{1}{\sqrt{2}} I_{2^{K-1}} \otimes
    \begin{pmatrix}
        0 & -i \\ i & 0
    \end{pmatrix}, \\
    \psi_1^{(1)} = \frac{1}{\sqrt{2}}
    \begin{pmatrix}
        0 & -i \\ i & 0
    \end{pmatrix}, \quad
    \psi_2^{(1)} = \frac{1}{\sqrt{2}}
    \begin{pmatrix}
        0 & 1 \\ 1 & 0
    \end{pmatrix}.
\end{gathered}
\end{equation}

$I_d$ is the $d \times d$ identity matrix, the superscript indicates the number of fermions $N = 2K$ and clearly all the $\psi_i^{(K)}$ are $2^K \times 2^K$ matrices. In order to treat this system analytically, we will always consider the $N \to \infty$ limit: this is also the case of interest when considering the holographic properties of the model.

\section{Large \texorpdfstring{$N$}{N} and Perturbation Theory}
In this section, we want to compute the observables of the system such as its Green function. We will do this in a perturbative approach, assuming our couplings $J_{i_1 \dots i_p}$ (hence the typical scale $J$ that appears in the variance) to be small. Indeed, the couplings have dimension $1$, so we expect the theory to be asymptotically free in the UV. From now on, we will always consider the Euclidean theory.

The two point functions are:
\begin{gather}
    G_{ij}(\tau) = \langle \Omega| \mathrm{T}[\psi_i(\tau) \psi_j(0)] | \Omega \rangle \equiv \Theta(\tau) \langle \Omega| \psi_i(\tau)\psi_j(0) |\Omega \rangle - \Theta(-\tau) \langle \Omega| \psi_j(0) \psi_i(\tau) | \Omega \rangle.
\end{gather}

$|\Omega\rangle$ is the vacuum of the interacting theory, $\Theta$ is the Heaviside theta function and $\psi_i(\tau)$ is the Euclidean time evolution $e^{H\tau}\psi_i(0) e^{-H\tau}$. Because of the symmetry under time translations, these functions only depend on the time distance $\tau$. The two point function that we will usually consider is the normalized trace of the expression above:
\begin{gather}
    G(\tau) \equiv \frac{1}{N} \sum_{i=1}^N G_{ii}(\tau).
\end{gather}

In the free theory, $H=0$ and the time evolution is trivial, hence the two point functions are constrained by the Clifford algebra:
\begin{gather}
    G^{(0)}_{ij}(\tau) = \frac{1}{2} \mathrm{sgn}(\tau) \delta_{ij} \implies G^{(0)}(\tau) = \frac{1}{2} \mathrm{sgn}(\tau), \quad \mathrm{sgn}(\tau) = \Theta(\tau)-\Theta(-\tau).
\end{gather}

If $i \neq j$, from $(\psi_i \psi_j)^T = -\psi_i \psi_j$ we deduce that all diagonal elements are null, while if $i = j$ we have $\psi_i^2 = 1/2$. This result is only true at zero temperature: $\beta = 1/T = + \infty$. If that is not the case, we have to require that our Green function satisfies $G(\tau+\beta) = -G(\tau)$ due to the antiperiodic boundary conditions of fermions at finite temperature. We can assume, without loss of generality, that the periodic time $\tau$ lies in $[-\beta/2, \beta/2)$, since times outside of this interval are related to it through translations of $k \beta, \, k \in \mathbb{Z}$.

For $\tau > 0$, we have \cite{Kapusta_Gale_2006}:
\begin{align}
    \nonumber G_{ij}(\tau,0) & = \tr [e^{-\beta H} \psi_i(\tau) \psi_j(0)]/\tr[e^{-\beta H}] \\ 
    \nonumber & = \tr [\psi_j(0) e^{-\beta H} \psi_i(\tau)]/\tr[e^{-\beta H}] \\
    & = \tr [e^{-\beta H} \psi_j(\beta) \psi_i(\tau)]/\tr[e^{-\beta H}] \label{eq:AntisymProof} \\
    \nonumber & = - \tr [e^{-\beta H} \mathrm{T}[\psi_i(\tau) \psi_j(\beta)]]/\tr[e^{-\beta H}] \\
    \nonumber & = -G_{ij}(\tau,\beta) = -G_{ij}(\tau-\beta,0).
\end{align}

A similar computation can be done for $\tau < 0$, yielding the same result. Notice that, if we use $G_{ij}(\tau-\tau')$, then $\tau-\tau' \in [-\beta, \beta)$. A natural way to match the zero temperature result to thermal $\tau \in [-\beta/2, \beta/2)$ and to give it the correct periodicity is to use a tangent function \cite{Trunin_2021} that links the time on the line to the one on the circle:
\begin{equation}
\begin{gathered}
    \label{eq:circlemap} \tau_{\rm line} = \tan \left( \frac{\pi \tau_{\rm circle}}{\beta} \right), \\
    G^{(0)}(\tau-\tau')_{\beta} = \frac{1}{2} \mathrm{sgn}\left( \tan \left( \frac{\pi \tau}{\beta} \right) - \tan \left( \frac{\pi \tau'}{\beta} \right) \right) = \frac{1}{2} \mathrm{sgn}\left( \frac{\sin \left( \frac{\pi (\tau-\tau')}{\beta} \right)}{\cos \left( \frac{\pi\tau}{\beta} \right) \cos \left( \frac{\pi\tau'}{\beta} \right)} \right), \\
    \partial_\tau G^{(0)}(\tau)_\beta = \sum_{k \in \mathbb{Z}} (-1)^k \delta(\tau-k\beta).
\end{gathered}
\end{equation}

We can drop the cosines in the sign function because they are always positive. This map has the nice property of mapping monotonically the real line to our interval of interest $[-\beta/2, \beta/2)$. Inside the time circle, the free Green function is indeed the inverse of the kernel of the Euclidean theory $\delta(\tau-\tau') \partial_\tau$. We have used the known property of Dirac's delta:
\begin{gather}
    \delta(f(x)) = \sum_{\{ x_i | f(x_i) = 0 \}} \frac{\delta(x-x_i)}{|f'(x_i)|}.
\end{gather}

We will usually drop the $\beta$ subscript.

We now proceed to build the Feynman rules of this model. The drawings refer to the case $p = 4$, but the generalization to other $p$ is obvious.
\begin{itemize}
    \item Assume a given sampling of the couplings $J_{i_1 \dots i_p}$. We can build diagrams from $p$ leg vertices with indices \{$i_1,\dots,i_p$\}:
    \begin{center}
        \includegraphics[width = 2 cm]{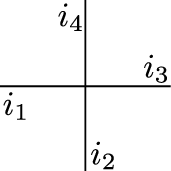}
    \end{center}
    
    \item For each realization of the model, Feynman diagrams can be computed. The next step is to take an ensemble average of each one of them. Since the couplings have a Gaussian distribution, one can use Wick's theorem and the known variance to evaluate the expectation values:
    \begin{gather}
        \langle J_{i_1 \dots i_p} J_{j_1 \dots j_p} \rangle_J = \frac{J^2(p-1)!}{N^{p-1}} \delta_{i_1 j_1} \dots \delta_{i_p j_p}.
    \end{gather}
    
    \item Let us focus on the two point functions of the fermions. The first non-null diagram is the ``melon'':
    \begin{center}
        \includegraphics[width = 7 cm]{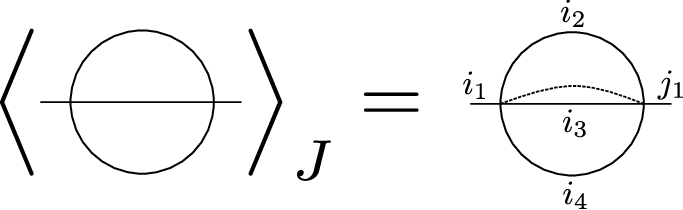}
    \end{center}
    The solid lines are the free fermionic propagators, while the dashed lines denote the Wick pairings of the ensemble average. All the indices are contracted except for the two external lines, so this diagram is equal to (repeated indices are summed over):
    \begin{gather}
        \langle J_{i_1 i_2 \dots i_p} J_{j_1 i_2 \dots i_p} \rangle_J \, G^{(0)}_{i_2 i_2} \dots G^{(0)}_{i_p i_p} = \frac{J^2(p-1)!}{N^{p-1}} G^{(0)}_{i_2 i_2} \dots G^{(0)}_{i_p j_p} \delta_{i_1 j_1} = (p-1)!J^2(G^{(0)})^{p-1} \delta_{i_1 j_1}.
    \end{gather}
    Note that every vertex lies at some time $\tau_i$ over which we integrate, with the free propagators connecting the times at their extremes: we have omitted this structure of the diagram for now.

    \item The $J^4$ diagrams exhibit different possible contractions, with different dependencies on $N$. For example, consider these two distinct possibilities:
    \begin{center}
        \includegraphics[width = 0.45 \textwidth]{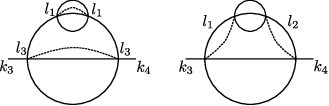}
    \end{center}
    In the first diagram, five of the seven propagators are diagonal and have their indices summed over, so they give a total $N^5$ contribution. The remaining two yield $G^{(0)}_{l_1 l_3}G^{(0)}_{l_3 l_1} \propto N$, so $N^6$ cancels out with the $N^{-6}$ coming from the two ensemble contractions and the diagram scales as $N^0$. In the second diagram, instead, the propagators associated to $l_1, l_2$ each give a $N$ contribution. The disorder average requires that the two internal lines that link the vertices at the bottom have the same index as two of the three lines connecting the upper vertices, so each pair of internal lines only contributes with a factor of $N$. Lastly, the average also requires the missing upper line to have both the indices $k_3, k_4$, thus yielding a $N^0 \delta_{k_3,k_4}$ factor. Globally, this diagram scales as $N^{-2}$.
    
    The bottom line of these observations is that the only diagrams that survive in the $N \to \infty$ limit are the ones where the ensemble averages connect vertices inside the same melon: by connecting vertices in different melons, more indices need to be contracted while summing them over. In doing so, one loses powers of $N$ that would otherwise be sourced by adjacent traces of the Kronecker deltas, thus the suppression due to the variance of the couplings prevails.
\end{itemize}

In the $N \to \infty$ limit, our two point function can be graphically written as:
\begin{center}
    \includegraphics[width = 1 \textwidth]{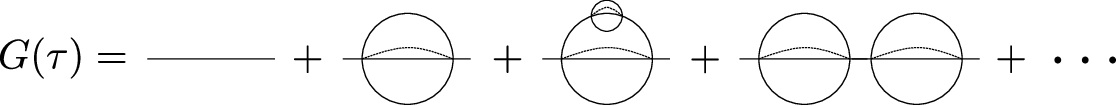}
\end{center}

We now define $\Sigma$ to be the ``self energy'', that is, the sum of all 1PI (one-particle-irreducible) diagrams or, equivalently, all the iterated melon diagrams. 1PI diagrams are connected diagrams that are still connected after cutting any single line. In terms of $G$ and $\Sigma$ we obtain the following closed set of consistency equations (the \textit{Schwinger-Dyson equations}):
\begin{center}
    \includegraphics[width = 1 \textwidth]{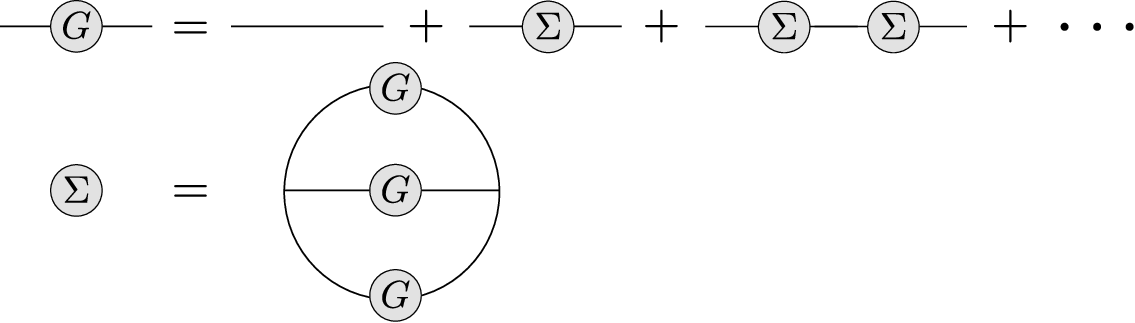}
\end{center}

Let us write the first one explicitly:
\begin{align}
    \nonumber G & = G^{(0)} + G^{(0)}\Sigma G^{(0)} + G^{(0)}\Sigma G^{(0)}\Sigma G^{(0)} + \dots \\
    & = G^{(0)}[1+\Sigma G^{(0)} + \Sigma G^{(0)}\Sigma G^{(0)} + \dots] \\
    \nonumber & = G^{(0)}[1 - \Sigma G^{(0)}]^{-1} = [(G^{(0)})^{-1} - \Sigma]^{-1}.
\end{align}

The above equation has to be interpreted as a product of matrices with two continuous indices (the time coordinates):
\begin{gather}
    (AB)(\tau,\tau') \equiv \int d\tau'' \: A(\tau,\tau'')B(\tau'',\tau').
\end{gather}

In this language, the ``$1$'' that appears above is a Dirac delta and the last equality in particular follows trivially from the properties of matrix products. The inverse of a function (or operator in general) is defined coherently by the following equation:
\begin{gather}
    \int d\tau'' A^{-1}(\tau,\tau'')A(\tau'',\tau') = \delta(\tau-\tau').
\end{gather}

It follows by definition of the free Green function that $(G^{(0)})^{-1}(\tau,\tau') = \delta(\tau-\tau') \partial_{\tau'}$. A complementary approach would be to derive these results in Fourier space, where matrix products are substituted by simple products of functions.

Finally, we have derived our Schwinger-Dyson equations when $N \to \infty$:
\begin{gather}
\label{eq:SDeqs}
\begin{cases}
    G(\tau,\tau') = [\delta \partial_\tau - \Sigma]^{-1}(\tau,\tau') \\
    \Sigma(\tau,\tau') = J^2 [G(\tau,\tau')]^{p-1}
\end{cases}
\end{gather}

Our boundary conditions for the Green function are the following:
\begin{equation}
\begin{gathered}
    G(\tau,\tau') = G(\tau-\tau',0) = G(0,\tau'-\tau) = -G(\tau'-\tau,0) = -G(\tau',\tau), \\
    G(\tau,\tau')_{\tau' \to \tau^\mp} = \pm \frac{1}{2}. \label{eq:SYKGBound}
\end{gathered}
\end{equation}

The first condition follows from the fermionic time ordering and time translation invariance, while the second follows from continuity and the equal time Clifford algebra $\{ \psi_i(\tau), \psi_j(\tau) \} = \delta_{ij}$. One could also interpret this condition by noting that dimensionful couplings become irrelevant in the UV, so we must recover the free theory Green function as the times coincide.

\section{The \texorpdfstring{$(G, \Sigma)$}{(G,S)} Formalism} \label{sec:GSFormal}
In this section, we show a powerful formalism that will be useful in later studies and that re-derives the previous result by employing the path integral. We have to be careful, since the actual Grassmann variables that we know how to handle from the usual Quantum Field Theory courses are the $c_i, c^\dagger_i$ defined in \eqref{eq:cicidag} and not the Majorana $\psi_i$ \cite{Trunin_2021}. Our reference \cite{Sarosi_2018} doesn't elaborate on this subtlety, so we will now show how we have deduced the way one should handle them. We will use $c^\dagger_i, \bar c_i$ interchangeably since in $(0+1)$ dimensions the representation of the Gamma matrices $\{ \gamma_\mu, \gamma_\nu \} = 2 \delta_{\mu \nu}$ is trivial: $\gamma_0 = 1$. Let us consider the partition function of the free theory first, in order to understand what happens with these variables:
\begin{equation}
\begin{aligned}
    \mathcal{Z}^{(0)} & = \int_{\psi_i(\beta/2) = -\psi_i(-\beta/2)} \left( \prod_{i=1}^{2K} \mathcal{D}\psi_i \right) \exp \left( -\int d\tau \: \sum_{i=1}^{2K} \frac{1}{2} \psi_i(\tau) \partial_\tau \psi_i(\tau) \right) \\
    & = \int \left( \prod_{i=1}^K \mathcal{D} \bar c_i \, \mathcal{D} c_i \left| \det \frac{\partial (\psi_{2i-1}, \psi_{2i})}{\partial (\bar c_i, c_i)} \right| \right) \exp \left( - \int d\tau \: \sum_{i=1}^K \frac{1}{2}(\bar c_i(\tau) \partial_\tau c_i(\tau) + c_i(\tau) \partial_\tau \bar c_i(\tau)) \right) \\
    & = \int [\mathcal{D}\bar c \mathcal{D} c] \exp \left( - \int d\tau \, d\tau' \: \sum_{i=1}^K \bar c_i(\tau) \delta_{ij} \delta(\tau-\tau') \partial_{\tau'} c_j(\tau') \right) = \left( \det (\delta(\tau-\tau')\partial_{\tau'}) \right)^{N/2} \\ 
    & = \sqrt{\det A(\tau,\tau')_{ij}}, \quad A(\tau,\tau')_{ij} \equiv \delta_{ij} \delta(\tau-\tau') \partial_{\tau'}, \quad i,j = 1,\dots,N.
\end{aligned}
\end{equation}

The antiperiodic boundary conditions for the fermions will be implicit from now on. We have used both integration by parts and the fact that $c_i, \bar c_i$ anticommute from the second to the third line. In the third line, we have used the known Berezin integral:
\begin{gather}
    \int [\mathcal{D}\bar \psi \mathcal{D}\psi] \exp \left( - \bar \psi A \psi \right) = \det A.
\end{gather}

By defining $\Psi \equiv (\psi_1, \dots, \psi_N)$, $A(\tau,\tau')$ is the matrix such that the action is $S = \int d\tau \, d\tau' \: \frac{1}{2} \Psi^T A(\tau,\tau') \Psi$. We have found out that the Gaussian integral of these new variables does not yield a determinant, but rather its square root:
\begin{gather}
    \label{eq:Majoint} \int \mathcal{D}\Psi \exp \left( -\frac{1}{2} \Psi A \Psi \right) = \sqrt{\det A}.
\end{gather}

The general result that one obtains is the Pfaffian of a matrix \cite{Berezin:1966nc}, but for any $2\ell \times 2\ell$ skew-symmetric matrix (such that $A^T = -A$) it is true that $(\mathrm{Pf} A)^2 = \det A$. In our case, $A$ is indeed skew-symmetric since $\partial_\tau^T = -\partial_\tau$ and we have $N = 2K$ fermions. Since our Majorana variables are certain combinations of Grassmann variables, we also deduce that they are Grassmann variables themselves such that $\psi_i \psi_j = - \psi_j \psi_i$.

We now consider the partition function of a fixed set of couplings:
\begin{gather}
    \mathcal{Z}(J) = \int \mathcal{D}\Psi \exp \left( -\int d\tau \: \left( \frac{1}{2} \psi_i \partial_\tau \psi_i + i^{p/2} J_{i_1 \dots i_p} \psi_{i_1} \dots \psi_{i_p} \right) \right).
\end{gather}

Repeated indices are summed over, with the convention that $J_{i_1 \dots i_p} \neq 0$ only when the indices are in strictly increasing order. Note that the Euclidean Lagrangian flips the sign of the couplings when compared to the Lorentzian one in \eqref{eq:SYKLag}. We want to compute a meaningful mean over the ensemble. We could either compute $\langle \mathcal{Z} \rangle$ or $\langle \log \mathcal{Z} \rangle = -\beta \langle F \rangle$, with $F$ being the free energy of the system. All the ensemble means of thermodynamic quantities would require that we carry out the latter computation, yet it turns out that the two only differ by $\mathcal{O}(1/N^{p-2})$ terms \cite{Sachdev_2015, Kitaev_2018}.

The main idea of the second approach would be to employ the so-called ``replica trick'', which stems from the observation that:
\begin{gather}
	\log \mathcal{Z} = \lim_{n \to 0} \partial_n \mathcal{Z}^n \implies \langle \log \mathcal{Z} \rangle = \lim_{n \to 0} \left[ \frac{\log (1 + (\langle \mathcal{Z}^n \rangle - 1))}{n} + \mathcal{O}(n) \right] = \lim_{n \to 0} \frac{\log \langle \mathcal{Z}^n \rangle}{n}.
\end{gather}

One can compute $\mathcal{Z}^n$ by taking $n$ identical copies of the system: this is done by considering $\psi_i^a$ with $a$ ranging from $1$ to $n$ inside the action. The result will clearly depend on $n$: by analytically continuing the result to non-integers, one can take the $n \to 0$ limit and find the desired mean.

Because of the equivalence between the two computations in the $N \to \infty$ limit (unless possibly $p = 2$, which we will never consider anyway) and because of simplicity, we will actually compute the first quantity. In the following, we use $I = \{i_1, \dots, i_p \}$, with $i_1 < \dots < i_p$.

\begin{align}
    \nonumber \langle \mathcal{Z} \rangle_J & = \int \prod_I \mathcal{D}J_I \: \mathcal{Z}(J) \exp \left( - \sum_I \frac{J_I^2}{2 \frac{J^2 (p-1)!}{N^{p-1}}} \right) \\
    & = \int \mathcal{D}\Psi \prod_I \mathcal{D}J_I \exp \left( -\int d\tau \: \frac{1}{2} \psi_i \partial_\tau \psi_i - i^{p/2} \int d\tau \: J_I \psi_{i_1} \dots \psi_{i_p} - \frac{J_I J_I}{2 \frac{J^2 (p-1)!}{N^{p-1}}} \right) \\
    \nonumber & = \mathrm{const.} \int \mathcal{D}\Psi \: \exp \left( -\int d\tau \: \frac{1}{2} \psi_i \partial_\tau \psi_i + J^2 \frac{(p-1)!}{2 N^{p-1}} i^p \int d\tau \, d\tau' \: \sum_I (\psi_{i_1} \dots \psi_{i_p})(\tau) (\psi_{i_1} \dots \psi_{i_p})(\tau') \right).
\end{align}

We have used that the integral over the couplings is quadratic and can be performed exactly:
\begin{gather}
    \label{eq:gaussint} \int \prod_i d\alpha_i \: \exp \left(-\frac{1}{2}\alpha^t A \alpha + B^t \alpha \right) = \frac{1}{\sqrt{\det \left(\frac{A}{2\pi} \right)}} \exp \left( \frac{1}{2}B^t A^{-1} B \right), 
\end{gather}

where $\alpha$ and $B$ are $N \times 1$ vectors and $A$ is a $N \times N$ positive definite matrix.
By using that in this case\footnote{When considering two different orderings $\{ i_1, \dots, i_p \}$ and $\{ i_{\sigma(1)}, \dots, i_{\sigma(p)} \}$ of the \textit{same} choice of $p$ indices inside a string of fermions, we don't have to worry about signs due to the anticommutations that achieve the increasing order of indices. This is thanks to the exponential having a product of two identical strings, and allows us to use $\sum_{i_1 \neq i_2 \dots \neq i_p}$ without caring whether the strings are ordered or not.}
\begin{gather}
    \psi^2 = 0, \quad \sum_I = \frac{1}{p!} \sum_{i_1 \neq i_2 \dots \neq i_p }
\end{gather}

and by introducing two new fields
\begin{gather}
    G(\tau,\tau') = \frac{1}{N} \sum_{i=1}^N \psi_i(\tau) \psi_i(\tau'), \quad \Sigma(\tau,\tau'),
\end{gather}

which can be inserted through a smart way of rewriting $1$ (up to a constant):
\begin{equation}
\begin{aligned}
    \label{eq:GSTrick} 1 & = \int \mathcal{D}G \: \delta \left( G(\tau,\tau') - \frac{1}{N} \psi_i(\tau) \psi_i(\tau') \right) \\
    & = \int \mathcal{D}G \, \mathcal{D}\Sigma \: \exp \left(- \frac{N}{2} \int d\tau \, d\tau' \left( G(\tau,\tau') - \frac{1}{N} \psi_i(\tau) \psi_i(\tau') \right) \Sigma(\tau,\tau') \right),
\end{aligned}
\end{equation}

we can also rewrite our partition function. The result of this procedure is:
\begin{equation}
\begin{split}
    \langle \mathcal{Z} \rangle_J = \int \mathcal{D}G \, \mathcal{D}\Sigma \, \mathcal{D}\Psi \: \exp \Bigg( & -\int d\tau \: \frac{1}{2} \psi_i \partial_\tau \psi_i + J^2 \frac{i^p N}{2p} (-1)^{\frac{p(p-1)}{2}} \int d\tau \, d\tau' G(\tau,\tau')^p \\
    & - \frac{N}{2} \int d\tau \, d\tau' \left( G(\tau,\tau') - \frac{1}{N} \psi_i(\tau) \psi_i(\tau') \right) \Sigma(\tau,\tau') \Bigg).
\end{split}
\end{equation}

Equation \eqref{eq:GSTrick} follows from the Wick rotation of the known relation $2\pi \delta(x) = \int dk \: e^{ikx}$. At this point, the field $\Sigma(\tau,\tau')$ is nothing more than a Lagrange multiplier that imposes the definition of the field $G(\tau,\tau')$. We observe that $i^p (-1)^{p(p-1)/2} = 1$ for even $p$ and that the fermions' action is quadratic, so we can integrate them out using \eqref{eq:Majoint}:
\begin{equation}
\begin{aligned}
    \langle \mathcal{Z} \rangle_J & = \int \mathcal{D}G \, \mathcal{D}\Sigma \left[ \det \left( \delta(\tau-\tau')\partial_{\tau'} - \Sigma(\tau,\tau') \right)\right]^{N/2} \exp \left( -\frac{N}{2} \int d\tau \, d\tau' \left( G(\tau,\tau')\Sigma(\tau,\tau') - \frac{J^2}{p} G(\tau,\tau')^p \right) \right) \\
    \label{eq:effaction} & = \int \mathcal{D}G \, \mathcal{D}\Sigma \: e^{-NI[G,\Sigma]}, \quad I[G,\Sigma] = -\frac{1}{2}\log \det (\delta\partial_\tau - \Sigma) + \frac{1}{2} \iint \left(G\Sigma - \frac{J^2}{p}G^p \right).
\end{aligned}
\end{equation}

In the $N \to \infty$ limit, the partition function is computed by simply considering its saddle point, whose equations of motion for $(G, \Sigma)$ can be shown to be exactly the Schwinger-Dyson equations \eqref{eq:SDeqs} by using that $\log \det A = \tr \log A$. The variation with respect to $G$ is trivial, so let us focus on the variation with respect to $\Sigma$:
\begin{equation}
\begin{gathered}
    \delta I = -\frac{1}{2} \int d\tau \, d\tau' \: \left\{ -[\delta \partial_\tau - \Sigma]^{-1}(\tau,\tau') \delta \Sigma(\tau',\tau) - G(\tau',\tau) \delta \Sigma(\tau',\tau) \right\} \\
    \implies G(\tau,\tau') = [\delta \partial_\tau - \Sigma]^{-1}(\tau,\tau').
\end{gathered}
\end{equation}

We have used that $G(\tau,\tau') = -G(\tau',\tau)$. If we had proceeded with the replica trick, we would have introduced fields analogous to $G$ and $\Sigma$ with two replica indices such as $G_{ab}$, linking different copies of the system among themselves. Ansatzing replica diagonal solutions (such that we always have $\delta_{ab}$ dependencies) would have solved the saddle equations, thus showing that in the $N \to \infty$ limit the two approaches coincide:
\begin{gather}
	\langle \log \mathcal{Z} \rangle = \lim_{n \to 0} \frac{\log \langle \mathcal{Z}^n \rangle}{n} \stackrel{\langle \mathcal{Z}^n \rangle = \langle \mathcal{Z} \rangle^n}{=} \log \langle \mathcal{Z} \rangle.
\end{gather}

\section{Large \texorpdfstring{$p$}{p} Expansion}
In this section, we will solve the Schwinger-Dyson equations in the $p \to \infty$ limit. This will come in handy later, when we will focus on the double-scaled limit of the model:
\begin{gather}
    N, p \to \infty, \quad \lambda = \frac{2p^2}{N} = \mathrm{const.}
\end{gather}

Our claim is that this limit yields a free theory with deviations that can be linked to a $1/p$ expansion. To show this, we ansatz a parametrization of the solution:
\begin{gather}
    \label{eq:BigpG} G(\tau) = \frac{1}{2} \mathrm{sgn}(\tau) e^{\frac{g(\tau)}{p-1}} = \frac{1}{2} \mathrm{sgn}(\tau) \left( 1 + \frac{g(\tau)}{p} + \dots \right),
\end{gather}

from which we also obtain ($\mathrm{sgn}(\tau)^{p-1} = \mathrm{sgn}(\tau)$):
\begin{gather}
    \Sigma(\tau) = J^2 \frac{1}{2^{p-1}} \mathrm{sgn}(\tau) e^{g(\tau)}.
\end{gather}

In the finite temperature case we use $\mathrm{sgn}\left( \sin\left( \frac{\pi \tau}{\beta} \right)\right)$ instead. Boundary conditions \eqref{eq:SYKGBound} now require: 
\begin{gather}
    g(\tau) = g(-\tau), \quad g(0) = g(\beta) = 0.
\end{gather}

In order to find $g(\tau)$, we turn to Fourier space and neglect $\mathcal{O}(1/p^2)$ terms. First, we transform \eqref{eq:BigpG}:
\begin{equation}
\begin{gathered}
    \tilde G(\omega) = (-i\omega)^{-1} + \frac{1}{2p}[\mathrm{sgn} \times g](\omega), \\
    \frac{1}{\tilde G(\omega)} = -i\omega + \frac{\omega^2}{2p}[\mathrm{sgn} \times g](\omega) + \mathcal{O}\left( \frac{1}{p^2} \right). 
\end{gathered}
\end{equation}

On the other hand, the first Schwinger-Dyson equation yields:
\begin{gather}
    \nonumber \frac{1}{\tilde G(\omega)} = -i\omega - \tilde \Sigma(\omega) \\
    \implies \tilde \Sigma(\omega) = -\frac{\omega^2}{2p}[\mathrm{sgn} \times g](\omega) 
    \implies \Sigma(\tau) = \frac{1}{2p} \partial^2_\tau \left( \mathrm{sgn}(\tau) g(\tau) \right) \\
    \nonumber \implies \partial^2_\tau \left( \mathrm{sgn}(\tau) g(\tau) \right) = 2\mathcal{J}^2 \mathrm{sgn}(\tau) e^{g(\tau)}. \label{eq:SYKg(p)}
\end{gather}

We have substituted $\mathcal{J}^2$ as defined in \eqref{eq:JvsJ}. The solutions of this equation are:
\begin{gather}
    e^{g(\tau)} = \frac{c^2}{\mathcal{J}^2} \frac{1}{[\sin(c(|\tau|+\tau_0))]^2}.
\end{gather}

By imposing the boundary conditions, we finally obtain:
\begin{gather}
    e^{g(\tau)} = \left( \frac{\cos \frac{\pi v}{2}}{\cos \left[ \pi v \left( \frac{1}{2} - \frac{|\tau|}{\beta} \right) \right]} \right)^2, \quad \beta \mathcal{J} = \frac{\pi v}{\cos \frac{\pi v}{2}}. \label{eq:thermalSYKprop}
\end{gather}

As $v$ runs from $0$ to $1$, $\beta \mathcal{J}$ runs from 0 to $+\infty$. The $\beta \to +\infty$ limit then reduces to:
\begin{gather}
    e^{g(\tau)} = \lim_{\beta \to +\infty} \left( \frac{1}{\cos \frac{\pi v \abs{\tau}}{\beta} + \tan \frac{\pi v}{2} \sin \frac{\pi v \abs{\tau}}{\beta}} \right)^2 = \frac{1}{(1+\mathcal{J}|\tau|)^2}.
\end{gather}

\section{The Conformal Limit}
In this section, we will discuss another interesting regime in which it is possible to analytically solve the Schwinger-Dyson equations: the IR limit or, equivalently, $|\tau-\tau'| \gg J^{-1}$. Consider the first equation in Fourier space once again: for small frequencies compared to $J$, the $-i\omega$ term can be neglected with respect to $\Sigma(\omega)$ since the latter is linked to $J$ through the second equation. Going back to time space, the equations are now:
\begin{gather}
\begin{cases}
    \int d\tau'' \: G(\tau,\tau'')\Sigma(\tau'',\tau') = -\delta(\tau-\tau') \\
    \Sigma(\tau,\tau') = J^2 [G(\tau,\tau')]^{p-1}
\end{cases}
\end{gather}

Consider a time reparametrization $\tau \to \phi(\tau)$ and impose that the two functions transform as follows:
\begin{gather}
\label{eq:Conftrasf}
\begin{cases}
    G(\tau,\tau') \to [\phi'(\tau) \phi'(\tau')]^\Delta G(\phi(\tau),\phi(\tau')) \\
    \Sigma(\tau,\tau') \to [\phi'(\tau) \phi'(\tau')]^{\Delta(p-1)} \Sigma(\phi(\tau), \phi(\tau'))
\end{cases}
\end{gather}

We argue that, if $G$ and $\Sigma$ are solutions, then their transformed counterparts are too. The second equation is trivially still satisfied after a reparametrization, whereas for the first one we have:
\begin{align}
    \nonumber & \int d\tau'' \: [\phi'(\tau) \phi'(\tau'')]^\Delta G(\phi(\tau),\phi(\tau'')) [\phi'(\tau'') \phi'(\tau')]^{\Delta(p-1)}\Sigma(\phi(\tau''),\phi(\tau'))  \\
    = & \int d\phi(\tau'') (\phi'(\tau''))^{\Delta p-1} (\phi'(\tau'))^{\Delta p} \left[\frac{\phi'(\tau)}{\phi'(\tau')} \right]^\Delta G(\phi(\tau),\phi(\tau'')) \Sigma(\phi(\tau''),\phi(\tau')) \\
    \nonumber = & - \phi'(\tau') \left[\frac{\phi'(\tau)}{\phi'(\tau')} \right]^{\frac{1}{p}} \delta(\phi(\tau)-\phi(\tau')) = -\delta(\tau-\tau').
\end{align}

In the third line we have chosen $\Delta = 1/p$, but we will see that this is absolutely necessary below. Notice that $-\delta(\tau-\tau')$ has to be seen as a fixed external source for the system of equations, so the integral of the reparametrized functions should yield exactly this source and not, for example, $-\delta(\phi(\tau)-\phi(\tau'))$. We have thus discovered a time reparametrization symmetry that is emergent in the IR, and explicitly broken by the $\delta(\tau-\tau')\partial_\tau$ term in the first Schwinger-Dyson equation. We have also found out that, under such transformations, $G$ and $\Sigma$ transform as conformal two point functions of primary fields with scaling dimension $\Delta = 1/p$ and $(p-1)\Delta$ respectively. We recall that, in a CFT$_1$, \textit{every} transformation of the time coordinate is conformal, so a field is a primary if and only if it transforms correctly under any $\phi(\tau)$. This is exactly the case for $G$ and $\Sigma$, as shown. An arbitrary transformation is characterized by its derivatives $\phi^{(n)}(\tau_0)$, therefore by an infinite amount of coefficients\footnote{This also happens in Euclidean CFT$_2$s in flat space, where all and only analytic functions of $\alpha = t+ix$ are conformal transformations.}: in this language, a primary has to be well-behaved under any $\tau \to \tau + \varepsilon, \: \varepsilon = \sum_{n \geq 0} \varepsilon_n \tau^n$, generated by some charge operator $Q_\varepsilon$.

We now look for solutions to these simplified equations at zero temperature. A clever idea is to use the previous observation by ansatzing a conformal two point function for $G$ (and consequently for $\Sigma$) that obeys $G(\tau,\tau') = -G(\tau',\tau)$:
\begin{gather}
    \label{eq:confGS}
    G(\tau) = \frac{b}{|\tau|^{2\Delta}} \, \mathrm{sgn}(\tau) = bd_\Delta (\tau), \quad \Sigma(\tau) = J^2 b^{p-1} d_{\Delta(p-1)} (\tau).
\end{gather}

Note that a general reparametrization might break the time translation invariance, but we are using the initial time coordinate for now. The easiest way to solve the first equation is to turn to Fourier space once again, hence we need to compute:
\begin{equation}
\begin{aligned}
    \tilde d_\Delta (\omega) & = \int_{-\infty}^{+\infty} d\tau \, e^{i\omega \tau} \frac{\mathrm{sgn}(\tau)}{|\tau|^{2\Delta}} \\
    & = - \int_{-\infty}^0 d\tau \, e^{i\omega \tau}(-\tau)^{-2\Delta} + \int_0^{+\infty} d\tau \, e^{i\omega\tau} \tau^{-2\Delta} \\
    & = 2i \Im \left[ \int_{0}^{+\infty} d\tau \, e^{i\omega\tau} \tau^{-2\Delta} \right].
\end{aligned}
\end{equation}

From the last equation, it is clear that $\tilde d_\Delta(-\omega) = -\tilde d_\Delta(\omega)$. For $\omega > 0$, we can use Jordan's lemma and the absence of poles in the first quadrant for the function $f(\tau \in \mathbb{C}) = e^{i\omega\tau} \tau^{-2\Delta}$ to compute the integral on the positive imaginary time axis $\tau = i \tilde \tau, \, \tilde \tau > 0$:
\begin{gather}
    \tilde d_\Delta(\omega) = 2i \Im \left[ i^{1-2\Delta} \int_{0}^{+\infty} d\tilde\tau \, e^{-\omega\tilde\tau} \tilde\tau^{-2\Delta} \right] = 2i \cos(\pi\Delta) \Gamma(1-2\Delta) \frac{1}{\omega^{1-2\Delta}}.
\end{gather}

We can use this result in $\tilde G(\omega) = -1/\tilde\Sigma(\omega)$. Matching the dependence on $\omega$ on both sides fixes $\Delta = 1/p$, so that we are left with:
\begin{gather}
    \nonumber -4J^2 b^p \cos^2\left( \frac{\pi}{p} \right) \Gamma\left(1-\frac{2}{p}\right) \Gamma\left(-1+\frac{2}{p} \right) = 1, \\
    \Gamma\left(1-\frac{2}{p}\right) \Gamma\left(-1+\frac{2}{p} \right) = \Gamma\left(1-\frac{2}{p}\right) \label{eq:bIR} \left(-1+\frac{2}{p}\right)^{-1} \Gamma\left(\frac{2}{p} \right) = -\left(1-\frac{2}{p}\right)^{-1} \frac{\pi}{\sin \left( \frac{2\pi}{p} \right)} \\
    \nonumber \implies b^p = \frac{1}{J^2 \pi} \left( \frac{1}{2} - \frac{1}{p} \right) \tan \left( \frac{\pi}{p} \right).
\end{gather}

In the specific case of $p = 2$, we can find $b$ in the following way:
\begin{gather}
    b^2 = \lim_{p\to 2} \frac{1}{4\pi J^2} \frac{p-2}{\cos\left( \frac{\pi}{2 \left( 1+\frac{p-2}{2} \right)} \right)} = \frac{1}{\pi^2 J^2}.
\end{gather}

The references only claim the final result, but we can actually determine the reparametrizations under which the Green function is invariant:
\begin{gather}
    [\phi'(\tau)\phi'(\tau')]^\Delta \, \frac{b}{|\phi(\tau)-\phi(\tau')|^{2\Delta}} \, \mathrm{sgn}(\phi(\tau)-\phi(\tau')) = \frac{b}{|\tau-\tau'|^{2\Delta}} \, \mathrm{sgn}(\tau-\tau').
\end{gather}

Since reparametrizations have to be injective, we can assume without loss of generality $\phi'(\tau) > 0$, so that $\mathrm{sgn}(\phi(\tau)-\phi(\tau')) = \mathrm{sgn}(\tau-\tau')$ and the equation that we need to solve is:
\begin{gather}
    \phi'(\tau)\phi'(\tau') = \left( \frac{\phi(\tau)-\phi(\tau')}{\tau-\tau'} \right)^2. \label{eq:SL2eq}
\end{gather}

Since this has to be true for all $\tau,\tau'$, we can fix $\tau' = 0$. The differential equation can then be easily solved and yield:
\begin{equation}
\begin{gathered}
    \phi(\tau) = \frac{\phi'(0)(1+\phi(0)c)\tau + \phi(0)}{\phi'(0)c \tau + 1} = \frac{\sqrt{\phi'(0)}(1+\phi(0)c)\tau + \frac{\phi(0)}{\sqrt{\phi'(0)}}}{\sqrt{\phi'(0)}c \tau + \frac{1}{\sqrt{\phi'(0)}}} = \frac{\alpha \tau + \beta}{\gamma \tau + \delta}, \quad \alpha\delta-\beta\gamma = 1, \\
    \phi'(\tau) = \frac{1}{\left( \sqrt{\phi'(0)}c \tau + \frac{1}{\sqrt{\phi'(0)}} \right)^2} = \frac{1}{(\gamma \tau + \delta)^2} > 0.
\end{gathered}
\end{equation}

$c$ is an integration constant. We have found out that invariant reparametrizations are all and only Möbius transformations. This means that \eqref{eq:confGS} breaks the full reparametrization invariance of the IR theory down to $SL(2,\mathbb{R})$. It is easy to generalize the zero temperature result to a generic $\beta$, since we just need to use the transformation properties \eqref{eq:Conftrasf} of the Green function in the case of \eqref{eq:circlemap}:
\begin{gather}
    G_{\rm circle}(\tau-\tau') = b \left( \frac{\pi}{\beta \left| \sin \left( \frac{\pi(\tau-\tau')}{\beta} \right) \right|} \right)^{2\Delta} \mathrm{sgn}\left( \sin \left( \frac{\pi(\tau-\tau')}{\beta} \right) \right).
\end{gather}

\section{IR Effective Action}
In the previous section, we have observed that a given solution of the Schwinger-Dyson equations in the IR can be mapped to other solutions through reparametrizations of the time. This means that the action \eqref{eq:effaction} without the derivative term has a saddle manifold parametrized by $\phi(\tau)$ instead of a single saddle point. Corrections due to the presence of the derivative then lift the degeneracy and identify our saddle solution in a unique way. The conformal limit requires $\abs{\theta} = \abs{\frac{2\pi \tau}{\beta}} \gg (\beta J)^{-1}$, so that we expect the action of the reparametrizations to be suppressed by powers of the dimensionless parameter $\beta J$ and to spawn in the first place due to the derivative corrections, as they are exact $0$-modes of the conformal theory. We can thus attempt to perform a strong coupling expansion (assuming $\beta J$ to be big) and separate the dynamics of the reparametrizations from the other fluctuations of the conformal $G_c(\tau)$ found in the previous section, which is responsible for a symmetry breaking from all reparametrizations down to $SL(2,\mathbb{R})$. The idea is that, although all fluctuations around the saddle are suppressed by the $N \to \infty$ limit, reparametrizations counter this suppression through an extra $\beta J$ dependence of the action in such a way that $\sim \beta J/N$ corrections induced by them could in principle be arbitrarily more relevant than those of any other excitation of the conformal saddle. The symmetry breaking is a strong hint that these dynamics will be governed by the Schwarzian derivative:
\begin{gather}
    S(\phi(\tau),\tau) \equiv \frac{\phi'''(\tau)}{\phi'(\tau)} - \frac{3}{2} \left( \frac{\phi''(\tau)}{\phi'(\tau)} \right)^2,
\end{gather}

which is the unique lowest order in derivatives Lagrangian that is $SL(2,\mathbb{R})$ invariant. This and several other properties of this mathematical object are proven in Appendix \ref{app:Schwarz}. Returning to \eqref{eq:effaction}, the only relevant term for the reparametrizations is the first one, since the others always obey the IR conformal symmetry. We can normalize it by adding a $\phi$-independent term:
\begin{gather}
    I_{\rm eff}[\phi] = -\frac{1}{2}\tr \log(\delta\partial_\tau - \Sigma_*^\phi) + \frac{1}{2}\tr \log(\delta\partial_\tau - \Sigma_*).
\end{gather}

$\Sigma_*$ is the true saddle of \eqref{eq:effaction} and $\Sigma_*^\phi$ is its reparametrization \eqref{eq:Conftrasf}:
\begin{gather}
    \Sigma_*^\phi(\tau,\tau') = [\phi'(\tau)\phi'(\tau')]^{1-\frac{1}{p}} \Sigma_*(\phi(\tau),\phi(\tau')).
\end{gather}

We are basically considering a very specific kind of fluctuations around the saddle of the action and determining an approximated Lagrangian that describes their behavior. We observe the following scalings:
\begin{gather}
    \delta(\tau-\tau')\partial_{\tau'} = \frac{4\pi^2}{\beta^2}\delta(\theta-\theta')\partial_{\theta'}, \quad \Sigma_* \propto J^2.
\end{gather}

This means that a strong coupling expansion is equivalent to a series expansion of the logarithms while assuming the derivative term to be much smaller than $\Sigma$. Since the $0$th order of the \textit{entire} effective action is reparametrization invariant, the leading order correction comes from expanding $\log (1 - \delta\partial_\tau(\Sigma)^{-1})$:
\begin{equation}
\begin{aligned}
    I_{\rm eff}[\phi] & = \frac{1}{2} \left[ \tr(\delta\partial_\tau (\Sigma_*^\phi)^{-1}) - \tr(\delta\partial_\tau (\Sigma_*)^{-1}) \right] + \mathcal{O}(\partial_\tau^2) \\
    & = -\frac{1}{2} \left[ \tr(\delta\partial_\tau G_*^\phi) - \tr(\delta\partial_\tau G_*) \right] + \mathcal{O}(\partial_\tau^2), \\
    (\delta \partial_\tau G_*)(\tau-\tau') & = \int d\tau'' \: \delta(\tau-\tau'') \partial_{\tau''} G_*(\tau''-\tau') = \partial_\tau G_*(\tau-\tau').
\end{aligned}
\end{equation}

We recall that $\Sigma^{-1} = -G + \mathcal{O}(\partial_\tau)$ as a consequence of the Schwinger-Dyson equations. Unfortunately, a naive computation of the traces yields Dirac deltas evaluated in 0 since they require to derive:
\begin{gather}
    G_*(\tau-\tau')|_{\tau'\to\tau} \approx \frac{1}{2}\mathrm{sgn}(\tau-\tau').
\end{gather}

It is therefore interesting to try to regularize the SYK model in order to make computations possible, then remove the regulator. We define a new derivative that substitutes $\partial_\tau$ in \eqref{eq:effaction}:
\begin{equation}
\begin{gathered}
    \partial_\tau^\varepsilon f(\tau) \equiv -\int d\tau' \: [\partial_{\tau'} \delta_\varepsilon(\tau-\tau')] f(\tau'), \\
    \delta_\varepsilon(\tau) = \frac{e^{-|\tau|/\varepsilon}}{2\varepsilon}.
\end{gathered}
\end{equation}

$\delta_\varepsilon(\tau)$ converges to the Dirac delta in distributional sense as $\varepsilon \to 0$. We prove this by taking a regular function $f(\tau)$ and considering:
\begin{gather}
    \lim_{\varepsilon \to 0} \int_{-\frac{\beta}{2}}^{+\frac{\beta}{2}} d\tau \: \frac{e^{-|\tau|/\varepsilon}}{2\varepsilon} f(\tau) = \lim_{\varepsilon \to 0} \frac{1}{2} \int_{-\frac{\beta}{2\varepsilon}}^{+\frac{\beta}{2\varepsilon}} d\tau \: e^{-|\tau|} f(\varepsilon \tau) = f(0) = \int_{-\frac{\beta}{2}}^{+\frac{\beta}{2}} d\tau \: \delta(\tau) f(\tau).
\end{gather}

We have assumed that $f(\tau)$ is such that we can commute the limit and the integral. If we take the space of the test functions to be the Schwartz space, for example, the result follows from Lebesgue's dominated convergence theorem applied to $|f_\varepsilon(\tau)| \equiv |e^{-|\tau|} f(\varepsilon \tau)| \leq e^{-|\tau|} |\mathrm{sup}_{\tau} \, f(\tau)|$, which is an integrable function. We could have chosen any regulator of the Dirac delta, but the nice properties of the chosen one are that it is invertible on $L^2(\mathbb{R})$, namely there are functions $f^{(n)}_\varepsilon(\tau) \in L^2(\mathbb{R})$ such that $\lim_{n \to \infty} \int d\tau' \: \delta_\varepsilon(\tau-\tau') f^{(n)}_\varepsilon(\tau') = \delta(\tau)$ (hence there exists a free propagator), and it is suppressed exponentially for big $\tau$. Let us look for the inverse of the new derivative in the infinite temperature case:
\begin{gather}
    \nonumber - \int_{-\infty}^{+\infty} d\tau' \: [\partial_{\tau'} \delta_\varepsilon(\tau-\tau')] H(\tau') = \int_{-\infty}^{+\infty} d\tau' \: \delta_\varepsilon(\tau-\tau') \partial_{\tau'} H(\tau') = \delta(\tau) \implies \tilde \delta_\varepsilon(\omega) \tilde {H'}(\omega) = 1, \\
    \tilde \delta_\varepsilon(\omega) = \int d\tau \: e^{i\omega\tau} \delta_\varepsilon(\tau) = \frac{1}{1+\varepsilon^2\omega^2} \\ 
    \nonumber \implies H'(\tau) = \delta(\tau) - \varepsilon^2 \delta''(\tau) \implies H(\tau) = \frac{1}{2} \mathrm{sgn}(\tau) - \varepsilon^2 \delta'(\tau).
\end{gather}

We have chosen the integration constant for $H(\tau)$ in order to match the known free propagator when $\varepsilon = 0$. Indeed, the invertibility of the regularized kernel was crucial to obtain this result, in particular $\lim_{n\to \infty} f^{(n)}_\varepsilon(\tau) = H'(\tau)$. An example of functions in $L^2(\mathbb{R})$ that converge to $H'(\tau)$ is obtained by taking the Gaussian regularization of the delta:
\begin{gather}
	f^{(n)}_\varepsilon(\tau) = \frac{1}{\sqrt{2\pi\sigma_n^2}} \left( 1 - \varepsilon^2 \, \frac{\tau^2-\sigma_n^2}{\sigma_n^4} \right) e^{-\tau^2/(2\sigma_n^2)}, \quad \sigma_n = \frac{1}{n J}.
\end{gather}

The only time scale of the system is $J$, so it is natural to fix our cutoff $\varepsilon$ to be $a_0/J, \, a_0 = \mathcal{O}(1)$: this way, we can say that the IR limit emerges when $\tau \gg \varepsilon$ and that it is going to be identical to the one in the initial SYK formulation.

The UV limit, on the other hand, is going to differ and to depend on the specific choice of our regulator. This is already clear when considering the saddle point of our new theory:
\begin{equation}
\begin{gathered}
    G_*^\varepsilon(\tau) = b \frac{1}{|\tau|^{\frac{2}{p}}} \mathrm{sgn}_\varepsilon(\tau), \\
    \label{eq:sgneps} \mathrm{sgn}_\varepsilon(\tau) = \begin{cases}
        \mathrm{sgn}(\tau) \quad \mathrm{when} \, |\tau| \gg \varepsilon \\
        |\tau|^{\frac{2}{p}} \frac{1}{b} [\delta_\varepsilon \partial_\tau]^{-1}(\tau) \quad \mathrm{when} \, |\tau| \ll \varepsilon
    \end{cases}.
\end{gathered}
\end{equation}

The new function $\mathrm{sgn}_\varepsilon(\tau)$ interpolates the two known behaviors smoothly when $|\tau| \approx \varepsilon$ and clearly depends on $\delta_\varepsilon(\tau)$ in the UV. A naive reparametrization of the saddle would now be
\begin{gather}
    G_*^{\phi,\varepsilon}(\tau,\tau') = [\phi'(\tau)\phi'(\tau')]^{\frac{1}{p}} G_*^{\varepsilon}(\phi(\tau),\phi(\tau')) = \left[ \frac{\phi'(\tau)\phi'(\tau')}{(\phi(\tau)-\phi(\tau'))^2} \right]^{\frac{1}{p}} b \: \mathrm{sgn}_\varepsilon(\phi(\tau)-\phi(\tau')),
\end{gather}

but we choose $G_*^{\phi,\varepsilon}$ to be proportional to $\mathrm{sgn}_\varepsilon(\tau-\tau')$ instead. This new choice is still such that the approximated first Schwinger-Dyson equation is obeyed by the transformed functions in the IR ($|\tau-\tau'| \gg \varepsilon$), so it is viable even though, in principle, the two do not match unless it is also true that $|\phi(\tau)-\phi(\tau')| \gg \varepsilon$.

It is now time to evaluate the regularized action:
\begin{gather}
    I_{\rm eff}^\varepsilon[\phi] \approx -\frac{1}{2} \left[ \tr (\delta_\varepsilon \partial_\tau G_*^{\phi,\varepsilon} ) - \tr ( \delta_\varepsilon \partial_\tau G_*^{\varepsilon} ) \right] = \frac{1}{2} \int d\tau \, d\tau' \: [\partial_{\tau'} \delta_\varepsilon(\tau-\tau')] (G_*^{\phi,\varepsilon}(\tau'-\tau) - G_*^{\varepsilon}(\tau'-\tau)).
\end{gather}

To do so, we consider the following expansion:
\begin{gather}
    \left[ \frac{\phi'(\tau)\phi'(\tau')}{(\phi(\tau)-\phi(\tau'))^2} \right]^{\frac{1}{p}} - \frac{1}{|\tau'-\tau|^{\frac{2}{p}}} = |\tau'-\tau|^{2-\frac{2}{p}} \left( \frac{1}{6p} S(\phi,\tau) + \mathcal{O}(\tau'-\tau) \right).
\end{gather}

\cite{Sarosi_2018} reports a wrong result of $S/12p$, which propagates to the following steps: in fact, we disagree by a factor of $2$. This formula is useful because the derivative of the regularized delta has support approximately in the range $|\tau'-\tau| \lesssim \varepsilon$, hence we obtain:
\begin{equation}
\begin{split}
    I^\varepsilon_{\rm eff}[\phi] & = -\frac{b}{12p} \int d\tau \: S(\phi,\tau) \int_{-\infty}^{+\infty} da \: [-\partial_a \delta_\varepsilon(a)] |a|^{2-\frac{2}{p}} \mathrm{sgn}_\varepsilon(a) = \\
    & = -\frac{b}{12p} \left[ \int_{-\infty}^{+\infty} da \: \frac{\mathrm{sgn}(a)}{2\varepsilon^2} e^{-\frac{|a|}{\varepsilon}} |a|^{2-\frac{2}{p}} \mathrm{sgn}_\varepsilon(a) \right] \int d\tau \: S(\phi,\tau).
\end{split}
\end{equation}

As suggested by the conformal symmetry breaking induced by the IR saddle, we have derived a Schwarzian action for the time reparametrizations. An observation on our part is that we can integrate $a \equiv \tau'-\tau$ on the entire real axis (rather than the actual interval $[-\beta/2-\tau, \beta/2-\tau)$) even for finite $\beta$ as $\tau$ varies from $-\beta/2$ to $+\beta/2$, provided that $\beta \gg \varepsilon$. If this is true, in fact, we commit a negligible mistake in not considering the exact range for $a$ at fixed $\tau$: by not stopping the integration at $|a| \sim \beta$, we commit a mistake of order $\mathcal{O}(e^{-\beta/\varepsilon}) \sim \mathcal{O}(e^{-\beta J})$. There is also an error due to the region of integration where $|\tau| \sim \beta/2$, where one of the extremes for $a$ is $0$ rather than $\mathcal{O}(\beta)$, but its temporal extension is only $\mathcal{O}(\varepsilon) \ll \beta$. Note that in this case we should consider fluctuations with respect to the thermal Green function\footnote{We underline that the above computation assumed zero temperature.}, which we can obtain from the zero temperature one with the usual tangent mapping. Constant aside, the final result is that the Schwarzian action actually becomes \eqref{eq:SchfinT}.

In any case, before having concluded, we still need to consider the $\varepsilon \to 0$ limit of this result. Unfortunately, it is impossible to compute the coefficient of the Schwarzian action exactly unless one finds the Green function of the interacting theory, yet we can evaluate the $a$ integral in the IR and UV limits of \eqref{eq:sgneps}:
\begin{equation}
\begin{gathered}
    I_{\rm eff}^{\varepsilon, \rm IR}[\phi] = -\frac{b}{12p} \Gamma\left(3-\frac{2}{p}\right) \varepsilon^{1-\frac{2}{p}} \int d\tau \: S(\phi,\tau), \\
    I_{\rm eff}^{\varepsilon, \rm UV}[\phi] = -\frac{\varepsilon}{12p} \int d\tau \: S(\phi,\tau).
\end{gathered}
\end{equation}

We know from \eqref{eq:bIR} that $b \propto J^{-\frac{2}{p}}$ and that $\varepsilon \propto J^{-1}$, hence both limits scale as $J^{-1}$. Since $\mathrm{sgn}_\varepsilon(a)$ interpolates these two cases, we can expect the actual result to scale as:
\begin{gather}
    I_{\rm eff}^\varepsilon[\phi] \sim \frac{1}{J} \int d\tau \: S(\phi,\tau).
\end{gather}

What we would now want to do is to consider the $\varepsilon \to 0$ limit of these results, since it was an arbitrary regulator that we inserted inside our theory. The problem with this procedure is that, unless $p = 2$, the coefficients of both the IR and the UV effective actions (hence the actual result itself) tend to zero. This is rather troublesome, as it means that we should either approach the problem in a different way or consider the next term in the derivative expansion.
Regardless, a Schwarzian derivative is still expected to be the final result of a more thorough analysis because of two reasons:
\begin{itemize}
    \item All and only Möbius time reparametrizations are symmetries of the conformal saddle. Consequently, a time reparametrization $\phi(\tau)$ or its composition with a Möbius transformation $(M \circ \phi)(\tau)$ are the exact same fluctuation, so the effective action should be unable to distinguish between the two. As a corollary, a pure Möbius transformation $M(\tau)$ is equivalent to the absence of fluctuations of $G_c(\tau)$, hence its effective action should be null.
    
    \item As argued before, a $(\beta J)^{-1}$ expansion is equivalent to a derivative expansion of \eqref{eq:effaction}, so our effective action for the reparametrizations should have the least possible number of derivatives while also possessing a $SL(2,\mathbb{R})$ symmetry.
\end{itemize}

The Schwarzian derivative perfectly satisfies both requirements and appears naturally in different approaches to the problem. There are, in fact, various attempts in the literature to find a compelling derivation of this result. Let us take for example the arguments provided in \cite{Bagrets_2016,Maldacena_2016,jevicki2016bilocal}, and briefly discuss what happens there.
\begin{itemize}
    \item In \cite{Bagrets_2016}, the specific case $p = 4$ is studied. The chosen approach is to consider the next term in the derivative expansion, which turns out to be:
    \begin{gather}
        I_{\rm eff}[\phi] = -\frac{1}{16 \sqrt{4\pi} J} \int d\tau \, d\tau' \: \frac{\phi'(\tau)^{3/2} \phi'(\tau')^{3/2}}{|\phi(\tau)-\phi(\tau')|^3}.
    \end{gather}
    Similarly to the case that we studied here, this action requires a UV regularization, which they take to be a $\Lambda \sim J$ cutoff in the frequency space. By antitransforming the regularized action, its definition in time space turns out to be:
    \begin{gather}
        I_{\rm eff}[\phi] = -\frac{1}{16\sqrt{4\pi} J} \int d\tau \, d\tau' \: \frac{\phi'(\tau)^{3/2} \phi'(\tau')^{3/2}}{(|\phi(\tau)-\phi(\tau')|+\delta)^3}, \quad \delta \sim J^{-1}\phi'((\tau+\tau')/2).
    \end{gather}
    Clearly, this action is non-local as it links reparametrizations at different times. They manage to recover a local Schwarzian action by first assuming it to be a correct description of the system and then studying its results for two and four point functions. These suggest that the non-local theory should change behavior at $\tau \gtrsim M \gg J^{-1}$, after which the connected four point function $\langle G G \rangle$ should decay as $\tau^{-3/2}$ instead of $\tau^{-1}$ (the disconnected behavior for $\Delta = 1/4$). Interestingly enough, $M$ is linked to the action by
    \begin{gather}
        I_{\rm eff}[\phi] = -\frac{M}{N} \int d\tau \: S(\phi,\tau).
    \end{gather}
    This provides a modification to really small frequencies by further altering the UV-regulated kernel $\sim \omega^2 \log(J/|\omega|)$ of the non-local theory, turning it into a local kernel $\sim \omega^2 \log(J/\Delta), \: \Delta \sim M^{-1}$. By determining the self-consistent $M$ that matches the Schwarzian action to the local kernel, they finally derive:
    \begin{gather}
        I_{\rm eff}[\phi] \approx -\frac{\log(N\log N)}{32\sqrt{4\pi}} \frac{1}{J} \int d\tau \: S(\phi,\tau).
    \end{gather}
    Notice how this result scales with $J^{-1}$ just like ours, even though it comes from a higher order term in the derivatives: in fact, it has to be this way for dimensional reasons. The problems with this approach are both the applicability of the procedure to larger $p$\footnote{In the initial action, the non-local kernel is $|\phi(\tau)-\phi(\tau')|^{-2-4/p}$ in general, hence the non-local behavior becomes stronger as $p \to \infty$.} and the comparison of their result with the action for linearized reparametrizations derived in \cite{Maldacena_2016}.

    \item In \cite{Maldacena_2016}, small reparametrizations $\phi(\tau) = \tau + \varepsilon(\tau)$ are considered. First of all, \eqref{eq:effaction} is expanded at quadratic order with respect to perturbations of the saddle solutions, obtaining:
    \begin{equation}
    \begin{gathered}
        I = \frac{J^2(p-1)}{4} g \cdot(\tilde K^{-1} - 1) g, \quad g \equiv (G-G_*)|G_*|^{\frac{2}{2-p}}, \\
        \tilde K(\tau_1, \tau_2; \tau_3, \tau_4) = -J^2(p-1)|G_*(\tau_1,\tau_2)|^{\frac{p-2}{2}} G_*(\tau_1,\tau_3) G_*(\tau_2,\tau_4) |G_*(\tau_3,\tau_4)|^{\frac{p-2}{2}}.
    \end{gathered}
    \end{equation}
    The Gaussian integral of the perturbations of $\Sigma$ has already been performed above. By considering the IR limit (so that $G_*(\tau) = G_c(\tau)$) and evaluating the action in the case of $G-G_* = \delta_\varepsilon G_c$, that is, the first order variation of the Green's function due to a small time reparametrization, they find the following result:
    \begin{gather}
        I_{\rm eff}[\varepsilon] = \frac{\alpha_S}{\mathcal{J}} \int_0^\beta d\tau \: \frac{1}{2}\left[ (\varepsilon'')^2 - \left( \frac{2\pi}{\beta} \right)^2 (\varepsilon')^2 \right],
    \end{gather}
    where $\alpha_S$ is a constant. Notice that they are assuming $\tau \in [0,\beta)$. It vanishes for global $SL(2,\mathbb{R})$ transformations only, so this term in the action explicitly breaks the full conformal symmetry that was already spontaneously broken by $G_c$: reparametrizations can be thus thought of as Pseudo-Nambu-Goldstone bosons in the theory, just like pions in QCD. The next step is to generalize this action to finite transformations in the case of $\beta = \infty$:
    \begin{gather}
        \phi(\tau) = \phi(0) + \phi'(0) \left( \tau + \frac{1}{2} \frac{\phi''(0)}{\phi'(0)}\tau^2 + \dots \right).
    \end{gather}
    For small $\tau$, any finite transformation is characterized by $\varepsilon' = 0, \, \varepsilon'' = \phi''/\phi'$, followed by a scaling and a translation. The latter leave $G_c$ invariant, thus a natural extension of the action is $\varepsilon'' \to \phi''/\phi'$. Up to a total derivative, this match to the linearized case yields our desired result:
    \begin{gather}
        I_{\rm eff}[\phi] = -\frac{\alpha_S}{\mathcal{J}} \int d\tau \: S(\phi,\tau).
    \end{gather}
    In order to find the finite temperature case, one first picks $\phi(\tau) = \tan \left( \frac{\pi\tau}{\beta} \right)$ with $\tau$ on the circle, then considers a further reparametrization $\psi(\tau)$ there. The composition law for the Schwarzian derivative in Appendix \ref{app:Schwarz} yields:
    \begin{equation}
    \begin{gathered}
    	S(\phi \circ \psi, \tau) = S(\phi,\psi(\tau)) \, (\psi')^2 + S(\psi,\tau) = \frac{1}{2} \left( \frac{2\pi}{\beta} \right)^2 (\psi')^2 + S(\psi,\tau), \\
        \label{eq:SchfinT} I_{\rm eff}[\psi] = -\frac{\alpha_S}{\mathcal{J}} \int_0^\beta d\tau \: \left[ S(\psi,\tau) + \frac{1}{2} \left( \frac{2\pi}{\beta} \right)^2 (\psi')^2 \right] \to \frac{\alpha_S}{2\mathcal{J}} \int_0^\beta d\tau \: \left[ \left( \frac{\psi''}{\psi'} \right)^2 - \left( \frac{2\pi}{\beta} \right)^2 (\psi')^2 \right],
    \end{gathered}
    \end{equation}
    which correctly reduces to the linearized result when $\psi(\tau) = \tau + \varepsilon(\tau)$, after having removed a total derivative in the second line.
    
    \item In \cite{jevicki2016bilocal}, two different solutions are presented. Both compute a differently regularized action:
    \begin{gather}
        I_{\rm eff}[\phi] = -\frac{1}{2} \lim_{s \to \frac{1}{2}} \int d\tau \, d\tau' \: G_*^\phi(\tau,\tau') Q_s(\tau,\tau'), 
    \end{gather}
    with $Q_s(\tau,\tau')$ a specific function that they determine in the paper. The first solution is a series expansion in $0 \leq \varepsilon = 1-2/p < 1$. To this end, one considers the reparametrized conformal Green function (we do not report the sign):
    \begin{equation}
    \begin{split}
    G^\phi_*(\tau,\tau') & = b(p(\varepsilon)) \left( \frac{\sqrt{\phi'(\tau)\phi'(\tau')}}{|\phi(\tau)-\phi(\tau')|} \right)^{1-\varepsilon} \\
    & = b(p) \left( \frac{\sqrt{\phi'(\tau)\phi'(\tau')}}{|\phi(\tau)-\phi(\tau')|} \right) \Bigg[ 1- \varepsilon \log \left( \frac{\sqrt{\phi'(\tau)\phi'(\tau')}}{|\phi(\tau)-\phi(\tau')|} \right) \\
    & \hspace{2.2 cm} + \frac{\varepsilon^2}{2} \left( \log \left( \frac{\sqrt{\phi'(\tau)\phi'(\tau')}}{|\phi(\tau)-\phi(\tau')|} \right) \right)^2 + \dots \Bigg].
    \end{split}
    \end{equation}
    This expansion is then inserted inside the regularized action, thus obtaining:
    \begin{gather}
        I_{\rm eff}[\phi] = - \frac{\alpha}{24\pi J} \int d\tau \: S(\phi,\tau), \quad \alpha = 1-\varepsilon^2 + \mathcal{O}(\varepsilon^3).
    \end{gather}
    The conceptual issue with this approach is that $p = 2$ (the point around which the expansion is made) is a fundamentally different case with respect to all the others: it is, in fact, a quadratic theory for any given member of the ensemble. It follows that taking a parametric expansion around this theory to determine valid results for any other $p$ is risky at the very least. Luckily, the second solution is more general and never makes this assumption. The final result in this case has the exact same structure as the one above, but now $\alpha = -12 \pi B_1(p) \gamma(p)$, where $\gamma(p)$ is a function presented in the paper and $B_1$ is a function whose approximate form is determined numerically in \cite{Maldacena_2016}.
\end{itemize}

Technicalities aside, a really important result has been shown: the reparametrizations of the IR saddle are described by a Schwarzian action. Among these, notably, Möbius transformations should not be integrated over in the partition function as they are basically a gauge symmetry of the theory. This fact will be the bridge that will allow us to connect this seemingly innocent Condensed Matter model to gravitational setups. \\

With this notion in mind, we are finally ready to explore a useful $(1+1)$ model of gravity: the Jackiw-Teitelboim (JT) gravity.

\newpage
\chapter{JT Gravity}
Before describing JT gravity, it is useful to first digress and talk about AdS$_2$ and how it emerges from the metric of extremal black holes in four dimensions, as it is one way to connect this two-dimensional model to interesting gravitational setups. We will mainly follow \cite{Sarosi_2018}, which in turn is based on \cite{Maldacena_1999, 10.1093/ptep/ptw124}.

\section{Near-Horizon Region of Extremal Black Holes} \label{sec:NearHorizon}
For a Reissner-Nordström black hole with magnetic charge $Q$ and mass $M$, the metric and the electromagnetic field are given by \cite{carroll2004spacetime, HamiltonGR}:
\begin{equation}
\begin{gathered}
    ds^2 = -\frac{(r-r_+)(r-r_-)}{r^2} dt^2 + \frac{r^2}{(r-r_+)(r-r_-)} dr^2 + r^2 d\Omega_2^2, \\
    F = Q \sin \theta \, d\phi \wedge d\theta, \\
    r_\pm = Q\ell_P + E\ell_P^2 \pm \sqrt{2QE\ell_P^3 + E^2 \ell_P^4}, \\
    E = M - \frac{Q}{\ell_P} \geq 0.
\end{gathered}
\end{equation}

$\ell_P = \sqrt{G_N}$ is the Planck length. $E$ is the difference between the actual black hole's energy $M$ and the energy it would have if it were extremal $(M_{\rm extremal} = Q/\ell_P)$, namely it is an excitation energy. If $E < 0$, the black hole's singularity would be naked and the spacetime would violate the cosmic censorship hypothesis. We focus on the case of extremal black holes, whose two horizon radii $r_\pm$ coincide: they are characterized by $E = 0$ and their only independent dimensionful parameter is $\ell_P$. By now defining a new coordinate
\begin{gather}
    z = \frac{Q^2 \ell_P^2}{r-r_+} > 0,
\end{gather}

we can ``zoom in'' to $r_+$ by simply taking $\ell_P \to 0$ while keeping $z$ fixed. Our metric becomes:
\begin{gather}
    ds^2 = \ell_P^2 Q^2 \left( \frac{-dt^2+dz^2}{z^2} + d\Omega_2^2 \right). \label{eq:AdS2}
\end{gather}

It is evident that we have factorized out a $S^2$ from our spacetime, leaving us with a non-trivial metric only for the $(t,z)$ coordinates. Equivalently, near an extremal black hole the spacetime becomes the product space AdS$_2 \times S^2$, where we are describing AdS$_2$ through the use of the Poincaré coordinates:
\begin{gather}
    ds^2 = \ell^2_{\rm AdS} \frac{-dt^2+dz^2}{z^2}.
\end{gather}

\begin{figure}
    \centering
    \includegraphics[width = 0.7 \textwidth]{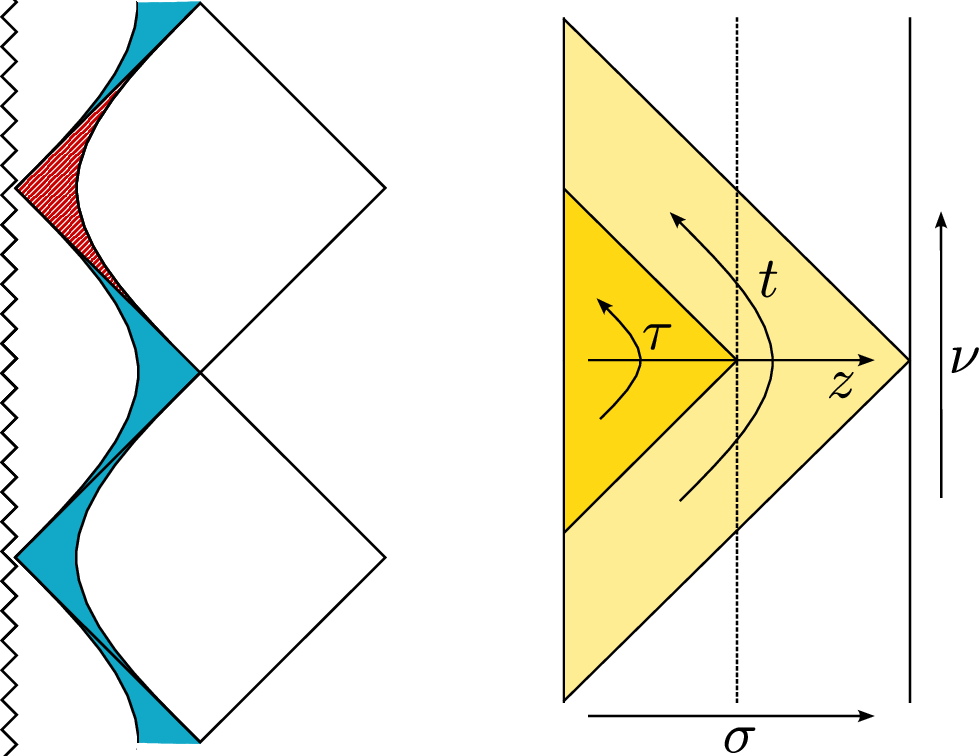}
    \caption{Left: Maximally extended Penrose diagram of an extremal black hole. The infinite chain signals the succession of black hole and white hole horizons that connect different universes. The vertical zigzagging line on the left is the singularity, spacelike infinities are on the right. The blue region is the AdS$_2$ near-horizon region, while the red dashed one is the patch covered by the Poincaré coordinates. \\ Right: Penrose diagram and coordinates of global AdS$_2$, which possesses two boundaries. The Poincaré patch is the light yellow region, while the dark yellow one is the Rindler patch.}
    \label{fig:AdSPen}
\end{figure}

$\ell_{\rm AdS}$ is the curvature radius of the spacetime. Refer to the left panel in Figure \ref{fig:AdSPen} for the Penrose diagram of an extremal black hole. As shown there, the spacetime can be extended to an infinite sequence of exterior and interior regions. The original coordinates $(t,r)$ only cover one exterior and one interior region (where they exchange their role of being a time and a space coordinate) of the global AdS$_2$ (in blue), which can instead be obtained from the Poincaré coordinates by taking the $(\nu,\sigma)$ coordinates:
\begin{equation}
\begin{gathered}
    ds^2 = \ell^2_{\rm AdS} \frac{-d\nu^2 + d\sigma^2}{\sin^2 \sigma}, \\
    \nu = \arctan(t+z) + \arctan(t-z), \quad \sigma = \arctan(t+z) - \arctan(t-z).
\end{gathered}
\end{equation}

It is now possible to decompactify the timelike coordinate $\nu$, thus extending its range from $[-\pi,\pi]$ to $[-\infty,+\infty]$. The same cannot be done to the spacelike coordinate $\sigma$, whose range is $[0,\pi]$ because of the $(\sin\sigma)^{-2}$ factor in the metric.

The global AdS$_2$ metric is conformally equivalent to a $(1+1)$ Minkowski spacetime. Its Penrose diagram is also in Figure \ref{fig:AdSPen} and shows that, unlike AdS$_{d > 2}$, there are two disconnected timelike boundaries at $\sigma = 0$ ($z = 0$) and $\sigma = \pi$ ($z=\infty$). There is another choice of coordinates that covers the so-called Rindler patch and is connected to the Poincaré coordinates in the following way:
\begin{equation}
\begin{gathered}
    z \pm t = \frac{(1 \pm \cosh \rho) \, e^{\tau/2} \mp \sinh \rho \, e^{-\tau/2}}{(1\pm \cosh \rho) \, e^{\tau/2} \pm \sinh \rho \, e^{-\tau/2}}, \label{eq:PoinToRind} \\
    ds^2 = \ell^2_{\rm AdS}(d\rho^2  \sinh^2 \hspace{-0.04 in} \rho \, d\tau^2).
\end{gathered}
\end{equation}

In the case of Minkowski spacetime in any number of dimensions, the Rindler coordinates describe an observer with constant proper acceleration (which coincides with the acceleration seen by the succession of inertial observers that have its same velocity at every instant).

We could have considered extremal black holes with electric charge $Q_e = \ell_P M$ and we would have again obtained an AdS$_2 \times S^2$ geometry near the horizon \cite{Anninos_2021}. This is a trivial consequence of the EM duality, which allows us to exchange $F_{\mu\nu}$ with $\tilde F_{\mu\nu}$ ($\vec{E} \to -\vec{B}, \: \vec{B} \to \vec{E}$) and electric charges with magnetic ones ($Q_{\rm el} \to -Q_{\rm mag}, \: Q_{\rm mag} \to Q_{\rm el}$) without altering the geometry ($T_{\mu\nu} \to T_{\mu\nu}$). Because of this, we stick to a magnetic charge in the following.

\section{Dilaton-Gravity Models} \label{sec:Dilatons}
In this section, we will show what is the connection of the previous study to our target two-dimensional models. We start from the Einstein-Hilbert action with an electromagnetic field:
\begin{gather}
    S = \frac{1}{16\pi \ell_P^2} \int d^4x \, \sqrt{-g} \left( R - \ell_P^2 F_{\mu \nu}F^{\mu \nu} \right).
\end{gather}

The normalization of the electromagnetic tensor is compatible with our choice of the radial magnetic field being $Q/r^2$, meaning that the action in \cite{Sarosi_2018} has a wrong factor of $1/4$ in front of the electromagnetic term. We can integrate out the angles by considering a static, spherically symmetric ansatz:
\begin{equation}
\begin{gathered}
    ds^2 = h_{ij} dx^i dx^j + e^{2\psi(r,t)}d\Omega_2^2, \\
    F = Q \sin \theta \, d\phi \wedge d\theta,
\end{gathered}
\end{equation}

with $i,j = 1,2, \, x^1 = t, \, x^2 = r$. We can use the following formula for our warped product space \cite{HeadrickCompendium}:
\begin{gather}
    R = R_h + 2e^{-2\psi} - 4 \frac{1}{\sqrt{-h}} \partial_i (\sqrt{-h} \, h^{ij} \partial_j \psi) - 6h^{ij}\partial_i \psi \partial_j \psi, \label{eq:Warped}
\end{gather}

with $R_h$ the Ricci scalar associated to the metric $h_{ij}$. Let us proceed with the computation:
\begin{equation}
\begin{split}
    S & = \frac{1}{16\pi \ell_P^2} \int dt \, dr \, d\theta \, d\phi \: \sqrt{-h} \, e^{2\psi} \sin \theta \: (R - 2 e^{-4\psi} Q^2 \ell_P^2) \\
    & = \frac{1}{4\ell_P^2} \int dt \, dr \, \sqrt{-h} \, [e^{2\psi} (R_h + 2 h^{ij} \partial_i \psi \partial_j \psi) + 2 - 2 e^{-2\psi} Q^2 \ell_P^2] \\
    & = \frac{1}{4\ell_P^2} \int dt \, dr \, \sqrt{-h} \, [\Phi^2 R_h + 2(\partial \Phi)^2 + 2 - 2 \Phi^{-2} Q^2 \ell_P^2]. \label{eq:BHReduce}
\end{split}
\end{equation}

In the second line, we have integrated by parts the third term appearing in \eqref{eq:Warped}, while in the third line we have defined $\Phi = e^\psi$. This result is an example of \textit{dilaton-gravity model}, which in a more general case is described by the following action \cite{almheiri2015models}:
\begin{gather}
    S = \frac{1}{16\pi G_N} \int d^2x \, \sqrt{-h} \, [\Phi^2 R_h + \lambda (\partial \Phi)^2 - U(\Phi^2/d^2)].
\end{gather}

$U$ is a dimensionless scalar potential, $d$ is a length parameter, $\lambda$ is a dimensionless coefficient. $\Phi^2$ is the dilaton field and has dimension (length)$^2$: it can be seen as a modulation of $G_N$, so that the Ricci scalar is linked to a dimensionless coupling $G_N \Phi^{-2}(r,t)$. What we have performed is a dimensional reduction of an initial four-dimensional theory, so $G_N$ has dimension (length)$^{-2}$ and the overall action is dimensionless. We can always assume $\lambda = 0$ because of the following Weyl transformation:
\begin{gather}
    h_{ij} \to \Phi^{-\alpha/2} h_{ij}, \quad \lambda \to \lambda - \alpha, \quad U(\Phi) \to \Phi^{-\alpha/2}U(\Phi),
\end{gather}

which follows from how the Ricci scalar transforms under this Weyl rescaling:
\begin{gather}
	\sqrt{-h} \, R \to \left( \Phi^{-\alpha/2} \sqrt{-h} \right) \left[ \Phi^{\alpha/2}R + \frac{\alpha}{2} \Phi^{\alpha/2} \nabla_\mu \nabla^\mu \log \Phi \right] = \sqrt{-h} \left[R + \frac{\alpha}{2} \frac{1}{\sqrt{-h}} \partial_\mu \left( \sqrt{-h} \, \frac{\partial^\mu \Phi}{\Phi} \right) \right].
\end{gather}

One can in fact integrate the second term by parts, whose result is that $\lambda$ is shifted by $-\alpha$.

Consider now the presence of matter fields and an action $S = I + S_{\rm matter}$. In two dimensions, our metric has three independent entries and we are free to fix two of them through diffeomorphisms of the two coordinates. This means that we can always pick the ``conformal gauge'':
\begin{gather}
    ds^2 = -e^{2\omega(r,t)}(dt^2-dr^2) = -e^{2\omega(u^+,u^-)}du^+ du^-,
\end{gather}

where $u^+ = t+r = \mathrm{constant}$ and $u^- = t-r = \mathrm{constant}$ are the trajectories of massless objects. For global AdS$_2$, for example, $t$ and $r$ here are $\nu$ and $\sigma$ respectively. One of the equations of motion is \cite{almheiri2015models}:
\begin{gather}
    -e^{2\omega} \partial_+(e^{-2\omega} \partial_+ \Phi^2) = G_N T_{++}^{\rm matter}, \label{eq:blowupeq}
\end{gather}

where the kinetic term of the dilaton is included in the stress tensor. Before we proceed, we first go back to \eqref{eq:BHReduce}: drawing inspiration from \eqref{eq:AdS2}, we look for solutions with constant $\Phi = \ell_{\rm AdS} = Q\ell_P$, so that the action greatly simplifies to:
\begin{gather}
    S = \frac{Q^2}{4} \int dt \, dr \, \sqrt{-h} \: R_h.
\end{gather}

If we stick to the conformal gauge and look for static solutions that only depend on $r$, the differential equation that we need to solve is the following \cite{almheiri2015models}:
\begin{equation}
\begin{gathered}
    \omega'' = \frac{1}{\ell_{\rm AdS}^2} e^{2\omega} \\
    \implies e^{2\omega} = \frac{\ell_{\rm AdS}^2}{r^2}, \; \frac{\ell_{\rm AdS}^2}{\sinh^2 r}, \; \frac{\ell_{\rm AdS}^2}{\sin^2 r}.
\end{gathered}
\end{equation}

This means that the warped product of a global AdS$_2$ with $S^2$ is indeed a full solution to Einstein's equations, and not only a near-horizon limit of the Reissner-Nordström geometry. Indeed, if we consider the radial proper length of this spacetime, it is infinite, so the ``near-horizon'' region is actually pretty ample:
\begin{gather}
    L = \int_0^\pi dr \: \sqrt{g_{rr}} = \ell_{\rm AdS} \int_0^\pi dr \: \frac{1}{\sin r} = +\infty.
\end{gather}

In terms of the extremal black hole, $\Phi = \ell_{\rm AdS}$ means sitting exactly at the horizon, which appears to be an infinitely long ``throat''. Having found this solution from our action and not the one in \cite{Sarosi_2018} proves that we are right: taking $\Phi = Q\ell_P$ in their case would not have set the potential term to $0$, yet it should as required by one of the equations of motion in \cite{almheiri2015models}. Going back to the case of matter being present, we can assume at first that the resulting spacetime is still at least asymptotically AdS$_2$ and check for consistency. By integrating \eqref{eq:blowupeq} along a null line $u^- = 0 \: (t = r = u^+/2)$ from one boundary to the other, we obtain:
\begin{gather}
    \int_0^{2\pi} du^+ \: e^{-2\omega} \, G_N T_{++}^{\rm matter} = [e^{-2\omega}\partial_+ \Phi^2]|_{u^+ \to 0} - [e^{-2\omega}\partial_+ \Phi^2]|_{u^+ \to 2\pi}. \label{eq:T++}
\end{gather}

For classical stress tensors, $T_{++} \geq 0$ locally due to the null energy condition $k^\mu k^\nu T_{\mu\nu} \geq 0$ for any $k^\mu k_\mu = 0$. For quantum stress tensors this is not true anymore, but the so-called ``averaged null energy condition'' (ANEC) usually holds \cite{Hartman_2017,Rosso_2020,kontou2015averaged}:
\begin{gather}
    \int d\lambda \: \langle \mathcal{E} | T_{++}^{\rm matter}| \mathcal{E} \rangle \geq 0,
\end{gather}

with $\lambda$ an affine parameter along the null geodesic of the lightlike vector $k^\mu$. This means that, although there may be counterexamples, it is reasonable to assume that in most situations the l.h.s. of \eqref{eq:T++} is going to be strictly positive. Apart from the case where it is exactly 0, $\Phi^2$ has to diverge near at least one of the boundaries to balance the $e^{-2\omega}$ factor:
\begin{equation}
\begin{aligned}
    e^{2\omega} \sim \frac{1}{\sin^2 r} & \sim \frac{4}{(u^+)^2} \quad && \mathrm{for} \: u^+ \to 0, \\
    & \sim \frac{4}{(2\pi - u^+)^2} \quad && \mathrm{for} \: u^+ \to 2\pi. \\
    \Phi^2|_{u^+ \to 0} & \sim \frac{4\alpha}{u^+} + \mathrm{regular}, \\
    \Phi^2|_{u^+ \to 2\pi} & \sim \frac{4\beta}{2\pi-u^+} + \mathrm{regular}.
\end{aligned}
\end{equation}

Higher order poles would give divergences and the regular parts are such that their first derivative is suppressed by $e^{-2\omega}$. If we substitute these functions, we obtain:
\begin{gather}
    -\alpha-\beta = \int_0^{2\pi} du^+ \: e^{-2\omega} \, T_{++}^{\rm matter} \geq 0.
\end{gather}

The presence of matter is thus responsible for two issues: $\Phi^2$ has to diverge at least on one boundary, and it also has to become infinitely negative there. This means that our assumption of an asymptotic AdS$_2$ region is absolutely inconsistent and we expect any kind of matter content (up to pathological cases) to heavily deform our initial spacetime.

This situation is exactly what we had described in Chapter 2, when we talked about irrelevant deformations of the CFT in the context of the holographic renormalization group. If we view our model from the perspective of extremal black holes, we can give an explicit interpretation of what is going on. As we will show in Section \ref{sec:BHgap}, there is a gap between the ground states of an extremal black hole as a quantum system and its first energy excitations, which give us a near-extremal black hole instead ($E > 0$). The CFT$_1$ that we obtain in the case of constant dilaton and empty AdS$_2$ in the near-horizon region of an extremal black hole is therefore only able to holographically describe a theory made of the ground states of this object. In order to see beyond the gap and distinguish near-extremality from extremality, we cannot perform an exact $\ell_P \to 0$ limit of the metric and we cannot decouple from the asymptotically flat part of the spacetime completely. In particular, recall the form of $z$:
\begin{gather}
    z = \frac{Q^2 \ell_P^2}{r-r_+}.
\end{gather}

The AdS$_2$ boundary where the CFT lives is at $z = 0$, that is, away from the black hole horizon. Looking for excitations above the gap means destroying the UV region of AdS$_2$, which is exactly what a blowing up dilaton (whose variations from a constant value, we recall, means accounting for variations of $e^{2\psi}$ and leaving the near-horizon region) achieves. This makes perfect sense, because for a near-extremal black hole the AdS$_2$ throat is not infinitely long and it connects to the flat asymptotic spacetime after a finite proper length \cite{Brown_2018}, while this does not happen for an extremal one: this is the main, essential difference between near-extremality and extremality, the effects of which we will investigate through JT gravity. Another difference is that, near the horizon, the metric is actually approximately Rindler when $E > 0$, but this is not an issue as it simply translates to a slightly modified bulk region (near $z = +\infty$) in AdS, whose extension is arbitrarily small as $E \to 0$: we will not be able to see this variation in the following, since it would require to perform a dimensional reduction in the case of $E > 0$, and is not particularly important anyway.

Looking for a varying dilaton is therefore equivalent to coupling the theory back to the asymptotic region. Not only that, but \textit{any} excitation wants to deflect the RG flow from the CFT$_1$ fixed point and couple it back to the asymptotic region. The outcome of this process in the UV depends on the chosen parameters and potential for the dilaton-gravity model. However, since a varying dilaton appears to be associated with an irrelevant operator\footnote{We have determined that $\Phi^2(r \to 0) \sim r^{-1} = r^{d-\Delta}$, therefore $\Delta = 2 > d = 1$.}, the IR part of the flow is expected to be universal and the deep interior geometry to always be close to AdS$_2$. In the following, we will cut off the RG flow sufficiently close to the IR fixed point in order to capture this universal dynamics, and study cut off AdS$_2$ spaces.

\section{Jackiw-Teitelboim Theory}
Due to what we have argued in the previous section, we are going to study cut off AdS$_2$ spaces with blowing up boundary conditions for the dilaton. Let us go back to the dilaton-gravity action:
\begin{gather}
    I = \frac{1}{16\pi G_N} \int d^2x \, \sqrt{-h} \, [\Phi^2 R_h + \lambda (\partial \Phi)^2 - U(\Phi^2/d^2)].
\end{gather}

We can look for AdS$_2$ solutions with a constant value for the dilaton $\Phi^2 = \phi_0$. Extremality of the action with respect to $\Phi^2$ then gives us:
\begin{gather}
    \frac{2}{\ell^2_{\rm AdS}} + \frac{1}{d^2} U'\left( \frac{\phi_0}{d^2} \right) = 0.
\end{gather}

$d$ is an external input scale (it was $\ell_P Q$ when we dimensionally reduced the black hole), so we have a relation between $\ell_{\rm AdS}$ and the free parameter $\phi_0$. If we deform this solution:
\begin{gather}
    \Phi^2 = \phi_0 + \phi,
\end{gather}

we expect from the previous section that $\phi \sim 1/r$ near the $r \to 0$ boundary in Poincaré coordinates. We choose to cut off our space at a value $r = \varepsilon > 0$ such that
\begin{gather}
    \frac{\phi(\varepsilon)}{d^2} \equiv \eta \ll 1.
\end{gather}

This lets expand our action around $\Phi^2 = \phi_0$:
\begin{gather}
    I = \frac{1}{16\pi G_N} \int d^2x \, \sqrt{-h} \left[ \phi_0 R_h - U \left( \frac{\phi_0}{d^2} \right) + \phi \left( R_h + \frac{2}{\ell^2_{\rm AdS}} \right) + \frac{\lambda}{4} \frac{(\partial \phi)^2}{\phi_0 + \phi} \right] + \mathcal{O}(\eta^2).
\end{gather}

Let us analyze this action term by term:
\begin{itemize}
    \item The first two terms are the usual Einstein-Hilbert action in two dimensions. In two dimensions, one must have a null cosmological constant (it is implied by tracing Einstein's equations), so the second term has to be removed through a local counterterm in case it is necessary. We expect that this operation is not needed for a well-defined theory, as was the case for the dimensionally reduced black hole. We can therefore impose $U(\phi_0/d^2) = 0$ as a requirement for a theory to make sense. In our cut off space, we will also need the GHY term and a boundary counterterm, so that the sum of all these pieces will give us the Euler characteristic of the manifold due to the Gauss-Bonnet theorem. In two dimensions, in fact, the Ricci scalar is two times the Gaussian curvature of the manifold.
    \item The third term is the Jackiw-Teitelboim theory that we will study in the following.
    \item The last term is a derivative that we have to expand in $\eta$, even though we could directly set $\lambda$ to $0$ in principle. To this end, we can make an estimate on the behavior of $\phi$ near the boundary. We know that $\phi \sim 1/r$, but $\phi$ has dimension (length)$^2$: a factor $\ell_P^2 E^2$ comes from the equation of motion \eqref{eq:blowupeq}, with $E$ being the excitation energy and the scale of the stress tensor\footnote{In two dimensions, the stress tensor has dimension (energy)$^2$. In \cite{Sarosi_2018} this factor is claimed to be $\ell_P^2 E^1$, which is at odds with this fact. Luckily, this mistake has no consequences on the final result.}. The missing factor has to be $\ell^3_{\rm AdS}$, which is the length scale of the derivatives. We obtain:
    \begin{gather}
        \frac{(\partial \phi)^2}{\phi_0 + \phi} \sim \frac{g^{rr}}{\phi_0 + \phi} \left( \frac{\ell_P^2 \ell_{\rm AdS}^3 E^2}{r^2} \right)^2 = \frac{\phi^2}{\ell_{\rm AdS}^2} \frac{1}{\phi_0 + \phi}.
    \end{gather}
    If we pick $\phi_0 = d^2$, we find that $2\ell^{-2}_{\rm AdS} = -d^{-2}U'(1)$ and finally:
    \begin{gather}
        \frac{(\partial \phi)^2}{\phi_0 + \phi} \sim \frac{|U'(1)|}{1+\frac{\phi}{d^2}} \left( \frac{\phi}{d^2} \right)^2 = \mathcal{O}(\eta^2).
    \end{gather}
\end{itemize}

The conclusion we draw from the previous reasoning is that, inside the cutoff surface, the action that governs the dynamics is universal (up to $\mathcal{O}(\eta^2)$ terms, that we neglect):
\begin{gather}
    I = \frac{\phi_0}{16\pi G_N} \int d^2x \, \sqrt{-h} \: R_h + \frac{1}{16\pi G_N} \int d^2x \, \sqrt{-h} \: \phi \left( R_h + \frac{2}{\ell^2_{\rm AdS}} \right). \label{eq:JTaction}
\end{gather}

Notice that we have removed the ``cosmological constant''. We will set $\ell_{\rm AdS} = 1$ from now on.

\section{Nearly \texorpdfstring{AdS$_2$}{AdS2} spaces}
Let us study the action \eqref{eq:JTaction} in the Euclidean signature. When studying holography and aiming at correlators in the CFT, it is a natural choice that one can then extend to the Lorentzian signature through analytic continuation. It is also easier to study in general. The embedding definition of Euclidean AdS$_2$ is equivalent to the hyperbolic disk, which is fully covered by both the Euclidean Poincaré and Rindler coordinates:
\begin{equation}
\begin{aligned}
    ds^2 & = \frac{dt^2 + dz^2}{z^2} \quad && (\mathrm{Poincar\acute e}), \\
    & = d\rho^2 + \sinh^2 \hspace{-0.04 in} \rho \, d\tau^2 && (\mathrm{Rindler}).
\end{aligned}
\end{equation}

\begin{figure}
    \centering
    \includegraphics[width = 0.8 \textwidth]{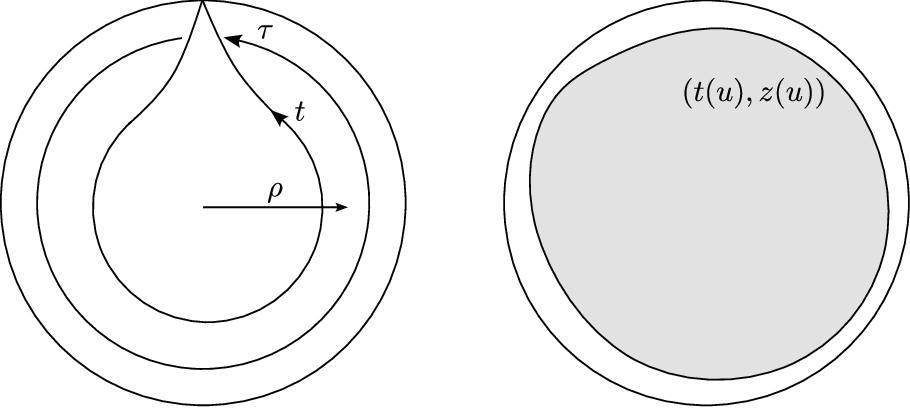}
    \caption{Left: Coordinates on the hyperbolic disk. $z$ is not shown, but it is $0$ on the boundary of the disk and increases towards $+\infty$ when going inwards. Right: A cutout from the hyperbolic disk.}
    \label{fig:NAdS}
\end{figure}

$t$ ranges on the whole real axis (touching the boundary at $\pm \infty$, as shown in Figure \ref{fig:NAdS}), while $\tau$ has period $2\pi$ and is an angle on the hyperbolic disk. The euclidean action with the addition of the GHY term is:
\begin{equation}
\begin{aligned}
    I & = -\frac{\phi_0}{16\pi G_N} \left[ \int_M d^2x \, \sqrt{h} \: R_h + 2 \int_{\partial M} dy \, \sqrt{\sigma} \: K \right] \\
     & - \frac{1}{16\pi G_N} \left[ \int_M d^2x \, \sqrt{h} \: \phi (R_h + 2) + 2 \int_{\partial M} dy \, \sqrt{\sigma} \: \phi_b K \right]. \label{eq:JTgrav}
\end{aligned}
\end{equation}

$\sigma$ is the induced $1 \times 1$ metric on the cutoff surface, $\phi_b$ is the value of $\phi$ on $\partial M$. This action is what defines the JT gravity model. Note that solving the Dirichlet problem in the presence of extra fields requires that we also fix their value on the boundary: the GHY term then naturally multiplies the trace of the extrinsic curvature $K$ by $(\phi_0+\phi)|_{\partial M} = \phi_0 + \phi_b$. The need for this modification is evident from the computations in Appendix \ref{app:2dDilaton}. Finally, we will later show that we actually need to substitute $K$ with $K-1$ in the last term in order to cancel a divergence, in a manner similar to what we have seen in Chapter 3.

Given a fixed cutoff of the disk, the first term is always proportional to its Euler characteristic: this means that deformations of the cut off manifold from one simply connected chunk of the disk to another keep this quantity invariant and are therefore zero modes of (this part of) the action. The degeneracy is lifted by the second term, that depends on the dilaton $\phi$. A way to describe simply connected cutouts is to fix the use of Poincaré coordinates (the equation of motion for $\phi$ enforces $R = -2$) and to parametrize the boundary $(t(u),z(u))$ with a parameter $u$. If we want $u$ to be a uniform time coordinate in the boundary theory, the induced metric has to be uniform too:
\begin{gather}
    \sigma = \frac{1}{\varepsilon^2}.
\end{gather}

$\varepsilon$ is the small parameter whose inverse plays the role of a UV cutoff in the boundary theory. If we write the line element on $\partial M$ from the bulk and the boundary point of views, the choice of $\sigma$ implies:
\begin{gather}
        \frac{d(t(u))^2 + d(z(u))^2}{z(u)^2} = \frac{du^2}{\varepsilon^2} \implies z = \varepsilon \sqrt{(t')^2 + (z')^2} = \varepsilon t' + \mathcal{O}(\varepsilon^3),
\end{gather}

so fixing $t(u)$ determines the cutout completely. This function can be thought of as the dynamical variable (a field) in our gravitational model along with the dependent $z(u)$, with $u$ being our ``time''. The Einstein-Hilbert part is the same for all cutouts, hence all $t(u)$: this is equivalent to a symmetry under reparametrizations of the boundary time $u \to f(u)$, with $f(u)$ such that the initial transformation $t(u) \to g(t(u))$ is matched to the $t(u) \to t(f(u))$ picture, which describes another surface since the $u$ coordinate is always the one with $\sigma$ as the boundary metric. Chunks that only differ by translations and rotations in the hyperbolic disk are virtually the same and are connected by a $SL(2,\mathbb{R})$ subgroup of all reparametrizations:
\begin{gather}
    t(u) \to \frac{at(u)+b}{ct(u)+d}, \quad ad-bc = 1.
\end{gather}

Indeed, the Poincaré metric exhibits a $SL(2,\mathbb{R})$ symmetry, as is clear when using the complex coordinate $\alpha = z+it$ and imposing $\omega(\alpha)$ to preserve it:
\begin{equation}
\begin{gathered}
    ds^2 = \frac{4}{(\omega+\bar\omega)^2} d\omega \, d\bar\omega \xrightarrow[\bar\omega = f(\bar\alpha)]{\omega = f(\alpha)} \frac{4 f'(\alpha)f'(\bar\alpha)}{(f(\alpha)+f(\bar\alpha))^2} d\alpha \, d\bar\alpha, \\
    \frac{f'(\alpha)f'(\bar\alpha)}{(f(\alpha)+f(\bar\alpha))^2} = \frac{1}{(\alpha+\bar\alpha)^2} \iff \omega = f(\alpha) = \frac{a\alpha + b}{c\alpha + d}, \quad ad-bc = 1.
\end{gathered}
\end{equation}

We know from Euclidean CFTs in two dimensions that $\omega(\alpha)$ has to be holomorphic, so $\omega$ cannot depend on $\bar\alpha$, otherwise we would not have $g_{\mu\nu} \to \Omega^2 g_{\mu\nu}$ and $d\omega \, d\bar\omega \propto d\alpha \, d\bar \alpha$. We also need $f$ to be a function with real coefficients for the Wick rotation back to the Lorentzian signature to make sense. The differential equation in the second line matches Equation \eqref{eq:SL2eq}, provided we consider $\alpha$ and $\bar\alpha$ as independent variables\footnote{This is clear from the Lorentzian case, where $\alpha = z-t$ and $\bar\alpha = z+t$ are the two independent lightcone coordinates.}, so we recover Möbius transformations once again. Note that this proof is absent in the references, which only claim the existence of this $SL(2,\mathbb{R})$ symmetry. When considering the boundary, then, the invariance for $\omega$ translates into the invariance for $t(u)$ that we claimed since $z(u) = \mathcal{O}(\varepsilon)$. This means that the choice of a specific cutout of AdS$_2$ spontaneously breaks the full reparametrization symmetry of the topological term in the action down to $SL(2,\mathbb{R})$. The ``Goldstone modes'' of this symmetry breaking pattern are the $t(u)$, which in fact have null Einstein-Hilbert action.

The discussion changes once we account for the terms that depend on $\phi$, which break the symmetry explicitly and give a finite action for $t(u)$. In terms of the extremal black hole, deviations of $\phi$ from zero means remembering that $e^{2\psi}$ is not a constant and therefore moving away from the very near-horizon region, towards the completion of the spacetime. In terms of the holographic renormalization group, the reparametrization symmetry is the conformal symmetry of the fixed point CFT$_1$, which is explicitly broken once we move to the UV along some irrelevant direction (the operator dual to $\phi$, with $\Delta = 2$). The main point is that the holographic description of this irrelevant deformation is captured by \ref{eq:JTgrav} for a large class of UV completions (the various dilaton-gravity models), which reduce to this same IR theory.

From a path integral or even classical perspective, as mentioned earlier, one can integrate $\phi$ out, since it is simply a Lagrange multiplier that fixes $R = -2$: the geometry is, as a matter of fact, AdS$_2$, and we are allowed to use the Poincaré coordinates. The only non-trivial term of our JT action is then:
\begin{gather}
    I = -\frac{1}{8\pi G_N} \int_{\partial M} \frac{du}{\varepsilon} \phi_b K = -\frac{1}{8\pi G_N} \int_{\partial M} \frac{du}{\varepsilon^2} \phi_r(u) K,
\end{gather}

where we have factored out the expected linear blow-up factor of $\phi_b$. Under reparametrizations $\psi(u)$ of the boundary time, i.e. the most general conformal transformation of the boundary CFT, the action we have written tells us that $\phi_r(u)$ transforms into $\psi(u)^{-1} \phi_r(\psi(u))$, that is, like a primary field with scaling dimension $\tilde \Delta = -1$. This means that it is the source for the operator dual to the dilaton in the CFT, which has scaling dimension $\Delta = d - \tilde \Delta = 2$. Recall in fact \ref{eq:sourcescale} and apply it to this scenario:
\begin{gather}
    \phi(z \to 0) = z^{1-2}(t'(u)\phi_r(u)) + \mathrm{subleading} \approx \frac{1}{\varepsilon} \phi_r(u),
\end{gather}

provided we have $\Delta = 2$ for the dual operator. While, given this picture, it would be tempting to imagine $t'(u) \phi_r(u)$ to be the source of the dual operator, this combination does not have the correct scaling dimension $\tilde \Delta = d - \Delta$, but rather it is invariant. Also, unlike the usual AdS/CFT scenario, our UV regulator is not $z_{\rm min}$, which is \textit{not} taken to be some fixed $\delta \to 0$ here, but rather the $\varepsilon$ parameter. This is the quantity that is correctly invariant under conformal transformations in the boundary (while $z$ transforms non-trivially) and the one that we should factor out of our regular functions (such as $\phi_r(u)$).

Let us now expand on the references by explicitly computing the trace of the extrinsic curvature through the use of a different formula:
\begin{gather}
    K = \frac{h_{ab}T^a \nabla_T n^b}{T_a T^a},
\end{gather}

where $T^a$ is the vector tangent to the boundary and $n^a$ is the usual orthogonal outgoing one. We show our proof starting from the definition, $K = h^{ab} \nabla_a n_b$. We claim that on the boundary, where all the quantities appearing below are defined:
\begin{gather}
    h^{ab} = \frac{T^a T^b}{T^2} + n^a n^b.
\end{gather}

To be the inverse of the metric, it must be true that:
\begin{gather}
    h^{ac} h_{cb} = \frac{T^a T_b}{T^2} + n^a n_b \stackrel{?}{=} \delta^{a}_b.
\end{gather}

It is immediate to check that the action of this linear operator on an arbitrary vector field $v^a = A T^a + B n^a$ localized on the boundary acts exactly as the identity due to $n_a n^a = 1, \: n_a T^a = 0$. The proof of the alternative formula for $K$ is then completed by noting that only the part of $h^{ab}$ that ``projects'' on $T$ survives, because $n_a n^a = 1$ implies $\nabla_b (n_a n^a) = 2 n_a \nabla_b n^a = 0$. The interesting aspect of using this formula is that we only have to consider the covariant derivative of $n^a$ along the boundary, so knowing the dependence of $(z,t)$ on $u$ only (and not on an extra ``radial'' coordinate $v$) is enough. The necessary ingredients are the following:
\begin{equation}
\begin{gathered}
    T = (T^u, T^v) = \partial_u (u,v) = (1,0), \quad T_a T^a = g_{uu} (T^u)^2 = \frac{1}{\varepsilon^2}, \\
    n^a = (n^t,n^z) = \frac{z}{\sqrt{(t')^2+(z')^2}} (z',-t'), \quad (T^t, T^z) = (t'(u),z'(u)), \\
    \Gamma^t_{tz} = \Gamma^t_{zt} = \Gamma^z_{zz} = - \Gamma^z_{tt} = - \frac{1}{z}.
\end{gathered}
\end{equation}

We need to compute the Christoffel symbols of the Poincaré coordinates because we do not know the metric in terms of $(u, \, v)$ nor any derivative of $n^a$ besides $\partial_u n^a$, so we cannot write the covariant derivative in terms of $(\partial_u, \partial_v)$ nor $(\partial_t, \partial_z)$, respectively. That being said, we can now determine the extrinsic curvature:
\begin{equation}
\begin{aligned}
    K & = \varepsilon^2 T^a T^b \nabla_a n_b = \varepsilon^2 T^a T^b (\partial_a n_b - \Gamma^c_{ab} n_c) = \varepsilon^2 (T^b \partial_u n_b - \Gamma^c_{ab} T^a T^b n_c) \\
    & = \frac{\varepsilon^2}{z\sqrt{t'^2+z'^2}} \left( t' z'' - z' t'' - (2\Gamma^t_{tz} t' z'^2 - \Gamma^z_{zz} z'^2 t' - \Gamma^z_{tt} t'^3) \right) \\
    & = \frac{t'((t')^2+(z')^2+zz'')-zz't''}{((t')^2+(z')^2)^{3/2}} = 1 + \varepsilon^2 S(t(u),u) + \mathcal{O}(\varepsilon^4).
\end{aligned}
\end{equation}

The first observation is that this result tells us that we need a boundary counterterm, that is, our JT action must actually contain a $\propto \phi_b(K-1)$ term in order to cancel the $\mathcal{O}(\varepsilon^{-2})$ divergence. The second one is that the action of this model is described by a Schwarzian derivative:
\begin{gather}
    I = -\frac{\phi_0 \chi(M)}{4 G_N} - \frac{1}{8\pi G_N} \int du \, \phi_r(u) S(t(u),u). \label{eq:JTSCHWA}
\end{gather}

$\chi(M)$ is the Euler characteristic of the cutout. $\phi_r(u)$ and $t(u)$ are connected by the equations of motion: we can either fix the first and determine the second or vice versa. If we take $\phi_r(u) = \bar \phi_r$ a constant, we see that our $(1+1)$ gravitational model is completely described by a boundary action with a dynamical field $t(u)$ living in only one dimension. The functional form of the action was to be expected due to the symmetry breaking pattern we had described earlier, just like in the SYK model.

There are several observations on our part regarding the links and analogies between the two theories.
\begin{itemize}
    \item In the SYK model, we had found a reparametrization invariance of the action that was explicitly broken by the derivative term. Here, we have that very same invariance that is instead explicitly broken by the varying dilaton part.
    \item In the SYK model, the biggest contribution to the action came from the conformal two point function, whose choice implied a spontaneous breaking of the approximate reparametrization symmetry down to $SL(2,\mathbb{R})$. Here, the topological term is the biggest contribution to the action and is completely determined by the choice of the cutout, which also spontaneously breaks the reparametrization symmetry down to $SL(2,\mathbb{R})$.
    \item In both models, the terms in the action that explicitly lift the degeneracy are the source of the Schwarzian derivative term, which is the first contribution possible to the action that preserves the residual $SL(2,\mathbb{R})$ symmetry by being unable to distinguish between configurations linked by a Möbius transformation.
    \item In both models, indeed, the $SL(2,\mathbb{R})$ symmetry group is to be considered a gauge symmetry of the theory: in the SYK model these transformations do not modify the conformal saddle, while here they preserve the Poincaré metric and modify our chosen cutout trivially.
\end{itemize}

All the computations we have done so far have shown us that indeed JT gravity is dual to the soft IR sector of the SYK model. The match is obtained (up to prefactors of the action) by identifying $u$ as the time $\tau$ in the boundary model and the dynamical field $t(u)$ as the soft reparametrization mode $\phi(\tau)$. This is an explicit realization of the nAdS$_2$/nCFT$_1$ correspondence (n stands for ``near''), where the nCFT$_1$ is the conformal limit of SYK with the first derivative correction and the bulk theory is JT gravity in a cutout of AdS$_2$. So: we have linked the SYK model, something that apparently has nothing to do with gravity, to the near-horizon region of a four-dimensional black hole. This can be considered an extraordinary result, and a way to showcase that approaching the study of quantum theories of gravity through the holographic principle has the potential to give us a great amount of information. In the next chapter, we will talk about the double-scaled SYK (DSSYK) model and we will stick to this limit for the rest of the thesis, with the objective of searching for more gravitational dualities. Before we do that, though, let us analyze the properties of JT gravity a bit more.

The Schwarzian action with a fixed $\bar \phi_r$ has the following equation of motion:
\begin{gather}
    \delta S(t,u) = \frac{1}{t'} \delta t''' - \frac{t'''}{(t')^2} \delta t' - 3 \frac{t''}{(t')^2} \delta t'' + 3 \frac{(t'')^2}{(t')^3} \delta t' \implies -\left[ \frac{1}{t'} \left( \frac{t''}{t'}\right)' \right]' \delta t = -\frac{[S(t,u)]'}{t'} \delta t = 0.
\end{gather}

We have integrated by parts and used that $t(u_{\rm max}) = t(u_{\rm min})$ on a closed contour. The classical solutions are those with constant Schwarzian. A smart way to look for non-trivial solutions is to apply the composition law \eqref{eq:SchComp} of the Schwarzian derivative:
\begin{gather}
    t(u) = \tan \frac{\tau(u)}{2} \implies S(t,u) = S(\tau,u) + \frac{1}{2} (\tau')^2.   
\end{gather}

A class of solutions with constant Schwarzian are $\tau(u) \propto u$. Note that this coordinate change is the Wick rotation of Equation \eqref{eq:PoinToRind} at the boundary $\rho = +\infty$ ($z = 0$):
\begin{gather}
	t = \tanh \frac{\tau}{2} \quad \stackrel{\mathrm{Euclidean}}{\longrightarrow} \quad t = \tan \frac{\tau}{2} \label{eq:PoinToRind2},
\end{gather}

so $\tau$ is the Euclidean Rindler time with period $2\pi$ and these solutions can be written as:
\begin{gather}
    \tau(u) = \frac{2\pi}{\beta} u \implies S(t(u),u) = \frac{2\pi^2}{\beta^2}.
\end{gather}

The idea is that \eqref{eq:PoinToRind2} follows from the $\mathcal{O}(\varepsilon^0)$ part of \eqref{eq:PoinToRind}, thus giving us an interpretation for $\tau$. The $2\pi$ periodicity of $\tau$ translates to a $\beta$ periodicity of $u$, so these solutions are thermal solutions with inverse temperature $\beta$ and $t(u) = t(u+\beta)$. This further estabilishes a link with the SYK model, since the transformation we have just used is exactly the one that we had already employed in the previous chapter to map the zero temperature solutions to the thermal ones, and the one that gave us the finite temperature action \eqref{eq:SchfinT}. Here, too, the use of $t(u) = \tan(\pi \psi(u)/\beta)$ (which generalizes to an arbitrary deformation the case of $\psi(u) = u$) gives us:
\begin{gather}
    I_\beta = -\frac{\phi_0 \chi(M)}{4 G_N} - \frac{\bar \phi_r}{8\pi G_N} \int_{-\beta/2}^{\beta/2} du \, \left[ S(\psi(u),u) + \frac{1}{2} \left( \frac{2\pi}{\beta} \right)^2 (\psi')^2 \right].
\end{gather}

\section{Entropy and Black Hole Gap} \label{sec:BHgap}
In this section, we will use the gravitational path integral to determine the thermodynamic properties of the spacetime and, because of our duality, also of the large $N$ SYK model. These computations, along with the link to the original four-dimensional black holes, is a focus on our part that is absent in the main source \cite{Sarosi_2018}. Remember that the Schwarzian action arose when considering a $(\beta J)^{-1} \ll 1$ expansion, namely a ``strong coupling'' regime. As usual, we will evaluate the $G_N \to 0$ limit, which in terms of the SYK model is equivalent to the $N \to \infty$ limit. We also want to suppress the  quantum fluctuations of the Schwarzian piece of the action, whose prefactor scales as $N/{\beta J}$: requiring this coefficient to be big is compatible with what we have done in the previous chapter, where $N \gg \beta J \gg 1$ (recall that we had taken the $N \to \infty$ limit before everything else). In terms of JT gravity, $(\beta J)^{-1}$ is mapped to $\bar \phi_r/\phi_0$, which is much smaller than $1$ by construction.

This is usual in the AdS/CFT correspondence, where a \textit{semiclassical} gravitational theory is dual to a ``large $N$'' CFT and is often able to describe regimes of strong coupling on the boundary: in AdS$_5 \times S^5/\mathcal{N} = 4$ SYM, for example, the strength of the coupling determines only whether the gravitational theory is influenced by string excitations or not. The interesting thing here is that the strong coupling limit of the SYK model is not hard to study, since it reduces to the conformal limit. Just like in the dual theory we have not found an infinite amount of ``string excitations'', the conformal limit suppresses every degree of freedom that is not a reparametrization of the time, which is then mapped to a quantum field in JT.

The partition function of JT gravity is simply the exponential of minus the classical action:
\begin{gather}
    \mathcal{Z}(\beta) = \exp \left( \frac{\phi_0 \chi(M)}{4G_N} + \frac{\pi \bar\phi_r}{4G_N \beta} \right).
\end{gather}

We obtain:
\begin{equation}
\begin{gathered}
    E = -\partial_\beta \log \mathcal{Z} = \frac{\pi \bar\phi_r}{4G_N} T^2, \\
    C = \partial_T E = \frac{\pi \bar\phi_r}{2G_N} T, \\
    S = \log \mathcal{Z} + \beta E = S_0 + \frac{\pi \bar\phi_r}{2G_N} T, \quad S_0 \equiv \frac{\phi_0 \chi(M)}{4G_N}. \label{eq:JTthermo}
\end{gathered}
\end{equation}

We can interpret $S_0$ as the big ground state entropy. For $T = 0$ and small $J$, for example, this degeneracy in the SYK model is clearly linked to all the $2^{N/2}$ states of the Hilbert space being equivalent. The correspondence tells us that a degeneracy is present for all values of the coupling. In the case of the extremal black hole, we expect $S_0$ to be the Bekenstein-Hawking term and the area of the horizon to vary for non-null temperatures.

In the rest of this section, we will interpret these quantities from the point of view of the (near) extremal black hole. First of all, we remember that deviations of $\Phi^2$ from $\phi_0$ (hence all the terms with $\bar\phi_r$) are associated with moving away from the very near-horizon region towards the outer part of the spacetime.
If we use the usual formula for the temperature, we obtain:
\begin{gather}
    T = \frac{f'(r_+)}{4\pi} = \frac{r_+-r_-}{4\pi r_+^2} = \frac{1}{2\pi} \left( \frac{2}{\ell_P Q^3} E \right)^{1/2} + \mathcal{O}(E^{3/2}) \implies E \approx 2\pi^2 Q^3 \ell_P T^2.
\end{gather}

We find that indeed $E \propto T^2$ and $C \propto T$ for small energies (temperatures). We remember that $\phi$ scales near the boundary as:
\begin{gather}
    \phi \sim \frac{\ell_P^2 \ell_{\rm AdS}^3 \epsilon^2}{\varepsilon} \implies \bar\phi_r \sim \ell_{\rm AdS}^3 = \ell_P^3 Q^3.
\end{gather}

The choice of the stress-energy tensor scaling as $\epsilon \sim \ell_P^{-1}$ gives us a match between the energies found through these two different approaches: at the same time, this is a natural scaling in a theory that has still not been coupled to matter. The fact that we have a match only for small energies may appear as worrisome, but our JT theory was obtained in the first place under the hypothesis that $\bar\phi_r \ll \varepsilon \phi_0$: for the black hole, $\phi_0 = \ell_P^2 Q^2$, therefore $\varepsilon \gtrsim \ell_{\rm AdS}$.

The entropy of a near-extremal black hole is easily computed with the area formula:
\begin{gather}
    S = \frac{4\pi r_+^2}{4G_N} = \frac{\pi}{G_N} \left[ Q^2 \ell_P^2 + \ell_P (2Q\ell_P)^{3/2} E^{1/2} + \mathcal{O}(E) \right] \approx \frac{\pi Q^2 \ell_P^2}{G_N} + 4\pi^2 Q^3 \ell_P T.
\end{gather}

The result reproduces the structure of the JT entropy, with a large contribution from the ground states (the extremal black hole) and a linear coefficient that is twice the one of the quadratic energy, like in \eqref{eq:JTthermo}.

The near-horizon limit of the metric was the result of taking $\ell_P \to 0$, so we also have to consider $Q \to \infty$ in order to allow for strictly positive energies. The alternative is to limit ourselves to the description of ground states, i.e. the microstates of the extremal black hole. Describing a black hole in terms of general relativity leads to exact thermodynamic laws, which suggests that this description only applies in a thermodynamic limit. A non-extremal black hole radiates Hawking quanta of energy $\sim T$: a thermodynamic description only makes sense if $E \gg T$, so that the emission of a quantum can be considered a quasi-equilibrium process, and consequently breaks down when $E \sim T$, that is:
\begin{gather}
    E_{\rm gap} \sim \frac{1}{\ell_P Q^3}.
\end{gather}

The semiclassical thermodynamic description we have given up until now makes no sense for near-extremal black holes with $0 < E < E_{\rm gap}$. $E_{\rm gap}$ can be seen as the scale of the gap above the ground states in the microscopic spectrum of the black hole \cite{Maldacena_1996}. If $Q$ doesn't blow up as $\ell_P$ is sent to 0, this gap goes to infinity and we are only allowed to describe the extremal black hole through the CFT$_1$ that lives on the boundary of the near-horizon AdS$_2$ geometry that emerges. Otherwise, our discussion in Section \ref{sec:Dilatons} holds.

To conclude, as already observed in the third chapter, an extremal black hole is such that its temperature is not fixed by the absence of conical singularities. On the other hand, such an object emits no radiation at all and has null surface gravity, which were the reasons that led us to associate a null temperature to this object. The absence of radiation emitted, in a way, makes its temperature quite arbitrary and meaningless, since there is nothing we can measure to determine it, yet any other choice that is not $T=0$ is misleading at best. It should now be clear to the reader that the $E \to 0$ limit of near-extremal black holes, which would also imply $T = 0$, is not obviously describing an extremal one\footnote{We have more elementary examples of this phenomenon in physics. A known one is the comparison between massless spin-$1$ fields and the $m \to 0$ limit of massive ones.}. Extremal and near-extremal spacetimes, in fact, are two manifolds with very different properties (consider, for example, their Penrose diagrams) and, as discussed above, there is actually a quantum energy gap separating them which contrasts a smooth limit between these two ``classical'' objects, so one should be extremely careful in this regard.

\section{Linearized Theory}
In this section, we will see what happens when we quantize the JT action \eqref{eq:JTSCHWA} with constant boundary dilaton. As an example, we will study the $\beta = 2\pi$ case, for which we can write $\tau(u) = u + \varepsilon(u)$. We will assume $\varepsilon(u)$ to be small, i.e. we will consider small deviations from the classical saddle point, just like it is usually done in QFT. We will stop at the quadratic terms, which give us the leading perturbative result in $G_N$. We start from the Schwarzian derivative:
\begin{gather}
    S(\tau,u) + \frac{1}{2}(\tau')^2 = \frac{1}{2} + \varepsilon' + \varepsilon''' + \frac{1}{2} (\varepsilon')^2 - \frac{1}{2} (\varepsilon'')^2 - (\varepsilon'' \varepsilon')' + \mathcal{O}(\varepsilon^3).
\end{gather}

We can drop total derivatives on the thermal circle, so we are left with:
\begin{gather}
    I_{\rm Sch} = \frac{C}{2} \int_0^{2\pi} du \,\left( (\varepsilon'')^2 - (\varepsilon')^2 \right), \quad C = \frac{\bar \phi_r}{8\pi G_N}.
\end{gather}

The two point function is simply the inverse of the Gaussian kernel. We can perform a Fourier transform:
\begin{gather}
    \varepsilon(u) = \frac{1}{\sqrt{2\pi}} \sum_{n \in \mathbb{Z}} \varepsilon_n e^{inu} \implies I = \frac{C}{2} \sum_{n \in \mathbb{Z}} (n^4-n^2) \varepsilon_n \varepsilon_{-n}.
\end{gather}

The kernel is zero for $n = 0, \pm 1$, but we have to remember that $SL(2, \mathbb{R})$ is a gauge symmetry of the theory we should not path integrate over. Indeed, imagine an infinitesimal transformation belonging to the gauge group:
\begin{equation}
\begin{gathered}
    t \to \frac{(1+\alpha) \, t + \beta}{\gamma t + (1-\alpha)} \approx t + (\beta + 2\alpha t - \gamma t^2) \equiv t + (a + bt + ct^2) = t + \Delta t, \\
    t + \Delta t = \tan \left( \frac{u+\varepsilon}{2} \right) \approx t + \frac{\varepsilon}{2 \cos^2 \left( \frac{u}{2} \right)}, \\
    \varepsilon_{SL(2,\mathbb{R})} = 2a \cos^2 \left( \frac{u}{2} \right) + 2b \sin \left( \frac{u}{2} \right) \cos \left( \frac{u}{2} \right) + 2c \sin^2 \left( \frac{u}{2} \right) = A + B e^{iu} + C e^{-iu}.
\end{gathered}
\end{equation}

In other words, we have to drop the $\varepsilon_0, \varepsilon_{\pm 1}$ modes as they are nothing more than the linearized $SL(2,\mathbb{R})$ transformations. The above proof is absent in \cite{Sarosi_2018}, which only states that the $\varepsilon_0, \varepsilon_{\pm 1}$ modes should be dropped without explicitly showing their connection to $SL(2,\mathbb{R})$. The propagator is therefore:
\begin{gather}
    \langle \varepsilon(u) \varepsilon(0) \rangle = \sum_{n \neq 0, \pm 1} \frac{e^{inu}}{2\pi} \langle \varepsilon_n \varepsilon_{-n} \rangle = \frac{1}{2\pi C} \sum_{n \neq 0, \pm 1} \frac{e^{inu}}{n^2(n^2-1)}.
\end{gather}

Again, we disagree with \cite{Sarosi_2018} when it comes to the prefactor of the propagator throughout this section, but we will agree with the results in the next section that will make use of said propagator: we assume, therefore, that these are typos on their part. These $\varepsilon(u)$ are quantum fluctuations of the Rindler time field $\tau$ around its vev $\langle \tau(u) \rangle = u$. We can rewrite the sum as a contour integral on the complex plane:
\begin{gather}
    \langle \varepsilon(u) \varepsilon(0) \rangle = \frac{1}{2\pi C} \oint_{\mathcal{C}} \frac{ds}{e^{2\pi i s}-1} \, \frac{e^{isu}}{s^2(s^2-1)}.
\end{gather}

\begin{figure}
    \centering
    \includegraphics[width = 0.85 \textwidth]{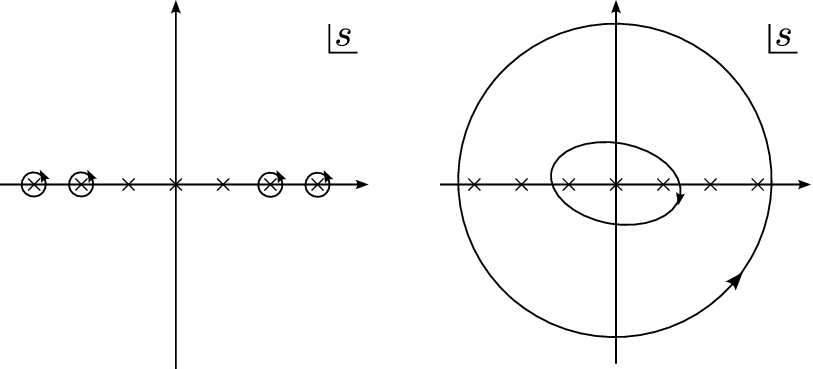}
    \caption{Left: Original contour of integration. Right: Deformed contour.}
    \label{fig:JTpoles}
\end{figure}

The contour $\mathcal{C}$ is the union of small circles running counter-clockwise around integer values of $s$, except $s = 0, \pm 1$. Indeed, these integers $s_0$ are simple poles of the function $\left( e^{2\pi i s} - 1 \right)^{-1}$, which can be expanded near them as $(2\pi i(s-s_0))^{-1}$. This contour integral can be deformed as shown in Figure \ref{fig:JTpoles}. We take the radius of the circle that we send to infinity to be $R = k + 1/2$, $k \in \mathbb{N}$ in order to avoid poles. The integral on the big circle goes to $0$ as $k \to +\infty$: to see this, let us write $s = R e^{i\theta}$ and distinguish the upper semicircle $(0 \leq \theta < \pi)$ from the lower semicircle $(\pi \leq \theta < 2\pi)$.
\begin{itemize}
    \item We have a $R^{-3}$ suppression coming from $ds/(s^2(s^2-1))$ for every value of $\theta$.
    \item On the upper semicircle we have $\sin \theta > 0$, therefore:
    \begin{equation}
    \begin{gathered}
        e^{isu} = e^{iuR \cos \theta - u R \sin \theta}, \quad e^{2 \pi i u} = e^{i 2\pi R \cos \theta - 2\pi R \sin \theta}, \\
        \frac{e^{isu}}{e^{2\pi i s} - 1} \sim - e^{-u R \sin \theta + iuR \cos \theta},
    \end{gathered}
    \end{equation}
    which further suppresses the integral.
    \item On the lower semicircle, instead, we have $\sin \theta < 0$, hence:
    \begin{gather}
        \frac{e^{isu}}{e^{2\pi i s} - 1} \sim e^{-(2\pi - u) R |\sin \theta| -i(2\pi - u) R \cos \theta},
    \end{gather}
    which again further suppresses the integral, since our time coordinate has range $0 \leq u < 2\pi$. Even if our range is, for example, $-\pi \leq u < \pi$, we can still take the result from positive $u$ and extend it to negative $u$ thanks to the $u \to -u$ $(n \to -n)$ symmetry of the propagator.
\end{itemize} 

The bottom line of this reasoning is that we only need to pick up the residues at $s = 0, \pm 1$ with a minus sign. We remember that for a pole $s_0$ of order $n$ of a function $f(s)$, the residue can be found through:
\begin{gather}
    \mathrm{Res}(f,s_0) = \frac{1}{(n-1)!} \lim_{s \to s_0} \frac{d^{n-1}}{dz^{n-1}}[(s-s_0)^n f(s)].
\end{gather}

$s_0 = 0$ is a pole of order 3, while $s_0 = \pm 1$ are poles of order 2. Putting everything together, we finally find our propagator:
\begin{gather}
    \langle \varepsilon(u) \varepsilon(0) \rangle = \frac{1}{2\pi C} \left( -\frac{(u-\pi)^2}{2} + (u-\pi) \sin u + 1 + \frac{\pi^2}{6} + \frac{5}{2} \cos u \right). \label{eq:JTpropag}
\end{gather}

We generalize the result to negative $u$ by substituting $u \to |u|$.

\section{Coupling to Matter}
As a final discussion, it is interesting to briefly investigate what happens when this model is coupled to matter. Imagine adding a minimally coupled free scalar to the action:
\begin{gather}
    I_{\rm matter} = \frac{1}{2} \int d^2x \, \sqrt{h} \left( h^{ab} \partial_a \chi \partial_b \chi + m^2 \chi^2 \right). \label{eq:matteract}
\end{gather}

The AdS/CFT dictionary tells us that the partition function of the dual CFT acquires a dependence on the boundary value $\tilde \chi_r(t)$, which acts as a source for a dual scalar operator. The Green's function for a massive scalar in Euclidean AdS$_{d+1}$ with Poincaré coordinates $(z,\vec x)$ is \cite{Freedman_1999}:
\begin{equation}
\begin{gathered}
    K_\Delta(z,\vec x - \vec y) = \frac{\Gamma(\Delta)}{\pi^{d/2} \Gamma \left( \Delta - \frac{d}{2} \right)} \left( \frac{z}{z^2 + (\vec x - \vec y)^2} \right)^\Delta, \\
    \lim_{z \to 0} z^{\Delta - d} K_\Delta(z, \vec x - \vec y) = \delta(\vec x - \vec y), \\
    z^\Delta (\nabla^2 - m^2) K_\Delta(z,\vec x) = \frac{2\Delta - d}{\sqrt{h}} \delta(z) \, \delta(\vec x), \quad \nabla^2 = z^{d+1} \partial_a (z^{-d+1} \partial_a), \\
    \chi(z,\vec x) = \int d^d y \: K_\Delta(z, \vec x - \vec y) \, \tilde \chi_r(\vec y) \implies (\nabla^2-m^2)\chi(z,\vec x) = (2\Delta - d) z^{d-\Delta+1} \delta(z) \tilde \chi_r(\vec x) \stackrel{z > 0}{=} 0.
\end{gathered}
\end{equation}

The massless case ($\Delta = d/2$) would require some adjustments, but we do not concern ourselves with it. The second line follows from observing that the limit on the left is zero ($\Delta > d/2$) unless $\vec x = \vec y$, then integrating both sides over $\mathbb{R}^d$ and checking that the result is the same. The third line is harder to obtain, but also requires to integrate both sides (multiplied by $\sqrt{h}$) after having verified by hand that the l.h.s is null for $m^2 = \Delta(\Delta-d)$ everywhere, apart from the singular point $(z,\vec x) = (0,0)$.

We can use these formulas to evaluate the on-shell action by first integrating Equation \eqref{eq:matteract} by parts, then substituting the third line for both fields:
\begin{equation}
\begin{aligned}
    I_{\rm matter} & = - \frac{1}{2} \int dz \, dt \: \chi(z,t) \left( \partial_a \left(\sqrt{h} h^{ab} \partial_b \right) - \sqrt{h} \, m^2 \right) \chi(z,t) \\
    & = -\frac{1}{2} \int dz \, dt \, dt' \, dt'' \: K_\Delta(z,t-t') \, \tilde \chi_r(t') \left( \partial_a \left(\sqrt{h} h^{ab} \partial_b \right) - \sqrt{h} \, m^2 \right) K_\Delta(z,t-t'') \, \tilde \chi_r(t'') \\
    & = -\frac{1}{2} \int dz \, dt \, dt' \, dt'' \: K_\Delta(z,t-t') \, \tilde \chi_r(t') \, 2\left( \Delta - \frac{1}{2} \right) z^{-\Delta} \delta(z) \, \delta(t-t'') \, \tilde \chi_r(t'') \\
    & = - D \int dt \, dt' \: \frac{\tilde \chi_r(t) \tilde \chi_r(t')}{|t-t'|^{2\Delta}}, \quad D = \frac{\left( \Delta - \frac{1}{2} \right) \Gamma(\Delta)}{\sqrt{\pi} \, \Gamma \left( \Delta - \frac{1}{2} \right)}.
\end{aligned}
\end{equation}

Since the quadratic piece in the sources generates the two point function of the dual primary in the boundary theory, conformal invariance alone determines the form of $I_{\rm matter}$, up to the constant $D$. We observe a subtlety that \cite{Sarosi_2018} doesn't emphasize. This computation seems to assume an AdS$_2$ background. If we ignore the backreaction of the field on the geometry, indeed, our spacetime is simply a subset of AdS$_2$ and we can use Poincaré coordinates. Even if we do not neglect backreaction, though, the condition $R = -2$ is enforced by the dilaton as a Lagrange multiplier, regardless of the matter content of the theory, so we always have a spacetime with a $(z(u),t(u))$ cutout, with $z(u)$ being determined by both the Dirichlet boundary conditions (think of the induced metric $\sigma$) and $t(u)$. We can make the dependence on $t(u)$ explicit by writing:
\begin{gather}
    \chi(z,t) = z(u)^{1-\Delta} \tilde \chi_r(t(u)) + \dots = \varepsilon^{1-\Delta}(t'(u))^{1-\Delta} \tilde \chi_r(t(u)) + \dots \equiv \varepsilon^{1-\Delta} \chi_r(t(u)) + \dots
\end{gather}

While $\tilde \chi(t)$ is the source in a theory with boundary time $t$, $\chi_r(t(u))$ is the object that correctly transforms under reparametrizations of the CFT$_1$ time $u$ (we recall that every transformation is conformal in one dimension) like a primary with scaling dimension $1-\Delta$, therefore it couples to an operator with scaling dimension $\Delta$. As already discussed in the case of the dilaton $\phi$, our situation is different from the basic AdS/CFT scenario, but we should always factor out the UV regulator $\varepsilon$. Again, then, $\chi_r(u)$ is the source that is useful for our analysis, which we can use to rewrite our on-shell action:
\begin{gather}
    I_{\rm matter} = - D \int du \, du' \: \left[ \frac{t'(u)t'(u')}{(t(u)-t(u'))^2} \right]^\Delta \chi_r(u) \chi_r(u').
\end{gather}

This formula describes how $\chi$ couples to the metric degree of freedom $t(u)$. At leading order in $G_N$, the boundary partition function with source $\chi_r(u)$ coupling to a dimension $\Delta$ operator is:
\begin{gather}
    \mathcal{Z}[\chi_r(u)] = \exp \left( -S_0 - I_{\rm Sch} - I_{\rm matter} \right),
\end{gather}

where $S_0 + I_{\rm Sch}$ is the action \eqref{eq:JTSCHWA}. $t(u)$ is obtained by extremizing $I_{\rm Sch} + I_{\rm matter}$ and depends on $\chi_r(u)$, so the exponential is not actually quadratic in $\chi_r(u)$ unless one ignores this backreaction. The Schwarzian term scales as $G_N^{-1}$, while the matter term scales as $G_N^0 \Delta^{3/2}$, so we can indeed do this if $\Delta$ grows slower than $G_N^{-2/3}$ as $G_N \to 0$. As a consequence, the CFT scalar $V(u)$ dual to $\chi_r$ is free, with only a non-trivial connected two point function:
\begin{gather}
    \langle V(u) V(u') \rangle \sim \left[ \frac{t'(u)t'(u')}{(t(u)-t(u'))^2} \right]^\Delta,
\end{gather}

with $t(u)$ a saddle of the Schwarzian action only. Corrections to this behavior can come from $\chi$ self-interacting or coupling to other bulk fields, from the backreaction or from loop corrections to the saddle. We will show the effects of the latter in the case of $\beta = 2\pi$. We make an expansion around the thermal saddle:
\begin{equation}
\begin{gathered}
    t(u) = \tan \left( \frac{u + \varepsilon(u)}{2} \right), \\
    \left[ \frac{t'(u)t'(u')}{(t(u)-t(u'))^2} \right]^\Delta = \frac{1}{\left( 2 \sin \frac{u_{12}}{2} \right)^{2\Delta}} [1 + \mathcal{B}(u_1,u_2) + \mathcal{C}(u_1,u_2) + \mathcal{O}(\varepsilon^3)].
\end{gathered}
\end{equation}

We have defined $u_{ij} \equiv u_i - u_j$ and we have denoted the linear and quadratic contributions with:
\begin{equation}
\begin{aligned}
    \mathcal{B}(u_1,u_2) & = \Delta \left( \varepsilon'(u_1) + \varepsilon'(u_2) - \frac{\varepsilon(u_1)-\varepsilon(u_2)}{\tan \frac{u_{12}}{2}} \right), \\
    \mathcal{C}(u_1,u_2) & = \frac{\Delta}{\left( 2 \sin \frac{u_{12}}{2} \right)^2} \Big[ (1+\Delta+\Delta \cos u_{12})(\varepsilon(u_1)-\varepsilon(u_2))^2 \\
    & - 2\Delta \sin u_{12} (\varepsilon(u_1) - \varepsilon(u_2))(\varepsilon'(u_1)+\varepsilon'(u_2)) \\ 
    & + (1-\cos u_{12})((\Delta-1)(\varepsilon'(u_1)^2+\varepsilon'(u_2)^2)+2\Delta \varepsilon'(u_1) \varepsilon'(u_2)) \Big].
\end{aligned}
\end{equation}

If we focus on the scalar field $\chi$, we can find the generator $W_{\rm matter}[\chi_r]$ of the connected $n$ point functions by taking the logarithm of the partition function $\mathcal{Z}_{\rm matter}[\chi_r] = \langle e^{-I_{\rm matter}} \rangle \approx \langle 1 - I_{\rm matter} + \frac{1}{2} I^2_{\rm matter}\rangle$:
\begin{equation}
\begin{aligned}
    W_{\rm matter}[\chi_r] = & \log \mathcal{Z}_{\rm matter}[\chi_r] \\ 
    = & \: D \int du_1 \, du_2 \: (1+\langle C(u_1,u_2) \rangle) \, \frac{\chi_r(u_1) \chi_r(u_2)}{\left( 2 \sin \frac{u_{12}}{2} \right)^{2\Delta}} \\
    & + \frac{1}{2} D^2 \int du_1 \, du_2 \, du_3 \, du_4 \: \frac{\chi_r(u_1)\chi_r(u_2)\chi_r(u_3)\chi_r(u_4)}{\left( 2 \sin \frac{u_{12}}{2} \right)^{2\Delta} \left( 2 \sin \frac{u_{34}}{2} \right)^{2\Delta}} \, \langle \mathcal{B}(u_1,u_2) \mathcal{B}(u_3, u_4) \rangle + \mathcal{O}(G_N^2).
\end{aligned}
\end{equation}

The second term is the result of the difference between $\langle V(u_1) \dots V(u_4) \rangle$ and its disconnected piece, so it is indeed the connected part of the four point function. We have used repeatedly that odd point functions in the quadratic theory of the free scalar $\varepsilon(u)$ vanish. We can see this field as carrying a $G_N^{1/2}$ factor, so working at one loop means that we are not allowed to have it appear more than twice in $W_{\rm matter}$ and that only its quadratic action should be considered in the path integral, as we have effectively done.

We should be wary of the time ordering of the $u_i$, especially when considering the four point function. In the case of the two point function, however, we can simply assume $u_1 > u_2$. If we use the invariance under time translations of \eqref{eq:JTpropag}, we obtain:

\begin{equation}
\begin{aligned}
    \langle \mathcal{C}(u_1,u_2) \rangle & = \frac{\Delta}{\left( 2 \sin \frac{u_{12}}{2} \right)^2} \Big[ 2(1+\Delta+\Delta \cos u_{12})(\langle \varepsilon(0) \varepsilon(0) \rangle - \langle \varepsilon(u_{12})\varepsilon(0) \rangle) \\
    & + 4\Delta \sin u_{12} \, \partial_{u_{12}} \langle \varepsilon(u_{12}) \varepsilon(0) \rangle \\
    & - 2(1-\cos u_{12})((\Delta-1)\partial^2_u \langle \varepsilon(u) \varepsilon(0) \rangle |_{u = 0} + \Delta \, \partial^2_{u_{12}} \langle \varepsilon(u_{12}) \varepsilon(0) \rangle) \Big] \\
    & = \frac{1}{2\pi C} \frac{\Delta}{\left( 2 \sin \frac{u_{12}}{2} \right)^2} \Big[ 2 + 4\Delta - u_{12}(2\pi - u_{12})(\Delta + 1) \\
    & - (\Delta u_{12}(2\pi - u_{12}) + 4\Delta + 2) \cos u_{12} +2 (\pi - u_{12})(2\Delta+1)\sin u_{12} \Big].
\end{aligned}
\end{equation}

Since $1/C \propto G_N^1$, this correction is subleading in powers of Newton's constant (as expected of a loop correction). We perfectly agree with \cite{Sarosi_2018}, which only reports the final result. The way we have evaluated the two point functions containing derivatives is the following:
\begin{equation}
\begin{gathered}
    \langle \varepsilon(u) \varepsilon'(u') \rangle = \partial_{u'} \langle \varepsilon(u-u') \varepsilon(0) \rangle = -\partial_u \langle \varepsilon(u-u') \varepsilon(0) \rangle, \\
    \langle \varepsilon'(u) \varepsilon'(u) \rangle = \lim_{u' \to u^-} \partial_u \partial_{u'} \langle \varepsilon(u) \varepsilon(u') \rangle = \lim_{u' \to u^-} \partial_u (-\partial_u) \langle \varepsilon(u-u') \varepsilon(0) \rangle = -\partial^2_u \langle \varepsilon(u) \varepsilon(0) \rangle |_{u = 0}, \\
    \langle \varepsilon'(u) \varepsilon'(u') \rangle = -\partial^2_u \langle \varepsilon(u-u') \varepsilon(0) \rangle.
\end{gathered}
\end{equation}

This concludes our overview of JT gravity. As stated earlier, we now go back to the SYK model: by exploring its double-scaled limit and the mathematical structure that arises, we will showcase its potential as a model where more computations can be carried out explicitly.

\newpage
\chapter{The DSSYK Model}
In this chapter, we will present a very interesting limit of the SYK model, which is being considered at the time of writing as a candidate to holographically describe quantum theories of de Sitter or, more weakly, spacetimes with inner de Sitter regions. This limit, along with some modifications, will be the protagonist of the novelties presented in this thesis.

This subfield is very active, with several physicists trying to unveil the connection between the DSSYK model and gravity. One of the pioneers of this link is Susskind, who has argued why this model (at infinite temperature) is expected to live at the ``stretched horizon'' of a static patch of de Sitter, i.e. the region that sits at a proper distance equal to a Planck length from the cosmological horizon, and holographically spawn this spacetime. His relevant works on this subject are \cite{susskind2021entanglement,susskind2022scrambling,lin2022infinite,susskind2022sitter,Susskind:2023hnj,Rahman:2024vyg}. His ideas have also received further input by Rahman in \cite{Rahman:2022jsf,Rahman:2024vyg}. The idea that, in order to spawn a low dimensional de Sitter spacetime, one should actually consider two copies of the DSSYK model and restrict the physical Hilbert space to states with equal energy on both sides is presented and developed in \cite{Narovlansky:2023lfz,verlinde2024doublescaled,Verlinde:2024zrh}. The conceptual differences in this approach is the presence of two copies of the system and the idea that they do not live at the stretched horizon of a static patch, but rather at the north and south pole of three-dimensional de Sitter, following the timelike trajectories of comoving observers. Interestingly, there has been a direct answer from Susskind and Rahman to this proposal \cite{Rahman:2023pgt}. Looking for a direct match of the partition function of a single copy of the DSSYK model to a two-dimensional gravitational theory has led to \cite{Blommaert_2024,Blommaert:2023wad,Blommaert:2024ydx,Goel:2023svz}. As stated at the beginning of the thesis, the latter is the approach we will follow ourselves. In particular, we will draw both inspiration and some results from these papers in Chapters 7 and 8.

This chapter will be dedicated to showing how the double-scaled limit allows us to do much more than the standard $N \to \infty$ limit, from building the Hilbert space of chords to using it to compute correlation functions of operators. The works we will base ourselves on are \cite{Berkooz:2018jqr, Berkooz:2018qkz}.

\section{Chord Diagrams}
Recall the Hamiltonian of the SYK model with $N$ Majorana fermions:
\begin{gather}
    H = i^{p/2} \sum_{1 \leq i_1 < \dots < i_p \leq N} J_{i_1 \dots i_p} \psi_{i_1} \dots \psi_{i_p}.
\end{gather}

Here, we will take slightly different normalizations from before:
\begin{equation}
\begin{gathered}
    \{ \psi_i, \psi_j \} = 2 \delta_{ij}, \quad i,j = 1,\dots,N, \\
    \langle J^2_{i_1 \dots i_p} \rangle_J = \binom{N}{p}^{-1} \mathcal{J}^2, \\
    \tr \mathbb{1} = 1.
\end{gathered}
\end{equation}

We recall that the $\langle \sbullet \rangle_J$ symbol indicates a mean over the ensemble. In the double-scaled limit, it is not necessary for the random couplings to be Gaussian, as we will soon show. It is enough to assume that they are independent and with zero mean. Only their standard deviation matters here, so they need not be identically distributed. We also require that the moments of $\binom{N}{p}^{1/2} J_{i_1 \dots i_p}$ are uniformly bounded by a number independent of $N$, so that several theorems hold for the Hamiltonian operator \cite{Erd_s_2014}. We will take $\mathcal{J} = 1$ in the following.

The definition of the double-scaled limit is the following:
\begin{gather}
    N \to \infty, \quad \lambda = \frac{2p^2}{N} = \mathrm{fixed}, \quad q \equiv e^{-\lambda}.
\end{gather}

Both $N$ and $p$ are always even. The main instrument we will employ to perform computations are the chord diagrams, which reduce our problem to combinatorics. We can interpret them physically as ``information flow'' or ``correlation flow'' diagrams. A perturbation will only be correlated at later times with itself, and the chords in the diagram keep track of such correlations.

In general, a chord diagram is a segment or a circle with nodes on it that are connected in pairs by chords. The combinatorics arise when considering the intersections among them. To better understand what we are talking about, we start by considering the moments:
\begin{gather}
    m_k = \langle \tr H^k \rangle_J = i^{kp/2} \sum_{I_1, \dots, I_k} \langle J_{I_1} \dots J_{I_k} \rangle_J \, \tr(\psi_{I_1} \dots \psi_{I_k}).
\end{gather}

The $I_j = \{ i^{(j)}_1, \dots, i^{(j)}_p \}$ are sets of $p$ distinct and ordered indices which range from $1$ to $N$, while $\psi_{I_j} \equiv \psi_{i^{(j)}_1} \dots \psi_{i^{(j)}_p}$. Computing these moments is the first step towards determining the ensemble-averaged partition function. Clearly, they are zero whenever $k$ is odd. The ensemble average of the $J$s scales as $\binom{N}{p}^{-k/2}$ due to our hypotheses. Because of the independence of the random couplings, no distinct $I$ can appear only once. When $N \to \infty$, furthermore, only the case where each $I$ appears exactly twice survives. In any other case, in fact, we have at most $\ell = \floor*{(k-1)/2}$ distinct elements in $\{ I_j \}$ and there are at most $\binom{N}{p}^\ell \times$ (number of possible assignments of the chosen sets to $I_1, \dots, I_k$) terms in the sum. Note that this estimate is an upper bound, since it allows for two or more $I_j$ to coincide, and that the number of assignments doesn't depend on $N$. This scenario scales as $\binom{N}{p}^{\ell - k/2} \to 0$, so we can completely ignore it. Had the random couplings been Gaussian, the pairwise contractions between them would have been automatic and this discussion would have only led us to conclude that it is highly unlikely for two or more contractions to be due to the same set of indices $I_j$.

\begin{figure}
    \centering
    \includegraphics[width = 0.3 \textwidth]{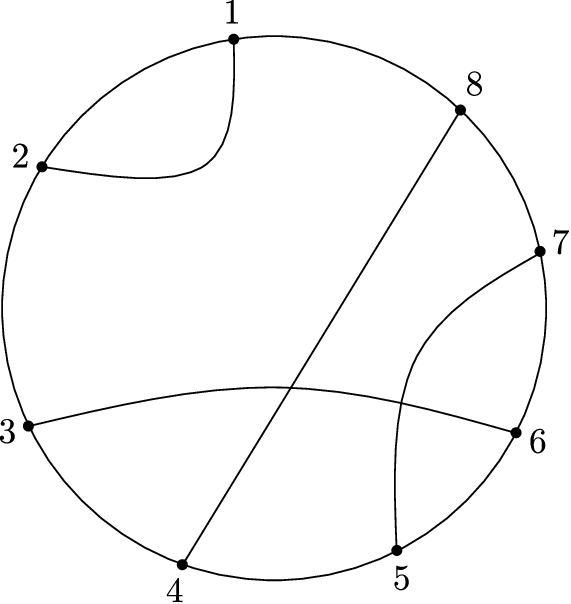}
    \caption{An example of chord diagram for $k = 8$. The chords tell us that $I_1 = I_2, \: I_3 = I_6, \: I_4 = I_8, \: I_5 = I_7$.}
    \label{fig:chords}
\end{figure}

This observation naturally leads us to the construction of chord diagrams. We represent $H^k$ as $k$ nodes on a circle (the trace is cyclic), labelled by an index $j = 1, \dots, k$. We connect the nodes in pairs with chords, signalling the pairs with the same $I_j$. We trivially obtain that $\langle J_{I_1} \dots J_{I_k} \rangle_J = \binom{N}{p}^{-k/2}$, so we are left with:
\begin{gather}
    m_k = i^{kp/2} \binom{N}{p}^{-k/2} \sum_{I_1, \dots, I_{k/2} \: \mathrm{distinct}} \tr(\psi_{I_{j(1)}} \dots \psi_{I_{j(k)}}),
\end{gather}

where the distribution of $I_1, \dots I_{k/2}$ among the $k$ strings of $\psi$ is described by each chord diagram, an example of which is in Figure \ref{fig:chords}.

In order to compute the trace, we need to disentangle every chord diagram by exchanging nodes, so that eventually the chords connect neighboring nodes. To do this, we need to commute the strings of $\psi$. The trace of a product of neighboring nodes is then $1$, because for every adjacent $\psi_{I_j} \psi_{I_j}$ we can pair the $\ell$-th fermion of the first string with the $\ell$-th fermion of the second string by performing $p-\ell$ anticommutations\footnote{Every fermion of the second string must cross $p-\ell$ fermions of the first one in order to be adjacent to its copy there, assuming that the ones before it in the second string have already performed the crossing.}, which yield an overall $(-1)^{(p-1)+(p-2)+\dots+1} = (-1)^{p(p-1)/2}$ prefactor. We obtain:
\begin{equation}
\begin{gathered}
    \psi_I \psi_I = (-1)^{p(p-1)/2} \psi_{i_1} \psi_{i_1} \dots \psi_{i_p} \psi_{i_p} = \frac{(-1)^{p/2}}{2^p} \{ \psi_{i_1}, \psi_{i_1} \} \dots \{ \psi_{i_p}, \psi_{i_p} \} = (-1)^{p/2} \\
    \implies i^{kp/2} \tr (\psi_{I_1} \psi_{I_1} \dots \psi_{I_{k/2}} \psi_{I_{k/2}}) = (-1)^{kp/4} \tr \, (-1)^{kp/4} = 1.
\end{gathered}
\end{equation}

Imagine we want to commute $\psi_{I_j}$ and $\psi_{I_k}$ and denote with $|I_j \cap I_k|$ the number of sites that the two sets have in common. If we start from the first fermion of $\psi_{I_k}$ and we move it in the first spot, in front of the entire $\psi_{I_j}$, we have performed $p$ anticommutations in the process and obtained a $(-1)^p = 1$ global factor, provided that it doesn't appear in $I_j$. The same reasoning applies to all the other fermions so, up to intersections, the outcome of the commutation yields a positive sign. Everytime there is an intersection, instead, commuting a fermion with its identical copy in the first string produces an obvious plus sign, so we get a $(-1)^{p-1} = -1$ factor at the end of the process. Finally, this reasoning tells us that:
\begin{gather}
    \psi_{I_j} \psi_{I_k} = (-1)^{|I_j \cap I_k|} \, \psi_{I_k} \psi_{I_j}.
\end{gather}

If we now interpret the counting of choices over all possible configurations as probabilities, we must ask ourselves how is $|I_j \cap I_k|$ distributed as a random variable when we extract $I_j, I_k$ uniformly from our $N$ Majorana fermions. We can do better and consider $A \subset \{ 1, \dots, N \}$ of size $p$ and $B$ of size $p'$, then determine the probability that $|A \cap B| = k$. We will assume $k \ll p, \: p' \ll N$. Indeed, in the double-scaled limit we have $p = p' \propto \sqrt{N}$ and we will show that the expected value $\sim k$ will be $\mathcal{O}(1) \ll \sqrt{N}$, so that this computation is consistent with the assumptions. Assuming $A$ to be fixed (since the $N$ elements are identical, it doesn't matter which ones exactly belong to this set), we first choose which are the $k$ elements in common, then we choose the remaining $p'-k$ elements for $B$. The probability we are looking for is simply the ratio with the total number of choices for $B$:
\begin{equation}
\begin{aligned}
    \mathcal{P}[|A \cap B| = k] & = \frac{\binom{p}{k} \binom{N-p}{p'-k}}{\binom{N}{p'}} = \frac{p(p-1)\dots(p-k+1)}{k!} \frac{p'!}{(p'-k)!} \frac{(N-p-p')!}{(N-p-p'+k)!} \frac{\binom{N-p}{p'}}{\binom{N}{p'}} \\
    & \approx \frac{p^k}{k!} p'^k \frac{1}{N^k} \prod_{j=0}^{p'-1} \left( 1 - \frac{p}{N-j} \right) \approx \frac{1}{k!} \left( \frac{pp'}{N} \right)^k \left( 1 - \frac{p}{N} \right)^{p'} \approx \frac{1}{k!} \left( \frac{pp'}{N} \right)^k e^{-pp'/N}.
\end{aligned}
\end{equation}

In the second line, we have approximated the first three fractions of the first line by using the assumed hierarchy between the quantities, then we have used it repeatedly to obtain the final result. What we obtain is interesting: the number of intersections is Poisson-distributed with parameter (mean of the distribution) $pp'/N$, which is indeed $\mathcal{O}(1)$ for sets of length $\sim \sqrt{N}$ and gives us the scale of the relevant $k$. Right now, we have $p = p'$, so this parameter is actually $\lambda/2$.

Let us elaborate on a detail that \cite{Berkooz:2018jqr} claims, but does not explicitly prove. We observe that intersections among strings are independent, i.e. it is impossible for a certain $\psi_k$ to appear in three or more $I_j$, so when disentangling the chords we can consider each step independently and the global process factorizes. To see that this is the case, we take three sets $A,B,C$ of lengths $p,p',p''$, and we determine the probability that $|A \cap B \cap C| = 0$. We can do this by taking the probability that $|A \cap B| = k$, multiplying it by the probability that none of these indices appear in $C$, then summing over $k$:
\begin{equation}
\begin{aligned}
    \mathcal{P}[|A \cap B \cap C| = 0] & = \sum_{k=0}^{\infty} \mathcal{P}[|A \cap B| = k] \frac{\binom{N-k}{p''}}{\binom{N}{p''}} \\
    & = \sum_{k=0}^{\infty} \mathcal{P}[|A \cap B| = k] \prod_{j=0}^{k-1} \left( 1 - \frac{p''}{N-j} \right) \\
    & \approx \sum_{k=0}^{\infty} \mathcal{P}[|A \cap B| = k] = 1.
\end{aligned}
\end{equation}

Note that we are summing up to infinity, but the error we commit is negligible and goes to $0$ as $N \to \infty$. We have used that the relevant $k$ are $\mathcal{O}(1)$ and that $p,p',p'' = \mathcal{O}(\sqrt{N})$, so that:
\begin{gather}
    \prod_{j=0}^{k-1} \left( 1 - \frac{p''}{N-j} \right) \approx \left( 1 - \frac{p''}{N} \right)^k \approx 1.
\end{gather}

When commuting two string of fermions inside the sum, then, thanks to the $\binom{N}{p}^{-k/2}$ prefactor that lets us convert the counting picture to the probabilistic one, the value of $(-1)^{|I_j \cap I_k|}$ we have to consider is simply its expectation value over the Poisson distribution. We have to repeat this operation independently for every intersection, until a given diagram is completely disentangled.

The final result is therefore the chord partition function:
\begin{gather}
    m_k = \sum_{\rm chord \: diagrams} \left( \sum_{j = 0}^\infty \frac{1}{j!} \left( \frac{p^2}{N} \right)^j e^{-p^2/N} (-1)^j \right)^{\# \: \mathrm{intersections}} = q^{\# \: \rm intersections}.
\end{gather}

\section{Observables}
The natural observables in the SYK model are polynomials of the $\psi$, i.e. operators of the form $O_I = \prod_{i \in I} \psi_i$, where $I$ is a set of indices of length $p_I$. In the following, we will consider lengths proportional to $\sqrt{N}$ and we will take random observables:
\begin{gather}
    M_A = i^{p_A/2} \sum_{1 \leq i_1 < \dots < i_{p_A} \leq N} J^{(A)}_{i_1 \dots i_{p_A}} \psi_{i_1} \dots \psi_{i_{p_A}}. \label{eq:randomOp}
\end{gather}

The $J^{(A)}_{i_1 \dots i_{p_A}}$ are independent, have zero mean, are such that the moments of $\binom{N}{p_A}^{1/2} J^{(A)}_{i_1 \dots i_{p_A}}$ do not depend on $N$ and 
\begin{gather}
    \langle J^{(A)}_{i_1 \dots i_{p_A}} J^{(B)}_{j_1 \dots j_{p_B}} \rangle_J = \binom{N}{p_A}^{-1} \delta^{AB}\delta_{i_1 j_1} \dots \delta_{i_{p_A} j_{p_B}}.
\end{gather}

These couplings are also independent of the random couplings in the Hamiltonian. We do not concern ourselves with different coupling strengths $\mathcal{J}^{(A)}$ since the way they appear is trivial, so we will also set them to $1$. If we see the DSSYK model as the dual of some gravitational theory, we can choose to probe the latter with the local stress-energy tensor in higher dimensions. This will be dual to some operator on the boundary theory (DSSYK precisely), which we can reasonably expect to be random and local, just like the Hamiltonian. The reasoning is explained in detail in \cite{Berkooz:2018qkz}, where the authors imagine the gravitational dual to contain a black hole: when acting on the states of its spectrum, it appears that a random Hamiltonian is a suitable effective description. The same applies to the stress-energy tensor and to any other single trace operator. This means that we have a plethora of operators that are expected to be of the form \eqref{eq:randomOp}, so this case is worth studying. Two point functions have the following form:
\begin{gather}
    \langle M(t_1) M(t_2) \rangle = \langle \tr e^{-\beta H} e^{iH t_1} M e^{-iH(t_1-t_2)} M e^{-i H t_2} \rangle_J, \label{eq:M2point}
\end{gather}

so, if we imagine to perform a series expansion of the exponentials, we are left with terms that look like
\begin{gather}
    \langle \tr H^{k_3} M H^{k_2} M H^{k_1} \rangle_J.
\end{gather}

\begin{figure}
    \centering
    \includegraphics[width = 0.3 \textwidth]{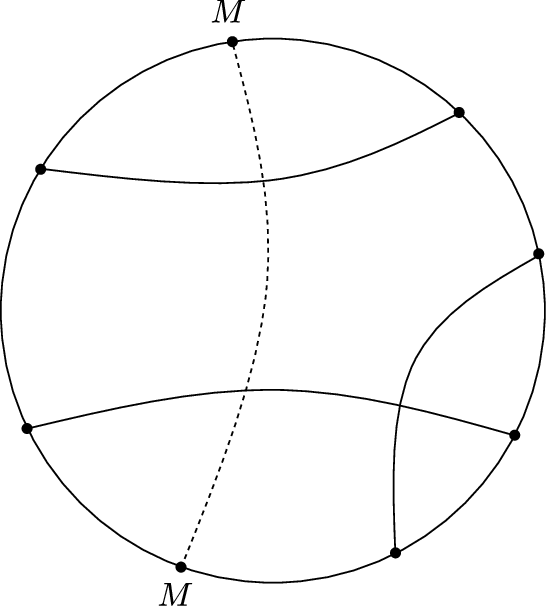}
    \caption{A chord diagram contributing to $\langle \tr M H^2 M H^4 \rangle_J$. The chord that connects the $M$ nodes has been represented as a dashed line. Nodes without a label are insertions of the Hamiltonian, which are connected by continuous lines.}
    \label{fig:chordsM}
\end{figure}

These are the quantities that we will compute in the following. To do this, we are going to generalize the chord diagrams presented in the previous section. The first observation is that we have two distinct and independent ensemble averages here: the one over the couplings of the Hamiltonian and the one over the random coefficients of the $M_A$. For the same reasons as before, even for $2n$ point functions of $M_A$ operators, only pairings of identical sets of coefficients survive in the $N \to \infty$ limit. Again, correlators of an odd number of operators are null because of the $J^{(A)}$ having zero mean. We need an even number of $M_A$ for every ``flavor'' $A$ and an even number of $H$, and all these different types of operators cannot be paired together. As a consequence, the chord diagrams generalize to a situation where we have different types of nodes that are all connected among themselves by different kinds of chords. An example of a chord diagram that contributes to \eqref{eq:M2point} is shown in Figure \ref{fig:chordsM}. The order of the nodes on the circle is fixed by the trace we are computing, and in this case prescribes us to insert $k_3$ $H$ nodes, an $M$ node, $k_2$ $H$ nodes, an $M$ node and finally $k_1$ $H$ nodes: note that the trace is cyclic and so is the circle, so we could have started from any operator appearing in the trace.

In order to compute the traces, we have to disentangle the diagrams once again. In this case there are different types of chords, so one needs not only to count the number of intersections, but also to check which chords intersect exactly. We have already shown that disentangling an $H-H$ intersection yields a factor $q = e^{-2p^2/N}$: in general, a similar computation would prove that disentangling an $H-M_A$ intersection yields a factor $q_A = e^{-2pp_A/N}$ and that disentangling an $M_A-M_B$ intersection yields a factor $q_{AB} = e^{-2p_A p_B/N}$.

In short:
\begin{gather}
    \langle \tr (\dots H^{k_3} M_B H^{k_2} M_A H^{k_1}) \rangle_J = \sum_{\rm chord \: diagrams} q^{\# \: H-H \: \mathrm{inters.}}  \prod_A q_A^{\# \: H-M_A \: \mathrm{inters.}} \prod_{A,B} q_{AB}^{\# \: M_A-M_B \: \mathrm{inters.}}. \label{eq:DSSYKnpoint}
\end{gather}

\section{Evaluating the Chord Partition Function}
In this section, we will develop the formalism that will let us compute the chord partition function:
\begin{gather}
    m_k = \langle \tr H^k \rangle_J = \sum_{\rm chord \: diagrams} q^{\# \: \rm intersections}.
\end{gather}

\begin{figure}
    \centering
    \includegraphics[width = 0.8 \textwidth]{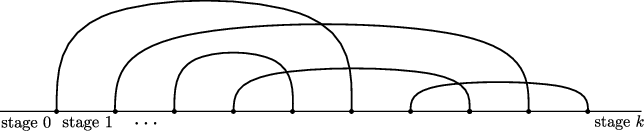}
    \caption{A chord diagram after it has been cut open. Note the height of the chords, which makes sure that the number of intersections between them is the same as the original diagram.}
    \label{fig:LineDiagram}
\end{figure}

The first step is to cut open the circular chord diagram at an arbitrary point and visualize it as a line, just like in Figure \ref{fig:LineDiagram}. An important thing to do when performing this operation is to order the chords, putting at the highest height those that appear earlier when reading the linearized diagram from left to right: this way, it is clear to see that the number of intersections of the ``new'' diagram is the same as the original one.

\begin{figure}
    \centering
    \includegraphics{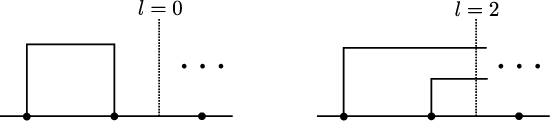}
    \caption{Possible partial chord diagrams at stage 2.}
    \label{fig:linestage2}
\end{figure}

We now introduce an auxiliary Hilbert space $\mathcal{H}_{\rm aux}$ with basis vectors $\ket{l}$, where the enumeration $l = 0,1,2,\dots$ represents the number of open chords. We interpret the diagram on the line as a system that evolves from left to right, with the nodes acting as a discrete ``time'' coordinate. To be more precise, we consider the number of open chords at every ``stage'', that we label with $i = 0,\dots,k$, between two adjacent nodes. $i = 0$ is the stage before the first node, while $i = k$ is the stage after the last one: there are always no open chords at both positions. Let us denote by $v_l^{(i)}$ the weighted sum over all possible open chord diagrams from stage 0 up to stage $i$ with $l$ open chords at stage $i$, where the weight for each diagram is $q$ to the power of the number of intersections until that point. We can compact this information into the infinite dimensional vectors $v^{(i)}$ that live in $\mathcal{H}_{\rm aux}$, where its components are the $v_l^{(i)}$. If we take, for example, $v^{(2)}$, we know that at stage $2$ we can either have $0$ or $2$ open chords: these are associated with two different diagrams with no intersections, as shown in Figure \ref{fig:linestage2}, so we have $v^{(2)} = (1,0,1,0,0,\dots)$.

We can write a recursion relation for the $v^{(i)}_l$. If we start from a situation with $l$ open chords at stage $i$, there are two possibilities for the number of chords in the previous stage.
\begin{itemize}
    \item The first possibility is that there are $l-1$ open chords at the previous stage, and then a new chord is opened in the following node.
    \item The second possibility is that in the previous stage there are $l+1$ open chords, and one is closed in the following node. Any open chord can close at that point, and in doing so it will intersect all the chords below it. These are $l+1$ different possibilities, which yield an extra factor of $q^0, q^1, \dots, q^l$, according to the choice of which chord is closed.
\end{itemize}

Summing over all the possibilities, we obtain:
\begin{gather}
    v^{(i+1)}_l = v^{(i)}_{l-1} + (1+q+q^2+...+q^l) v^{(i)}_{l+1} = v^{(i)}_{l-1} + \frac{1-q^{l+1}}{1-q} v^{(i)}_{l+1} \equiv v^{(i)}_{l-1} + \eta_l \, v^{(i)}_{l+1}.
\end{gather}

This can be rewritten in terms of a transfer matrix $T$ that ``evolves'' $v^{(i)}$ into $v^{(i+1)}$:
\begin{equation}
\begin{gathered}
    v^{(i+1)} = T v^{(i)}, \\
    T = \begin{pmatrix}
        0 & \frac{1-q}{1-q} & 0 & 0 & \dots \\
        1 & 0 & \frac{1-q^2}{1-q} & 0 & \dots \\
        0 & 1 & 0 & \frac{1-q^3}{1-q} & \dots \\
        0 & 0 & 1 & 0 & \dots \\
        \vdots & \vdots & \vdots & \vdots & \ddots 
    \end{pmatrix}.
\end{gathered}
\end{equation}

More compactly:
\begin{gather}
    v^{(i+1)}_{l_1} = T_{l_1 l_2} v^{(i)}_{l_2}, \quad T_{l_1 l_2} = \delta_{l_2, l_1-1} + \eta_{l_1} \delta_{l_2, l_1+1}.
\end{gather}
Our input is that at stages $i = 0$ and $i = k$ there are no open chords, so we start from $v^{(0)} = \ket{0}$ and, after evolving this state through our transfer matrix $k$ times, we should only consider those diagrams that have no open chords left. In other words, we simply have to project $v^{(k)}$ onto $\ket{0}$ in order to obtain $m_k$, since we have already accounted for the intersections:
\begin{gather}
    m_k = \langle \tr H^k \rangle_J = \langle 0| T^k |0 \rangle.
\end{gather}

This result tells us that $T$ is effectively the Hamiltonian of the auxiliary system: any function of $H$ inside a trace in the original Hilbert space can be replaced with the expectation value on $\ket{0}$ of same function of $T$. In a way, we could think of this dual system as the gravitational theory dual to DSSYK. The next step is to diagonalize $T$ and determine both its eigenvalues (which play the role of energies $E$) and eigenvectors. We first perform a similarity transformation on the matrix:
\begin{equation}
\begin{gathered}
    \hat T = PTP^{-1}, \quad P = \mathrm{diag}(P_0,P_1,P_2, \dots), \\
    P_0 = 1, \quad P_{l \geq 1} = \prod_{i=0}^{l-1} \sqrt{\eta_i} = \frac{\sqrt{(q;q)_l}}{(1-q)^{l/2}}, \\
    \hat T_{l_1 l_2} = P_{l_1} T_{l_1 l_2} P_{l_2}^{-1} = \sqrt{\eta_{l_2}} \, \delta_{l_2, l_1-1} + \sqrt{\eta_{l_1}} \, \delta_{l_2, l_1+1},
\end{gathered}
\end{equation}

where we have defined the $q$-Pochammmer symbol
\begin{gather}
    (a;q)_n \equiv \prod_{k=0}^{n-1} (1-aq^k).
\end{gather}

The matrix $\hat T$ is clearly symmetric (which means that $T$ can be diagonalized) and can be used instead of $T$ when computing chord partition functions:
\begin{gather}
    m_k = \langle 0| T^k |0 \rangle = \langle 0| P^{-1} \hat T^k P | 0 \rangle = \langle 0| \hat T^k | 0 \rangle.
\end{gather}

When looking for the spectrum, though, it is best to use the original $T$ matrix. It is an infinite-dimensional matrix with the following asymptotic behavior:
\begin{gather}
    T_{\rm asymp} = \begin{pmatrix}
        0 & \frac{1}{1-q} & 0 & 0 & \dots \\
        1 & 0 & \frac{1}{1-q} & 0 & \dots \\
        0 & 1 & 0 & \frac{1}{1-q} & \dots \\
        0 & 0 & 1 & 0 & \dots \\
        \vdots & \vdots & \vdots & \vdots & \ddots
    \end{pmatrix}.
\end{gather}

If we define $\delta T = T - T_{\rm asymp}$, our eigenvector problem has the same structure as a scattering problem in Quantum Mechanics:
\begin{gather}
    T \psi = (T_{\rm asymp} + \delta T) \, \psi = E \psi.
\end{gather}

We can interpret the components of $\psi$ as the values of a wave function at discrete positions, so that the scattering happens on a half-line with $\delta T$ being the potential close to the origin. This is an intuitive explanation as to why we can extract the spectrum of $T$ from $T_{\rm asymp}$, just like one does when considering plane waves at infinity, which are the eigenvectors of the kinetic term in the Hamiltonian that survives at long distances. However, it should be clear from a purely mathematical point of view that the need for $\psi$ to satisfy the eigenvector condition regardless of ``position'', coupled with the fact that $T$ only connects adjacent sites (that is, it is ``local''), leads us to the observation that, when restricting ourselves to the components at infinity, they must necessarily satisfy $T_{\rm asymp} \psi_{\infty} = E\psi_{\infty}$ with the same energy $E$. The only exception to this discussion are bound states, which in fact are characterized by $\psi_{\infty} = 0$. Fortunately, we have found a neat argument and a mathematical observation to explain why there are none in this case, so that the continuous part of the spectrum is a complete basis of the auxiliary Hilbert space. The original papers \cite{Berkooz:2018jqr,Berkooz:2018qkz}, in fact, do not seem to properly address this issue.

We first perform another similarity transformation on $\hat T$ using $S_{ij} = (-1)^i \delta_{ij}$:
\begin{gather}
    \tilde T_{ij} = (S \hat T S^{-1})_{ij} = -\hat T_{ij} \implies \tilde T_{\rm asymp} = \frac{1}{\sqrt{1-q}} 
    \begin{pmatrix}
        0 & -1 & 0 & 0 & 0 & \dots \\
        -1 & 0 & -1 & 0 & 0 & \dots \\
        0 & -1 & 0 & -1 & 0 & \dots \\
        0 & 0 & -1 & 0 & -1 & \dots \\
        \vdots & \vdots & \vdots & \vdots & \vdots & \ddots
    \end{pmatrix}.
\end{gather}

The scattering problem can be rewritten as:
\begin{gather}
    \sqrt{1-q} \left( \tilde T_{\rm asymp} + \frac{2}{\sqrt{1-q}} + \delta \tilde T \right) \psi = \sqrt{1-q} \, (E-E_0) \psi, \quad E_0 \equiv -\frac{2}{\sqrt{1-q}}.
\end{gather}

As we will see, $E_0$ is the minimum possible energy of the system. We now have that the first two terms in the l.h.s. are the discretization of the second derivative operator with a negative sign, while the third term acts like a potential:
\begin{gather}
    \sqrt{1-q} \, \delta \tilde T = \begin{pmatrix}
        0 & 1-\sqrt{1-q} & 0 & 0 & 0 & \dots \\
        1-\sqrt{1-q} & 0 & 1-\sqrt{1-q^2} & 0 & 0 & \dots \\
        0 & 1-\sqrt{1-q^2} & 0 & 1-\sqrt{1-q^3} & 0 & \dots \\
        0 & 0 & 1-\sqrt{1-q^3} & 0 & 1-\sqrt{1-q^4} & \dots \\
        \vdots & \vdots & \vdots & \vdots & \vdots & \ddots
    \end{pmatrix}.
\end{gather}

In general, the problem with this matrix is that it is not diagonal, but we can still see that it is ``local'' since it only connects adjacent sites of the vector $\psi$. In the continuum limit, for $\psi$ a ``continuous'' function, its components vary very slowly, so that we can effectively describe the behavior of $\delta \tilde T$ by using a diagonal matrix instead:
\begin{gather}
    \sqrt{1-q} \, (\delta \tilde T \psi)_{n \geq 1} = (1-\sqrt{1-q^n}) \psi_{n-1} + (1-\sqrt{1-q^{n+1}}) \psi_{n+1} \approx (2-\sqrt{1-q^n} - \sqrt{1-q^{n+1}}) \psi_n. 
\end{gather}

From this, we can clearly see that our potential term is always positive (i.e. it is repulsive) and decays exponentially for big $n$ and/or small $q$, where it behaves as:
\begin{gather}
    \sqrt{1-q} \, (\delta \tilde T \psi)_n \approx \frac{1+q}{2} q^n \psi_n.
\end{gather}

Such a potential does not allow for bound states, so we are only going to consider $E \geq E_0$ eigenstates from now on. This heuristic argument becomes more concrete when there is an obvious small parameter with which one can rescale the discrete ``position'' index $i$ to obtain a continuous coordinate. In the $q \to 1$ limit, for example, we can define the continuous coordinate in the following way:
\begin{gather}
    x = \abs{\log q} \, i.
\end{gather}

This way, the eigenvalue problem becomes:
\begin{gather}
    \left( - \abs{\log q}^2 \frac{d^2}{dx^2} + 2(1-\sqrt{1-e^{-x}}) \right) \psi(x) = \sqrt{1-q} \, (E-E_0) \psi(x).
\end{gather}

In the $q \to 0$ limit, similarly, we can take $x = q \, i$ and obtain:
\begin{gather}
    \left( - q^2 \frac{d^2}{dx^2} + \frac{1+q}{2} e^{-\abs{\log q} \, x/q} \right) \psi(x) = \sqrt{1-q} \, (E-E_0) \psi(x).
\end{gather}

At the end of the discussion, we will prove in a formal way that there are no bound states in the spectrum, but this picture has the property of being a nice physical interpretation of what is going on. 

Going back to the study of scattering states, $T_{\rm asymp}$ is a tridiagonal Toeplitz matrix. The most general $n \times n$ Toeplitz matrix is:
\begin{gather}
    A_n = \begin{pmatrix}
        a & c & 0 & 0 & 0 & \dots \\
        b & a & c & 0 & 0 & \dots \\
        0 & b & a & c & 0 & \dots \\
        0 & 0 & b & a & c & \dots \\
        \vdots & \vdots & \vdots & \vdots & \vdots & \ddots
    \end{pmatrix}.
\end{gather}

We will now determine the eigenvalues of said matrices through an explicit computation, whereas in \cite{Berkooz:2018jqr,Berkooz:2018qkz} they are simply reported. First, we can write a recursion relation for the determinants $\Delta_n$ of these matrices as a function of their size:
\begin{gather}
    \Delta_n = a \Delta_{n-1} - bc \Delta_{n-2}, \quad \Delta_0 = 1, \: \Delta_1 = a
    \implies \Delta_n = \frac{\left( \frac{a+\sqrt{a^2-4bc}}{2} \right)^{n+1} - \left( \frac{a-\sqrt{a^2-4bc}}{2} \right)^{n+1}}{\sqrt{a^2-4bc}}.
\end{gather}

The solution is found in the usual way, namely by ansatzing $\Delta_n = x^n$ in the recursion relation, finding the roots $x_{1,2}$ and fixing the coefficients of $\Delta_n = \alpha x_1^n + \beta x_2^n$ by requiring that the initial conditions are satisfied. When looking for the spectrum, we substitute $a \to a' = a - \lambda$ and find the values of $\lambda$ such that $\Delta'_n = 0$. This is achieved when:
\begin{gather}
    1 = \left( \frac{a' - \sqrt{a'^2-4bc}}{a' + \sqrt{a'^2-4bc}} \right)^{n+1} \implies \frac{1 - \sqrt{1-\frac{4bc}{a'^2}}}{1 + \sqrt{1-\frac{4bc}{a'^2}}} = \omega_k = \exp \left( \frac{2ik\pi}{n+1} \right), \quad 0 \leq k \leq n,
\end{gather}

where $\omega_k$ is an $(n+1)$th root of unity. Inverting the relation yields:
\begin{gather}
    a' = \pm \frac{\omega_k + 1}{2 \sqrt{\omega_k}} 2\sqrt{bc} \implies \lambda = a \pm 2 \sqrt{bc} \cos \left( \frac{k\pi}{n+1} \right).
\end{gather}

In the case of $T_{\rm asymp}$, since $n \to \infty$, the spectrum becomes continuous as the spacing between consecutive eigenvalues goes to $0$. The final result is:
\begin{gather}
    E(x) = \frac{2}{\sqrt{1-q}} \cos(\pi x), \quad x \in [0,1].
\end{gather}

We find that the density of states is uniformly distributed in the interval $\theta = \pi x \in [0,\pi]$, so that the insertion of a complete set of energy eigenstates has a flat measure in the continuum:
\begin{gather}
    \sum_E \to \int_0^\pi d\theta.
\end{gather}

The next step is determining the eigenvectors $v^{(\mu)}$ of $T$ (and \textit{not} of $T_{\rm asymp}$, whose usefulness was limited to finding which eigenvalues to input) with energy $E(\mu) = 2\mu/\sqrt{1-q}$. By definition:
\begin{gather}
    T v^{(\mu)}_l = \frac{2\mu}{\sqrt{1-q}} v^{(\mu)}_l = v^{(\mu)}_{l-1} + \frac{1-q^{l+1}}{1-q} v_{l+1}^{(\mu)}, \quad \quad v_0^{(\mu)} = 1, \: v_{-1}^{(\mu)} \equiv 0. 
\end{gather}

It is useful to rescale the components:
\begin{gather}
    v_l^{(\mu)} = \frac{(1-q)^{l/2}}{(q;q)_l} u_l^{(\mu)}.
\end{gather}

This way, the recursion relation simplifies to:
\begin{gather}
    2\mu u_l^{(\mu)} = (1-q^l) u_{l-1}^{(\mu)} + u_{l+1}^{(\mu)}, \quad \quad u_{-1}^{(\mu)}=0, \: u_{0}^{(\mu)} = 1.
\end{gather}

This recursion relation is exactly the one satisfied by the continuous $q$-Hermite polynomials $H_n(x|q)$, whose properties are reported in Appendix \ref{app:qHermite}. This means that our eigenvectors are:
\begin{gather}
    v_l^{(\mu)} = \frac{(1-q)^{l/2}}{(q;q)_l} H_l(\mu|q), \quad -1 \leq \mu = \cos \theta \leq 1.
\end{gather}

The fact that the eigenvectors are uniquely determined by $\mu$ (therefore by their energy) implies that there is no degeneracy in the spectrum. The components $\hat \psi_l(\cos\theta|q)$ of the normalized eigenvectors of $\hat T$ are those of $P v^{(\mu)}$, up to a normalization factor:
\begin{gather}
    \hat \psi_l(\cos\theta|q) = N(\theta,q) P_l v_l^{(\mu)} = N(\theta,q) \frac{H_l(\cos \theta|q)}{\sqrt{(q;q)_l}}. \label{eq:EigenThL}
\end{gather}

In Appendix \ref{app:qHermite}, we show that
\begin{gather}
    N(\theta,q) = \hat\psi_0(\cos\theta|q) = \frac{\sqrt{(q;q)_\infty} |(e^{2i\theta};q)_\infty|}{\sqrt{2\pi}}
\end{gather}

is the normalization that makes the following identities true:
\begin{equation}
\begin{gathered}
    \sum_{n=0}^\infty \hat\psi_n (\cos\theta|q) \hat \psi_n(\cos \phi|q) = \delta(\theta-\phi), \\
    \int_0^\pi d\theta \: \hat \psi_n(\cos\theta|q) \hat \psi_m(\cos\theta|q) = \delta_{nm}.
\end{gathered}
\end{equation}

This result tells us that we can take the eigenvectors $| \hat \psi(\theta) \rangle$ of $\hat T$, which are defined by the projections $\langle l| \hat \psi(\theta) \rangle = \hat \psi_l (\cos\theta|q)$, and use them to write the identity operator:
\begin{gather}
    \mathbb{1} = \int_0^\pi d\theta \: |\hat \psi(\theta)\rangle \langle \hat\psi(\theta)|.
\end{gather}

Clearly, this is the correct decomposition only if there are no bound states, which is a statement that we are finally able to prove formally. The key observation is that the trace of $T$ is:
\begin{gather}
	\tr T = \sum_n E_n + \int_0^\pi d\theta \: E(\theta) = 0. 
\end{gather}

The $E_n$ are the energies of the bound states, which are necessarily upper-bounded by the lowest scattering energy $E_0 < 0$. Since both the trace of $T$ and the contribution from the continuum are null, we are left with:
\begin{gather}
	\sum_n E_n = 0, \qquad E_n \leq E_0 < 0 \; \forall n,
\end{gather}

which is obviously only possible if there are no bound states in the theory.

A generic matrix element of $\hat T^k$ can then be computed through an insertion of this decomposition of the identity:
\begin{gather}
    \langle l|\hat T^k|m \rangle = \int_0^\pi d\theta \: \hat \psi_l(\cos\theta|q) \hat \psi_m(\cos\theta|q) \, E^k(\theta).
\end{gather}

Taking $|l\rangle = |m\rangle = |0\rangle$ leads to an expression for the moments:
\begin{gather}
    m_k = \int_0^\pi \frac{d\theta}{2\pi} \: (q, e^{\pm 2i\theta}; q)_\infty \left( \frac{2 \cos\theta}{\sqrt{1-q}} \right)^k. \label{eq:DSSYKmoment}
\end{gather}

We have used the shorthand notation $(a_1, a_2, \dots a_n; q)_\infty \equiv (a_1; q)_\infty (a_2; q)_\infty \dots (a_n;q)_\infty$. Note that $m_{2r+1} = 0$ since $\theta \in [0,\pi/2]$ and $\theta \in [\pi/2,\pi]$ give opposite contributions to the integral. We can use this result to obtain the ensemble average of the partition function of the theory:
\begin{gather}
    \langle \mathcal{Z} \rangle_J (\beta) = \langle \tr e^{-\beta H} \rangle_J = \sum_{k=0}^\infty \frac{(-\beta)^k}{k!} m_k = \int_0^\pi \frac{d\theta}{2\pi} \: (q, e^{\pm 2i\theta}; q)_\infty \exp \left( -\frac{2 \beta \cos\theta}{\sqrt{1-q}} \right). \label{eq:DSSYKpi}
\end{gather}

When studying the SYK model, we had managed to rewrite the partition function in terms of two fields $G(\tau,\tau')$ and $\Sigma(\tau,\tau')$, over which there was a path integral left to perform. In the double-scaled limit, instead, we have virtually managed to compute that path integral, and we are only left with an integral over a simple, real variable. All these integrals can actually be performed and several simplifying limits are possible, for details see Appendix \ref{app:ZDSSYK}. It is worth noting that the moments and the partition functions contain the density of energy eigenstates of the DSSYK model with respect to both the $\theta$ parameter (we call it $\Psi(\theta,q)$) and the energy itself (we call it $v(E(\theta)|q)$):
\begin{equation}
\begin{gathered}
    \Psi(\theta,q) = |\hat\psi_0(\cos\theta|q)|^2 = \frac{(q,e^{\pm 2i\theta};q)_\infty}{2\pi}, \\
    d\theta \: \Psi(\theta,q) = dE \: v(E(\theta)|q) \implies v(E(\theta)|q) = \frac{\sqrt{1-q}}{4\pi \sin\theta} (q,e^{\pm 2i\theta};q)_\infty.
\end{gathered}
\end{equation}

Note that this density is normalized to $1$ as a consequence of \eqref{eq:qHermOrt} applied to $n=m=0$:
\begin{gather}
    \int_0^{\pi} d\theta \: \Psi(\theta,q) = \int_{-\frac{2}{\sqrt{1-q}}}^{\frac{2}{\sqrt{1-q}}} dE \: v(E(\theta),q) = 1.
\end{gather}

\section{Evaluating Correlation Functions}
\begin{figure}
    \centering
    \includegraphics[width = 0.85 \textwidth]{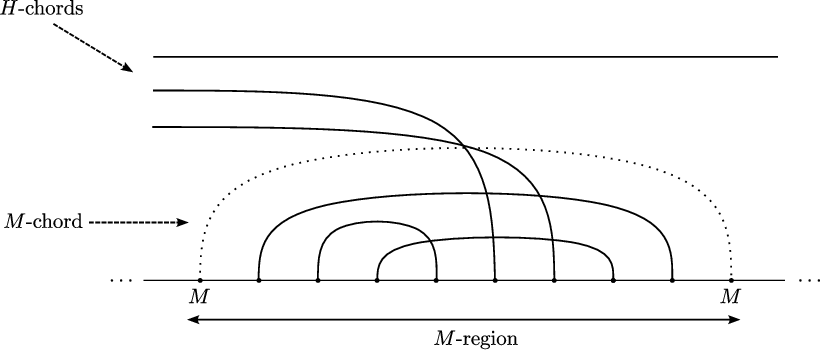}
    \caption{Example of $M$-region inside a chord diagram. Between two paired $M$ operators, only $H$ nodes appear. Notice again how the highest chords are those that have been created earlier, regardless of their type.}
    \label{fig:linediagM}
\end{figure}

In this section, we will describe the instruments needed to compute \eqref{eq:DSSYKnpoint}. We will focus on the case of a random operator $M$, which was defined in \eqref{eq:randomOp}. We had seen an example of a modified chord diagram in Figure \ref{fig:chordsM}. The first step is to once again cut the diagram open: thanks to the presence of a trace, we can cut at any point. The smartest choice is to cut right before an $M$ insertion, so that the diagram on the line starts with an $M$-chord that opens from an $M$ node: this is the choice that we will make in the next section, where we will obtain the two point function. Still, for an arbitrary cut, we obtain the situation in Figure \ref{fig:linediagM}. The technology that we have developed in the previous section is modified by the presence of the $M$-region, namely the region between two paired $M$ nodes: this situation can be generalized to uncrossed $2n$ point functions of $M_A$ operators, where there is a sequence of these regions in the diagram that do not interact among themselves, i.e. there are no $M$ nodes between two paired $M_A$ operators.

Just like in the previous section, we use the auxiliary Hilbert space $\mathcal{H}_{\rm aux}$: we only count $H$-chords and not $M$-chords. We now introduce a factor of $\tilde q = e^{-2pp'/N}$, with $p'$ the length of the $\psi$ strings in $M$, for every $H$-chord incoming into the $M$-region, and another factor of $\tilde q$ for every $H$-chord leaving the $M$-region. This is achieved by inserting a matrix
\begin{gather}
    S = \begin{pmatrix}
        1 & 0 & 0 & \dots \\
        0 & \tilde q & 0 & \dots \\
        0 & 0 & \tilde q^2 & \dots \\
        \vdots & \vdots & \vdots & \ddots \\
    \end{pmatrix}
\end{gather}

immediately before and after the $M$-region: if this region is made of $k$ stages, we substitute $T^k$ with $S T^k S$. Consider taking a fixed chord diagram with $l$ open chords before entering the region and $l'$ open chords when leaving it: the insertions of $S$ multiply this diagram by $\tilde q^{l+l'}$. These insertions of $S$ behave in the following way.
\begin{itemize}
    \item They give a $\tilde q$ factor to the open chords at the entrance that close inside the $M$-region and to the chords that open from $H$ nodes inside the region without closing inside it. This is the intended behaviour, as these do intersect the $M$-chord once.
    \item They do not give any $\tilde q$ factor to the chords that open and close inside the $M$-region. Again, this is intended as they do not intersect the $M$-chord.
    \item They give a $\tilde q^2$ factor to the open chords at the entrance that do not close inside the $M$-region, but are still open at the exit. This is not the correct factor, since these chords do not intersect the $M$-chord.
\end{itemize}

We need to fix the last situation. Before we do that, though, we observe that, if we limit ourselves to two point functions, the smart choice of starting from an $M$ node avoids this issue completely, since we are always in the situation of no incoming open chords at the start. Still, it is interesting to see how this problem can be solved. It is crucial, for example, when computing four point functions of operators, although we won't show how to do that in this thesis.

First, consider a diagram with one chord out of $l$ not closing in the $M$-region, which is multiplied by an extra $\tilde q^2$. Let this chord be above all the others and let $l$ be a fixed number throughout this discussion: indeed, we can always put it on top up to factors of $q$. One can recover the correct factor for such a diagram if one takes $l-1$ incoming chords, multiplies the outcome of their evolution by $1-\tilde q^2$ and sums it to the starting diagram. Unfortunately, the result of this operation is still wrong when applied to diagrams with more than one chord not closing in the $M$-region. We want to further modify the evolution inside the $M$-region by inserting more matrices that fix all the possible scenarios. In particular, they should also add all the chord diagrams with $l-i$ incoming chords and evolve them through the $M$-region, before adding back $i$ chords and multiplying their contribution by an appropriate factor, so that the sum of all of these extra diagrams yields the correct result. 

\begin{figure}
    \centering
    \includegraphics[width = 1 \textwidth]{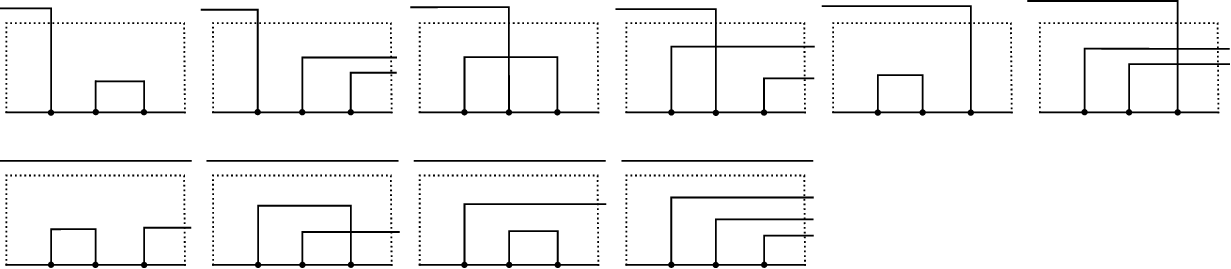}
    \caption{Chord diagrams with $l = 1$ open chords when entering an $M$-region that contains three $H$ nodes. These contribute to correlation functions of $M$.}
    \label{fig:MregionEx}
\end{figure}

We start from a simple example, with $l = 1$ and three $H$ nodes in the $M$-region. All the possible diagrams are shown in Figure \ref{fig:MregionEx}. The diagrams in the first row are assigned the correct value $\tilde q(1 + \tilde q^2 + q + q \tilde q^2 + 1 + q^2 \tilde q^2)$, while those in the second one are assigned $\tilde q^2 (\tilde q + q  \tilde q + \tilde q + \tilde q^3)$, with an extra $\tilde q^2$ factor. If we forget about the initial chord, the second row is also the list of all the diagrams with $l = 0$. This means that by adding $l = 0$ diagrams multiplied by $1 - \tilde q^2$, the final result would be the correct one.

The matrix that implements going from $l$ chords to $l-1$ chords is the following:
\begin{gather}
    U = \begin{pmatrix}
        0 & 1 & 0 & \dots \\
        0 & 0 & 1 & \dots \\
        0 & 0 & 0 & \dots \\
        \vdots & \vdots & \vdots & \ddots
    \end{pmatrix}, \quad U_{l,l'} = \delta_{l',l+1}.
\end{gather}

Likewise, the matrix that restores the missing chords (from $l$ to $l+1$) is:
\begin{gather}
    D = \begin{pmatrix}
        0 & 0 & 0 & \dots \\
        1 & 0 & 0 & \dots \\
        0 & 1 & 0 & \dots \\
        \vdots & \vdots & \vdots & \ddots
    \end{pmatrix}, \quad D_{l,l'} = \delta_{l,l'+1}.
\end{gather}

\begin{figure}
    \centering
    \includegraphics[width = 0.85 \textwidth]{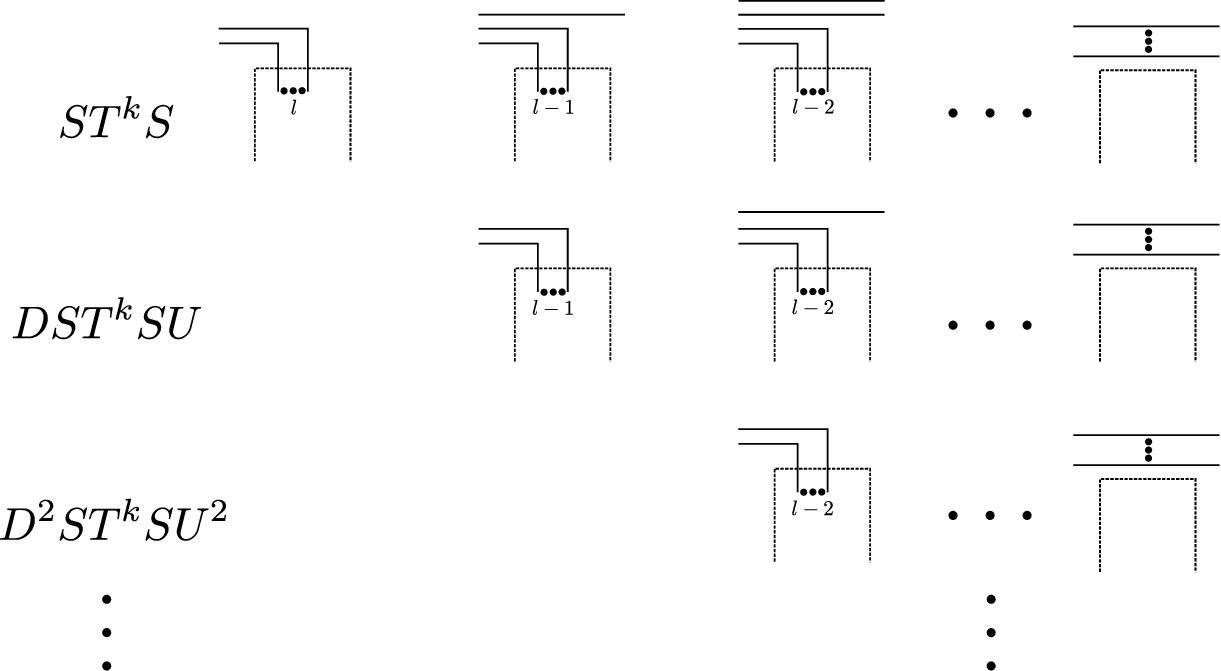}
    \caption{A schematic drawing of the different contributions to the transition matrix. Each diagram in the figure stands for the sum of the diagrams with the indicated number of chords closing in the $M$-region. The $i$-th row stands for the correcting diagrams with $l-i+1$ chords entering the $M$-region, to which $i$ additional chords are added in the end. In the $j$-th column, we see all the diagrams where $l-j+1$ chords close in the $M$-region. This upper triangular structure is such that the first $j$ $P_k^{(l)}$ can be determined by only ``repairing'' the diagrams with $j$ or less chords that do not intersect the $M$-region.}
    \label{fig:Mregionchain}
\end{figure}

Our discussion tells us that the propagation of $l$ incoming chords through an $M$-region containing $k$ $H$ nodes should be described by the following modified transition matrix:
\begin{gather}
    \sum_{j=0}^{l} P_j^{(l)} D^j S T^k S U^j, \label{eq:MTransition}
\end{gather}

with $P_j^{(l)}$ numerical coefficients that depend on $q, \tilde q$. The different contributions of the terms in \eqref{eq:MTransition}, according to the number of chords closing in the $M$-region, are shown in Figure \ref{fig:Mregionchain}.

Let us give our personal explanation of why Figure \ref{fig:Mregionchain} correctly describes what is going on, in particular what is the mechanism that ``kills'' diagrams with too many lines closing in the $M$-region in each row. The general idea is that, by linearity, the vector $v^{(i)} = (v^{(i)}_0, v^{(i)}_1, v^{(i)}_2, \dots)$ that we evolve with our transition matrix is made of all the possible chord diagrams, multiplied by some factor due to intersections. When we apply $U^{j}$ to a specific diagram, we are shifting its initial position inside $v^{(i)}$ by $j$ slots to the left. When we evolve it with $T^k$, then, once we arrive at the point where the $(l-j+2)$th $H$ chord needs to close, the subcomponent of $v$ due to this diagram is already in the leftmost position, that is, it is proportional to $(1, 0, 0, \dots)$ instead of $(0,1,0, \dots)$. As a consequence, it should be projected by $T$ on $v_{-1}$ rather than $v_0$, which is impossible, and is therefore ``killed'' by the evolution.

As a trivial example, let us consider the case of $l=1$ open chords that encounter an $M$-region with a single $H$ node, so that the incoming vector is $v = (0,1,0,\dots)$. We can forget about the insertions of $S$ here, as they are irrelevant for the argument. Clearly, $Tv = (1,0,1,\dots) = (1,0,0,\dots) + (0,0,1,\dots)$, namely the sum of the diagram where the chord closes on the $H$ node and of the one where another chord opens from the node. When we consider $(DTU)v$, we obtain $(0,0,1,\dots)$, namely only the diagram where the chord opens from the $H$ node: $U$ has brought the vector to $(1,0,0,\dots)$, so there is no chord that can close on the node as required by the first diagram. Indeed, the diagram that has survived is the one that contracts $l-1 = 0$ $H$-chords in the $M$-region, to which we have added a chord at the end through $D$. This example generalizes to the mechanism that we have described earlier, i.e. the absence of a $v_{-1}$ component to which one could project through $T$. All the other diagrams, instead, survive this evolution and are present in the end.

\begin{figure}
    \centering
    \includegraphics[width = 0.82 \textwidth]{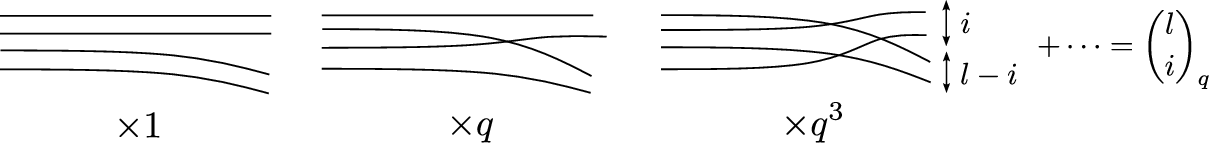}
    \caption{Weighted sum over the possibilities of choosing $i$ out of $l$ chords, with weight $q^{\# \: \mathrm{intersections}}$: the result is the $q$-binomial coefficient. We can imagine moving these $i$ chords at the top of the diagram because they are going to be crossed by the other $l-i$ $H$-chords when closing in the $M$-region.}
    \label{fig:intersectM}
\end{figure}

Up until now, we have assumed that the $i$ chords out of the initial $l$ that did not close in the $M$-region were above all the others, but this is not always the case. In a general situation, they are $i$ chords at any height. When the other $l-i$ chords close in the $M$-region, they actually cross the chords that are below them, so in assuming that all the $i$ chords are above the rest we are missing some factors of $q$. We can choose which of the $l$ chords do not close in the $M$-region in $\binom{l}{i}$ ways, multiply each choice by $q$ to the power of the number of intersections, then sum over all the possibilities, as shown in Figure \ref{fig:intersectM}. This actually yields the $q$-binomial coefficient, also called the \textit{Gaussian binomial coefficient}:
\begin{gather}
    \binom{l}{i}_q \equiv \frac{(1-q^l)(1-q^{l-1})\dots(1-q^{l-i+1})}{(1-q^i)(1-q^{i-1})\dots(1-q)} = \frac{(q;q)_l}{(q;q)_i (q;q)_{l-i}}, \quad \binom{l}{i}_q \equiv 0 \: (i > l).
\end{gather}

Let us show that this is true. Denote with $V(i,l-i)$ the result of the sum. Consider the chord at the bottom: the sum can be split into the terms where this chord belongs to the $i$ chords and those where it doesn't. The first batch of terms is equal to $q^{l-i} V(i-1,l-i)$, since this chord has to cross the other $l-i$ chords to be put above them but, apart from this, it is mapped in a ``full'' weighted sum with one fewer chord. The second batch of terms is instead equal to $V(i,l-i-1)$ for similar reasons, but now this chord should not be moved. We have found a recursion relation that we need to solve:
\begin{gather}
    V(i,j) = V(i,j-1) + q^j V(i-1,j), \quad V(0,j) = V(i,0) = 1, \: i,j \geq 0.
\end{gather}

It is trivial to verify that $V(i,j) = \binom{i+j}{i}_q$ is indeed the solution to this equation.

We can now determine the $P_j^{(l)}$. The way we do this is by considering the diagrams in which $l-j$ chords close inside the $M$-region. Each such diagram is part of the sets of diagrams represented in the $(j+1)$th column of Figure \ref{fig:Mregionchain}. The operator \eqref{eq:MTransition} associates the correct weight to this kind of diagrams, plus an excessive value that we set to zero:
\begin{gather}
    (\tilde q^{2j} -1) \binom{l}{l-j}_q + \tilde q^{2(j-1)} \binom{l-1}{l-j}_q P_1^{(l)} + \tilde q^{2(j-2)} \binom{l-2}{l-j}_q P_2^{(l)} + \dots + P_j^{(l)} = 0.
\end{gather}

This factor we are setting to $0$ multiplies the weight of all the diagrams with the chords not closing in the $M$-region placed above the rest. As already observed, we have a triangular system, which is solved by:
\begin{gather}
    P_j^{(l)} = \frac{(q;q)_l (\tilde q^2; q)_j}{(q;q)_{l-j} (q;q)_j} = \binom{l}{l-j}_q (\tilde q^2; q)_j.
\end{gather}

We now give our proof of the claim. We proceed by induction: assume that we have fixed $P_i^{(l)}$, \mbox{$1 \leq i \leq j-1$}, through the first $j-1$ equations of the triangular system, and that they are given by the above formula. The basis for the induction is that it is true for $P_0^{(l)} = 1$. The $j$th equation then gives us:
\begin{gather}
    P_j^{(l)} = \binom{l}{l-j}_q \left( 1 - \sum_{i=0}^{j-1} \binom{j}{i}_q \tilde q^{2(j-i)} (\tilde q^2;q)_i \right).
\end{gather}

The proof by induction is complete if we show the following identity:
\begin{gather}
    1 - \sum_{i=0}^{j-1} \binom{j}{i}_q \tilde q^{2(j-i)} (\tilde q^2;q)_i = (\tilde q^2;q)_j. \label{eq:PijProof}
\end{gather}

To this end, it is useful to employ the Gaussian binomial theorem, i.e. the $q$-analog of the usual Newton binomial theorem:
\begin{gather}
    (a;q)_n = \sum_{k \in \mathbb{Z}} \binom{n}{k}_q q^{k(k-1)/2} (-a)^k = \sum_{k=0}^n \binom{n}{k}_q q^{k(k-1)/2} (-a)^k.
\end{gather}

This theorem reduces to the known formula for $q=1$ and can also be proven through a simple induction, see \cite{ProofWiki} for details. We apply it to the l.h.s. of \eqref{eq:PijProof}:

\begin{equation}
\begin{aligned}
    & 1 - \sum_{i=0}^{j-1} \sum_{\ell=0}^{i} \binom{j}{i}_q \binom{i}{\ell}_q (-1)^\ell \tilde q^{2(j-i+\ell)} q^{\ell(\ell-1)/2} \\
    = \: & 1 - \sum_{\ell=0}^{j-1} \sum_{k \equiv i - \ell = 0}^{j-\ell-1} \binom{j}{\ell}_q \binom{j-\ell}{k}_q (-1)^\ell \tilde q^{2(j-k)} q^{\ell(\ell-1)/2} \\
    = \: & 1 - \sum_{\ell=0}^{j-1} \sum_{a \equiv j-k = \ell+1}^j \binom{j}{\ell}_q \binom{j-\ell}{j-a}_q (-1)^\ell \tilde q^{2a} q^{\ell(\ell-1)/2} \\
    = \: & 1 - \sum_{a=1}^j \binom{j}{j-a}_q \tilde q^{2a} \sum_{\ell = 0}^{a-1} \binom{a}{a-\ell}_q (-1)^\ell q^{\ell(\ell-1)/2} \\
    = \: & 1 - \sum_{a=1}^j \binom{j}{j-a}_q \tilde q^{2a} \left[ \sum_{\ell = 0}^a \binom{a}{a-\ell}_q (-1)^\ell q^{\ell(\ell-1)/2} - (-1)^a q^{a(a-1)/2} \right] \\
    = \: & 1 - \sum_{a=1}^j \binom{j}{j-a}_q \tilde q^{2a} \left[ (1;q)_a - (-1)^a q^{a(a-1)/2} \right] \\
    = \: & 1 + \sum_{a=1}^j \binom{j}{a}_q q^{a(a-1)/2} (-\tilde q^2)^a = (\tilde q^2;q)_j.
\end{aligned}
\end{equation}

In the last line, we have used that $(1;q)_a = 0$. This concludes our proof. With this result, we have fully determined how $l$ initial chords propagate through the $M$-region.

\section{Two Point Functions}
In this section, we compute \eqref{eq:M2point}. We start by considering:
\begin{gather}
    \langle \tr H^{k_1} M H^{k_2} M \rangle_J.
\end{gather}

We have used that the trace is cyclic to have an $M$ in the rightmost position: from a chord diagram perspective, this is equivalent to cutting it open next to an $M$ insertion and placing this $M$ node in the leftmost position on the line. Switching to the auxiliary $\mathcal{H}_{\rm aux}$ picture, we have to compute (recall that $P$ and $S$ are both diagonal):
\begin{gather}
    \langle \tr H^{k_1} M H^{k_2} M \rangle_J = \langle 0| T^{k_1} P_0^{(0)} D^0 ST^{k_2} S U^0 |0 \rangle = \langle 0| T^{k_1} S T^{k_2} |0 \rangle = \langle 0| \hat T^{k_1} (PSP^{-1}) \hat T^{k_2} |0 \rangle = \langle 0| \hat T^{k_1} S \hat T^{k_2} | 0 \rangle.
\end{gather}

We can insert several decompositions of the identity:
\begin{equation}
\begin{aligned}
    \langle 0| \hat T^{k_1} S \hat T^{k_2} | 0 \rangle & = \sum_{n,m=0}^\infty \int_0^\pi d\theta \, d\phi \: \langle 0| \hat T^{k_1} |\hat \psi(\theta) \rangle \langle \hat \psi(\theta) | n \rangle \langle n| S |m \rangle \langle m| \hat T^{k_2} | \hat \psi(\phi) \rangle \langle \hat \psi(\phi) | 0 \rangle \\
    & = \int_0^\pi d\theta \, d\phi \: \hat\psi_0(\cos\theta|q) \hat\psi_0(\cos\phi|q) E(\theta)^{k_1} E(\phi)^{k_2} \sum_{n=0}^\infty \tilde q^n \hat\psi_n(\cos\theta|q) \hat\psi_n(\cos\phi|q).
\end{aligned}
\end{equation}

We remember the form of the eigenvectors \eqref{eq:EigenThL}, so that we can use \eqref{eq:qHermOrtSum} with $t = \tilde q$:
\begin{gather}
    \sum_{n=0}^\infty \tilde q^n \hat\psi_n(\cos\theta|q) \hat\psi_n(\cos\phi|q) = \hat\psi_0(\cos\theta|q) \hat\psi_0(\cos\phi|q) \frac{(\tilde q^2;q)_\infty}{(\tilde q e^{i(\theta+\phi)}, \tilde q e^{i(\theta-\phi)}, \tilde q e^{-i(\theta-\phi)}, \tilde q e^{-i(\theta+\phi)} ; q)_\infty}.
\end{gather}

Finally, we obtain:
\begin{gather}
    \langle \tr H^{k_1} M H^{k_2} M \rangle_J = \int_0^\pi \prod_{j=1}^2 \left[ \frac{d\theta_j}{2\pi} \: (q,e^{\pm 2i\theta_j};q)_\infty \left( \frac{2 \cos\theta_j}{\sqrt{1-q}} \right)^{k_j} \right] \frac{(\tilde q^2;q)_\infty}{(\tilde q e^{i(\pm\theta_1 \pm \theta_2)};q)_\infty}.
\end{gather}

We can easily exponentiate this result to obtain the two point function \eqref{eq:M2point}:
\begin{equation}
\begin{aligned}
    \langle \tr e^{-\beta H} M(t_1) M(t_2) \rangle_J & = \langle \tr e^{iH((t_1-t_2)+i\beta)} M e^{-iH(t_1-t_2)} M \rangle_J \\
    & = \sum_{k_1 = 0}^{\infty} \sum_{k_2 = 0}^{\infty} \frac{(i(t_1-t_2)-\beta)^{k_1}}{k_1!} \frac{(-i(t_1-t_2))^{k_2}}{k_2!} \langle \tr H^{k_1} M H^{k_2} M \rangle_J \\
    & = \int_0^\pi \prod_{j=1}^2 \left[ \frac{d\theta_j}{2\pi} \: (q,e^{\pm 2i\theta_j};q)_\infty \exp \left( \frac{2 i t'_j \cos\theta_j}{\sqrt{1-q}} \right) \right] \frac{(\tilde q^2;q)_\infty}{(\tilde q e^{i(\pm\theta_1 \pm \theta_2)};q)_\infty}, \label{eq:M2pointfinal}
\end{aligned}
\end{equation}

where $t'_1 \equiv (t_1-t_2) + i\beta, \: t'_2 \equiv -(t_1-t_2)$. One could show that any other way of cutting the circle yields the same result through the technology developed in the previous section, hence showing its consistency with the cyclicity of the trace, but we will not do that here.

We can summarize this result by introducing diagrammatic rules. The ``skeleton'' chord diagrams that we introduce for this purpose only have the $M$-chords and no explicit insertions of the Hamiltonian. Here, the $M$ operators are inserted at Euclidean (and not Lorentzian) times. The circle in the diagram, on which we place our $M$ operators, is the thermal circle of circumference $\beta$. The correlation functions are then built from the following rules.
\begin{itemize}
    \item A segment along the circle is an evolution induced by the $T$ matrix for an Euclidean time $\theta = \Delta \tau$ between two paired $M$ operators. We can interpret this segment as a propagator of an eigenstate of $T$ with energy $E(\theta) = 2 \cos\theta/\sqrt{1-q}$, so that we obtain a factor of $e^{-\Delta \tau E(\theta)}$.
    \begin{center}
        \includegraphics[width = 0.35 \textwidth]{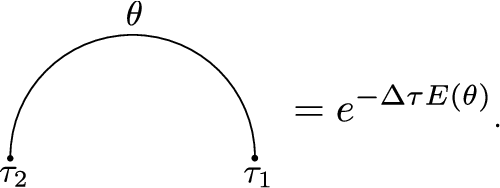}
    \end{center}
    
    \item Next, we sum over the energy eigenstates that can propagate in this segment, that is, over $\theta$. The measure is the one that appears in the orthogonality relation of the $q$-Hermite polynomials:
    \begin{gather}
        d\mu(\theta) = \frac{d\theta}{2\pi} \, (q,e^{\pm 2i\theta};q)_\infty.
    \end{gather}

    \item Each operator insertion (such as $M$) between two segments $\theta_1$ and $\theta_2$ is to be thought of as a matrix element $\langle \hat \psi(\theta_1) | \mathcal{O} |\hat \psi(\theta_2) \rangle$, since we are propagating $T$ eigenstates in the diagram. If we define $m_A$ through $\tilde q_A = q^{m_A}$, the corresponding vertex in the skeleton chord diagram is assigned the value:
    \begin{center}
        \includegraphics[width = 0.4 \textwidth]{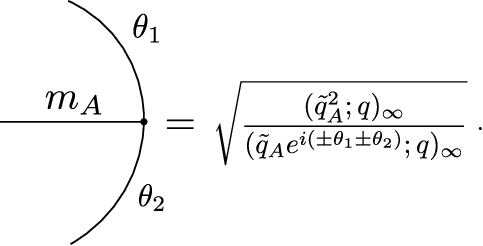}
    \end{center}

    \item A contracted pair of operators ``conserves the energy'', meaning that the same $\theta$ variable is used before and after such a contracted pair if there are no additional operator insertions in between. In other words, the propagation through an $M$-region (described by an operator $\mathcal{O}$ in $\mathcal{H}_{\rm aux}$ that acts on $| l \rangle$ like \eqref{eq:MTransition}) of the eigenvectors of $T$ does not change their initial energy at the ``exit'': equivalently, they are also eigenvectors of $\mathcal{O}$, regardless of the separation between the two $M$. We do not show the proof of this statement here, as it would require introducing a new type of special function along with its properties, but it is described in \cite{Berkooz:2018jqr}. We also integrate over a single $\theta$ variable inside the region between the paired operators, unless insertions of other operators appear.
\end{itemize}

\begin{figure}[H]
    \centering
    \includegraphics[width = 0.3 \textwidth]{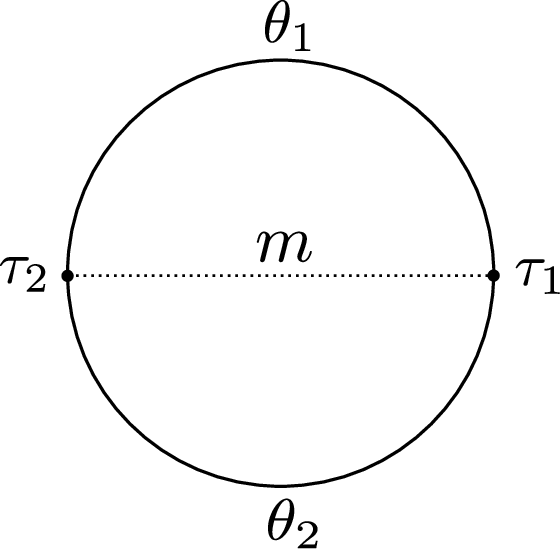}
    \caption{A skeleton chord diagram for the two point function.}
    \label{fig:skeletonchord}
\end{figure}

A full skeleton chord diagram for the Euclidean two point function is in Figure \ref{fig:skeletonchord}.

The $\theta_1$ variable is integrated over a segment of length $\tau_1-\tau_2$ (we assume $\tau_1 > \tau_2$), while the $\theta_2$ variable is integrated over a segment of length $\beta-(\tau_1-\tau_2)$, so these rules give us the following two point function:
\begin{gather}
    \langle \tr e^{-\beta H} M(-i\tau_2) M(-i\tau_1) \rangle_J = \int_0^\pi \prod_{j=1}^2 \left[ \frac{d\theta_j}{2\pi} \: (q,e^{\pm 2i\theta_j};q)_\infty \exp \left( -\frac{2 \tau'_j \cos\theta_j}{\sqrt{1-q}} \right) \right] \frac{(\tilde q^2;q)_\infty}{(\tilde q e^{i(\pm\theta_1 \pm \theta_2)};q)_\infty},
\end{gather}

where $\tau'_1 \equiv \tau_1-\tau_2$, $\tau'_2 \equiv \beta - (\tau_1-\tau_2)$. This result agrees with \eqref{eq:M2pointfinal} evaluated on the thermal circle. These rules are still valid when considering four point functions, but in that case an extra rule is needed in order to describe the crossing of two $M$-chords. We will not consider four point functions in this thesis, so the above set of rules is sufficient.

\section{Fermionic Propagators, Large \texorpdfstring{$N$}{N} and Large \texorpdfstring{$p$}{p}} \label{sec:Glambdavsp}
In this section, we will consider the differences between the double-scaled limit that we have studied throughout this chapter and the usual two-step limit, where one first takes $N \to \infty$ at fixed $p$, then $p \to \infty$. In particular, we will see what happens when considering the thermal two point function of a Majorana fermion:
\begin{gather}
    G(\tau) = \sum_{i=1}^N \frac{1}{N} \frac{\langle \tr \left( e^{-(\beta-\tau)H} \psi_i e^{-\tau H} \psi_i \right) \rangle_J}{\langle \tr e^{-\beta H} \rangle_J} = \mathrm{sgn}(\tau) \left( 1 + \frac{1}{p} \sum_{m,n=0}^\infty \sum_{k=1}^\infty \frac{\lambda^m P_k^{(m,n)}[\beta-\tau,\tau]}{p^n} \right). \label{eq:Glamvsp}
\end{gather}

$P_k^{(m,n)}$ are homogeneous polynomials of degree $k$ which are symmetric in their two arguments, as required from the structure of the trace. There is no $k=0$ polynomial because we know the UV behavior $G(\tau \to 0) \approx \mathrm{sgn}(\tau)$. Note that we do not have a prefactor of $1/2$, unlike \eqref{eq:BigpG}, because of the different normalization $\{ \psi_i, \psi_j \} = 2 \delta_{ij}$. We can expect such an expansion to exist, because both the fixed $p, \: \lambda \to 0$ and fixed $\lambda, \: p \to \infty$ limits are valid. Such an expansion is sure to converge at high temperatures: for small $\lambda$, the discussion in Appendix \ref{app:ZDSSYK} suggests that this condition translates to requiring that $\sqrt{\lambda} \beta < 1$. This expansion alone is able to tell us what happens in these two different limits:
\begin{itemize}
    \item In the usual two-step limit, only the leading order in $\lambda$ of the expansion survives, i.e. the $\mathcal{O}(\lambda^0)$ terms. On the other hand, this limit lets us easily determine the subleading terms in the $p^{-1}$ expansion. In principle, we could perform less approximated computations and also obtain the first few terms in a $\lambda$ expansion, for example due to $1/N$ corrections.
    \item In the double-scaled limit, only the $\mathcal{O}(p^{-1})$ terms survive, but we are able to determine the contributions from all powers of $\lambda$. Even in this case, a more precise approach would yield subleading corrections.
\end{itemize}

In the first case, we had determined the first correction of order $\mathcal{O}(p^{-1})$ to the propagator, which we had called $g(\tau)$. In fact, recall \eqref{eq:thermalSYKprop}:
\begin{gather}
    e^{g(\tau)} = \left( \frac{\cos \frac{\pi v}{2}}{\cos \left[ \pi v \left( \frac{1}{2} - \frac{|\tau|}{\beta} \right) \right]} \right)^2, \quad \beta \mathcal{J} = \frac{\pi v}{\cos \frac{\pi v}{2}}.
\end{gather}

We now want to determine the propagator that we obtain from a chord diagram expansion. If we consider $G_{ii}(\tau)$, the $\psi_i$ operators are again represented by nodes that are connected by a different kind of chord, which now carries a single fixed index $i$ rather than $p_A$ different ones (as was the case with the $M_A$). There are also no random couplings appearing in the operator. When considering intersections of the fermionic chord with the $H$-chords, we cannot use the Poisson distribution for the mean number of ``shared'' fermions in the two strings, since the fermionic one only has unitary length. Our determination of the exact result is pretty straightforward: given a fixed set $A$ of $p$ indices, the single index that we choose for the second set $B$ will appear in the first set with probability $p/N$, and won't with probability $1-p/N$. The expected value of $(-1)^{|A \cap B|}$ is therefore:
\begin{gather}
    \tilde q = \mathbb{E}\left[(-1)^{|A \cap B|}\right] = (+1) \times \left( 1-\frac{p}{N} \right) + (-1) \times \frac{p}{N} = 1 - \frac{\lambda}{p}. 
\end{gather}

We now give a more detailed explanation of what is said in \cite{Berkooz:2018jqr}. The above result for $\tilde q$ means that disentangling intersections between a fermionic chord and an $H$-chord gives $\mathcal{O}(p^{-1})$ deviations from the case where there are no such intersections (or, more precisely, with respect to the same diagrams with $\tilde q$ set to $1$). Every chord diagram appearing in the numerator is mapped to one appearing in the denominator and contributing to the partition function by first bringing the $\psi_i$ together and then removing them, which is a ``free'' operation at order $\mathcal{O}(p^0)$. This implies that our propagator should be equal to $1$ up to $\mathcal{O}(p^{-1})$ terms, thus matching the free theory in the same way prescribed by the expansion \eqref{eq:Glamvsp}. Another interesting observation is that ``UV'' diagrams with $\tau = 0$ have no $H$-chords in the $\psi_i$-region and no intersections being assigned factors of $\tilde q$, so this situation can only lead to the free theory result for any interacting scenario when taking the ratio with the partition function, as we already knew from our analysis in Chapter 4.

One could be worried about the presence of other $\mathcal{O}(p^{-1})$ corrections to the chord diagram approach, which could make the final result of the computation unreliable already at order $\mathcal{O}(p^{-1})$, hence completely useless. They actually are present, but they are not an issue. Recall how we obtained the chord diagram picture in the first place: we assumed $N \to \infty$ to keep only pairwise contractions and then assigned factors of $q$ to each intersection between $H$-chords by assuming a Poisson distribution for the number of common sites between two different sets of indices. This, in turn, also required that $p \to \infty$: the real distribution is actually not Poisson and receives $\mathcal{O}(p^{-1})$ corrections in general. The interesting thing, though, is that these effects clearly modify the numerator and denominator in \eqref{eq:Glamvsp} in the same way, so they only have a $\mathcal{O}(p^{-2})$ effect on $G(\tau)$ in the end:
\begin{gather}
    G = \frac{B_0 + B_1 p^{-1} + A_1 p^{-1} + B_2 p^{-2} + \dots}{B_0 + B_1 p^{-1} + \dots} = 1 + \frac{A_1}{B_0} p^{-1} + \mathcal{O}(p^{-2}).
\end{gather}

The effects due to the fermionic chords are the $A_{i \geq 1}$, while all the other contributions are represented by the $B_{i \geq 0}$. As we can see, we don't have to worry about the $B_1$ coefficient and we can use \eqref{eq:M2pointfinal} in the case of $M \to \psi_i$ reliably, provided we only keep the $\mathcal{O}(p^{-1})$ part of the result. Being able to perform such an integral would precisely give us the $\mathcal{O}(p^{-1})$ part of \eqref{eq:Glamvsp} non-perturbatively with respect to $\lambda$. \\

With this analysis, the review part of this thesis is finally over. Although we have already greatly expanded on the details with respect to the source material in all the previous chapters, it is now time for us to obtain completely new results in the remainder of this work.

\newpage
\chapter{The ``Charged'' SYK Model}
In this chapter, we will consider a modified version of the SYK model, where the Majorana fermions are substituted by Dirac fermions. We will call it the Complex SYK (cSYK) model. The properties of this model have been studied in several papers, notably by Sachdev himself and others in \cite{Sachdev_2015,Davison_2017,Gu:2019jub,Sachdev:2023try,Sachdev:2023fim} and by Susskind in \cite{Susskind:2020fiu}. Other examples of works on this subject are \cite{Bulycheva:2017uqj,Gaikwad:2018dfc,Chaturvedi:2018uov,Sorokhaibam:2019qho,Louw:2022njq,Louw:2022njq2}. Interestingly, the chord diagram formalism has been extended to the double-scaled limit of this model \cite{Berkooz:2020uly}. As is clear from these references, the link between the charged model and gravity is also of great interest to the community. What we will do in the following is to adapt the approach presented in \cite{Goel:2023svz} to the case of the complex model: in particular, we will be interested in the effects of a non-null chemical potential $\mu$ that is now allowed by the new $U(1)$ global symmetry. To our knowledge, this way of obtaining the relevant equations has not been followed in the literature already, nor has their consequent analysis been performed. This chapter is a preliminary study of the cDSSYK model, with the aim of gaining familiarity with the novelties of this model with respect to the Majorana one. The study of the gravitational dual will be carried out in the next chapter.

The model is described by the following Euclidean Lagrangian and the following Hamiltonian, with complex random couplings:
\begin{equation}
\begin{gathered}
    L = \sum_{i=1}^N \psi^\dagger_i \partial_\tau \psi_i + \sum_{\substack{1 \leq i_1 < \dots < i_{p/2} \leq N \\ 1 \leq j_1 < \dots < j_{p/2} \leq N}} J_{i_1 \dots i_{p/2},j_1 \dots j_{p/2}} \psi^{\dagger}_{i_1} \dots \psi^{\dagger}_{i_{p/2}} \psi_{j_1} \dots \psi_{j_{p/2}}, \\
    H = H^\dagger = \sum_{\substack{1 \leq i_1 < \dots < i_{p/2} \leq N \\ 1 \leq j_1 < \dots < j_{p/2} \leq N}} J_{i_1 \dots i_{p/2},j_1 \dots j_{p/2}} \psi^{\dagger}_{i_1} \dots \psi^{\dagger}_{i_{p/2}} \psi_{j_1} \dots \psi_{j_{p/2}}, \\
    \label{eq:herm} J_{i_1 \dots i_{p/2},j_1 \dots j_{p/2}} = J^*_{j_1 \dots j_{p/2},i_1 \dots i_{p/2}}, \\
    \langle |J_{i_1 \dots i_{p/2},j_1 \dots j_{p/2}}|^2 \rangle_J = 2^{p-1} \frac{[(p/2-1)!]^2}{N^{p-1}} \mathcal{J}^2, \\
    \{ \psi_i, \psi_j \} = 0, \quad \{ \psi^\dagger_i, \psi^\dagger_j \} = 0, \quad \{ \psi_i, \psi^\dagger_j \} = \delta_{ij},
\end{gathered}
\end{equation}

where the last mean is an ensemble average over many samplings of the couplings. From now on, we will use the indices $I = \{i_1, \dots, i_{p/2}\}, J = \{j_1, \dots, j_{p/2}\}$. The global $U(1)$ symmetry is simply $\psi_i \to e^{i\alpha} \psi_i, \: \alpha \in \mathbb{R}$ for all the $N$ fermions. As usual, $N$ and $p$ are assumed to always be even. We observe that the free theory actually has a larger $U(N)$ symmetry group, $\psi_i \to U_{ij} \psi_j, \: U_{ij} \in U(N)$, but that this is broken down to $U(1)$ by the interacting terms coupling all the fermions through random coefficients. Let us discuss this more in detail. First, we rewrite the Hamiltonian in the following way:
\begin{gather}
	H = \sum_{\substack{1 \leq i_1 < \dots < i_{p/2} \leq N \\ 1 \leq j_1 < \dots < j_{p/2} \leq N}} \frac{1}{[(p/2)!]^2} J_{i_1 \dots i_{p/2},j_1 \dots j_{p/2}} \sum_{\sigma_i, \sigma_j} \mathrm{sgn}(\sigma_i) \, \mathrm{sgn}(\sigma_j) \psi^{\dagger}_{\sigma_i(1)} \dots \psi^{\dagger}_{\sigma_i(p/2)} \psi_{\sigma_j(1)} \dots \psi_{\sigma_j(p/2)},
\end{gather}

where $\sigma_i$ and $\sigma_j$ are all the possible permutations of the ordered $I$ and $J$, respectively. Then, we perform an arbitrary $U(N)$ transformation of the fermions, which in turn modifies the Hamiltonian and the Lagrangian:
\begin{equation}
\begin{gathered}
	\hspace{-1 cm} H \to \sum_{\substack{1 \leq \ell_1 < \dots < \ell_{p/2} \leq N \\ 1 \leq k_1 < \dots < k_{p/2} \leq N}} \sum_{\sigma_\ell, \sigma_k} \frac{1}{[(p/2)!]^2} \hat J_{\sigma_\ell(1) \dots \sigma_\ell(p/2),\sigma_k(1) \dots \sigma_k(p/2)} \psi^{\dagger}_{\sigma_\ell(1)} \dots \psi^{\dagger}_{\sigma_\ell(p/2)} \psi_{\sigma_k(1)} \dots \psi_{\sigma_k(p/2)}, \\
	\hspace{-1.4 cm} \hat J_{\sigma_\ell(1) \dots \sigma_\ell(p/2),\sigma_k(1) \dots \sigma_k(p/2)} = \sum_{\substack{i_1 \neq i_2 \dots \neq i_{p/2} \\ j_1 \neq j_2 \dots \neq j_{p/2}}} \mathrm{sgn}(\sigma_i) \, \mathrm{sgn}(\sigma_j) U^\dagger_{\sigma_\ell(1) \sigma_i(1)} \dots U^\dagger_{\sigma_\ell(p/2) \sigma_i(p/2)} J_{i_1 \dots i_{p/2},j_1 \dots j_{p/2}} U_{\sigma_j(1) \sigma_k(1)} \dots U_{\sigma_j(p/2) \sigma_k(p/2)}.
\end{gathered}
\end{equation}

If we now reorder the fermions in the sum and we go back to the initial way of writing the Hamiltonian, we find how the couplings are affected by $U$:
\begin{gather}
	\tilde J_{\ell_1 \dots \ell_{p/2}, k_1 \dots k_{p/2}} = \frac{1}{[(p/2)!]^2} \sum_{\sigma_\ell, \sigma_k} \mathrm{sgn}(\sigma_\ell) \, \mathrm{sgn}(\sigma_k) \hat J_{\sigma_\ell(1) \dots \sigma_\ell(p/2),\sigma_k(1) \dots \sigma_k(p/2)}.
\end{gather}

This transformation is a symmetry if and only if $\tilde J_{I,J} = J_{I,J}$ for every $I, J$ and for every unitary transformation $U$. We can interpret these equations as constraints in the space of all the possible couplings, which are satisfied only for very specific choices. As a consequence, when integrating over all the possible extractions of the $J_{I,J}$ to take ensemble averages, the hypersurface of choices that do not break $U(N)$ has null measure and is therefore irrelevant. This reasoning applies not only to $U(N)$, but to every other subgroup that strictly contains the global $U(1)$: for $U(1)$ transformations and for these only, it is easy to verify that $\tilde J_{I,J} = J_{I,J}$ is trivially satisfied, therefore there are no constraints on the possible couplings and the entire configuration space obeys this symmetry. Ultimately, it is the ensemble average that breaks the free $U(N)$ symmetry down to the residual $U(1)$, to which we can associate a chemical potential $\mu$ that couples to the conserved charge $Q$:
\begin{gather}
	Q(\tau) = \frac{1}{N} \sum_{i=1}^N \psi_i^\dagger(\tau)\psi_i(\tau) - \frac{1}{2}, \quad \frac{dQ}{d\tau} = 0.
\end{gather}

In the following sections, we will work at increasing levels of complexity, starting from $\mu = 0, \: \beta = +\infty$ all the way to the most general case.

\section{Path Integral at \texorpdfstring{$\mu = 0, \: \beta = +\infty$}{mu = 0, beta = +inf}}
The Euclidean partition function of the system for a fixed set of couplings is:
\begin{gather}
    \mathcal{Z}(J) = \int [\mathcal{D}\bar\psi \, \mathcal{D}\psi] \exp\left( -\int d\tau \, \sum_{i=1}^N \bar\psi_i \partial_\tau \psi_i - \sum_{I,J} J_{I,J} \, \bar \psi_{i_1} \dots \bar \psi_{i_{p/2}} \psi_{j_1} \dots \psi_{j_{p/2}} \right). \label{eq:startingCSYK}
\end{gather}

We remember that in $(0+1)$ dimensions $\gamma_0 = 1$, hence we can use $\psi^\dagger$ and $\bar\psi$ interchangeably. The symbol $[\mathcal{D}\bar\psi \, \mathcal{D}\psi]$ indicates the product over the measure of the $N$ fermions. We now take the ensemble mean of the partition function:
\begin{equation}
\begin{aligned}
    \langle \mathcal{Z} \rangle_J & = \int \prod_{I,J} \mathcal{D}J_{I,J} \: \mathcal{Z}(J) \exp \left( - \sum_{I,J} \frac{|J_{I,J}|^2}{2^p \frac{[(p/2-1)!]^2}{N^{p-1}} \mathcal{J}^2} \right) \\
    \label{eq:meanZ} & = \int [\mathcal{D}\bar\psi \, \mathcal{D}\psi] \exp \Bigg( -\int d\tau \, \sum_{i=1}^N \bar\psi_i \partial_\tau \psi_i \\
    & + 2^{p-2} \frac{[(p/2-1)!]^2}{N^{p-1}}\mathcal{J}^2 \sum_{I,J} \int d\tau \, d\tau' \: (\bar \psi_{i_1} \dots \bar \psi_{i_{p/2}} \psi_{j_1} \dots \psi_{j_{p/2}})(\tau) (\bar \psi_{j_1} \dots \bar \psi_{j_{p/2}} \psi_{i_1} \dots \psi_{i_{p/2}})(\tau') \Bigg).
\end{aligned}
\end{equation}

We have used that the integral over the couplings is quadratic and can be performed exactly thanks to \eqref{eq:gaussint}. To recast our situation into something compatible with this formula, we first write:
\begin{equation}
\begin{split}
    \langle \mathcal{Z} \rangle_J = & \int \prod_{I>J} \mathcal{D}\Re(J_{I,J}) \, \mathcal{D}\Im(J_{I,J}) \prod_{I=J} \mathcal{D}\Re(J_{I,J}) \\
    & \exp \left( \sum_{I,J} -\frac{(\Re(J_{I,J}))^2}{2\sigma^2} -\frac{(\Im(J_{I,J}))^2}{2\sigma^2} + \Re(J_{I,J})B_{I,J} + \Im(J_{I,J})(iB_{I,J}) + C \right).
\end{split}
\end{equation}

We have used \eqref{eq:herm} to perform the integral only over the independent couplings. $I>J$ is any possible ordering, for example the lexicographic ordering. We rewrite the argument of the exponential as:
\begin{equation}
\begin{split}
    & \sum_{I>J} -\frac{(\Re(J_{I,J}))^2}{\sigma^2} -\frac{(\Im(J_{I,J}))^2}{\sigma^2} + \Re(J_{I,J})(B_{I,J}+B_{J,I}) + \Im(J_{I,J})(iB_{I,J}-iB_{J,I}) \, + \\
    & \sum_{I=J} -\frac{(\Re(J_{I,J}))^2}{2\sigma^2} + \Re(J_{I,J})B_{I,J} + C.
\end{split}
\end{equation}

It is now straightforward to compute the Gaussian integrals and check that the result is indeed \eqref{eq:meanZ}. By permuting the anticommuting Grassmann variables $\psi_i, \bar\psi_i$ and by using that, just like the SYK model in Chapter 4,
\begin{gather}
    \psi^2 = \bar \psi^2 = 0, \quad \sum_{I,J} = \frac{1}{[(p/2)!]^2} \sum_{\substack{i_1 \neq i_2 \dots \neq i_{p/2} \\ j_1 \neq j_2 \dots \neq j_{p/2}}},
\end{gather}
we obtain:
\begin{gather}
    \langle \mathcal{Z} \rangle_J = \int [\mathcal{D}\bar\psi \, \mathcal{D}\psi] \exp \left( -\int d\tau \, \sum_{i=1}^N \bar\psi_i \partial_\tau \psi_i + \frac{N\mathcal{J}^2}{p^2} \int d\tau \, d\tau' \: \left( -4 G(\tau,\tau')G(\tau',\tau) \right)^{p/2} \right).
\end{gather}

We have used the definition of the fermionic two point function of the system:
\begin{gather}
    G(\tau,\tau') = \frac{1}{N} \sum_{i=1}^N \psi_i(\tau)\bar\psi_i(\tau').
\end{gather}

The $(-1)^{p/2}$ factor emerges in the following way. We start from $\psi^\dagger_I \psi_J \psi^\dagger_J \psi_I$: permuting $\psi^\dagger_I$ and $\psi_J$ gives some sign that is equal to the one obtained by then permuting $\psi^\dagger_I$ and $\psi^\dagger_J$, so the starting point is equal to $\psi_J \psi^\dagger_J \psi^\dagger_I \psi_I$. In order to obtain $\psi_{j_1} \psi^\dagger_{j_1} \dots \psi_{j_{p/2}} \psi^\dagger_{j_{p/2}}$, we have to move $\psi^\dagger_{j_1}$ through $p/2-1$ fermions, then $\psi^\dagger_{j_2}$ through $p/2-2$ fermions, and so on. Similarly, in order to obtain $\psi_{i_1} \psi^\dagger_{i_1} \dots \psi_{i_{p/2}} \psi^\dagger_{i_{p/2}}$, we have to move $\psi_{i_1}$ through $p/2$ fermions, then $\psi_{i_2}$ through $p/2-1$ fermions, and so on. Summing all these terms, we have that moving through $0 \leq i \leq p/2-1$ fermions is an operation that we have to perform twice, so only $(-1)^{p/2}$ coming from moving $\psi_{i_1}$ doesn't cancel out.

Following what is usually done in the SYK model, it is now useful to switch to the ($G,\Sigma$) formalism. To do so, we insert the identity (up to a constant factor) inside the path integral:
\begin{equation}
\begin{aligned}
    1 & = \int \mathcal{D}G \; \delta \left( G(\tau,\tau') - \frac{1}{N} \sum_{i=1}^N \psi_i(\tau)\bar\psi_i(\tau') \right) \\
    & = \int \mathcal{D}G \, \mathcal{D}\Sigma \; \exp \left(-N \int d\tau \, d\tau' \: \left( G(\tau,\tau') - \frac{1}{N} \sum_{i=1}^N \psi_i(\tau)\bar\psi_i(\tau') \right) \Sigma(\tau',\tau) \right). \label{eq:SigmaIntroCSYK}
\end{aligned}
\end{equation}

Just like in Section \ref{sec:GSFormal}, this follows from the Wick rotation of $2\pi \delta(x) = \int dk \, e^{ikx}$. Note the use of $\Sigma(\tau',\tau)$ rather than $\Sigma(\tau,\tau')$. We can now perform the path integral over the fermions thanks to the formula $\int [\mathcal{D}\bar\psi \mathcal{D}\psi] \exp(-\bar\psi A \psi) = \det A$:
\begin{equation}
\begin{aligned}
    \langle \mathcal{Z} \rangle_J = \int \mathcal{D}G \, \mathcal{D}\Sigma \: \exp \Bigg( & N \log\det [\delta(\tau-\tau')\partial_{\tau'} + \Sigma(\tau,\tau')] \\
    & - N\int d\tau \, d\tau' \left( \Sigma(\tau',\tau)G(\tau,\tau') - \frac{\mathcal{J}^2}{p^2}(-4 G(\tau,\tau')G(\tau',\tau))^{p/2} \right) \Bigg). \label{eq:ZGS}
\end{aligned}
\end{equation}

We want to focus on the double-scaled limit of the SYK model which, we recall, is given by:
\begin{gather}
    N,p \to +\infty, \quad \lambda = \frac{2p^2}{N} = \mathrm{const.}
\end{gather}

Therefore, we follow \cite{Goel:2023svz} and insert the following zero temperature ansatz in the path integral, whose change of measure is trivial:
\begin{gather}
    \Sigma(\tau,\tau') = \frac{\sigma(\tau,\tau')}{p}, \quad G(\tau,\tau') = \left( \frac{1}{2}\mathrm{sgn}(\tau-\tau')+ \tilde C \right)\left(1+\frac{g(\tau,\tau')}{p} \right).
\end{gather}

The different choice of $G(\tau,\tau')$ here is due to the different free propagator in the regular and complex models. In this case, a path integral approach to the free theory easily yields that the propagator is:
\begin{equation}
\begin{gathered}
    G_{\rm free}(\tau,\tau') = (\delta \partial_\tau)^{-1}(\tau,\tau'); \\
    \partial_\tau G_{\rm free}(\tau,\tau') = \delta(\tau-\tau') \implies G_{\rm free}(\tau,\tau') = \frac{1}{2}\mathrm{sgn}(\tau-\tau')+\tilde C.
\end{gathered}
\end{equation}

We will see that the constant $\tilde C$ is connected to the electric charge of the system. We focus on $\log\det[\delta(\tau-\tau')\partial_{\tau'} + \Sigma(\tau,\tau')] = \tr\log [\delta(\tau-\tau')\partial_{\tau'} + \Sigma(\tau,\tau')]$ first. We can freely subtract $\tr\log [\delta(\tau-\tau')\partial_{\tau'}]$ as this only modifies the partition function by an irrelevant constant. This way, we get:
\begin{equation}
\begin{aligned}
    & \tr\log [\delta(\tau-\tau')\partial_{\tau'} + \Sigma(\tau,\tau')] - \tr\log [\delta(\tau-\tau')\partial_{\tau'}] \\
    = \: & \tr\log \left[ \delta(\tau-\tau') + \int d\tau'' \: \Sigma(\tau,\tau'') ( \delta \partial_\tau)^{-1}(\tau'',\tau') \right] \\
    \approx &  \int d\tau \, d\tau' \: \Sigma(\tau,\tau') ( \delta \partial_\tau)^{-1}(\tau',\tau) - \frac{1}{2} \int d\tau_1 \dots d\tau_4 \: \Sigma(\tau_1,\tau_2) ( \delta \partial_\tau)^{-1}(\tau_2,\tau_3) \Sigma(\tau_3,\tau_4) ( \delta \partial_\tau)^{-1}(\tau_4,\tau_1).
\end{aligned}
\end{equation}

In the last line, we have expanded the logarithm up to $\mathcal{O}(1/p^2)$ terms, since the rest appear to be subleading in \eqref{eq:ZGS}. The inverse $A(\tau,\tau')$ of $\delta(\tau-\tau')\partial_{\tau'}$ is again found by solving:
\begin{gather}
    \int d\tau'' \: \delta(\tau-\tau'') \, \partial_{\tau''} A(\tau'',\tau') = \partial_\tau A(\tau,\tau') = \delta(\tau-\tau') \implies A(\tau,\tau') = \frac{1}{2}\mathrm{sgn}(\tau-\tau') + C.
\end{gather}

Once more, we have a constant $C$ appearing in the solution. For now, we don't bother to fix it, but we notice that we can freely take the ansatz for $G(\tau,\tau')$ to have the same $C$ as here. We make an observation: if $\Sigma(\tau,\tau') = -\Sigma(\tau',\tau)$, then the trace of the logarithm is actually independent of $C$ and, if we flip our mental picture, we can imagine that we are fixing the constant of $(\delta \partial_\tau)^{-1}(\tau,\tau')$ to be the same as the one in the fermionic propagator. This is surely the case in the Majorana SYK model. $\Sigma(\tau,\tau')$ is introduced in a way similar to \eqref{eq:SigmaIntroCSYK} there, but for Majorana fermions the term it multiplies in the parentheses is antisymmetric in its arguments by construction: this means that any symmetric part of $\Sigma(\tau,\tau')$ is irrelevant and we can assume this function to be purely antisymmetric. What happens to the term inside the trace of the logarithm, then, is the following:
\begin{equation}
\begin{gathered}
    \int d\tau'' \: \Sigma(\tau,\tau'') ( \delta \partial_\tau)^{-1}(\tau'',\tau') = \int d\tau'' \: \Sigma(\tau,\tau'') \left( \frac{1}{2} \mathrm{sgn}(\tau''-\tau') + C \right), \\
    C \int_{-\infty}^{+\infty} d\tau'' \: \Sigma(\tau,\tau'') = C \int_{-\infty}^{+\infty} d\tau'' \: \mathrm{sgn}(\tau-\tau'') B(|\tau-\tau''|) = C \int_{-\infty}^{+\infty} dx \: \mathrm{sgn}(x) B(|x|) = 0.
\end{gathered}
\end{equation}

In the second line, we have rewritten the antisymmetric $\Sigma(\tau,\tau')$ as a sign function times a function of the modulus of the distance between its two arguments (we recall that the theory has a time translation symmetry).

In any case, putting this result back into \eqref{eq:ZGS}, we obtain:
\begin{equation}
\begin{aligned}
    \langle \mathcal{Z} \rangle_J = & \int \mathcal{D}g \, \mathcal{D}\sigma \: \exp \Bigg( \frac{2\mathcal{J}^2}{\lambda} \int d\tau \, d\tau' \: (1-4C^2)^{p/2} \exp \left( \frac{g(\tau,\tau')+g(\tau',\tau)}{2} \right) \\
    & - \frac{1}{\lambda} \int d\tau_1 \dots d\tau_4 \: \sigma(\tau_1,\tau_2) A(\tau_2,\tau_3) \sigma(\tau_3,\tau_4) A(\tau_4,\tau_1) -\frac{2}{\lambda} \int d\tau \, d\tau' \: \sigma(\tau,\tau')A(\tau',\tau)g(\tau',\tau) \Bigg). \label{eq:Zsigma}
\end{aligned}
\end{equation}

We have used that:
\begin{gather}
    A(\tau,\tau')A(\tau',\tau) = -\frac{1}{4}(1-4C^2).
\end{gather}

Picking the same constant for $G(\tau,\tau')$ and $(\delta \partial_\tau)^{-1}(\tau,\tau')$ turns out to be actually \textit{necessary} in order to cancel the $\mathcal{O}(1/p)$ terms and have a global, finite $1/\lambda$ prefactor in the exponential, although these terms would have been automatically null in the case of an antisymmetric $\Sigma(\tau,\tau')$.

The integral over $\sigma(\tau,\tau')$ can be performed exactly. Before we do that, though, let us study the saddle point equations for both $g$ and $\sigma$ that dominate in the $\lambda \to 0$ limit. What we expect to obtain from this is the $\mathcal{O}(\lambda^0/p)$ part of the fermionic propagator of the complex DSSYK model. Although we are performing the double-scaled limit, hence from the observations in Section \ref{sec:Glambdavsp} we would in principle be able to obtain the full $\lambda$ dependence of the $\mathcal{O}(1/p)$ part of the propagator, the saddle point approximation clearly gives us only the dominant piece.

Varying the action with respect to $\sigma$ gives us:
\begin{gather}
    - \int d\tau_3 \, d\tau_4 \: A(\tau_2,\tau_3) \sigma(\tau_3,\tau_4) A(\tau_4,\tau_1) = A(\tau_2,\tau_1) g(\tau_2,\tau_1). \label{eq:cDSSYKdsigma}
\end{gather}

Varying the action with respect to $g$, instead, gives us:
\begin{equation}
\begin{gathered}
    \mathcal{J}^2 (1-4C^2)^{p/2} \exp \left( \frac{g(\tau,\tau') + g(\tau',\tau)}{2} \right) = \sigma(\tau,\tau') A(\tau',\tau) \\
    \implies \sigma(\tau,\tau') = -4 \mathcal{J}^2 (1-4C^2)^{p/2-1} A(\tau,\tau') \exp \left( \frac{g(\tau,\tau') + g(\tau',\tau)}{2} \right).
\end{gathered}
\end{equation}

This $\sigma$ tells us that, for arbitrary $C$, $\Sigma$ is not antisymmetric and therefore which $C$ appears in $(\delta \partial_\tau)^{-1}$ is not irrelevant. Again, having the same $C$ for $(\delta \partial_\tau)^{-1}$ and $G$ is both the most natural choice and the only one that cancels out ``divergent'' ($\propto p^1$) terms in the partition function. If we plug this result in \eqref{eq:cDSSYKdsigma} and take derivatives on both sides with respect to $\tau_1$ and $\tau_2$, we obtain:
\begin{gather}
    \partial^2_\tau (A(\tau)g(\tau)) = 4\mathcal{J}^2 (1-4C^2)^{p/2-1} A(\tau) \exp\left( \frac{g(\tau) + g(-\tau)}{2} \right). \label{eq:cSYKg(p)}
\end{gather}

We have used that the theory is invariant under time translations, so that $g(\tau_2,\tau_1) = g(\tau \equiv \tau_2-\tau_1)$ and similarly for $A(\tau_2,\tau_1)$. In the case of an ``uncharged'' scenario with $C = 0$ we recover \eqref{eq:SYKg(p)}, up to a different factor of $2$.

In order to understand how to integrate $\sigma$ out in the partition function, we will use discrete indices. We have to be really careful in the following, because a naive approach would be unable to correctly account  for the sign functions. This is actually a mistake that is committed in \cite{Goel:2023svz}, whose end result is the lack of all the sign functions in \eqref{eq:cSYKg(p)} and a correction $g(\tau)$ to the free, ``uncharged'' propagator that only works for positive arguments. What we have here has the following structure:
\begin{gather}
    \int \mathcal{D}\sigma_{ij} \: \exp \left( -\frac{1}{2} \frac{2}{\lambda} \sigma_{ij} {B^{ij}}_{lm} \sigma^{lm} -\frac{2}{\lambda} \sigma_{ij} C^{ij} \right) = \exp \left( +\frac{1}{\lambda} C_{ij} {(B^{-1})^{ij}}_{lm} C^{lm} \right), \quad {B^{ij}}_{lm} = {A^j}_{l} {A_m}^{i}.
\end{gather}

The inverse matrix satisfies ${(B^{-1})^{ij}}_{lm} {B^{lm}}_{kr} = \delta^i_k \delta^j_r$. We claim that the inverse matrix is ${(\delta \partial_\tau)^i}_m {(-\delta \partial_\tau)_l}^j$. Indeed, going back to continuous indices, we obtain:
\begin{gather}
    \left( \int d\tau_m \: \delta(\tau_i-\tau_m)\partial_{\tau_m} A(\tau_m-\tau_k) \right) \left( -\int d\tau_l \: \delta(\tau_l-\tau_j) \partial_{\tau_l} A(\tau_r-\tau_l) \right) = \delta(\tau_i-\tau_k) \delta(\tau_r-\tau_j).
\end{gather}

It follows that the integration yields, after an additional integration by parts:
\begin{equation}
\begin{aligned}
    \langle \mathcal{Z} \rangle_J = \int \mathcal{D}g \: \exp \Bigg( \frac{2\mathcal{J}^2}{\lambda} \int d\tau \, d\tau' \: (1-4C^2)^{p/2} \exp \left( \frac{g(\tau,\tau')+g(\tau',\tau)}{2} \right) \\
    + \frac{1}{\lambda} \int d\tau \, d\tau' \: \partial_\tau (A(\tau,\tau')g(\tau,\tau')) \partial_{\tau'} (A(\tau',\tau)g(\tau',\tau)) \Bigg). \label{eq:Zmu0pre}
\end{aligned}
\end{equation}

We argue that, in principle, we cannot assume $g(\tau,\tau^\pm) = 0$ for arbitrary $\mu$ and $\beta$, which will be important later. First, $G(\tau,\tau')_{\tau'\to\tau}$ (almost) reduces to the free theory propagator: in the usual SYK model, this was due to the equal time Clifford algebra constraint. In this variation, although the fact that the couplings are dimensionful still holds, there is another deep reason for which this has to be the case. Our theory, in fact, possesses a $U(1)$ global symmetry $\psi_i \to e^{i\alpha} \psi_i$, with $\alpha$ a real constant. We recall that the conserved charge associated to this new symmetry is:
\begin{gather}
    Q(\tau) = \frac{1}{N} \sum_{i=1}^N \psi_i^\dagger(\tau)\psi_i(\tau) - \frac{1}{2}.
\end{gather}

Both in the interacting and non-interacting model, then, we find that by definition:
\begin{gather}
    G(\tau,\tau')_{\tau'\to\tau^+} = -\langle Q \rangle - \frac{1}{2},
\end{gather}

provided we are considering $G$ as the actual propagator of the theory and not as a variable of the path integral. The mean charge density $\langle Q \rangle$ is computed in the grandcanonical ensemble in the following way:
\begin{gather}
	\langle Q \rangle \equiv \frac{1}{N} \frac{\tr[e^{-\beta (H-\mu N Q)} \sum_{i=1}^N \psi_i^\dagger(\tau) \psi_i(\tau)]}{\tr[e^{-\beta(H-\mu N Q)}]} - \frac{1}{2}.
\end{gather}

Since this is a conserved charge of the theory, it is time independent and is a fixed parameter of the system that depends on $\mu$. By using definitions, we can conclude:
\begin{equation}
\begin{gathered}
	\langle Q \rangle = -\frac{1}{2} -\lim_{\tau' \to \tau^+} G(\tau,\tau') = -\frac{1}{2} - \left( -\frac{1}{2} + C \right) \left( 1 + \frac{g(0^-)}{p} \right) \implies \langle Q \rangle = -C + \frac{g(0^-)}{p} \left( \frac{1}{2} - C \right), \\
	-\frac{1}{2} \leq \langle Q \rangle \leq +\frac{1}{2}. \label{eq:QvsCg0}
\end{gathered}
\end{equation}

The range for $\langle Q \rangle$ is determined by the fact that it is the normalized sum of $N$ terms that range from $0$ to $1$ (the expectation value on the vacuum of the number of fermions of each kind), reduced by $1/2$. Our definition of the mean charge $\langle Q \rangle$ is the one that is null at half-filling, so that deviations from the ``uncharged'' results are clearly tied to the filling of the system. The picture that arises is that fermions and ``holes'' at each ``site'' (type of fermion) have opposite charges (equal to $\pm 1/2$ respectively), so only if they are equal in number we have $\langle Q \rangle = 0$. This relation is actually able to fix and give a physical interpretation to $C(\mu)$: since ultimately $G(\tau) = G_{\rm free}(\tau) (1+g(\tau)/p)$, $C(\mu)$ is minus the mean charge the system would have with the same $\mu$ if it were free:
\begin{gather}
	C(\mu) = -\langle Q \rangle_{\rm free}(\mu).
\end{gather}

This is further substantiated by the fact that $A(\tau,\tau')$ emerges in the path integral in a way that is completely independent of the $\propto \mathcal{J}^2$ part of the effective action, hence it has to be the same regardless of whether interactions are present or not. In the following, we rename $\langle Q \rangle \to \langle Q \rangle_{\rm true}$ and $\langle Q \rangle_{\rm free} \to \langle Q \rangle$ for convenience.

Equation \eqref{eq:QvsCg0} tells us that interactions alter the relation between $\langle Q \rangle$ and $\mu$, and in this sense $G(\tau,\tau')_{\tau'\to\tau}$ is not exactly the free theory propagator here. Interestingly, it has been determined in the literature that the $\langle Q \rangle (\mu)$ relation is only corrected at order $\mathcal{O}(1/p^2)$ \cite{Davison_2017}, from which we infer that $g(0^-) = 0$. In this section, in any case, we are setting the chemical potential $\mu$ to $0$, so both $\langle Q \rangle_{\rm true}$ and $\langle Q \rangle$ are null, which implies that $G(\tau,\tau') = -G(\tau',\tau)$ and $g(\tau,\tau') = g(\tau',\tau)$, again with $g(0) = 0$.

When evaluating the saddle point of the path integral and possibly expanding around it to compute corrections, one can refer to the following proof. We underline that this is actually a proof of the two point function's behavior (its vacuum expectation value), not of the $G$ variable inside the path integral, though this is consistent when considering the $\lambda \to 0$ saddle point of the final result. We know that, up to a normalization factor:
\begin{equation}
	\begin{gathered}
		G(\tau,\tau') = \frac{1}{N} \sum_{i=1}^{N} \bra{\Omega} \mathrm{T} [\psi_i(\tau) \bar\psi_i(\tau')] \ket{\Omega} \propto \int [\mathcal{D}\bar\psi \, \mathcal{D}\psi] \, \sum_{i=1}^N \psi_i(\tau) \bar\psi_i(\tau') \, \exp (-S_{\rm eff}[\psi,\bar\psi]), \\
		S_{\rm eff}[\psi,\bar\psi] = \int d\tau \, \sum_{i=1}^N \bar\psi_i \partial_t \psi_i - \frac{N\mathcal{J}^2}{p^2} \int d\tau \, d\tau' \: \left( -4 G(\tau,\tau')G(\tau',\tau) \right)^{p/2}.
	\end{gathered}
\end{equation}

Even in the case of finite temperature, we can integrate the kinetic term by parts and anticommute the variables. Since the measure is made up of an even ($N$) number of fermions, we can anticommute all of them and obtain $[\mathcal{D}\bar\psi \, \mathcal{D}\psi] = [\mathcal{D}\psi \, \mathcal{D}\bar\psi]$. We can also anticommute the fermionic variables in $G(\tau,\tau')G(\tau',\tau)$. If we now redefine our variables by exchanging $\psi \longleftrightarrow \bar\psi$, we obtain:
\begin{gather}
	G(\tau,\tau') \propto \int [\mathcal{D}\bar\psi \, \mathcal{D}\psi] \, \sum_{i=1}^N \bar\psi_i(\tau)\psi_i(\tau') \, \exp (-S_{\rm eff}[\psi,\bar\psi]) = -G(\tau',\tau).
\end{gather}

Going back to \eqref{eq:Zmu0pre}, using that $g(\tau,\tau) = 0$ naively would give us:
\begin{equation}
\begin{aligned}
    \langle \mathcal{Z} \rangle_J = \int \mathcal{D}g \: \exp \Bigg( \frac{2\mathcal{J}^2}{\lambda} \int d\tau \, d\tau' \: (1-4\langle Q \rangle^2)^{p/2} \exp \left( \frac{g(\tau,\tau')+g(\tau',\tau)}{2} \right) \\
    + \frac{1}{\lambda} \int d\tau \, d\tau' \: A(\tau,\tau') A(\tau',\tau) \partial_\tau g(\tau,\tau') \partial_{\tau'} g(\tau',\tau) \Bigg). \label{eq:Zmu0}
\end{aligned}
\end{equation}

This operation is performed in \cite{Goel:2023svz} but it is not correct, as we will soon show.

Using that $\langle Q \rangle = 0$ and $g(\tau,\tau')=g(\tau',\tau)$ to simplify \eqref{eq:Zmu0} would then yield:
\begin{align}
    \langle \mathcal{Z} \rangle_J = \int \mathcal{D}g \: \exp \Bigg( -\frac{1}{\lambda} \int d\tau \, d\tau' \: \left( \frac{1}{4} \partial_\tau g(\tau,\tau') \partial_{\tau'} g(\tau,\tau') - 2\mathcal{J}^2 e^{g(\tau,\tau')} \right) \Bigg).
\end{align}

Again, this partition function is incorrect. A sign of the underlying problem is the equation of motion of $g(\tau)$ that one obtains for the $\lambda \to 0$ saddle:
\begin{gather}
    \partial^2_\tau g(\tau) = 4\mathcal{J}^2 e^{g(\tau)}.
\end{gather}

We see a mismatch with \eqref{eq:cSYKg(p)}, namely the absence of the sign functions. Their presence is crucial, since their effect is making $g(\tau)$ a function of $|\tau|$, therefore granting that it is even. Without them, it is impossible to require that both $g(\tau)$ be even and satisfy $g(0) = g(\beta) = 0$ for an arbitrary inverse temperature $\beta$. We now check what goes wrong when going from \eqref{eq:Zmu0pre} to \eqref{eq:Zmu0} by performing all the derivatives:
\begin{equation}
\begin{aligned}
    & \int d\tau \, d\tau' \: \partial_\tau (A(\tau,\tau')g(\tau,\tau')) \partial_{\tau'} (A(\tau',\tau)g(\tau',\tau)) \\
    = & \int d\tau \, d\tau' \left( -\frac{1}{4} \partial_\tau g(\tau,\tau') \partial_{\tau'} g(\tau',\tau) + 2\delta(\tau-\tau')g(\tau,\tau')A(\tau',\tau)\partial_{\tau'} g(\tau',\tau) + \delta^2(\tau-\tau') g(\tau,\tau')g(\tau',\tau) \right) \\
    = & \int d\tau \, d\tau' \: \left( -\frac{1}{4} \partial_\tau g(\tau,\tau') \partial_{\tau'} g(\tau',\tau) \right) + 0 + \delta(0) \, g^2(0) \int d\tau.
\end{aligned}
\end{equation}

The issue is that throwing away the extra terms is getting rid of a contribution to the action that depends on $g(\tau)$ and is not clearly null. As a consequence, varying the second line with respect to $g(\tau,\tau')$ to find the equation of motion yields:
\begin{equation}
\begin{aligned}
    & -\frac{1}{2} \partial^2_\tau g(\tau) + 4\delta(\tau)A(-\tau)\partial_\tau g(\tau) +2 \delta'(\tau) A(-\tau) g(\tau) + 2\mathcal{J}^2 e^{g(\tau)} = 0 \\
    \implies & A(\tau) \partial^2_\tau g(\tau) + 2\delta(\tau) \partial_\tau g(\tau) + \delta'(\tau) g(\tau) = \partial^2_\tau (A(\tau)g(\tau)) = 4\mathcal{J}^2 A(\tau) e^{g(\tau)}.
\end{aligned}
\end{equation}

What we find is that the terms neglected in \cite{Goel:2023svz} give an important contribution to the equations of motion and are needed to ``reconstruct'' the derivatives of the sign function.

This subtlety aside, the result we have obtained for the partition function is similar to \cite{Goel:2023svz}, and a similar analysis of the saddle point solution and the one-loop corrections can be carried out here. Studying the $\mu = 0, \: \beta = +\infty$ case has been pretty useful, but there would be nothing new to learn by proceeding with these assumptions. We are now ready to see how the situation changes once we add a chemical potential and, therefore, a mean electric charge.

\section{Path Integral at \texorpdfstring{$\mu \neq 0, \: \beta = +\infty$}{mu != 0, beta = +inf}}
To study the effects of turning on a chemical potential $\mu$ that couples to the conserved $U(1)$ charge, we add a $+\int d\tau \: \mu \left( \sum_{i=1}^N \bar\psi_i(\tau)\psi_i(\tau) - N/2 \right)$ term to the exponential in the path integral. 

This addition implies that $\langle Q \rangle$ can be linked to $\mu$ by noting that:
\begin{gather}
	\langle Q \rangle = \frac{1}{\beta N} \frac{\partial \log \langle \mathcal{Z} \rangle_J (\beta,\mu)}{\partial \mu}. \label{eq:QdlogZ}
\end{gather}

This relation tells us that $\langle Q \rangle$ is an odd function of $\mu$ both in the free and in the interacting theory, since $\langle Z \rangle_J$ is an even function of $\mu$. To see this, we use the following equivalent definition of the charge:
\begin{gather}
	Q(\tau) = \frac{1}{2} - \frac{1}{N} \sum_{i=1}^N \psi_i(\tau) \psi_i^\dagger(\tau),
\end{gather}

so that the path integral becomes:
\begin{gather}
	\langle \mathcal{Z} \rangle_J(\beta, \mu) = \int [\mathcal{D}\bar\psi \, \mathcal{D}\psi] \,  \exp \left[ -S_{\rm eff}[\psi,\bar\psi] - \mu \int d\tau \left( \sum_{i=1}^N \psi_i(\tau) \bar\psi_i(\tau) - \frac{N}{2} \right) \right], \\
	S_{\rm eff}[\psi,\bar\psi] = \int d\tau \, \sum_{i=1}^N \bar\psi_i \partial_t \psi_i - \frac{N\mathcal{J}^2}{p^2} \int d\tau \, d\tau' \: \left( -4 G(\tau,\tau')G(\tau',\tau) \right)^{p/2}.
\end{gather}

If we now exchange $\psi \longleftrightarrow \bar\psi$, we have already determined that $S_{\rm eff}$ is mapped into itself and that the measure of integration does not change, so that the only effective variation is having mapped $\mu$ into $-\mu$. This is just a change of variables, so we infer that $\langle Z \rangle_J(\beta,\mu) = \langle Z \rangle_J(\beta, -\mu)$.

By repeating the same computations of the previous section, it is straightforward to see that the main difference lies in a different fermionic determinant:
\begin{gather}
    \det [\delta(\tau-\tau')\partial_{\tau'} + \Sigma(\tau,\tau')] \to \det [\delta(\tau-\tau')(\partial_{\tau'} - \mu) + \Sigma(\tau,\tau')].
\end{gather}

Our function $A(\tau,\tau')$ is now the solution of a new differential equation:
\begin{gather}
    \partial_\tau A(\tau,\tau') - \mu A(\tau,\tau') = \delta(\tau-\tau') \implies A(\tau,\tau') = \left( \frac{1}{2} \mathrm{sgn}(\tau-\tau') + C \right) e^{\mu(\tau-\tau')}.
\end{gather}

It is always true that $G(\tau,\tau') = A(\tau,\tau')(1 + g(\tau,\tau')/p)$.

Restarting from \eqref{eq:Zsigma} with the modified $A(\tau,\tau')$, we once again need to determine the inverse matrix ${(B^{-1})^{ij}}_{lm}$. Let us verify that it is ${(\delta(\partial_\tau-\mu))^i}_m {(-\delta(\partial_\tau+\mu))_l}^j$:
\begin{equation}
\begin{gathered}
    \left( \int d\tau_m \: \delta(\tau_i-\tau_m)(\partial_{\tau_m}-\mu) A(\tau_m-\tau_k) \right) \left( -\int d\tau_l \: \delta(\tau_l-\tau_j) (\partial_{\tau_l}+\mu) A(\tau_r-\tau_l) \right) \\
    = \delta(\tau_i-\tau_k) \delta(\tau_r-\tau_j).
\end{gathered}
\end{equation}

After integrating by parts, the partition function is now:
\begin{equation}
\begin{aligned}
    \langle \mathcal{Z} \rangle_J = \int \mathcal{D}g \: \exp \Bigg( \frac{2\mathcal{J}^2}{\lambda} \int d\tau \, d\tau' \: ( 1 - 4\langle Q \rangle^2 )^{p/2} \exp \left( \frac{g(\tau,\tau')+g(\tau',\tau)}{2} \right) \\
    + \frac{1}{\lambda} \int d\tau \, d\tau' \: (\partial_\tau - \mu) (A(\tau,\tau')g(\tau,\tau')) (\partial_{\tau'} - \mu) (A(\tau',\tau)g(\tau',\tau)) \Bigg). \label{eq:cSYKallincludZ}
\end{aligned}
\end{equation}

We want to study the structure of the saddle point solution and its dependence on $\mu$. Using the invariance under time translations, the equation of motion is:
\begin{gather}
    (\partial_\tau^2 - 2\mu \partial_\tau + \mu^2) (A(\tau)g(\tau)) = 4 \mathcal{J}^2 (1-4 \langle Q \rangle^2)^{p/2-1} A(\tau) \exp \left( \frac{g(\tau)+g(-\tau)}{2} \right). \label{eq:cSYKeom}
\end{gather}

From this equation, we deduce the important fact that the system experiences an effective coupling strength that depends on $\langle Q \rangle$:
\begin{gather}
    \mathcal{J}^2_{\rm eff} (\langle Q \rangle) = \mathcal{J}^2(1-4\langle Q \rangle^2)^{p/2-1}.
\end{gather}

As soon as $\langle Q \rangle \neq 0$, the effective coupling strength immediately drops to $0$ because of the $p \to \infty$ limit. We choose to slightly modify the system, then, by fixing $\mathcal{J}_{\rm eff}$ rather than $\mathcal{J}$ or, equivalently, by imposing that the coupling strength $\mathcal{J}$ scales as $\mathcal{J}(\langle Q \rangle) \propto (1-4\langle Q \rangle^2)^{-(p-2)/4}$. We will study this equation further in the next section, where we will consider the most general thermal scenario.

\section{Path Integral at \texorpdfstring{$\mu \neq 0, \, \beta \neq +\infty$}{mu not 0, beta not infinite}} \label{sec:finalPI}
The previous computation was performed at zero temperature, so we now turn to the finite temperature scenario. In this case, one significant relation for the fermionic two point function $G(\tau,\tau') = G(\tau-\tau')$ is the Kubo-Martin-Schwinger (KMS) relation:
\begin{gather}
    G(\tau+\beta) = -e^{-\beta\mu} G(\tau). \label{eq:KMSrel}
\end{gather}

There is an extremely important hypothesis underlying this result: we must evolve fermions with $H$ for this condition to be satisfied. The Green function we are studying is the one associated to fermions that are actually time-evolved with $H-\mu N Q$, which implies that it simply satisfies $G(\tau+\beta) = -G(\tau)$. The only difference between the two, though, is simply an extra $e^{\mu\tau}$ factor. All these statements are proven in Appendix \ref{app:KMSSYK}. Since our objective is to determine $g(\tau)$ and the propagator associated to $H-\mu N Q$ is the one that arises naturally from our path integral treatment of the system, we will keep on using this one, although a trivial relation connects the two.

What we have to do is find the finite temperature free propagator, then solve \eqref{eq:cSYKeom} with the appropriate boundary conditions. For small $\tau$, all functions are indistinguishable from their zero temperature counterpart, so all the observations that previously assumed $\tau' \to \tau^\pm$ still hold.

By recalling \eqref{eq:circlemap}, the thermal generalization of $A(\tau)$ (which solves the same differential equation regardless of temperature) is:
\begin{gather}
    A(\tau) = \left( \frac{1}{2} \mathrm{sgn} \left( \sin \left( \frac{\pi\tau}{\beta} \right) \right) - \langle Q \rangle \right) e^{\mu\tau}. \label{eq:AtauFreeMu}
\end{gather}

Consequently, the $p \to \infty$ limit of the two point function is:
\begin{gather}
    G(\tau) = A(\tau) \left( 1 + \frac{g(\tau)}{p} \right).
\end{gather}

The calculations we have performed in the previous section play out in the same way here, so we find \eqref{eq:cSYKeom} once again. The only difference lies in the different boundary conditions, so that the solution of the previous section is simply the $\beta \to +\infty$ limit of the general case.

In Appendix \ref{app:chargeSYK}, we prove the following important result:
\begin{gather}
	\langle Q \rangle (\mu) = \frac{1}{2} \tanh \left( \frac{\beta\mu}{2} \right).
\end{gather}

This relation correctly satisfies $\langle Q \rangle (-\mu) = - \langle Q \rangle (\mu)$ and $-1/2 \leq \langle Q \rangle \leq 1/2$. We recall that this is the relation between the charge and the chemical potential in the free theory, but that it coincides with the actual $\langle Q \rangle_{\rm true} (\mu)$ up to irrelevant $\mathcal{O}(1/p^2)$ corrections.

To solve \eqref{eq:cSYKeom}, we decompose $g(\tau)$ into a symmetric and an antisymmetric part:
\begin{equation}
\begin{gathered}
    g(\tau,\mu) = g_S(|\tau|,\mu) + \beta\mu \, \mathrm{sgn}(\tau) g_A(|\tau|,\mu). \label{eq:cSYKsymantisym}
\end{gathered}
\end{equation}

The dimensionless $\beta\mu$ prefactor in front of $g_A$ is a good ansatz, since the antisymmetric part is absent when $\mu = 0$.

If we integrate \eqref{eq:cSYKeom} in $[-\varepsilon,\varepsilon], \: \varepsilon \ll 1$, we obtain:
\begin{equation}
\begin{aligned}
    \partial_\tau(Ag)(0^+) -2\mu (Ag)(0^+) & = \partial_\tau(Ag)(0^-) -2\mu (Ag)(0^-) \\
    -\mu A(0^+) g(0^+) + A(0^+) \dot g(0^+) & = -\mu A(0^-)g(0^-) + A(0^-) \dot g(0^-).
\end{aligned}
\end{equation}

We have used that $A(\tau)$ satisfies $\partial_\tau A(\tau) = \delta(\tau) + \mu A(\tau)$. Note that $g(\tau)$ is usually not continuous because of the $g_A$ piece that arises when $\mu \neq 0$, with $g_A(0^+) \neq 0$ in general. The left hand side is
\begin{gather}
    \left( \frac{1}{2} - \langle Q \rangle \right) (-\mu g_S(0^+) - \beta\mu^2 g_A(0^+) + \dot g_S(0^+) + \beta\mu \dot g_A(0^+)),
\end{gather}

while the right hand side is
\begin{gather}
    \left( -\frac{1}{2} - \langle Q \rangle \right) (-\mu g_S(0^+) + \beta\mu^2 g_A(0^+) - \dot g_S(0^+) + \beta\mu \dot g_A(0^+)).
\end{gather}

We have used that $g_{S,A}(0^+) = \pm g_{S,A}(0^-), \: \dot g_{S,A}(0^+) = \mp \dot g_{S,A}(0^-)$.

Equating them yields:
\begin{gather}
    2\langle Q \rangle (\beta\mu^2 g_A(0^+) -\dot g_S(0^+)) = \mu(g_S(0^+)- \beta \dot g_A(0^+)). \label{eq:cSYKboundaryA}
\end{gather}
    
The differential equation in $\tau = 0$ has given us this boundary condition. Other boundary conditions come from the periodicity of the propagator $G(\tau) = -G(\tau-\beta)$, which implies $g(\tau) = g(\tau-\beta)$. Since $g(\beta) = g(0^-)$ has to be true for all $\mu$, it has to hold in particular for all powers of $\mu$ independently, that is:
\begin{gather}
	g_S(\beta) = g_S(0^-) = g_S(0^+), \qquad g_A(\beta) = g_A(0^-) = -g_A(0^+).
\end{gather}

If we perform a series expansion of the solution, in fact, we will obtain:
\begin{gather}
    g_S(\tau) = \sum_{k = 0}^{\infty} (\beta\mu)^k g_S^{(k)}(\tau), \qquad \beta \mu g_A(\tau) = \sum_{k=0}^{\infty} (\beta\mu)^{k+1} g_A^{(k)}(\tau), \label{eq:cSYKgAgSExp}
\end{gather}

The fourth and final boundary condition comes from \eqref{eq:QvsCg0}:
\begin{gather}
	g(0^-) = g_S(0^+) - \beta\mu g_A(0^+) = 2p \, \frac{\langle Q \rangle_{\rm true}(\mu) - \frac{1}{2} \tanh \frac{\beta\mu}{2}}{1+\tanh \frac{\beta\mu}{2}} = 0. \label{eq:cSYK4thBound}
\end{gather}

We have used that $\langle Q \rangle_{\rm true}$ is indistinguishable from $\langle Q \rangle$ at the relevant order $\mathcal{O}(1/p)$. This $g(0^-) = 0$ condition is formally not required in the degenerate case $\beta\mu = -\infty$, but we impose its validity everywhere by extrapolation. We have found the necessary four boundary conditions, which in principle apply to all the $g_S^{(k)}, \: g_A^{(k)}$ separately. We can now restrict ourselves to $\tau > 0$, as the extension to $\tau < 0$ is obvious. We obtain:
\begin{gather}
	(\mu^2 A(\tau)g(\tau) + 2\mu A(\tau) \dot g(\tau) + A \ddot g(\tau)) - (2\mu^2 A(\tau)g(\tau) + 2\mu A(\tau) \dot g(\tau)) + \mu^2 A(\tau)g(\tau) = 4\mathcal{J}^2_{\rm eff} A(\tau) e^{g_S(\tau)}
\end{gather}

We can get rid of $A(\tau)$ everywhere (for example by multiplying by $A(-\tau)$ on both sides) and then insert the decomposition \eqref{eq:cSYKsymantisym}. The resulting equation is actually a sum of two separate differential equations, which we can identify through the opposite behavior under a $\tau \to -\tau$ symmetry (that is, the part that switches sign for $\tau < 0$ and the part that doesn't). Before doing this, though, we first notice that we can use dimensionless variables $\tilde \mu \equiv \mu/\mathcal{J}_{\rm eff}, \: \tilde \tau \equiv \mathcal{J}_{\rm eff} \tau, \: \tilde \beta \equiv \mathcal{J}_{\rm eff} \beta$, and obtain:
\begin{gather}
	\partial^2_{\tilde \tau} g(\tilde \tau) = 4 e^{g_S(\tilde \tau)}.
\end{gather}

From now on, $\mu, \tau$ and $\beta$ will be used to refer to these dimensionless variables (this is equivalent to setting $\mathcal{J}_{\rm eff} = 1$). The system of differential equations is:
\begin{equation}
\begin{cases}
	\ddot g_S(\tau) = 4e^{g_S(\tau)} \\
	\beta \mu \ddot g_A(\tau) = 0
\end{cases}.
\end{equation}

The first equation is even under $\tau \to -\tau$, while the second one is odd. Our decomposition \eqref{eq:cSYKsymantisym} is therefore a solution for all $\tau$. This system can be easily solved analytically for any value of $\mu$:
\begin{equation}
\begin{gathered}
	g_S(\tau) = \log \left( \frac{c_1^2}{2 \cos^2 (c_1 \tau + c_2)} \right), \\
	g_A(\tau) = c_3 \left( \frac{\tau}{\beta} - \frac{1}{2} \right).
\end{gathered}
\end{equation}

We have already imposed the correct periodicity for $g_A(\tau)$. In the case of $g_S(\tau)$, the function needs to satisfy:
\begin{equation}
\begin{gathered}
	\cos^2(c_1 \abs{\tau} + c_2) = \cos^2(c_1 \abs{\tau-\beta} + c_2) \stackrel{\tau > 0}{=} \cos^2(c_1 \beta + c_2 - c_1 \tau) = \cos^2(c_1 \tau - c_2 - c_1 \beta) \\
	\implies c_1\beta = k\pi - 2c_2, \quad k \in \mathbb{Z}.
\end{gathered}
\end{equation} 

All these possibilities are ascribable to only two different scenarios, depending on whether $k$ is even or odd. If $k = 2r$ is even, we obtain:
\begin{gather}
	g_S(\tau) = \log \left( \frac{c_1^2}{2 \cos^2 \left( c_1 \abs{\tau} - c_1 \frac{\beta}{2} + r\pi \right)} \right) = \log \left( \frac{c_1^2}{2 \cos^2 \left( c_1 \abs{\tau} - c_1 \frac{\beta}{2} \right)} \right).
\end{gather}

If $k = 2r+1$ is odd, on the other hand, we obtain:
\begin{gather}
		g_S(\tau) = \log \left( \frac{c_1^2}{2 \cos^2 \left( c_1 \abs{\tau} - c_1 \frac{\beta}{2} + r\pi + \frac{\pi}{2} \right)} \right) = \log \left( \frac{c_1^2}{2 \sin^2 \left( c_1 \abs{\tau} - c_1 \frac{\beta}{2} \right)} \right).
\end{gather}

We know from \eqref{eq:thermalSYKprop} (the only difference with that study is $\mathcal{J} \to \sqrt{2} \mathcal{J}_{\rm eff}$) that $c_1\beta + 2c_2 = 0$ for $\mu = 0$, which implies the presence of the cosine at the denominator. This choice is the right one, because there is always a solution of the equation:
\begin{gather}
	\frac{2}{\beta^2} = \frac{\cos^2 c_2}{c_2^2},
\end{gather}

due to the codomain of the r.h.s. (as a function of $c_2$) being $\mathbb{R}^+$. On the other hand, since $\sin^2 x/x^2 \leq 1$, there are no real solutions of the following equation if $\beta < \sqrt{2}$:
\begin{gather}
	\frac{2}{\beta^2} = \frac{\sin^2 c_2}{c_2^2}.
\end{gather}

This is convincing evidence that we should always pick the cosine. In any case, if we reasonably assume that $(c_1\beta + 2c_2)(\mu)$ is a continuous function of $\mu$, it has to be $0$ for all values of $\mu$. As a consequence, we are left with two parameters $c_2(\mu)$ and $c_3(\mu)$, which must be fixed through the remaining two boundary conditions.

We now have:
\begin{gather}
	g_S(\tau) = \log \left( \frac{2c_2^2}{\beta^2 \cos^2 \left[ c_2 \left( 1-\frac{2 \abs{\tau}}{\beta} \right) \right]} \right), \quad \dot g_S(\tau) = - \mathrm{sgn}(\tau) \frac{4c_2}{\beta} \tan \left[ c_2 \left( 1-\frac{2 \abs{\tau}}{\beta} \right) \right]
\end{gather} 

Due to the $c_2 \longleftrightarrow -c_2$ symmetry of $g_S(\tau)$, we will always assume $c_2 > 0$. Plugging what we have found into \eqref{eq:cSYK4thBound} allows us to determine $c_3(\mu)$ as a function of $c_2(\mu)$:
\begin{gather}
	c_3(\mu) = - \frac{2}{\beta\mu} \log \left( \frac{2c^2_2(\mu)}{\beta^2 \cos^2 c_2(\mu)} \right). \label{eq:c3munumb1}
\end{gather}

On the other hand, plugging these functions into \eqref{eq:cSYKboundaryA} also yields:
\begin{gather}
	c_3(\mu) = \frac{1}{1-\frac{\beta\mu}{2} \tanh \frac{\beta\mu}{2}} \left[ \log \left( \frac{2c^2_2(\mu)}{\beta^2 \cos^2 c_2(\mu)} \right) - \frac{4\tanh \frac{\beta\mu}{2}}{\beta\mu} \, c_2(\mu) \tan c_2(\mu) \right].
\end{gather}

Finally, equating these two different expressions fixes $c_2(\mu)$:
\begin{gather}
	- \frac{2}{\beta\mu} \log \left( \frac{2c^2_2(\mu)}{\beta^2 \cos^2 c_2(\mu)} \right) = \frac{1}{1-\frac{\beta\mu}{2} \tanh \frac{\beta\mu}{2}} \left[ \log \left( \frac{2c^2_2(\mu)}{\beta^2 \cos^2 c_2(\mu)} \right) - \frac{4\tanh \frac{\beta\mu}{2}}{\beta\mu} \, c_2(\mu) \tan c_2(\mu) \right]. \label{eq:whatisc2mu}
\end{gather}

There are two limiting cases in which computations can be carried out analytically, namely the $\beta\abs{\mu} \ll 1$ and $\beta\abs{\mu} \gg 1$ limits.

We first determine the $\mathcal{O}(\beta\mu)$ corrections to $g(\tau)$ for small $\beta\mu$. In this case, we can write:
\begin{equation}
\begin{gathered}
	c_2(\mu) \approx c_2^{(0)} + (\beta\mu) \, c_2^{(1)}, \quad c_3(\mu) \approx - \frac{4(1+c_2^{(0)} \tan c_2^{(0)})}{c_2^{(0)}} \, c_2^{(1)}, \\
	c_2^{(0)} = \frac{\pi v}{2}, \qquad \frac{\frac{\pi v}{2}}{\cos \frac{\pi v}{2}} = \frac{1}{\sqrt{2}} \, \beta.
\end{gathered}
\end{equation}

$c_2^{(1)}$ satisfies the $\mathcal{O}((\beta\mu)^0)$ part of \eqref{eq:whatisc2mu}:
\begin{equation}
\begin{gathered}
	2 c_2^{(0)} \tan c_2^{(0)} - \frac{4(1 + c_2^{(0)} \tan c_2^{(0)})}{c_2^{(0)}} \, c_2^{(1)} = 0 \\
	\implies c_2^{(1)} = \frac{\left( c_2^{(0)} \right)^2 \tan c_2^{(0)}}{2 (1 + c_2^{(0)} \tan c_2^{(0)})}, \qquad c_3^{(0)} = -2 c_2^{(0)} \tan c_2^{(0)}.
\end{gathered}
\end{equation}

A systematic expansion of \eqref{eq:whatisc2mu} in powers of $\beta\mu$ can be carried out, which allows to determine the series expansion of $c_2(\mu) = \sum_{k=0}^{+\infty} (\beta\mu)^k c_2^{(k)}$ and, in turn, also the series expansion of $c_3(\mu) = \sum_{k=0}^{+\infty} (\beta\mu)^k c_3^{(k)}$.

We now turn to the $\beta\abs{\mu} \to +\infty$ limit of the above expressions, which depends on the sign of $\mu$. Up to exponentially suppressed terms that cannot be captured by a perturbative expansion, in fact, we have that $\tanh \frac{\beta\mu}{2} = \mathrm{sgn}(\mu)$. It is useful to redefine the antisymmetric coefficient according to $\beta \mu \, c_3(\mu) \to c_3(\mu)$, so that we do not have an artificial divergent prefactor in front anymore.
\begin{itemize}
	\item If $\mu > 0$, the leading order that survives in \eqref{eq:whatisc2mu} is $\mathcal{O}((\beta\mu)^{-2})$, which yields:
	\begin{gather}
		2 c_2(+\infty) \tan c_2(+\infty) = \log \left( \frac{2c^2_2(+\infty)}{\beta^2 \cos^2 c_2(+\infty)} \right). \label{eq:c2diagnoser}
	\end{gather}
	This equation implies a finite limit for $c_2(\mu \to +\infty)$, which in turn tells us that $c_3(+\infty) = \mathcal{O}((\beta\mu)^0)$:
	\begin{gather}
		c_3(+\infty) = -4 c_2(+\infty) \tan c_2(+\infty).
	\end{gather}
	
	\item If $\mu < 0$, the leading order that survives in \eqref{eq:whatisc2mu} is $\mathcal{O}((\beta\mu)^{-1})$, which yields:
	\begin{gather}
		\log \left( \frac{2c^2_2(-\infty)}{\beta^2 \cos^2 c_2(-\infty)} \right) = 0 \implies \frac{c_2(-\infty)}{\cos c_2(-\infty)} = \frac{1}{\sqrt{2}} \, \beta.
	\end{gather}
	This equation also implies a finite limit for $c_2(\mu \to -\infty)$, but now $c_3(-\infty) = \mathcal{O}((\beta\mu)^{-1})$ instead.
\end{itemize}

Just like before, we can perform a series expansion of $c_{2,3}(\mu)$ in powers of $(\beta\mu)^{-1}$: $c_{2,3}(\mu) = \sum_{k=0}^{+\infty} (\beta\mu)^{-k} c_{2,3}^{(k)}$. Note that the expansions will have different coefficients for different signs of $\mu$. As an application, we determine the first deviation of $c_2(\pm \infty)$ from its asymptotic constant value and the first correction to $c_3(\pm \infty)$.
\begin{itemize}
	\item If $\mu > 0$, the $\mathcal{O}((\beta\mu)^{-3})$ piece of \eqref{eq:whatisc2mu} implies:
	\begin{gather}
		c_2^{(1)} = 0.
	\end{gather}
	We now prove that $c_2^{(k \geq 1)} = 0$, which implies that the asymptotic value only receives $\mathcal{O}(e^{-\beta\mu})$ non-perturbative corrections. It is sufficient to check the consistency of this ansatz in \eqref{eq:whatisc2mu}, where we rename $x \equiv (\beta\mu)^{-1}$ and $c_2(+\infty) \equiv c_2$:
	\begin{equation}
	\begin{aligned}
		& -2x \log \left( \frac{2c^2_2}{\beta^2 \cos^2 c_2} \right) - \frac{1}{1-\frac{1}{2x}} \left[ \log \left( \frac{2c^2_2}{\beta^2 \cos^2 c_2} \right) - 4x c_2 \tan c_2 \right] \\
		= & -2x \log \left( \frac{2c^2_2}{\beta^2 \cos^2 c_2} \right) + \sum_{n=1}^{+\infty} (2x)^n \left[ \log \left( \frac{2c^2_2}{\beta^2 \cos^2 c_2} \right) - 4x c_2 \tan c_2 \right] \\
		= & \sum_{n=2}^{+\infty} (2x)^n \left[ \log \left( \frac{2c^2_2}{\beta^2 \cos^2 c_2} \right) - 2 c_2 \tan c_2 \right] = 0.
	\end{aligned}
	\end{equation}
	
	As for the antisymmetric part, we consequently find that $c_3^{(k \geq 1)} = 0$ and that $c_3(+\infty)$ only gets corrected at order $\mathcal{O}(e^{-\beta\mu})$.
	
	\item If $\mu < 0$, the $\mathcal{O}((\beta\mu)^{-2})$ piece of \eqref{eq:whatisc2mu} implies:
	\begin{gather}
		c_2^{(1)} = - \frac{\left( c_2^{(0)} \right)^2 \tan c_2^{(0)}}{1 + c_2^{(0)} \tan c_2^{(0)}}.
	\end{gather}
	
	Plugging this into \eqref{eq:c3munumb1} then yields:
	\begin{gather}
		c_3^{(1)} = 4 c_2^{(0)} \tan c_2^{(0)}.
	\end{gather}
\end{itemize}

Once again, a systematic expansion of the equations is possible: this procedure returns increasingly better approximations for $c_2(\mu)$ and $c_3(\mu)$ as one determines higher order terms.

\begin{figure}
	\centering
	\includegraphics[width = 0.6 \textwidth]{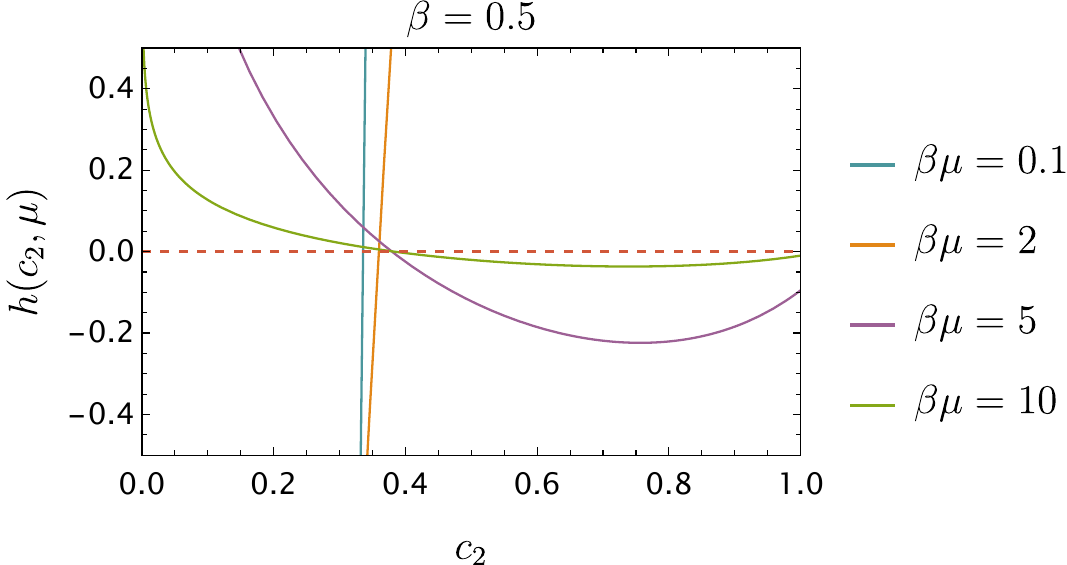} \\
	\includegraphics[width = 0.41 \textwidth]{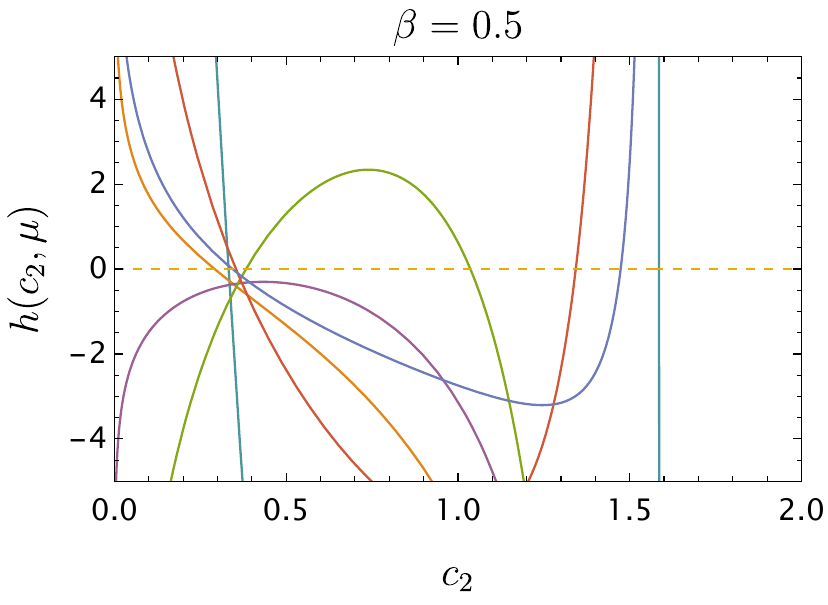} \quad
	\includegraphics[width = 0.55 \textwidth]{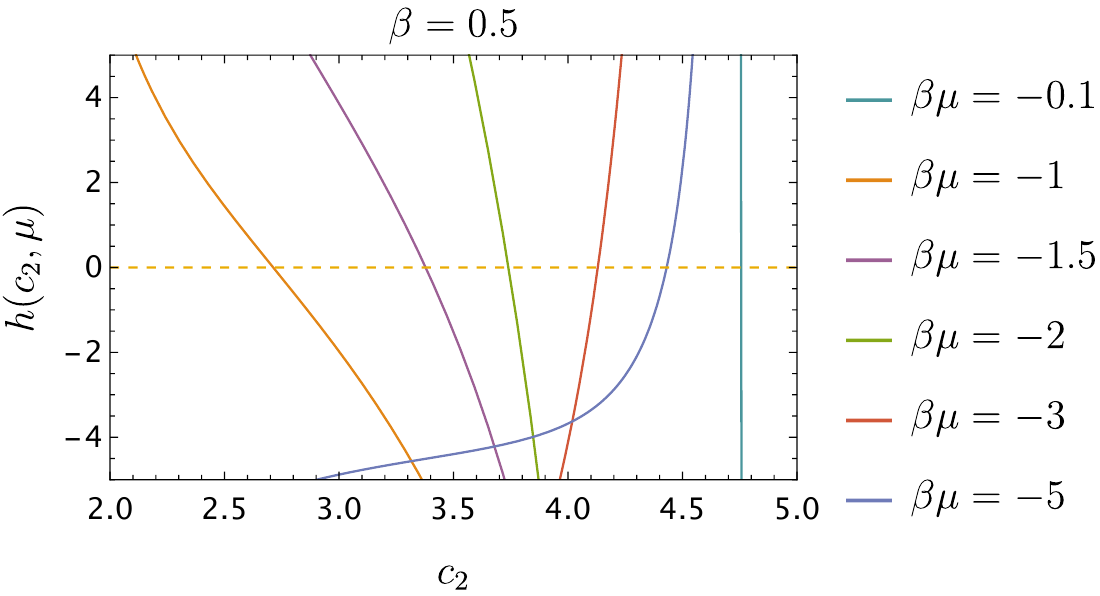} \\
	\includegraphics[width = 0.48 \textwidth]{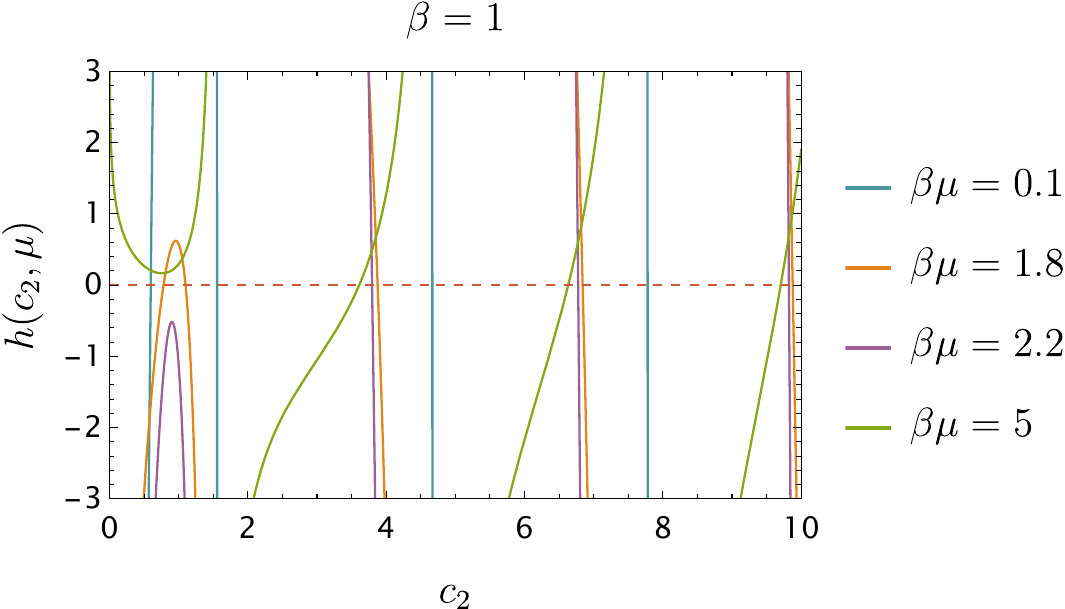} \quad
	\includegraphics[width = 0.49 \textwidth]{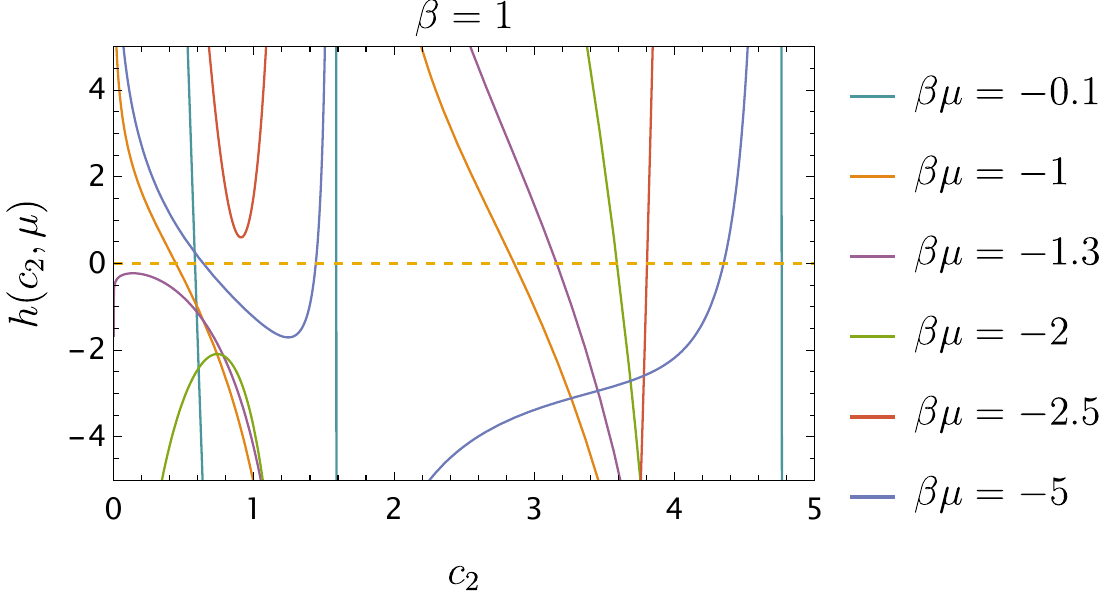}
	\caption{Plot of $h(c_2,\mu)$, which we define as the r.h.s. minus the l.h.s. of \eqref{eq:whatisc2mu}, as a function of $c_2$ for different $\mu$ and $\beta$. In principle, up to unknown mechanisms that uniquely determine a single one, every $c_2$ such that $h = 0$ is a valid solution of our system. We observe that there are infinitely many roots of $h = 0$, which merits further investigation. We can see that, for $\beta = 0.5$, every $\mu > 0$ allows a $c_2(\mu)$ that is continuously connected to the known $c_2(0)$, hence there is no ambiguity in our choice. For $\beta = 1$, on the other hand, when $\beta\mu$ is ``big'', the first branch does not intersect the $h = 0$ axis, therefore we are forced to pick a different branch and we have a discontinuity in the behavior of $c_2(\mu)$. For $\mu < 0$, regardless of the value of $\beta$, the first branch does not intersect the $h = 0$ line for a finite range of $\beta\mu$, before crossing it again with opposite concavity. We estimate that the concavity of the first branch flips for $\beta\mu \approx \pm 2.4$ (that is, the roots of $(\beta\mu/2) \tanh (\beta\mu/2) = 1$), which is where the denominator of the r.h.s. of $\eqref{eq:whatisc2mu}$ is singular.}
	\label{fig:cSYKmanysols}
\end{figure}

What one would miss from this asymptotic analysis alone is that, depending on the value of $\beta$ (or, equivalently, $\mathcal{J}_{\rm eff}$), $c_2(\mu)$ can be discontinuous multiple times as $\beta\mu$ varies from $-\infty$ to $+\infty$. The reason is that \eqref{eq:whatisc2mu} has infinitely many roots for $\beta\mu \neq 0$: for small $\beta\mu$, it is clear that the solution we need to consider is the one that is connected to $c_2(0)$. As $\beta\abs{\mu}$ grows, this type of solution may cease to exist depending on the value of $\beta$, and we are then forced to take a different ``branch'' for $c_2(\mu)$. We show what happens as we vary $\beta\mu$ in the case of $\beta = 0.5$ and $\beta = 1$ in Figure \ref{fig:cSYKmanysols}.

The minimal choice we have made, which implies the discussion below, is to always take the smallest positive root. At the end of the chapter, we discuss a possible way of determining the right one, which is left to future work. For the remainder, we will proceed by making this decision as a working hypothesis.

For $\mu > 0$, as we will soon show, this phenomenon only happens for $\beta > \beta_{\rm crit} \approx 0.72$. Our physical interpretation is that, regardless of which new root (out of the infinitely many) we are forced to choose for sufficiently large values of $\beta\mu$, this will result in an enhancement of the magnitude of $g(\tau)$. Since $g(\tau)$ is connected to correlations of a given fermion at different times, what we deduce is that $T < T_{\rm crit}$ allows for an ``ordered'' phase of highly charged states (big $\mu$). Therefore, we suspect a connection to some kind of phase transition of the system at hand. Since this transition happens for $\beta\mu \sim 1$, we also deduce that, as $T$ decreases, the minimum $\mu$ that induces this new phase also becomes increasingly smaller.

What is even more interesting is the behavior for $\mu < 0$, which is always the same even for extremely low $\beta$. While, for $\mu > 0$, there may or there may not be an enhancement of $c_2(\mu)$ at some point, what we find here is that $c_2(\mu)$ transitions to a higher value in a certain range of $\beta\mu$ (which becomes narrower as $\beta$ decreases), then it transitions back to the initial branch and connects smoothly to $c_2(-\infty) = c_2(0)$. We do not yet have a clear interpretation of this phenomenon, but we notice the following: if a big $g(\tau)$ is indeed related to an ordered system, it is not surprising that increasing the temperature reduces the range in which this phase manages to exist.

\begin{figure}
	\centering
	\includegraphics[width = 0.475 \textwidth]{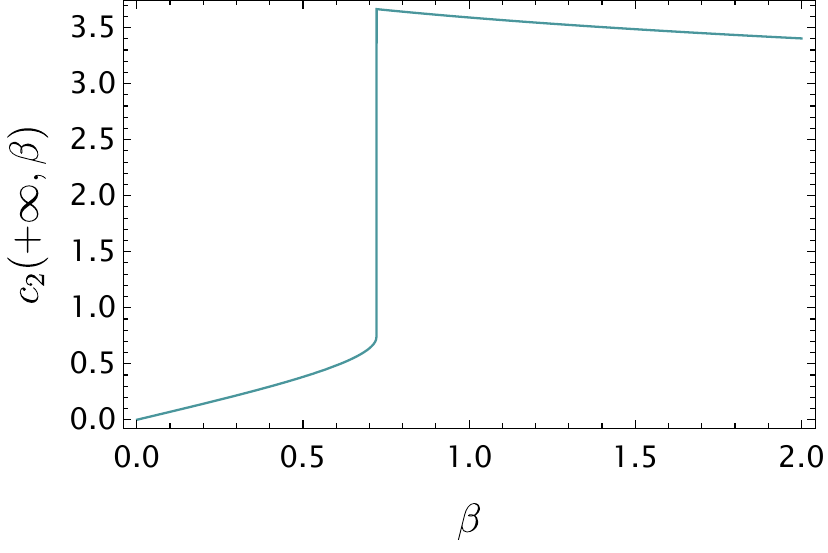} \quad
	\includegraphics[width = 0.495 \textwidth]{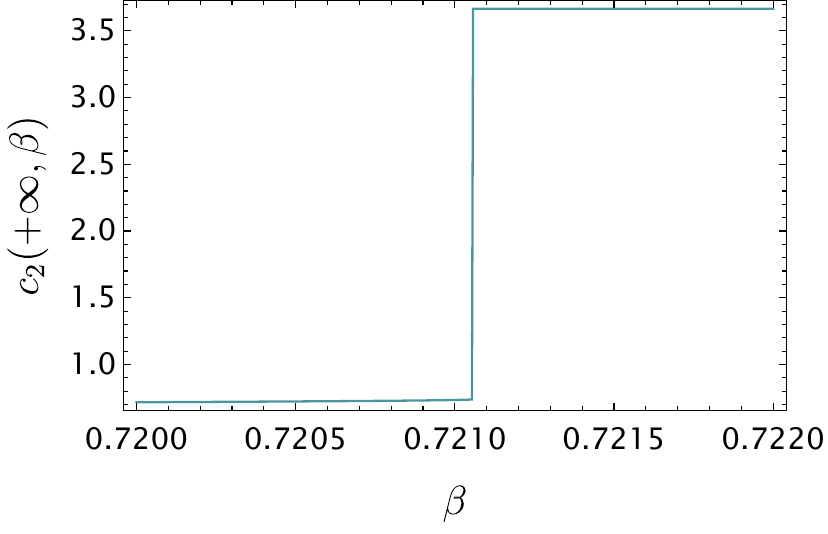}
	\caption{Left: Plot of $c_2(+\infty)$ as a function of $\beta$. Right: Zoom of the plot near $\beta_{\rm crit}$, where $c_2(+\infty)$ exhibits a discontinuous increase. This behavior is caused by the first branch of $h(c_2,\mu)$ (as defined in Figure \ref{fig:cSYKmanysols}) not intersecting the $h = 0$ line for sufficiently big $\beta\mu$, thus forcing $c_2$ to be taken from the second branch of the function. We find $\beta_{\rm crit} \approx 0.72$.}
	\label{fig:cSYKbetacrit}
\end{figure}

A clever way to determine the value of $\beta_{\rm crit}$ is by considering the behavior of $c_2(+\infty)$ (which solves \eqref{eq:c2diagnoser}), as it increases steeply if and only if the class of roots continuously connected to $c_2(0)$ ceases to exist for a fixed value of $\beta$. We plot the dependence of $c_2(+\infty)$ on $\beta$ in Figure \ref{fig:cSYKbetacrit}, from which we extract that $\beta_{\rm crit} \approx 0.72$.

\begin{figure}
	\centering
	\includegraphics[width = 0.48 \textwidth]{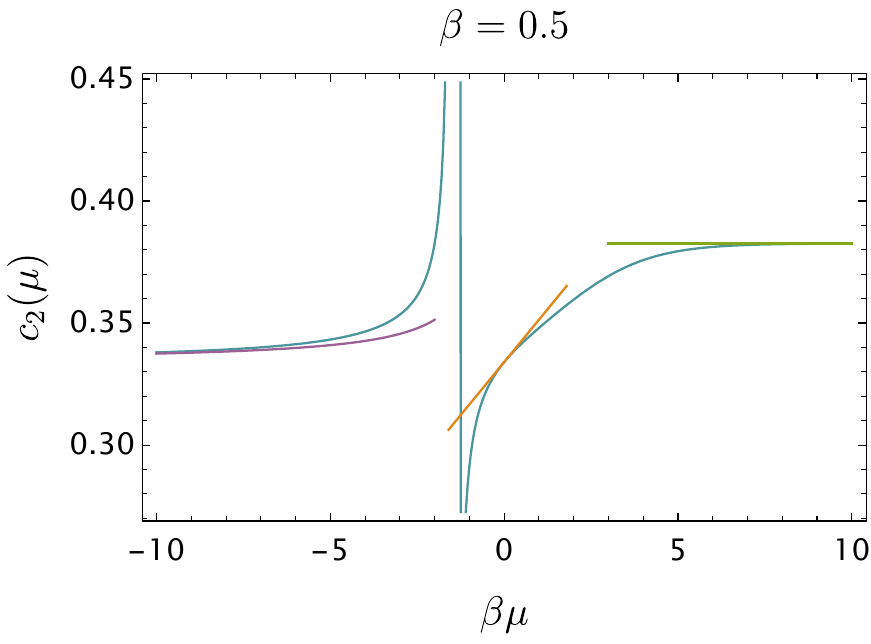} \quad
	\includegraphics[width = 0.48 \textwidth]{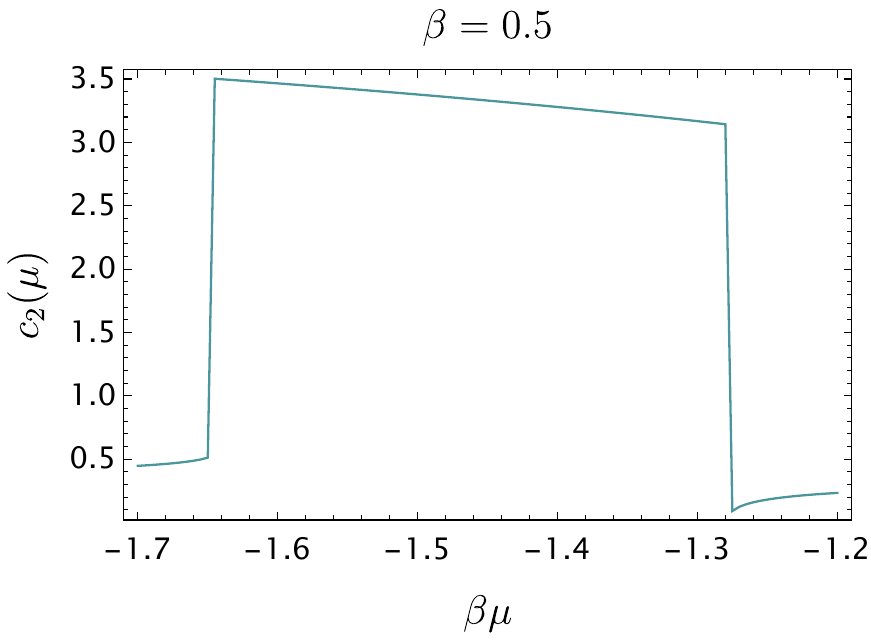} \\
	\includegraphics[width = 0.48 \textwidth]{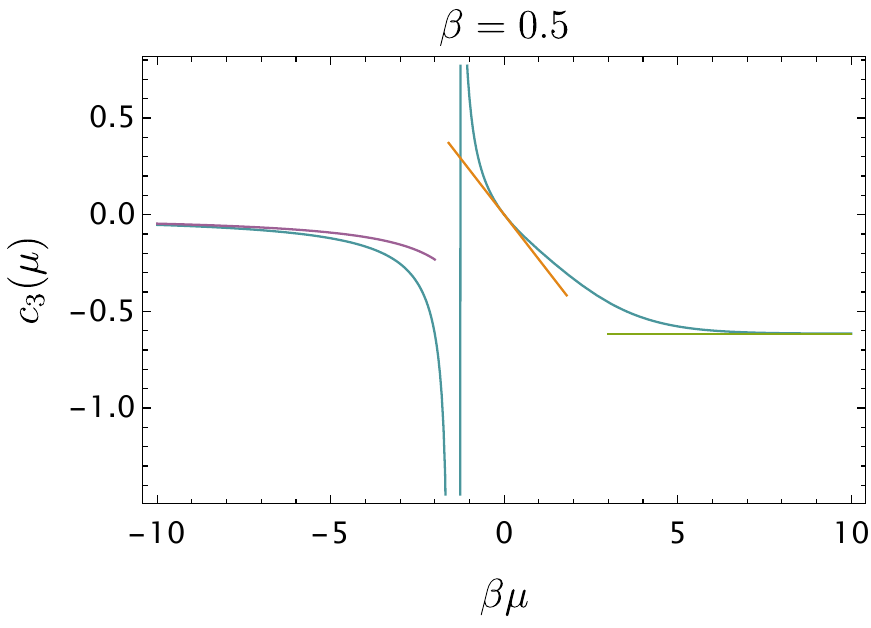} \quad
	\includegraphics[width = 0.48 \textwidth]{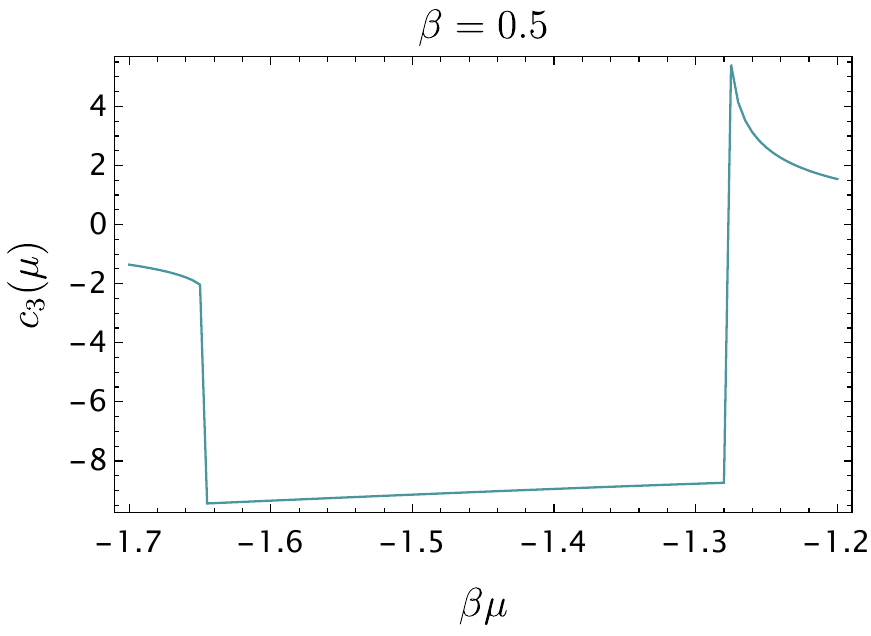} \\
	\includegraphics[width = 0.48 \textwidth]{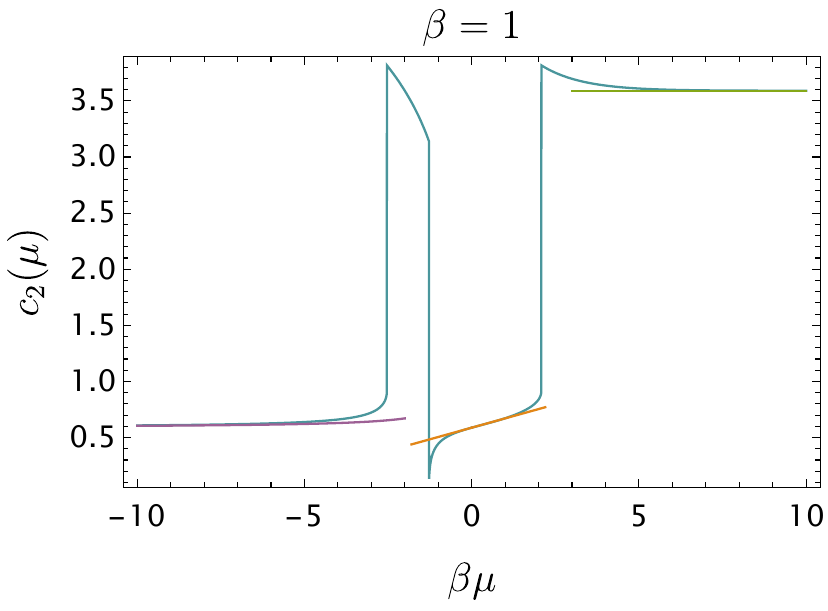} \quad
	\includegraphics[width = 0.48 \textwidth]{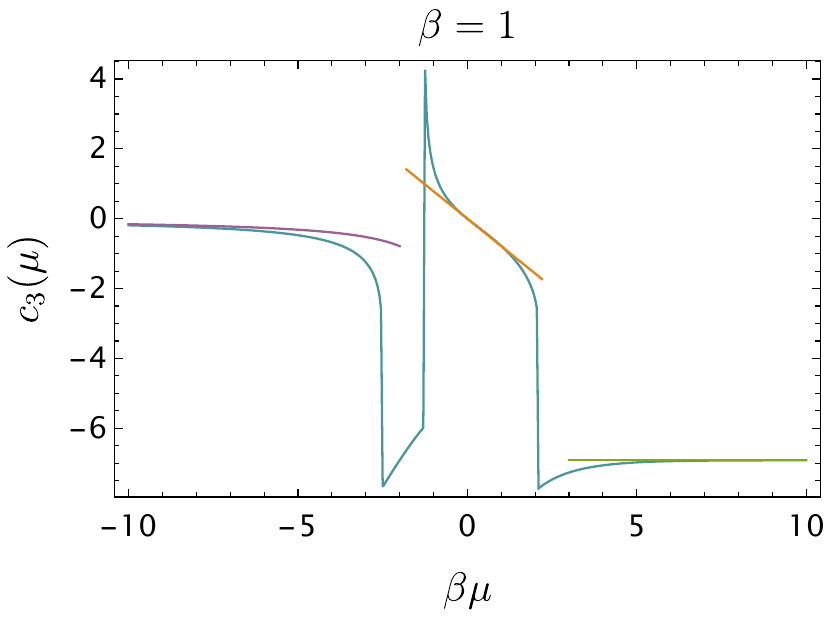}
	\caption{Plot of $c_2(\mu)$ and $c_3(\mu)$ for $\beta = 0.5$ and $\beta = 1$, along with our analytical approximations. We are using the rescaled $\beta \mu \, c_3(\mu) \to c_3(\mu)$. The light blue lines are the numerical solutions, the orange lines are the linear expansions of the functions around $\beta\mu = 0$, the green lines are their asymptotic values at $+\infty$ and the violet lines are their asymptotic values at $-\infty$ plus their first correction. We see that there are always two discontinuities for negative $\mu$, while a discontinuity for positive $\mu$ is only present if $\beta > \beta_{\rm crit}$. For small $\beta$, the ``anomalous region'' between the two $\mu < 0$ discontinuities is much narrower than the case of big $\beta$.}
	\label{fig:cSYKc2c3numb1}
\end{figure}

\begin{figure}
	\centering
	\includegraphics[width = 0.485 \textwidth]{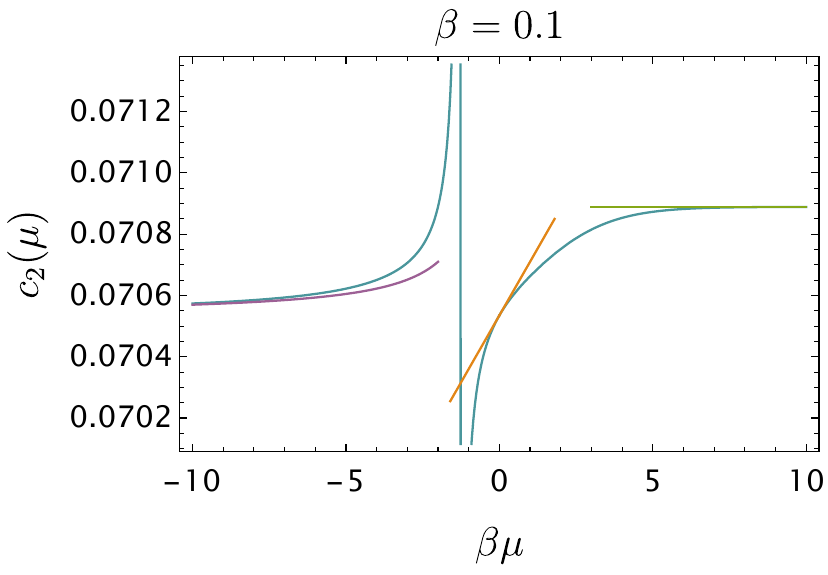} \quad
	\includegraphics[width = 0.475 \textwidth]{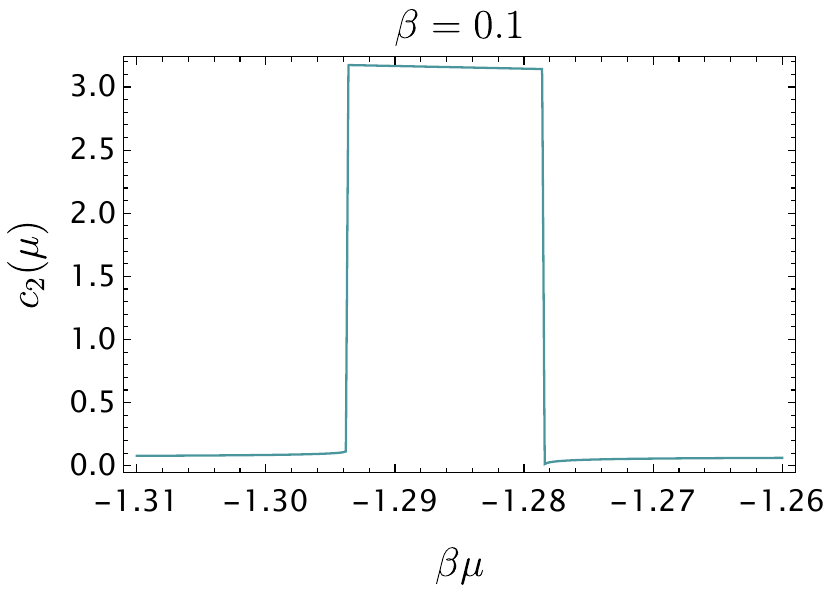} \\
	\includegraphics[width = 0.48 \textwidth]{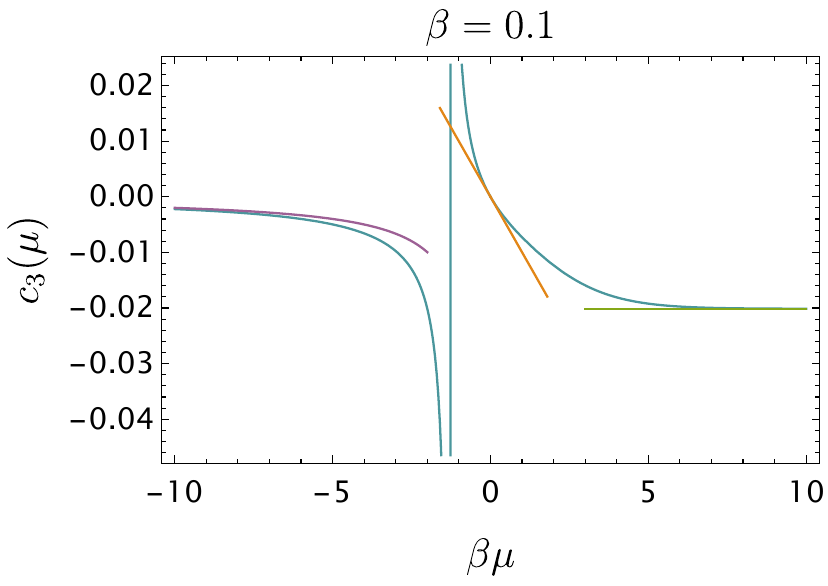} \quad
	\includegraphics[width = 0.48 \textwidth]{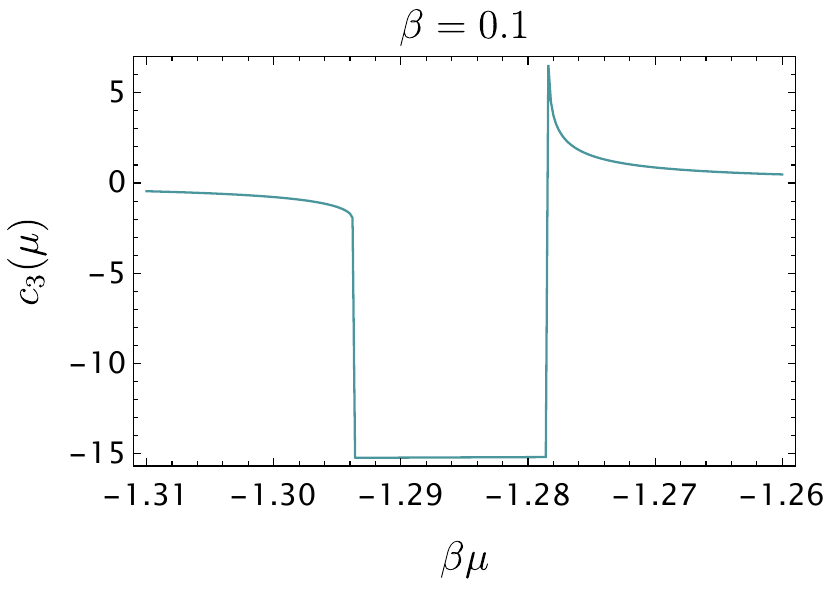}
	\caption{Plot of $c_2(\mu)$ and $c_3(\mu)$ for $\beta = 0.1$, along with our analytical approximations. We are using the rescaled $\beta \mu \, c_3(\mu) \to c_3(\mu)$. The light blue lines are the numerical solutions, the orange lines are the linear expansions of the functions around $\beta\mu = 0$, the green lines are their asymptotic values at $+\infty$ and the violet lines are their asymptotic values at $-\infty$ plus their first correction. We see that there are still two discontinuities for negative $\mu$, while the $\mu > 0$ part of the functions is continuous. In this case, the ``anomalous region'' between the two $\mu < 0$ discontinuities is even narrower than $\beta = 0.5$, coherently with our expectations.}
	\label{fig:cSYKc2c3numb2}
\end{figure}

\begin{figure}
	\centering
	\includegraphics[width = 0.395 \textwidth]{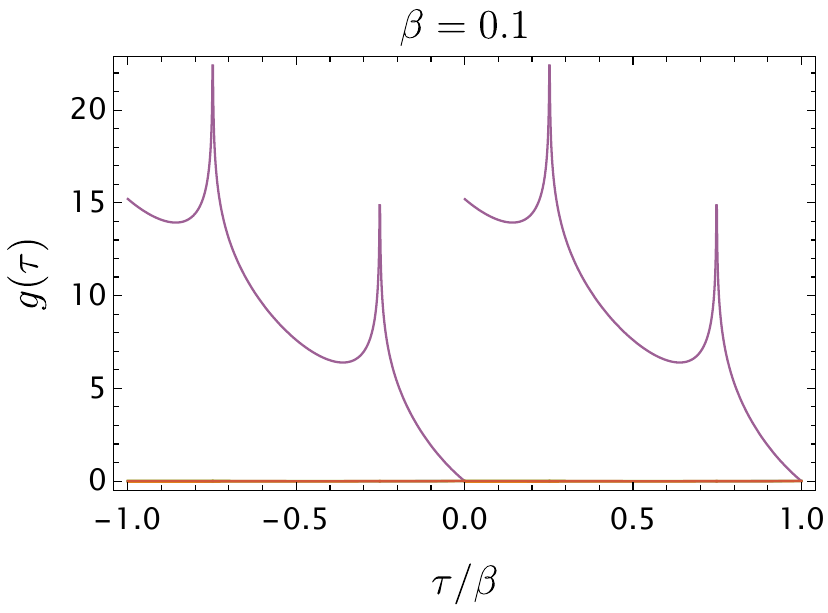} \quad
	\includegraphics[width = 0.575 \textwidth]{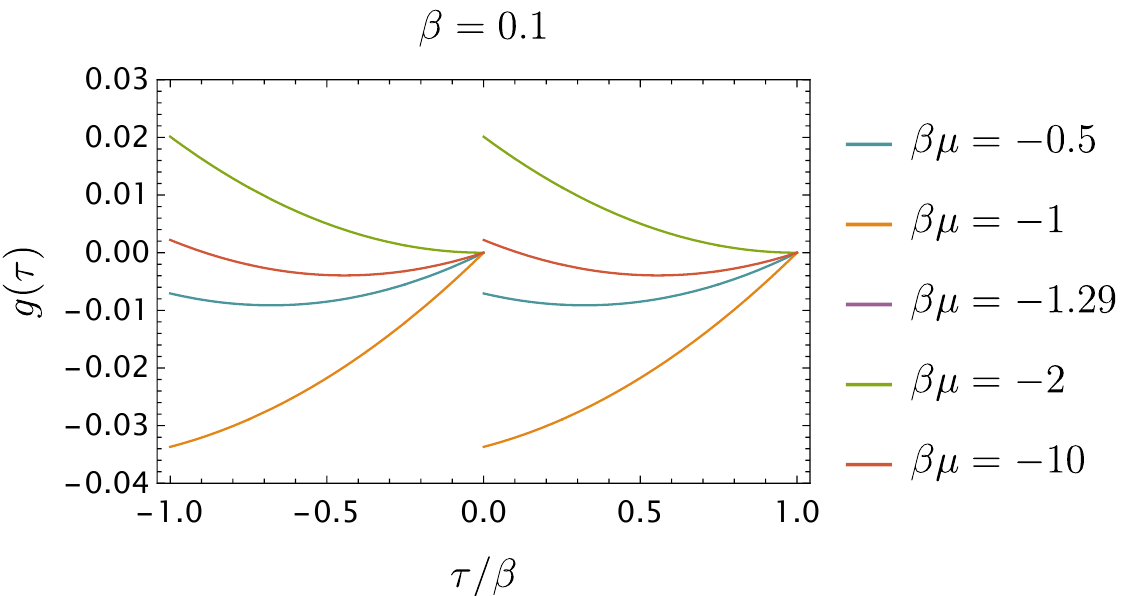} \\
	\includegraphics[width = 0.48 \textwidth]{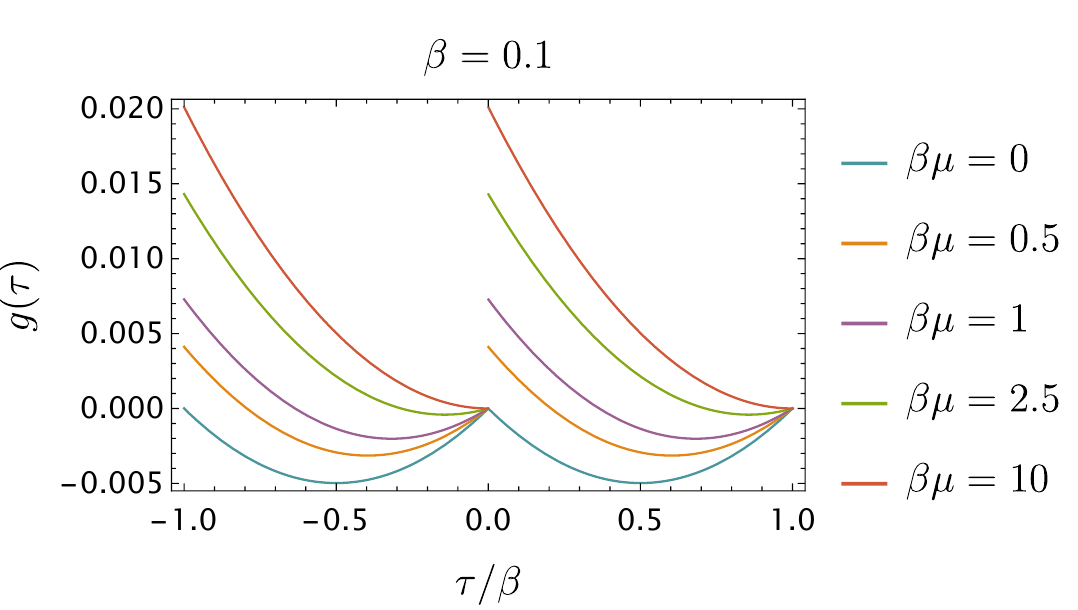} \quad
	\includegraphics[width = 0.48 \textwidth]{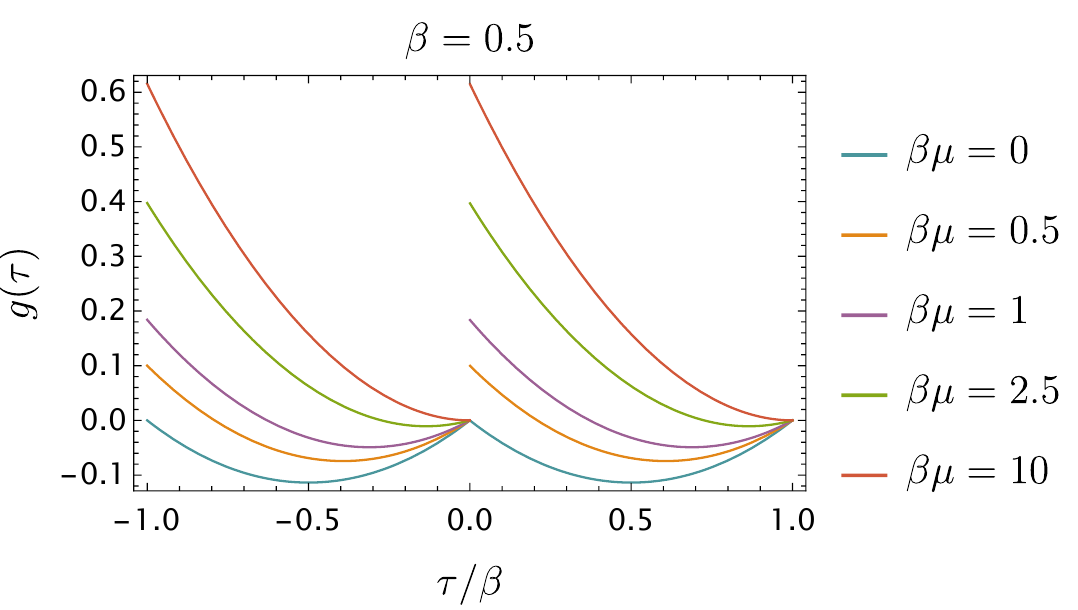} \\
	\includegraphics[width = 0.415 \textwidth]{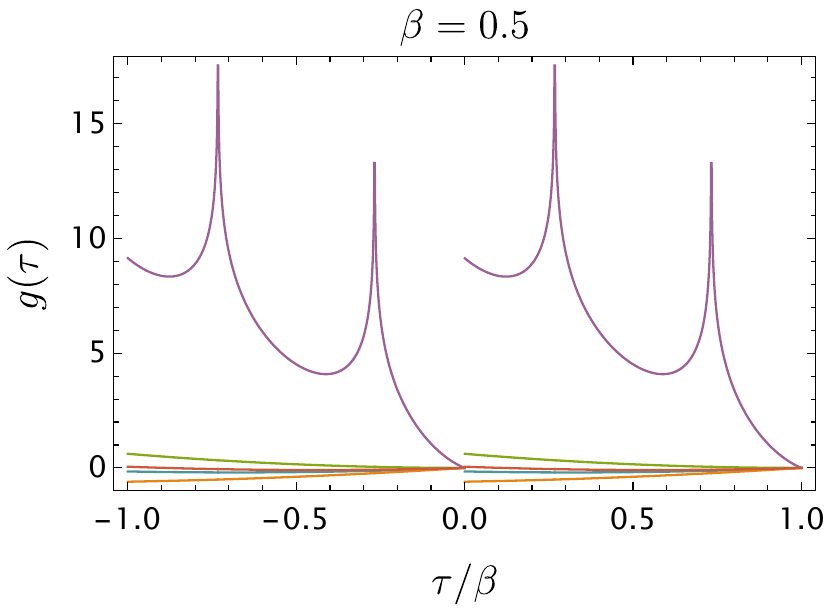} \quad
	\includegraphics[width = 0.555 \textwidth]{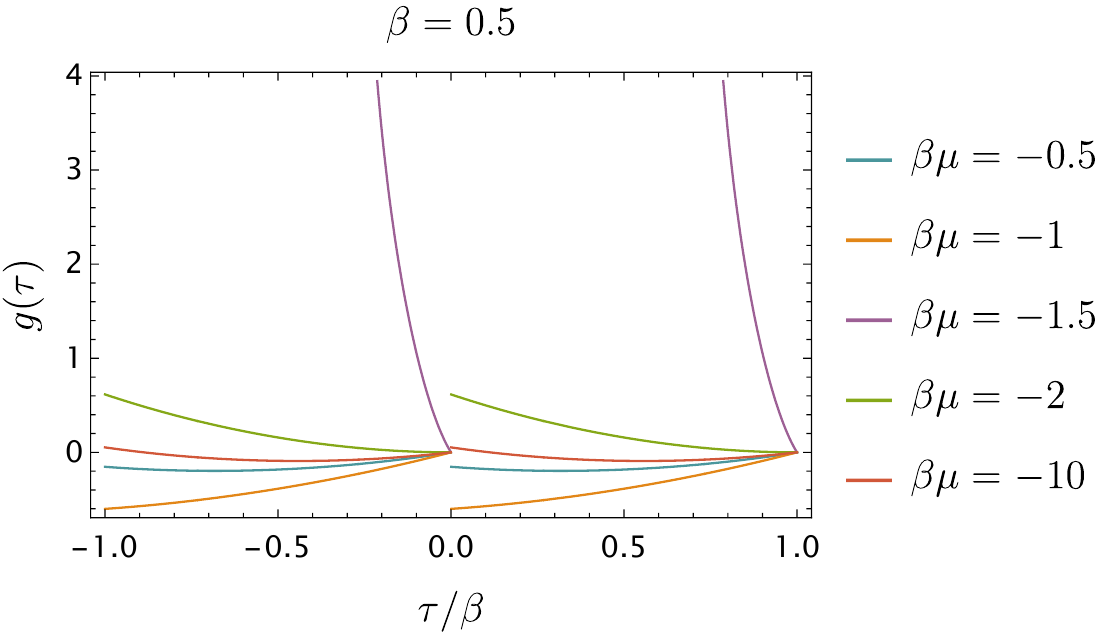} \\
	\includegraphics[width = 0.48 \textwidth]{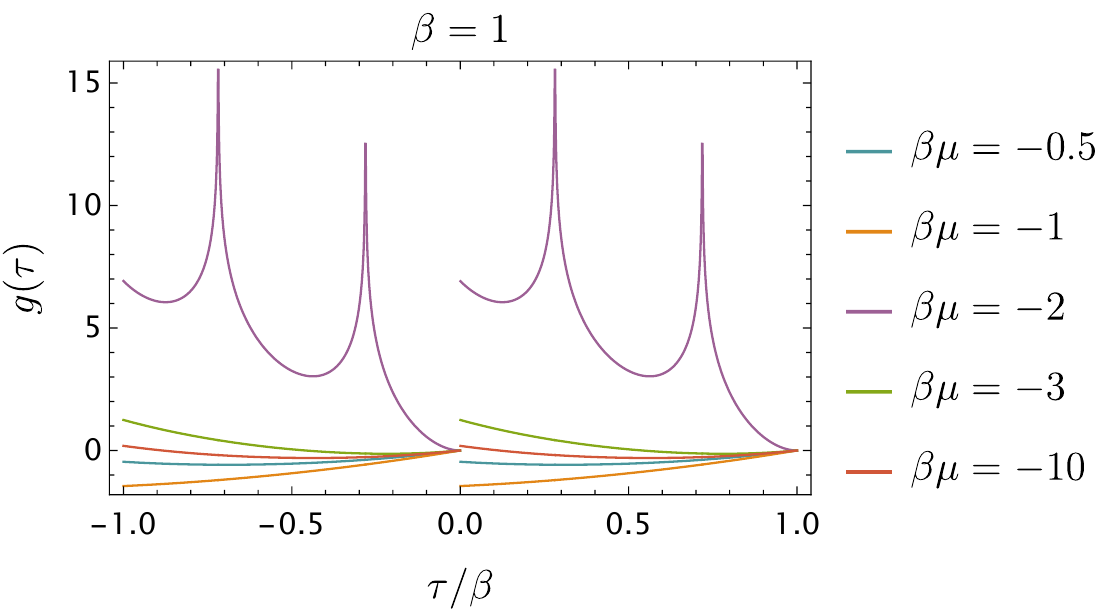} \quad
	\includegraphics[width = 0.48 \textwidth]{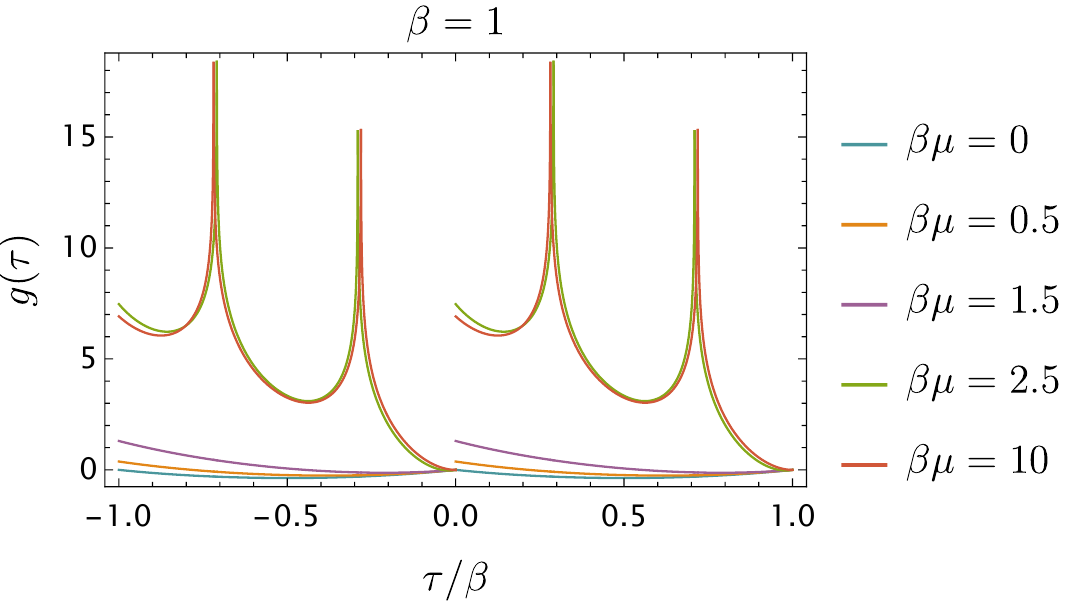}
	\caption{Plot of $g(\tau)$ for different values of $\beta$ and $\mu$. The violet lines in the $\beta\mu < 0$ plots capture the behavior of $g(\tau)$ in the anomalous region, where its value is greatly enhanced. Similarly, for $\beta = 1 > \beta_{\rm crit}$ and $\mu > 0$, the very same enhancement takes place for sufficiently big $\beta\mu$. The function is always periodic with period $\beta$, as expected.}
	\label{fig:cSYKfinalplot}
\end{figure}

In Figures \ref{fig:cSYKc2c3numb1} and \ref{fig:cSYKc2c3numb2}, we compare our approximated results to the full numerical solution for $\beta = 0.1$ and $\beta = 0.5$, which are below $\beta_{\rm crit}$, and for $\beta = 1$, which is above $\beta_{\rm crit}$. In Figure \ref{fig:cSYKfinalplot}, finally, we plot $g(\tau)$ for different values of $\beta$ and $\mu$.

The discontinuous behavior of $g(\tau)$ with respect to $\mu$ translates into discontinuities of $\langle \mathcal{Z} \rangle_J(\beta,\mu)$, which, at this level, is simply the saddle of \eqref{eq:cSYKallincludZ} with the $g(\tau)$ we have just analyzed. This property of the partition function estabilishes a clearer link with the idea of different possible phases of our system. It would be interesting to study the saddle $\langle Z \rangle_J(\beta,\mu)$ for the infinitely many possible values of $c_2(\mu)$, and their associated thermodynamics. Determining the heat capacity, for example, would determine the stability of each root. Better yet, it is reasonable to expect that the preferred root is the one that minimizes the free energy of the system with respect to all the others: therefore, computing the partition function should give us a definitive answer. \\

With these calculations, we conclude our study of the properties of $g(\tau)$ in the cDSSYK model. In the next chapter, we will review the bulk dual of the DSSYK model and study the propagation of scalar fields there, before turning to the study of the gravitational theory associated to the charged variation.

\newpage
\chapter{The Bulk Dual of (c)DSSYK}
As discussed at the beginning of the previous chapter, there are numerous ongoing attempts of trying to determine the dual theory of the DSSYK model. In this chapter, we will investigate the properties of one particular gravitational theory that has been proposed in \cite{Blommaert_2024,Blommaert:2023wad,Blommaert:2024ydx}, then we will extend it to the case of the complex DSSYK model. In these papers and the most recent one in particular, it is argued that the bulk dual of the DSSYK model is given by a sine dilaton model, whose euclidean action is:
\begin{equation}
\begin{gathered}
    \label{eq:bulkdualSYK} S_E = -\frac{1}{16\pi G_N} \int d^2x \, \sqrt{g} \, \left( \Phi R + V(\Phi) \right) - \frac{1}{8\pi G_N} \int dt \, \sqrt{h} \, \left( \Phi K -i \ell \frac{e^{-i \abs{\log q} \Phi/\ell^2}}{2 \abs{\log q}} \right), \\
    V(\Phi) = \frac{\sin(2\abs{\log q} \Phi/\ell^2)}{\abs{\log q}}, \quad \log q = -\frac{p^2}{N}.
\end{gathered}
\end{equation}

As usual, $K$ is the extrinsic curvature of the boundary, $h$ is the induced metric there and $\Phi$ is the dilaton field. $\ell$ is a length constant. The last term in the action is a boundary term that does not modify the equations of motion when solving the Dirichlet problem of fixed induced metric and boundary value of the dilaton: it is needed to have a finite action. In the case of JT gravity, the term that served the same purpose was simply $-(8\pi G_N)^{-1} \int dt \, \sqrt{h} \: (-\Phi/\ell)$. We will come back to what the boundary conditions are later.

\section{(Naive) Gravitational Semiclassics}
We will now study the properties of the semiclassical solution of this theory. First, we rescale the dilaton field $2\abs{\log q}\Phi \to \ell^2 \Phi$, thus making it dimensionless:
\begin{gather}
    (8\pi G_N) S_E = \frac{1}{2\abs{\log q}} \left( -\frac{1}{2} \int d^2x \, \sqrt{g} \, \left( \ell^2 \Phi R + 2\sin \Phi \right) - \int dt \, \sqrt{h} \, \left( \ell^2 \Phi K -i \ell e^{-i \Phi/2} \right) \right). \label{eq:BulkDSSYK}
\end{gather}

We see from this rescaling that the saddle point is valid when $\abs{\log q} \ll 1 \: (\lambda \ll 1)$, series expansion in $G_N$ aside. Note that this limit is not JT gravity (as the initial $V(\Phi)$ would suggest) because we are keeping the rescaled dilaton fixed, which means that the initial dilaton field is scaling as $\abs{\log q}^{-1}$. Given this potential $V(\Phi) = 2\sin(\Phi)$ and taking the dimensionless spatial coordinate to be $r = \Phi$, the Lorentzian metric that solves Einstein's equations is \cite{Cavagli_1999,Anninos:2017hhn,Anninos_2022}:
\begin{equation}
\begin{gathered}
    ds^2/\ell^2 = - f(r) \, dt^2 + \frac{dr^2}{f(r)}, \quad f(r) = 2 \cos r_h - 2 \cos r, \\
    \ell^2 R = -2\cos \Phi = -V'(\Phi). \label{eq:notansatzgmn}
\end{gathered}
\end{equation}

In Appendix \ref{app:2dDilaton} we show how we have thoroughly derived this result, in particular how this is not simply an ansatz, but rather the most general solution. We see that the metric has a black hole horizon at $r_h$ and a cosmological horizon at $r_c = 2\pi - r_h$. Up to an exchange between these two values and because of a $r_h \longleftrightarrow -r_h$ symmetry of the metric, we can always assume that $0 \leq r_h \leq \pi$, at least for a single connected patch of the spacetime. In this regard we observe that, if this theory is a dimensional reduction of a higher dimensional spacetime, it is possible in principle that two apparently disconnected patches (the one with positive $r$ and the one with negative $r$) are actually connected through the dimensions that we have integrated out here. To be more precise, because of the periodicity of the cosine, there is actually an infinite sequence of disconnected patches and horizon radii given by with:
\begin{equation}
\begin{gathered}
    r_{\pm}^{(k)} = \pm r_h + 2k\pi, \quad k \in \mathbb{Z}, \\
    r_+^{(-k)} = -r_-^{(k)}.
\end{gathered}
\end{equation}

\begin{figure}
    \centering
    \includegraphics[width = 1 \textwidth]{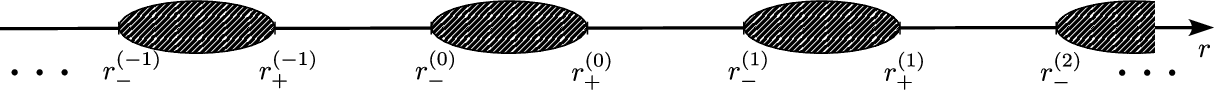} \\
    \vspace{0.7 cm}
    \includegraphics[width = 0.45 \textwidth]{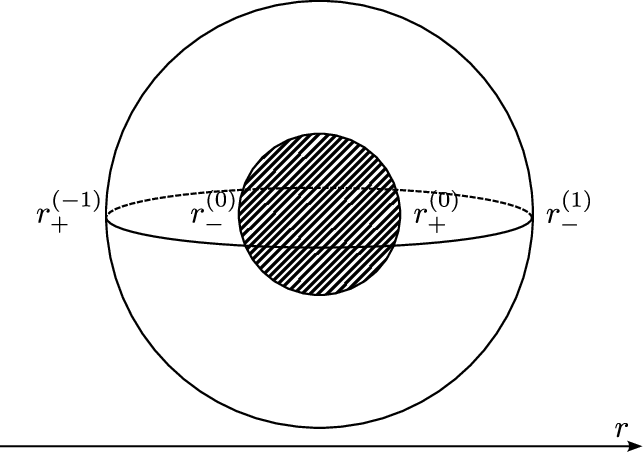}
    \caption{The first figure shows the infinite sequence of horizons on the real axis. The prohibited, ``inner horizon'' regions are depicted as shaded ovals. The disconnected patches in two dimensions are given by $r_+^{(k)} \leq r \leq r_-^{(k+1)}$. The region between what we have defined as $r_h$ and $r_c$ is $r_+^{(0)} \leq r \leq r_-^{(1)}$. \\ The second figure shows that if, for example, our two-dimensional picture was just the projection on a line of a black hole surrounded by a cosmological horizon, the starting four-dimensional manifold would also include the region $r_+^{(-1)} \leq r \leq r_-^{(0)}$, which would not be disconnected from the first one.}
    \label{fig:BulkPatches}
\end{figure}

A sketch of this sequence of patches and how to possibly connect them is shown in Figure \ref{fig:BulkPatches}. As long as we stick to two dimensions, we do not need to worry about the patches that are disconnected from the one under study. In the Euclidean signature, in fact, the manifold closes at the horizons, so what lies beyond is not of any relevance.

Although the coordinate range of the spatial direction tends to 0 as $r_h$ tends to $r_h = \pi$, its proper length has a finite limit. In fact, imagine taking:
\begin{equation}
\begin{gathered}
    \label{eq:dS2} r_h = \pi - \varepsilon, \quad r_c = \pi + \varepsilon, \quad \varepsilon \ll 1, \\
    r = \pi + \delta, \quad \delta \in (-\varepsilon, \varepsilon). 
\end{gathered}
\end{equation}

In this situation, we obtain:
\begin{equation}
\begin{gathered}
    f(\delta) \approx \varepsilon^2-\delta^2, \\
    L = \ell \int_{-\varepsilon}^{+\varepsilon} d\delta \: f(\delta)^{-1/2} = \pi \ell, \quad R \approx +2/\ell^2. \label{eq:varepsto0SYK}
\end{gathered}
\end{equation}

As $\varepsilon$ tends to 0, the classical solution becomes a static patch of dS$_2$. This can actually also be interpreted as a dimensionally reduced near-Nariai geometry, which describes the near-horizon limit of the Schwarzschild-de Sitter spacetime with near-coincident horizons \cite{Anninos_2022}. To see how this metric emerges from a four-dimensional black hole, we start from the general metric \cite{Galante:2023uyf}:
\begin{gather}
    ds^2 = -\left( 1 - \frac{r^2}{\ell^2} - \frac{2G_N M}{r} \right) dt^2 + \frac{dr^2}{1 - \frac{r^2}{\ell^2} - \frac{2G_N M}{r}} + r^2 d\Omega_2^2.
\end{gather}

\begin{figure}
    \centering
    \includegraphics[width = 0.65 \textwidth]{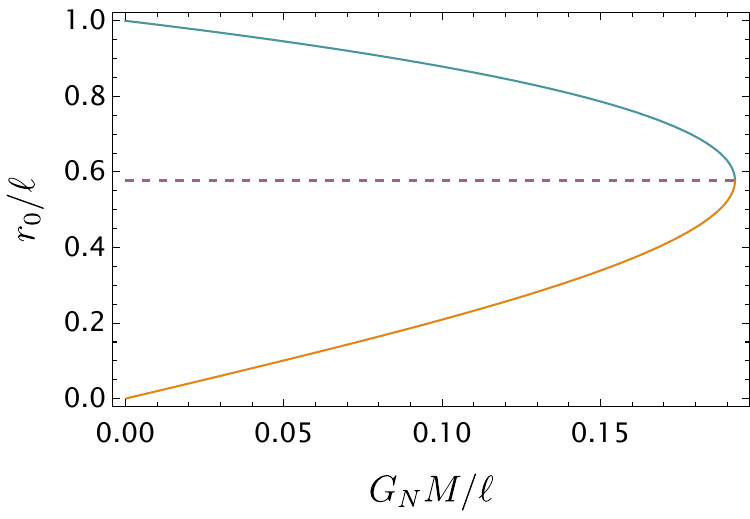}
    \caption{Real positive solutions of the cubic equation $f(r_0)=0$. The blue line is the cosmological horizon, the orange line is the black hole horizon, the dashed line is the value $r_0^*$ that both curves reach for the maximum value of the mass $M^* = 3^{-3/2} \ell/G_N$.}
    \label{fig:SdSBlack}
\end{figure}

Looking for the horizons $r_0$ of the metric yields three roots, out of which one is negative for all values of the black hole mass $M$ and the other two are plotted in Figure \ref{fig:SdSBlack}.

An interesting fact about black holes emerges in de Sitter: they cannot have an arbitrarily big mass, but there is a maximum achievable value $M^* = 3^{-3/2} \ell/G_N$ such that the black hole and cosmological horizons coincide and are equal to $r_0^* = 3^{-1/2} \ell$. If we now consider a black hole with mass $M = M^* - \delta M$, such that the cosmological horizon $r_c$ and the black hole horizon $r_h$ are given by
\begin{gather}
    r_h = r_0^* - \sqrt{\frac{2G_N \delta M}{\sqrt{3} \ell}} \, \ell + \mathcal{O}(\delta M), \quad r_c = r_0^* + \sqrt{\frac{2G_N \delta M}{\sqrt{3} \ell}} \, \ell + \mathcal{O}(\delta M), \quad \sqrt{\frac{2G_N \delta M}{\sqrt{3} \ell}} \, \ell \equiv \varepsilon,
\end{gather}

we can expand the metric between the horizons at leading order in $\varepsilon$ by defining new coordinates $\{ \tau, \rho \}$:
\begin{gather}
    \tau \equiv \frac{\varepsilon}{\ell} t, \quad \rho \equiv \ell \frac{r-r_0^*}{\varepsilon}, \quad \tau \in \mathbb{R}, \: \rho \in [-\ell,\ell].
\end{gather}

We expand $f(r)$ first:
\begin{equation}
\begin{aligned}
    f(r) & = 1 - \frac{r^2}{\ell^2} - \frac{2G_N M^*}{r} + \frac{2G_N \delta M}{r} \\
    & = f \left( r_0^* + \varepsilon \frac{\rho}{\ell}, M^* \right) + \frac{2G_N \delta M}{r_0^*} + \mathcal{O}(\varepsilon^3) \\
    & = \frac{1}{2} \varepsilon^2 \frac{\rho^2}{\ell^2} \, \partial^2_r f(r_0^*,M^*) + 3 \frac{\varepsilon^2}{\ell^2} + \mathcal{O}(\varepsilon^3) \\
    & = \frac{3\varepsilon^2}{\ell^2} \left( 1 - \frac{\rho^2}{\ell^2} \right) + \mathcal{O}(\varepsilon^3).
\end{aligned}
\end{equation}

Plugging this result into the metric yields:
\begin{gather}
    ds^2 \approx -3 \left( 1- \frac{\rho^2}{\ell^2} \right) \, d\tau^2 + \frac{d\rho^2}{3 \left( 1- \frac{\rho^2}{\ell^2} \right)} + \frac{\ell^2}{3} d\Omega_2^2.
\end{gather}

The final change of coordinates that reproduces the static patch is $\rho \to 3^{1/2} \rho, \: \tau \to 3^{-1/2} \, \tau$. In this limit, we see that the the spacetime factorizes into dS$_2 \times S^2$ and that the dS$_2$ part is exactly one of its static patches. If one performs a dimensional reduction of this manifold as $\varepsilon \to 0$, then, we obtain exactly \eqref{eq:varepsto0SYK}. This is a hint of this geometry being a dimensional reduction of a four-dimensional manifold, since we obtain the same metric through the same procedure of moving the cosmological horizon close to the black hole one. From this point of view, the first ``negative'' patch is relevant in the four-dimensional case, as it is connected to our initial patch by circumnavigating the event horizon of the black hole at the center of de Sitter. One can look at the thermodynamic quantities in order to substantiate this claim.

\begin{figure}
    \centering
    \includegraphics[width = 1 \textwidth]{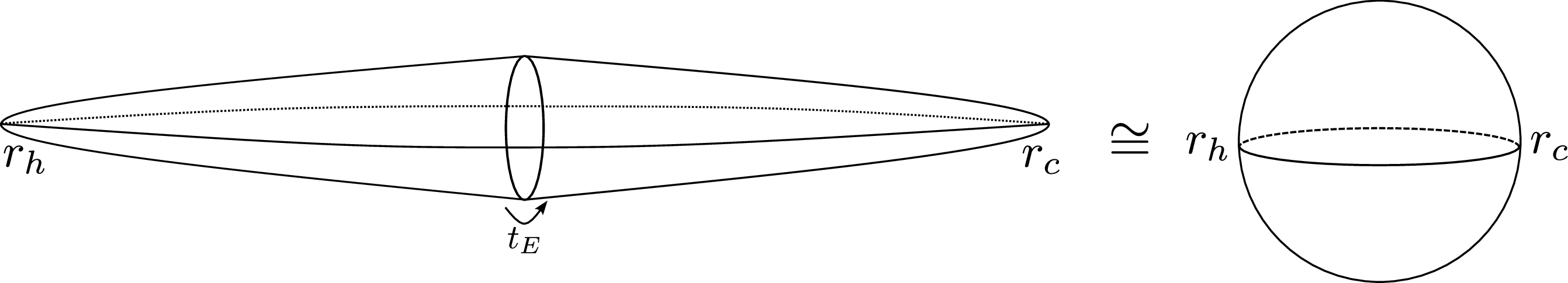}
    \caption{Representation of the Euclidean two-manifold $M$ that solves Einstein's equations. The absence of conical singularities implies that it is topologically equivalent to a two-sphere, with the two horizons lying at opposite poles.}
    \label{fig:DoubleSigaro}
\end{figure}

To study the thermodynamic properties of the semiclassical solutions, it is useful to turn to the Euclidean action. The resulting two-dimensional manifold $M$ has the topology of $S^2$ and no boundary, since we require the absence of conical defects at both the black hole horizon and the cosmological one. The consequence of this requirement is that the manifold caps smoothly at both points, so that the the end result is two cigars (like the one in Figure \ref{fig:Sigaro}), glued together as shown in Figure \ref{fig:DoubleSigaro}. Requiring the correct periodicity in the Euclidean time $\beta$ at both $r = r_h, \: r = r_c$ fixes the inverse temperature of the spacetime:
\begin{gather}
    \beta = \frac{4\pi\ell}{|f'(r_h)|} = \frac{4\pi \ell}{|f'(r_c)|} = \frac{2\pi \ell}{\sin(r_h)}.
\end{gather}

There is no mismatch between the required periodicities of the Euclidean time at the two horizons, so a smooth $S^2$ saddle is indeed possible in this theory. From a thermodynamic point of view, this result tells us that a thermal equilibrium (hence a canonical ensemble) is possible for any value of $r_h$.

\section{Gravitational Boundary Conditions}

We could go on and study the thermodynamics of the two-dimensional system that we have just described, looking for hints of higher-dimensional spacetimes, and this would surely be an interesting topic. Surprisingly, however, the way we need to proceed when matching this gravitational theory to the DSSYK model is radically different and is described in \cite{Blommaert:2024ydx}. We now review their results while completing missing passages in their computations, before turning back to our original contributions. The naive way of proceeding would be to consider the patches in Figure \ref{fig:BulkPatches}. What needs to be done, though, is to actually complexify the $r$ coordinate (and the dilaton field $\Phi$) and to take the ``boundary'' of the manifold not to be the cosmological horizon $r_c$ or any other cutoff radius between $r_h$ and $r_c$, but to lie at
\begin{gather}
    \Phi_{\rm bdy} = \frac{\pi}{2} + i\infty.
\end{gather}

In short, the subtlety of this situation is that only the right boundary conditions of the same starting action are holographically connected to the DSSYK model. After all, the holographic principle suggests that the DSSYK model lives at the boundary of the dual geometry, but \textit{where is it, exactly}? The answer is $r_0 \equiv \Phi_{\rm bdy}$. This is only one of the boundary conditions of the correct Dirichlet problem, the other involving the boundary metric:
\begin{gather}
    \sqrt{f(\Phi_{\rm bdy})} \, e^{i \Phi_{\rm bdy}/2} = i,
\end{gather}

which is clearly satisfied by $f(r)$ once we have determined that the DSSYK model should live at $\Phi_{\rm bdy}$. If one takes this theory as is and computes the curvature of the spacetime at $\Phi_{\rm bdy}$, one finds the really odd result $R = 2i \sinh(\infty)/\ell^2$. Building the holographic dictionary between the two theories requires an extra step. If one forgets for a moment that the chord Hilbert space that we have introduced in Chapter 6 only has states with a non-negative number of chords, one can show that the DSSYK model can be rewritten as a one-dimensional quantum theory that is called the $q$-Schwarzian. ``Semiclassical'' ($\abs{\log q} \ll 1$) correlators of scalar primaries with conformal weight $\Delta$ in this boundary theory are dual to correlators of bulk massive scalars in the fixed background geometry that do not couple minimally to the sine dilaton metric, but rather also to the dilaton field through the following action:
\begin{gather}
    S = \frac{1}{2} \int d^2x \, \sqrt{g} \left( g^{\mu\nu} \partial_\mu \phi \, \partial_\nu \phi + m^2 e^{-i\Phi} \phi^2 \right) = \frac{1}{2} \int d^2x \, \sqrt{g_{\rm eff}} \left( g^{\mu\nu}_{\rm eff} \partial_\mu \phi \, \partial_\nu \phi + m^2\phi^2 \right)
\end{gather}

That is, the probe experiences an \textit{effective} metric that is a Weyl rescaling of the starting sine dilaton one:
\begin{equation}
\begin{aligned}
    & g_{\mu\nu, \rm eff} = e^{-i\Phi} g_{\mu\nu} \\
    \implies & R_{\rm eff}(\Phi) = e^{i\Phi} R(\Phi) - 2e^{i\Phi} g^{\mu\nu} \nabla_\mu \nabla_\nu (-i \Phi/2) \\
    = \: & e^{i\Phi} R(\Phi) + ie^{i\Phi} \partial_\Phi f(\Phi)/\ell^2 = -2/\ell^2.
\end{aligned}
\end{equation}

This means that the particle actually behaves as if in AdS$_2$ everywhere, so we recover the usual setup of the AdS/CFT correspondence.

One can also say that the particle's boundary-to-boundary renormalized geodesic length in this effective metric
\begin{gather}
	L = \int \frac{ds}{\ell} \: e^{-i\Phi/2} = \int \frac{ds_{\rm eff}}{\ell},
\end{gather}

after having quantized the gravitational theory, is an operator with a discrete spectrum that is dual to the chord number operator $\mathbf{n}$ in the $q$-Schwarzian (DSSYK) theory:
\begin{gather}
	\mathbf{L} = 2 \abs{\log q} \mathbf{n}.
\end{gather}

This choice is the one that matches the Hamiltonian operators of the quantum bulk and boundary theories and tells us that particles in the quantized sine dilaton model have discretized trajectories. Better yet, since $\mathbf{L}$ is a canonical variable of the gravitational quantum mechanics that exists a priori from any scalar probe, it tells us that spacetime itself is discretized. At the semiclassical level, coherently, this is the dictionary that emerges from the previous comparison between the two point function on the boundary and in the bulk. The renormalized boundary-to-boundary propagator $\langle \phi_r(x_2) \phi_r(x_1) \rangle$ can in fact be computed both from a field perspective and from a particle one through the use of the worldline formalism, whose path integral's saddles select the geodesics \cite{Worldline}:
\begin{gather}
	\langle \phi_r(x_2) \phi_r(x_1) \rangle = \int_{X(0) = x_1}^{X(\ell L) = x_2} \mathcal{D}X \: \exp \left( -m \int_{X(s)} ds_{\rm eff} \right) \approx e^{-\Delta L}. \label{eq:worldstart}
\end{gather}

This approximation requires a heavy particle, so that $m^2\ell^2 = \Delta(\Delta-1) \implies \Delta \approx m \ell \gg 1$. We prove this statement in Appendix \ref{app:worldline}. Although the computation of two point functions in DSSYK and in the $q$-Schwarzian theory in the $q \to 1$ limit does not require $\Delta \gg 1$, this way of computing propagators in the bulk theory does: for these values, the two theories yield the exact same result (up to a slight mismatch with the DSSYK model, where the ``negative chord number states'' are absent), provided that $r_0$ is exactly $\pi/2 + i\infty$. Going back to the operator equality, keeping only the subspace of states where $\mathbf{L} \geq 0$ is both required (because of $\mathbf{n} \geq 0$) and natural, since semiclassical Euclidean trajectories always have $ds_{\rm eff} \geq 0$.

This discussion has addressed the boundary value of the coordinate $r$ (although a clearer picture is given below), but what about its other extremal value? A full match of the partition functions of the DSSYK model and of the gravitational theory tells us that we should sum over all geometries that start at any horizon $r_\pm^{(k)}$ and end at $\Phi_{\rm bdy}$ while following an arbitrary path on the complex $r$ plane. This is a clear statement at the semiclassical level: these are all and only the geometries that have both the right time periodicity (the same one is imposed by any of the horizons) and the right boundary conditions for the metric and the dilaton.

Turning back to the effective metric, we can introduce:
\begin{gather}
    r(\rho) = \frac{\pi}{2} + i \log(\rho + i\cos\theta),
\end{gather}

so that the effective geometry becomes:
\begin{gather}
    ds_{\rm eff}^2/\ell^2 = F_{\rm AdS}(\rho) \, dt^2 + \frac{d\rho^2}{F_{\rm AdS}(\rho)}, \quad F_{\rm AdS}(\rho) = \rho^2-\sin^2\theta.
\end{gather}

\begin{figure}
    \centering
    \includegraphics[width=0.43 \textwidth]{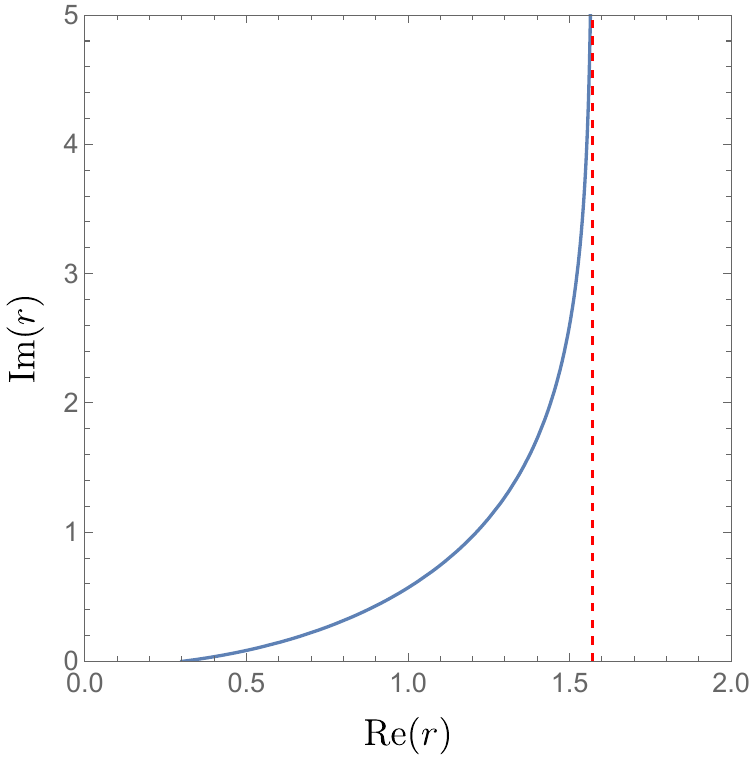} \qquad \includegraphics[width=0.45 \textwidth]{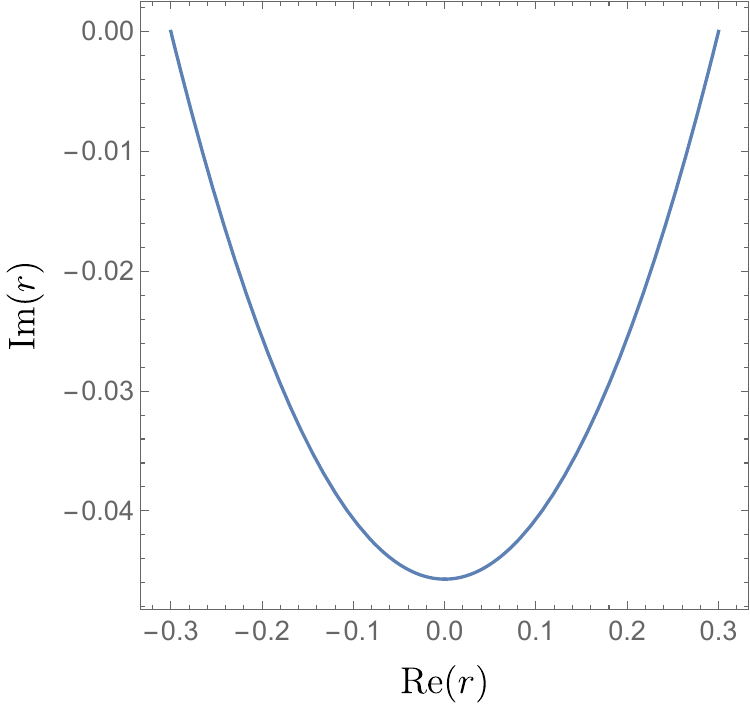}
    \caption{Trajectories on the complex $r$ plane spanned by $\rho \in [\sin\theta,+\infty]$ (left figure) and \mbox{$\rho \in [-\sin\theta,+\sin\theta]$} (right figure) in the case of $\theta = 0.3$. We see that this parametrization starts from an horizon on the real axis and terminates at $\Phi_{\rm bdy}$, as it should. The red dashed line in the plot on the left is $\Re(r) = \pi/2$.}
    \label{fig:DSSYKComplexR}
\end{figure}

We have renamed $r_h \to \theta$ and we will use this notation throughout the rest of the chapter. We have taken the cut of the logarithm to lie on the negative semiaxis, so that the angle of complex numbers has range $\vartheta \in [-\pi,\pi]$ and $r(\sin \theta) = \theta$. We had already determined that the particle perceives an AdS$_2$ geometry. With this choice of coordinates we see, in particular, that it is an AdS$_2$ black hole \cite{Cadoni_1995} with an event horizon at $\rho = \sin\theta$ and inverse temperature $\beta = 2\pi\ell/\sin\theta$. The trajectory in the complex $r$ plane spanned by $\rho \in [\sin\theta,+\infty]$ is shown in Figure \ref{fig:DSSYKComplexR}. It is exactly a trajectory that starts from an horizon on the real axis and terminates at $\Phi_{\rm bdy}$. Considering the range $\rho \in [-\sin\theta,+\infty]$ is equivalent to starting from the cosmological horizon; from the AdS$_2$ point of view, the interval $[-\sin\theta,+\sin\theta]$ explores the inner horizon region of the black hole, while on the $r$ plane it extends the trajectory as also shown in Figure \ref{fig:DSSYKComplexR}. If we were to consider any other integration contour starting from any other horizon and reaching $\Phi_{\rm bdy}$, since $R_{\rm eff} = -2/\ell^2$ everywhere, we would have other AdS$_2$ black hole solutions with $(g_{\mathrm{eff}})_{tt}(r_\pm^{(k)}) = g_{tt}(r_\pm^{(k)}) = 0$.

This picture is the reason why we need to place the DSSYK model at $\Phi_{\rm bdy}$: it is just the position of the Weyl rescaled AdS$_2$ boundary $\rho = +\infty$, where boundary operators dual to the bulk ones need to be inserted, translated in terms of the sine dilaton metric.

The reference \cite{Blommaert:2024ydx} does not explicitly prove that the boundary counterterm in the action should be the one appearing in \eqref{eq:BulkDSSYK} so, now that we have described the origin of the boundary conditions, we will do it ourselves for completeness. We will consider the most general case and show that, for a generic dilaton gravity model with potential $V(\Phi)$, the boundary counterterm should be:
\begin{gather}
	(8\pi G_N)(2\abs{\log q}) \delta S_E = - \int dt \, \sqrt{h} \, \left( - \ell \sqrt{G(\Phi)} \right), \quad G(\Phi_{\rm bdy}) = \int^{\Phi_{\rm bdy}} d\Phi \: V(\Phi). \label{eq:DilatonCounter}
\end{gather}

This term does not modify the equations of motion when considering the Dirichlet problem with fixed $h$ and $\Phi$. We report the $2\abs{\log q}$ factor to make contact with the case of interest and the rescaled dilaton, but obviously this proof is unrelated to this. We first write the on-shell action \eqref{eq:BulkDSSYK} without the counterterm:
\begin{equation}
\begin{aligned}
	-(8\pi G_N/\ell^2) (2\abs{\log q}) S_E & = \int dt \left[ \frac{1}{2} \int_{\Phi_h}^{\Phi_{\rm bdy}} d\Phi \: (\ell^2 \Phi R + V(\Phi)) + \sqrt{f(\Phi_{\rm bdy})} \, \ell K \Phi_{\rm bdy} \right] \\
	& = \int dt \: \frac{1}{2} \left[ \int_{\Phi_h}^{\Phi_{\rm bdy}} d\Phi \: (-\Phi V'(\Phi) + V(\Phi)) + V(\Phi_{\rm bdy}) \Phi_{\rm bdy} \right].
\end{aligned}
\end{equation}

The lower bound $\Phi_h$ is an horizon of the metric, $f(\Phi_h)$ = 0, or any other lower limit for the value of the dilaton. We have used that:
\begin{gather}
	\ell^2 R = -V'(\Phi), \quad f(\Phi) = \int_{\Phi_h}^{\Phi} d\Phi' \: V(\Phi'), \quad \ell K = \partial_{\mu} n^\mu = \partial_{\Phi} \sqrt{f(\Phi)} = \frac{V(\Phi_{\rm bdy})}{2\sqrt{f(\Phi_{\rm bdy})}}.
\end{gather}

We can integrate the first term of the action by parts, thus obtaining:
\begin{gather}
	-(8\pi G_N/\ell^2) (2\abs{\log q}) S_E = \int dt \left[ \int_{\Phi_h}^{\Phi_{\rm bdy}} d\Phi \: V(\Phi) + \frac{1}{2} V(\Phi_h) \Phi_h \right].
\end{gather}

We now assume that the volume integral of the potential diverges and that we do need to renormalize the action in the first place. This is clearly the case for the usual dilaton potential in JT gravity, $V(\Phi) = 2\Phi$, and for the situation at hand, where $V(\pi/2 + i\alpha) = 2 \cosh \alpha$ diverges as $\alpha \to \infty$ and the integration measure is asymptotically $d\Phi = i \, d\alpha$. Therefore, we have to consider the counterterm \eqref{eq:DilatonCounter}, where we can take any lower limit of integration for $G(\Phi)$, which we call $\Phi_c$. On-shell, this term becomes:
\begin{equation}
\begin{aligned}
	-(8\pi G_N/\ell^2) (2\abs{\log q}) \delta S_E & = -\int dt \: \sqrt{\int_{\Phi_h}^{\Phi_{\rm bdy}} d\Phi \: V(\Phi)} \sqrt{\int_{\Phi_c}^{\Phi_{\rm bdy}} d\Phi \: V(\Phi)} \\
	& = -\int dt \int_{\Phi_h}^{\Phi_{\rm bdy}} d\Phi \: V(\Phi) \left[ 1 - \int_{\Phi_h}^{\Phi_c} d\Phi \: V(\Phi) \left( \int_{\Phi_h}^{\Phi_{\rm bdy}} d\Phi \: V(\Phi) \right)^{-1} \right]^{1/2} \\
	& \approx -\int dt \left[ \int_{\Phi_h}^{\Phi_{\rm bdy}} d\Phi \: V(\Phi) - \frac{1}{2} \int_{\Phi_h}^{\Phi_c} d\Phi \: V(\Phi) \right]. 
\end{aligned}
\end{equation}

The renormalized on-shell action is therefore:
\begin{gather}
	S_E + \delta S_E = -\frac{\beta \ell}{16\pi G_N} \frac{1}{2\abs{\log q}} \left[ V(\Phi_h) \Phi_h + \int_{\Phi_h}^{\Phi_c} d\Phi \: V(\Phi) \right], \label{eq:RenormDilatonAction}
\end{gather}

which is finite. In the presence of an horizon $\Phi_h$, we also recall that $\beta = 4\pi \ell/|V(\Phi_h)|$. If we finally take $\Phi_c = \pi/2$, which will be necessary later, and apply this formula to the sine dilaton model, we obtain:
\begin{gather}
	\sqrt{G(\Phi_{\rm bdy})} = \sqrt{- 2 \cos \Phi_{\rm bdy}} \stackrel{e^{i\Phi_{\rm bdy}} = 0}{=} \sqrt{-2 e^{-i\Phi_{\rm bdy}}/2} = i e^{-i\Phi_{\rm bdy}/2}.
\end{gather}

\section{The Duality of Thermodynamics} \label{sec:thermodual}
Let us work through the duality on both sides, starting from the gravitational one. Evaluating the on-shell renormalized action \eqref{eq:RenormDilatonAction} on the saddles associated to $\theta_{\pm}^{(k)}$ leads to the following semiclassical partition function:
\begin{equation}
\begin{aligned}
    \label{eq:fakeZgrav} \mathcal{Z}_{\rm grav}(\beta) & \approx \sum_{\pm, \: k \in \mathbb{Z}} \exp \left[ \frac{\ell^2}{8\pi G_N} \left( \frac{\pi \theta_\pm^{(k)}}{\abs{\log q}} + \frac{\beta \ell^{-1} \cos\theta}{2\abs{\log q}} \right) \right] \\
	& \approx \int dE(\theta) \: \exp \left[ \frac{\ell^2}{8\pi G_N} \left( \frac{\pi \theta}{\abs{\log q}} + \frac{\beta \ell^{-1} \cos\theta}{2\abs{\log q}} \right) \right].
\end{aligned}
\end{equation}

In the last line, we have used that our infinite sum can be seen as the result of performing a saddle point approximation of an integral over a $\theta$ variable. Indeed, if one looks for the saddles of the integral $\bar\theta$, one finds:
\begin{gather}
	\frac{\partial}{\partial\theta} ( 2\pi\theta + \beta \ell^{-1} \cos\theta ) = 0 \implies \beta = \frac{2\pi \ell}{\sin\bar\theta}.
\end{gather}

Notice an important feature that will be crucial in the match: we have decided to pick $\beta = 4\pi\ell/V(\Phi_h)$, which implies a negative temperature for the cosmological horizon saddles (the $\theta_-^{(k)}$). Although this choice may seem strange at first, it is suggested by black hole mechanics in de Sitter. We will come back to this topic in more detail in Section \ref{sec:Susskind}.

We observe that, in a way, the integral on the last line is the result that we actually get from computing the semiclassical partition function: we start from a path integral over the metric and the dilaton field and the classical solutions that emerge exhibit a free parameter $\theta$. Since a path integral is virtually a sum over all possible field configurations, it is clear that the semiclassical one has to integrate over this parameter. The integration measure $dE(\theta)$ is of no concern to us at this point since, ultimately, the saddle point approximation of the integral gives us the infinite sum regardless. As we will see from the DSSYK side, the measure of integration is actually modified (``renormalized'') by the 1-loop terms. For convenience, we are going to fix 
\begin{gather}
    \ell = \sqrt{8\pi G_N} \equiv \ell_P,
\end{gather}

then set $\ell$ to $1$ from now on.

For a given saddle spacetime, the energy, heat capacity and entropy are given by:
\begin{equation}
\begin{aligned}
	E = -\partial_{\beta} \log\mathcal{Z}_{\rm grav} = -\frac{\cos\theta}{2\abs{\log q}}, \\
	C = -\beta^2 \partial_\beta E = \frac{\pi \sin \theta}{\abs{\log q} \cos\theta}, \\
	S = (1-\beta \partial_\beta) \log\mathcal{Z}_{\rm grav} = \frac{\pi \theta}{\abs{\log q}}.
\end{aligned}
\end{equation}

It is important to observe that we have taken derivatives with respect to $\beta$. We recall our discussion in Section \ref{sec:firstex}, namely that every observer perceives a different temperature, so one should take derivatives with respect to the right one. As we will see, the $\beta$ of the gravitational theory is going to be the same $\beta$ of DSSYK, so this is the inverse temperature that is experienced by the boundary theory and the one with respect to which we should take derivatives. 

If $0 \leq \theta_+^{(0)} < \pi/2$, black hole saddles have a positive heat capacity, while cosmological ones are unstable when assigned a negative temperature. The situation flips if $\pi/2 < \theta_+^{(0)} \leq \pi$. The entropies of all the $\theta < 0$ horizons are negative, although their moduli satisfy the Bekenstein-Hawking formula with area \mbox{$A = \theta/(2\abs{\log q})$}. This is an extremely weird result but, for better or for worse, we will soon see that this is not the end of the story. We will need to modify this theory to perfectly match it to the DSSYK model. The point is that, as we have said earlier, this putative bulk theory is actually dual to the $q$-Schwarzian Quantum Mechanics, whose difference with respect to DSSYK is that its Hilbert space also possesses ``negative chord number'' states that have yet to be removed. Since the chord Hilbert space is dual to the gravitational one and is therefore connected to spacetime itself and its discretization, we observe that the unwanted presence of these negative chord states may be connected to negative entropies. We will substantiate this claim in Section \ref{sec:DefectOp}. A final observation is that only the saddles with the lowest free energy dominate: these are clearly $\theta_+^{(+\infty)}$ and $\theta_-^{(+\infty)}$ for all $\beta$, whose entropy is positive.

We now turn to the DSSYK model in order to see what we have missed in the previous picture and how to fix it at the semiclassical level. Ironically enough, at the quantum level, it is sufficient to remove the undesired states, but what are the consequences of this operation on the emergent spacetime? Recall the partition function of the DSSYK model \eqref{eq:DSSYKpi}:
\begin{gather}
	\mathcal{Z}_{\rm DSSYK}(\beta) = \int_0^\pi \frac{d\theta}{2\pi} \: (q^2, e^{\pm 2i\theta}; q^2)_\infty \exp \left( -\frac{2 \beta \mathcal{J} \cos\theta}{\sqrt{1-q^2}} \right).
\end{gather}

Recall that we are now using $q = e^{-p^2/N}$. For consistency with the previous discussion in this chapter, we adopt the same conventions of \cite{Blommaert:2024ydx}, namely we drop the constant $(q^2;q^2)_\infty/(2\pi)$, we take $\theta \to \pi-\theta$ and we redefine the variance of the couplings:
\begin{gather}
    \langle J^2_{i_1 \dots i_p} \rangle_J = \frac{1}{4\lambda} \binom{N}{p}^{-1} \mathcal{J}^2, \quad \lambda = 2\abs{\log q}.
\end{gather}

The final result is:
\begin{gather}
	\mathcal{Z}_{\rm DSSYK}(\beta) = \int_0^\pi d\theta \: (e^{\pm 2i\theta}; q^2)_\infty \exp \left( \frac{\beta \mathcal{J} \cos\theta}{\sqrt{2\abs{\log q}(1-q^2)}} \right).
\end{gather}

We now set $\mathcal{J} = 1$ and we consider the $\abs{\log q} \ll 1$ limit, which allows us to perform the following expansion of the $q$-Pochhammer symbol:
\begin{equation}
\begin{aligned}
	\log(x;q^2)_\infty & = \log \prod_{k=0}^\infty (1-xq^{2k}) = \sum_{k=0}^\infty \log(1-xq^{2k}) \\
	& = - \sum_{k=0}^\infty \sum_{n=1}^\infty \frac{x^n}{n} q^{2nk} = - \sum_{n=1}^\infty \frac{x^n}{n} \frac{1}{1-q^{2n}} \\
	& = - \sum_{n=1}^\infty \frac{x^n}{n} \left( \frac{1}{2n\abs{\log q}} + \frac{1}{2} \right) + \mathcal{O}(\abs{\log q}) \\
	& \approx -\frac{\mathrm{Li}_2(x)}{2\abs{\log q}} + \frac{1}{2} \log(1-x).
\end{aligned}
\end{equation}

Using this expansion and the fact that the dilogarithm satisfies the identity
\begin{gather}
	\mathrm{Li}_2(x) + \mathrm{Li}_2(1/x) = -\frac{\pi^2}{6} - \frac{1}{2}\log^2(-x),
\end{gather}

we obtain, up to $\mathcal{O}(\abs{\log q})$ terms:
\begin{equation}
\begin{aligned}
	\log(e^{\pm 2i\theta};q^2)_\infty & \approx \frac{1}{2\abs{\log q}} \left( \frac{\pi^2}{6} + \frac{1}{2} (i(2\theta-\pi))^2  \right) + \frac{1}{2} \log[(1-e^{2i\theta})(1-e^{-2i\theta})] \\
	& = -\frac{\pi^2}{6\abs{\log q}} + \frac{\pi\theta-\theta^2}{\abs{\log q}} + \log(2\sin\theta). \label{eq:loge2ithapprox}
\end{aligned}
\end{equation}

Recall that $\theta \in [0, \pi]$ and that the cut of the logarithm lies on the negative semiaxis, so $\log(-e^{2i\theta})$ is equal to $i(2\theta-\pi),$ with $2\theta-\pi \in [-\pi,\pi]$.

These approximated formulas are basically a 1-loop expansion, where we keep the first subleading terms. Interestingly enough, the only relevant effect of the 1-loop terms is a renormalization of the integration measure $d\theta \to \sin\theta \, d\theta = d\cos\theta$, so that the partition function becomes:
\begin{gather}
	\mathcal{Z}_{\rm DSSYK} \approx \int_0^\pi d\cos\theta \: \exp \left( \frac{\pi\theta - \theta^2}{\abs{\log q}} + \frac{\beta \cos\theta}{2\abs{\log q}} \right). \label{eq:SYKpartitionapp}
\end{gather}

$\mathcal{O}(\abs{\log q}^0)$ corrections would also affect the second term in the exponential, but we neglect them. An alternative would be a different redefinition of the variance of the couplings, such that the second term has that form for every value of $\abs{\log q}$. We find a first, important mismatch between the actual DSSYK model and the putative bulk theory. The saddle point approximation of the integral, in fact, gives us the following inverse temperature and entropy:
\begin{gather}
	\beta = \frac{2\pi-4\theta}{\sin\theta}, \quad S = \frac{\pi\theta-\theta^2}{\abs{\log q}}.
\end{gather}

A particular feature is that the entropy is not monotonic in $\theta$, but rather it cancels at the extremes $\theta = 0, \: \pi$ and has a maximum at $\theta = \pi/2$. The same result is obtained in a different way in Appendix \ref{app:ZDSSYK}, since $S \propto \log \Psi(\theta,q)$ (see \eqref{eq:ApproxDistrDSSYK}) and here we have the pointwise limit of the distribution, i.e. the Gaussian limit. As a consequence, the temperature of the system is negative when the mean energy is above \mbox{$E = 0$} $(\theta = \pi/2)$, since $\beta = dS/dE = 2 \abs{\log q} \, (\sin\theta)^{-1} \, dS/d\theta < 0$ there. This system has a truncated spectrum, so negative temperatures are not problematic and are actually needed to achieve $\langle E \rangle > 0$. The inverse temperature covers the whole positive semiaxis in the range $[0,\pi/2]$, decreasing monotonically from $+\infty$ to $0$. Similarly, it covers the negative semiaxis in the range $[\pi/2,\pi]$, decreasing monotonically from $0$ to $-\infty$.

This is one of the interesting characteristics of the DSSYK model, namely the difference between the actual temperature $\beta^{-1}$ and the so-called ``tomperature'', which is a concept that was already introduced by Lin and Susskind \cite{lin2022infinite}. By tomperature, Lin and Susskind mean the change in energy due to the removal of one qubit (i.e. two fermionic degrees of freedom) while keeping the couplings involving all the other fermions fixed. From a holographic point of view, the qubit is a quantum that leaves the horizon and transports energy equal to the tomperature, so this quantity is basically the Hawking temperature of the dual horizon.

If one takes the real time $T \to \infty$ limit of a two point function of a dimension $\Delta$ operator, at least in the $\abs{\log q} \ll 1$ scenario, one can show that there is an exponential decay of the type:
\begin{gather}
	\lim_{T \to \infty} \tr \left( e^{-(\beta/2+iT)H} M_\Delta e^{-(\beta/2-iT)} M_\Delta \right) \sim \exp \left( - \Delta \frac{2\pi}{\beta_{\rm BH}} T \right).
\end{gather}

For clarity, a random operator of size $p'$ has dimension $\Delta = p'/p$. Recall, in fact, the conformal limit of the SYK model, where the Green function \eqref{eq:confGS} had scaling dimension $1/p$ and is the propagator of a single fermion: a random operator is a product of $p'$ fermions and behaves effectively as a primary of dimension $\Delta$. The decay of DSSYK correlators is not driven by the temperature $\beta^{-1}$, but rather by the Hawking temperature that we had found earlier: $\beta_{\rm BH} = 2\pi/\sin\theta$. So: tomperature also drives the decay of correlators, in agreement with \cite{lin2022infinite}. From now on, we will use the term ``tomperature'' to refer to both the effective decay temperature of correlators and the Hawking temperature, since they are equal.

This feature has a profound meaning, especially if we consider the case $\theta = \pi/2$, where $\beta = 0$ and $\beta_{\rm BH} = 2\pi$. The DSSYK model is, in fact, a system with a finite number of degrees of freedom, so its thermodynamic observables are not expected to diverge in any way at infinite temperature: since the spectrum truncates at the ``UV'' energy $E_{\rm max} = 2/\sqrt{1-q^2}$, the average energy of the system has to remain finite. Indeed, the correlators have a thermal decay with a finite tomperature, as shown.

\section{The Missing Step: The Defect Operator} \label{sec:DefectOp}
We have shown that there is a mismatch between the thermodynamics of the two models, which is caused at the quantum level by the presence of the negative chord number states. In this section, we will show how to modify the semiclassical bulk theory in order to correct this issue. The quantum picture of the sine dilaton gravity is obtained in \cite{Blommaert:2024ydx}, but it gives us no intuition on the structure of the spacetime. Therefore, reviewing it is not of interest to us. The two most prominent aspects of the quantum picture is that the basis of the Hilbert space are the eigenvectors of $\mathbf{L}$, $| \mathbf{L} \geq 0 \rangle$, which are dual to the chord states of DSSYK, and that as a consequence the following equalities hold:
\begin{equation}
\begin{gathered}
	\tr(e^{-\beta H_{\rm SYK}}) = \langle \mathbf{L} = 0| e^{-\beta H_{\rm grav}} | \mathbf{L} = 0 \rangle, \\
	\langle E|L \rangle = H_{L/(2\abs{\log q})}(-2\abs{\log q}E | q^2), \quad E(\theta) = -\frac{\cos \theta}{2\abs{\log q}}.
\end{gathered}
\end{equation}

The variance of the couplings in DSSYK has been chosen in such a way that the energies are proportional to $1/\abs{\log q}$ even when $\abs{\log q}$ is not much smaller than $1$. The most important requirement of the semiclassical modification of the bulk theory is that the result must have inverse temperature $\beta = (2\pi-4\theta)/\sin\theta$, but black holes with inverse temperature $\beta_{\rm BH}$. In order for this to happen, there has to be some modification of gravity sitting exactly at the horizon. The solution is to introduce a ``defect'' operator with opening angle $\gamma$ in the partition function:
\begin{gather}
	V_\gamma = \int d^2x \, \sqrt{g} \: e^{-(2\pi-\gamma) \Phi/(2\abs{\log q})}.
\end{gather}

We will now be more detailed with respect to the source material, and investigate the consequences of this term in the partition function. The insertion of the operator yields:
\begin{gather}
    \mathcal{Z}_{\rm grav} = \int d^2 x_0 \int \mathcal{D}g \, \mathcal{D}\Phi \: \exp \left[ \int d^2x \: \left( \frac{1}{2} \log g - \frac{(2\pi-\gamma)\Phi}{2\abs{\log q}} \right) \delta^{(2)}(x-x_0) - S_E \right]
\end{gather}

We can forget about the $\log g$ term as it is $\mathcal{O}(\abs{\log q}^0)$, while the other one is responsible for introducing a delta source for the Ricci scalar:
\begin{gather}
    \sqrt{g} R - 2(2\pi-\gamma) \delta^{(2)}(x-x_0) + \sqrt{g} \, V'(\Phi) = 0.
\end{gather}

A saddle approximation of the position $x_0$ of the defect tells us that it is actually located where $\Phi$ is an extremum, namely at the horizon (where $\Phi = \theta$). Regardless of how the geometry is exactly modified by a defect sitting at the horizon, the semiclassical partition function is, up to a sum over saddles:
\begin{gather}
    \mathcal{Z}_{\rm grav} \sim \exp \left( \frac{\gamma \theta}{2 \abs{\log q}} + \frac{\beta \cos\theta}{2\abs{\log q}} \right).
\end{gather}

Simply, the new term has shifted the old action by a term proportional to $(\gamma-2\pi) \theta$. The choice of $\beta < 0$ for cosmological horizons is necessary for this result to also apply to them. Were we not to do this, the defect contribution to \eqref{eq:fakeZgrav} would not cancel $-2\pi\theta_-^{(k)}/(2\abs{\log q})$ in the action, which is the term we would find instead of its opposite.

If we now restore the sum over saddles, we correctly obtain:
\begin{gather}
    \mathcal{Z}_{\rm grav}(\gamma) = \int dE(\theta) \: \exp \left( \frac{\gamma\theta}{2\abs{\log q}} + \frac{\beta\cos\theta}{2\abs{\log q}} \right).
\end{gather}

$\theta$ is an integration variable of the path integral, so we are not allowed to simply fix $\gamma = 2\pi-2\theta$ (\cite{Blommaert:2024ydx} wrongly reports $2\pi-4\theta$), which would be a step further towards a match with $\mathcal{Z}_{\rm DSSYK}$. What we can do is integrate the opening angles with a Gaussian weight (a ``wave function'' $\psi(\gamma)$ of the opening angle). Let us consider the most general function and let us initially assume that $\gamma \in \mathbb{R}$:
\begin{gather}
    \mathcal{Z}_{\rm grav} = \int_{-\infty}^{+\infty} d\gamma \, \exp \left( -\frac{(a+\gamma)^2}{2b\abs{\log q}} \right) \mathcal{Z}_{\rm grav}(\gamma) \propto \int dE(\theta) \: \exp \left( -\frac{b}{8} \frac{\left( \frac{8a}{b} - 2\theta \right)\theta}{2\abs{\log q}} + \frac{\beta\cos\theta}{2\abs{\log q}} \right).
\end{gather}

To obtain the desired result, we would need $b = -8, \: a = -2\pi$. The integral would then diverge because of the growing exponential, but we can take the analytic continuation of the result where it exists. This is the approach in the paper, but we prefer a different interpretation, namely we take $\gamma$ to be purely imaginary:
\begin{gather}
    \mathcal{Z}_{\rm grav} = \int_{-\infty}^{+\infty} i \, d\gamma \, \exp \left( \frac{(a+i\gamma)^2}{2b\abs{\log q}} \right) \mathcal{Z}_{\rm grav}(i\gamma) \propto \int dE(\theta) \: \exp \left( \frac{b}{8} \frac{\left( -\frac{8a}{b} - 2\theta \right)\theta}{2\abs{\log q}} + \frac{\beta\cos\theta}{2\abs{\log q}} \right).
\end{gather}

We now need $b = 8, \: a = -2\pi$, which make the integral well-defined. Interestingly enough, this tells us that the opening angles of the defect over which we integrate are not real, but the final result is equivalent to choosing a real $\theta$-dependent angle $\gamma = 2\pi-2\theta$.

Let us go back to the semiclassical partition function of the DSSYK model and rewrite it in a form that allows us to conclude that the previous prescription is the correct one. Although this result is stated in \cite{verlinde2024doublescaled}, no derivation is presented. From Appendix \ref{app:ZDSSYK}, we recover that the partition function can be written in terms of a Jacobi theta function (see Equation \eqref{eq:civitavecchia}):
\begin{gather}
    \mathcal{Z}_{\rm DSSYK} \approx \int_0^\pi d\theta \: \Psi(\theta,q) \, \exp \left( \frac{\beta \cos\theta}{2\abs{\log q}} \right), \quad \Psi(\theta,q) = \frac{\sin \theta}{\pi q^{1/4}} \: \theta_1 \left( \frac{\theta}{\pi}, \frac{i\abs{\log q}}{\pi} \right).
\end{gather}

Since $\abs{\log q} \ll 1$, we can write:
\begin{gather}
    \theta_1 \left( \frac{\theta}{\pi}, \frac{i\abs{\log q}}{\pi} \right) \approx 2 \sqrt{\frac{\pi}{\abs{\log q}}} e^{- \theta^2/\abs{\log q} - \pi^2/(4\abs{\log q})} \sinh \left( \frac{\pi\theta}{\abs{\log q}} \right) \prod_{m=1}^{\infty} \left( 1- 2 \cosh \left( \frac{2\pi\theta}{\abs{\log q}} \right) e^{-2m\pi^2/\abs{\log q}} \right)
\end{gather}

Note that we have restored part of the infinite product, as opposed to what we do in the appendix. We want to rewrite the Jacobi theta function for an arbitrary real $\theta \neq k\pi, \: k \in \mathbb{Z}$: if $\theta = k\pi$, in fact, the $\sin\theta$ function sets $\Psi(\theta,q)$ to $0$ anyway, but also the $m = |k|$ term in the product suppresses the value of $\theta_1$ exponentially if $k \neq 0$ (or the hyperbolic sine cancels it completely if $k = 0$). We will check at the end that our approximated result for $\theta_1$ behaves well around these values. If we define $\bar m$ as the non-negative integer such that $\bar m \pi < |\theta| < (\bar m + 1)\pi$, then the infinite product will be dominated by the hyperbolic cosines for $m \leq \bar m$ and by $1$ for $m > \bar m$:
\begin{gather}
    \theta_1 \left( \frac{\theta}{\pi}, \frac{i\abs{\log q}}{\pi} \right) \approx \sqrt{\frac{\pi}{\abs{\log q}}} e^{- \theta^2/\abs{\log q} - \pi^2/(4\abs{\log q})} \mathrm{sgn}(\theta) e^{\pi|\theta|/\abs{\log q}} (-1)^{\bar m} \prod_{m=1}^{\bar m} e^{2\pi(|\theta|-\bar m \pi)/\abs{\log q}}.
\end{gather}

If we merge $\mathrm{sgn}(\theta)$ with $\sin\theta$ in $\Psi(\theta,q)$, we obtain an even extension of the distribution to negative angles in a natural way. Forgetting about the sign and the constant prefactor, then, we have:
\begin{equation}
\begin{aligned}
     \theta_1 \left( \frac{\theta}{\pi}, \frac{i\abs{\log q}}{\pi} \right) & \propto (-1)^{\bar m} \exp \left[ \frac{1}{\abs{\log q}} \left( -\theta^2-\frac{\pi^2}{4} + \pi|\theta| + 2\pi \bar m |\theta| - \pi^2 \bar m(\bar m + 1) \right) \right] \\
     & = (-1)^{\bar m} \exp \left[ -\frac{1}{\abs{\log q}} \left(|\theta|-\pi \left( \bar m + \frac{1}{2} \right) \right)^2 \right].
\end{aligned}
\end{equation}

Note that the result is also correct for $\bar m = 0$. We have found that the distribution is a Gaussian around $\pi/2 + \bar m \pi$ with variance $\sigma^2 = \abs{\log q}/2$, so this approximation also suppresses the case $\theta = k\pi$ correctly. In the $\abs{\log q} \to 0$ limit, for a fixed value of $\theta$, we can substitute the actual distribution with the sum of all the aforementioned Gaussians, plus those with peaks at negative values of $|\theta|$ (which are unattainable). The consequence of this is an absolutely negligible error, so that up to a constant we can finally write:
\begin{gather}
    \mathcal{Z}_{\rm DSSYK} = \int dE(\theta) \: \rho(\theta) \, \exp \left( \frac{\beta \cos\theta}{2\abs{\log q}} \right), \quad \rho(\theta) = \sum_{n=-\infty}^\infty (-1)^n \exp \left[ -\frac{1}{\abs{\log q}} \left(\theta - \pi \left(n + \frac{1}{2} \right) \right)^2 \right].
\end{gather}

Let us now complete the computation on the gravitational side, but with a particular prescription: the cosmological saddles have to be taken as negative contributions to the total partition function. The DSSYK partition function is integrating only over $\theta \in [0,\pi]$, while the gravitational one is over the whole real axis. We can match the extremes of integration in the following way\footnote{We take all the possible $2\pi n$ translations of $\theta \in [0,\pi]$ and $\theta \in [-\pi, 0]$, thus covering the real axis.}:
\begin{equation}
\begin{aligned}
    \mathcal{Z}_{\rm grav} & = \int_{-\infty}^{+\infty} i \, d\gamma \, \exp \left( \frac{(2\pi-i\gamma)^2}{16\abs{\log q}} \right) \int_0^\pi dE(\theta) \sum_{n=-\infty}^{\infty} 2\sinh \left( \frac{i \gamma(\theta+2\pi n)}{2\abs{\log q}} \right) \exp \left( \frac{\beta\cos\theta}{2\abs{\log q}} \right) \\
    & \propto \int_0^\pi dE(\theta) \: \tilde \rho(\theta) \, \exp \left( \frac{\beta \cos\theta}{2\abs{\log q}} \right),
\end{aligned}
\end{equation}

where the density that appears here is
\begin{gather}
    \tilde \rho(\theta) = \sum_{n=-\infty}^{\infty} \left[ \exp \left( -\frac{(\theta + \pi(2n-1/2))^2}{\abs{\log q}}\right) - \exp \left( -\frac{(\theta + \pi(2n+1/2))^2}{\abs{\log q}}\right) \right].
\end{gather}

To show that $\tilde \rho(\theta) = \rho(\theta)$, we rewrite the sum in $\rho(\theta)$ using $\{ n \in \mathbb{Z} \} = \{ -2m, \: -2m-1 \}_{m \in \mathbb{Z}}$:
\begin{gather}
    \rho(\theta) = \sum_{m=-\infty}^{\infty} \left[ \exp \left( -\frac{(\theta + \pi(2m-1/2))^2}{\abs{\log q}} \right) - \exp \left( -\frac{(\theta + \pi(2m+1/2))^2}{\abs{\log q}} \right) \right] = \tilde \rho(\theta).
\end{gather}

With this computation, we have successfully shown how the two theories can be matched at the semiclassical level. If we restore $\ell$ and $\mathcal{J}$, in particular, we find that these quantities are related by:
\begin{gather}
    \mathcal{J} = \ell^{-1}.
\end{gather}

In Section \ref{sec:thermodual}, we had found that all the $\theta < 0$ saddles were characterized by negative entropies and we had observed that negative chord number states may have been the culprits. If we add $\theta-$independent terms to $\rho(\theta)$ to recover the structure shown in \eqref{eq:SYKpartitionapp} for all saddles, we find that the entropy is now:
\begin{equation}
\begin{gathered}
    S_n(\theta) = \frac{1}{\abs{\log q}} \left[ \pi^2 \left( n + \frac{1}{2} \right)^2 - \left( \theta - \pi \left( n + \frac{1}{2} \right) \right)^2 \right] = \frac{\theta \, (\pi(2n+1)-\theta)}{\abs{\log q}}, \\
    \pi \left(n + \frac{1}{2} \right) - \frac{\pi}{2} \leq \theta \leq \pi \left(n + \frac{1}{2} \right) + \frac{\pi}{2}.
\end{gathered}
\end{equation}

This result gives the correct entropy for $n = 0$. It is also easy to show that $S_n(\theta) \geq 0$ for all $n$ and for all $\theta$, which points towards the correctness of our claim that negative entropies are connected to negative chord number states.

We notice a possible inconsistency of this procedure. In deriving the $\propto \pi\theta$ term in the Euclidean action, we have used that $\beta = 4\pi/V(\theta)$. On the other hand, the insertion of the defect has modified the saddle $\bar\theta(\beta)$ relation to match the one in DSSYK, so it is unclear that the end result is unaffected by never performing this substitution in the middle of the calculations. Another possible issue is that the entire $S_E$ action has been computed using the unperturbed geometry, to which the defect contribution has been added while neglecting its backreaction on spacetime. It may be possible that a consistent computation yields the same result, or that a modification to the way the defect is introduced is necessary to cure the matching procedure. In this thesis, we will assume that this step is not problematic, but it should be explicitly checked that this is indeed the case.

There is another observation on our part. The gravitational action we have presented at the beginning has a prefactor of $\ell_P^{-2}$, which was chosen to estabilish a connection with the dilaton-gravity models that one can expect from a dimensional reduction of four-dimensional theories. We have then set $\ell = \ell_P$ to cancel the prefactor. We could have introduced a more agnostic theory with an a priori prefactor of $\ell^{-2}$, which would then have given us the freedom to set $\ell$ to whatever value we wanted and, consequently, $\mathcal{J}$ on the DSSYK side. Conversely, though, the fact that we expect to find $\ell = \ell_P$ from our physical reality tells us that the boundary theory of interest is expected to have couplings of order $\ell_P^{-1} = M_P$.

This makes things even more interesting when one recalls that the conformal limit of the SYK model kicks in at $t \gtrsim \mathcal{J}^{-1} = t_P$, and that this IR limit describes energy scales $\omega \lesssim M_P$. This means that the full SYK model completes JT gravity (which is dual only to its IR part) exactly to energies above $M_P$. Then, if we consider the specific case of the double-scaled limit on one side, we find that the UV completion of JT gravity that it induces in the bulk is exactly our sine dilaton model. In fact, by considering the sine potential of the initial dilaton field:
\begin{gather}
    V(\Phi) = \frac{\sin(2\abs{\log q} \Phi/\ell^2)}{\abs{\log q}},
\end{gather}

this clearly reduces to the JT potential $V_{\rm JT} = 2\Phi/\ell^2$ as long as $\abs{\Phi} \ll \ell^2/\abs{\log q}$. JT gravity was found in the first place by placing a cutoff on the value of $\Phi$ and consequently working in the ``IR'' region, where every UV theory reduced to the same universal dynamics. This requirement on $\Phi$ implies $\theta \ll 1$, which is exactly the IR part of the DSSYK energy spectrum.

To conclude, the unusual operations that were required in order to perform the match were the use of an imaginary opening angle in the defect operator, a negative temperature for cosmological horizons and a negative contribution to the partition function coming from the cosmological saddles. It may be that the minus sign in front of the cosmological saddles is due to the deteterminant of the quadratic expansion of the action, which is not guaranteed to be real given the presence of the complex $r$ coordinate, around these solutions. Verifying this hypothesis would require a one-loop computation on the gravitational side, which is outside the scope of this thesis and is left to future work.

\section{Static Patch Holography?} \label{sec:Susskind}
Having shown that the duality between the two theories holds, it is natural to ask how this result ties with the several claims about static patch holography discussed by Lin, Rahman and Susskind in the references reported at the beginning of Chapter 6.

We start with the claim that, to be dual to de Sitter spacetimes, the DSSYK model should be taken at infinite temperature $(\beta = 0)$. This is to ensure that the resulting dual spacetime has the maximum possible entropy: this is, in fact, a known property of empty de Sitter static patches \cite{Maeda_1998,Bousso_2001,Bousso_2000}. The main argument is the following: if one adds matter with its own entropy inside the static patch, the cosmological horizon will shrink in such a way that the new total entropy is at most the empty starting one. This happens because matter tends to escape beyond the cosmological horizon\footnote{To see this immediately, consider timelike trajectories in the Penrose diagram of dS. Matter leaves the causal diamond of a comoving observer asymptotically in the future unless its position coincides with theirs, with the diamond being exactly the static patch.} and entropy should never decrease in any physical process: since an observer tends to experience empty de Sitter asymptotically, its entropy should be maximum. We can take, as an example of this fact, the four-dimensional Schwarzschild-de Sitter black hole, and check that the entropy is maximum when the black hole is not present.

For a generic value of the mass $M$, the total entropy is:
\begin{gather}
    S(M) = S_{\rm c}(M) + S_{\rm b}(M) = \frac{\pi}{G_N}(r^2_c(M)+r^2_b(M)).
\end{gather}

\begin{figure}
    \centering
    \includegraphics[width = 0.65 \textwidth]{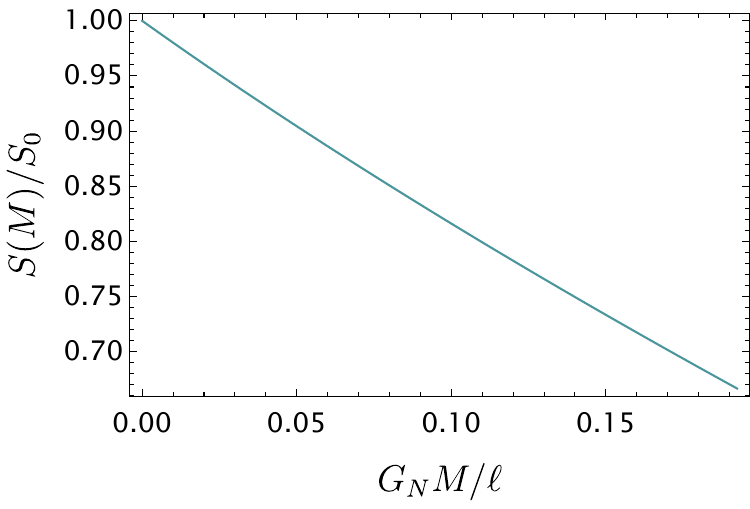}
    \caption{Ratio between SdS$_4$ entropy and empty dS$_4$ entropy as a function of the black hole mass $M$, up to $M^*$. Clearly, the entropy decreases monotonically as more and more matter is added inside the static patch.}
    \label{fig:SdSEntropy}
\end{figure}

$r_c$ and $r_b$ are the radii of the cosmological horizon and of the event horizon, which were already determined as a function of $M$ in Figure \ref{fig:SdSBlack}. The ratio between the total entropy $S(M)$ and the empty de Sitter entropy $S_0$ is shown in Figure \ref{fig:SdSEntropy}, which proves the expected behavior.

As a consequence of this, if we start from empty de Sitter and we imagine forming a small black hole with mass $\delta M$, the second law of black hole mechanics is actually:
\begin{gather}
    \delta M = -T_{\rm dS} \delta S, \quad T_{\rm dS} = \frac{1}{2\pi \ell}.
\end{gather}

Entropy is reduced when adding matter, so there is a surprising minus sign that appears above \cite{PhysRevD.15.2738}. We choose to view this result as a negative temperature for the cosmological horizon, which is what we have assigned to the $\theta_-^{(k)}$ in the sine dilaton model. Finally: this assignment is not only needed for the match to be possible, but it is also supported by this phenomenon on the gravitational side.

Next, we recall that the temperature and inverse entropy in the $\abs{\log q} \to 0$ DSSYK model are:
\begin{gather}
   \beta = \frac{2\pi-4\theta}{\sin\theta}, \quad S = \frac{\pi\theta-\theta^2}{\abs{\log q}}.
\end{gather}

The choice of $\theta = \pi/2$ maximizes the entropy and gives us $\beta = 0$. There seems to be a mismatch with the fact that empty de Sitter has finite temperature $T_{\rm dS} = (2\pi \ell)^{-1}$, but this is solved by the finite tomperature of DSSYK, which for $\theta = \pi/2$ is exactly equal to $(2\pi \mathcal{J}^{-1})^{-1}$ and is more suitable for an association with the horizon temperature. These match for $\ell_{\rm dS} \equiv \ell = \mathcal{J}^{-1}$, which is compatible with what we have estabilished in the previous section.

For $\theta = \pi/2$, the sine dilaton geometry is:
\begin{gather}
    ds^2 = -(-2\cos r) \, dt^2 + \frac{dr^2}{-2\cos r}, \quad R(r) = -2 \cos r.
\end{gather}

The curvature on the real line is always non-negative and ranges from $0$ at the two horizons to a maximum of $R = +2$ when $r = \pi$. Although it is not exactly the static patch, this geometry does have a ``near de Sitter'' inner region. Susskind claims that the boundary theory should be positioned at the stretched horizon of the static patch, but here we see that it lives at $\Phi_{\rm bdy} = \pi/2 + i\infty$. The interpretation of the imaginary part of the radial coordinate is unclear outside of the effective geometric picture but, by focusing on the real part, different intuitions emerge from different integration contours for $\Phi$. In any case, the real part of $\Phi_{\rm bdy}$ is always $\pi/2$, which is exactly the position of the horizon for $\theta = \pi/2$.
\begin{itemize}
    \item If one takes the integration contour to be $\Phi = \pi/2 +i\alpha, \: \alpha \in [0,+\infty]$, the geometry never lies on the real axis. Since the real part is always $\pi/2$, the intuition is that the DSSYK model does live on the horizon, in a way.
    \item If one takes the integration contour to first link $\pi/2$ to $\pi$ while staying on the real axis, then follow a contour that eventually reaches $\Phi_{\rm bdy}$, the geometry actually has a first part of positive curvature. $r = \pi$ is an interesting value, because $R = +2$ and locally a static patch is experienced:
    \begin{gather}
        ds^2 \approx -2 \left( 1-\frac{\delta^2}{2} \right) \, dt^2 + \frac{d\delta^2}{2 \left( 1-\frac{\delta^2}{2} \right)}, \quad r = \pi + \delta, \: |\delta| \ll 1.
    \end{gather}
    This was basically what we had found for $\theta = \pi$ in the first section, where the local static patch was everything that was there to experience. $r = \pi$ is also a pode geodesic, namely a geodesic followed by a comoving observer in de Sitter. This choice of contour emphasizes the possible role of an observer in a quantum theory of gravity in de Sitter \cite{Chandrasekaran_2023,witten2023background}. The intuition here is that this geometry appears to have a ``centaur''-like structure, with a de Sitter IR part and something else in the UV that ties to the boundary, glued together at the pode $r = \pi$. This kind of geometry (where the UV part is AdS$_2$) has been explored in \cite{Anninos:2017hhn,Anninos_2021,Anninos_2022}.
\end{itemize}

Note that we have used the metric of the geometry without defects, since the discussion does not change in that case. It would be interesting to find observables that are able to probe this geometry and not the effective AdS$_2$ one, since this would allow us to gain insight into the quantum nature of de Sitter. Operationally, one would need a probe that couples non-trivially to the dilaton and to the metric (and not to the specific combination that yields $(g_{\mu\nu})_{\rm eff}$). As for Susskind's claims, they appear to have successfully caught all the interesting physics of DSSYK and its holographic dual (particularly its possibility of describing a dS spacetime), but they are slightly off when it comes to details.

\section{Scalar Waves}
In this section, we demonstrate how probes that couple minimally to the sine dilaton geometry would possess entirely different properties with respect to the ones that couple to the effective metric, hence giving us the chance to study the de Sitter region of the spacetime. To do this, we will investigate the behavior of a minimally coupled real scalar field $\psi$ of mass $m$ in the naive Lorentzian bulk geometry, then the behavior of a non-minimally coupled scalar in the effective AdS$_2$ one. The equation of motion of a Fourier mode with frequency $\omega$ in the first case, $\psi(r,t) = e^{i\omega t} \psi(r)$, is:
\begin{equation}
\begin{aligned}
    & \partial_\mu (g^{\mu\nu} \partial_\nu \psi(r,t)) - m^2 \psi(r,t) = 0 \\
    \implies & \omega^2 \frac{1}{f(r)} \psi(r) + \partial_r (f(r) \partial_r \psi(r)) = m^2 \psi(r).
\end{aligned}
\end{equation}

This equation can be recast in a Schrödinger form by using an auxiliary coordinate that maps $[\theta, 2\pi-\theta]$ to $[-\infty, +\infty]$:
\begin{equation}
\begin{gathered}
    y(r) = \int_{\pi}^r dr' \: \frac{1}{f(r')} = \frac{1}{2 \sin \theta} \log \left( \frac{\sin((r-\theta)/2)}{\sin((r+\theta)/2)} \right), \\
    \tan \left( \frac{r}{2} \right) = \frac{1 + e^{2 \sin\theta \, y}}{1 - e^{2 \sin\theta \, y}} \tan \left( \frac{\theta}{2} \right).
\end{gathered}
\end{equation}

This way, we obtain:
\begin{equation}
\begin{gathered}
    \left( -\frac{d^2}{dy^2} + m^2 f(r(y)) \right) \psi(y) = \omega^2 \psi(y), \\
    V(y) = m^2 f(r(y)) = m^2 \frac{2\sin^2\theta}{\cosh (2\sin\theta \, y) - \cos\theta}.
\end{gathered}
\end{equation}

\begin{figure}
    \centering
    \includegraphics[width = 0.7 \textwidth]{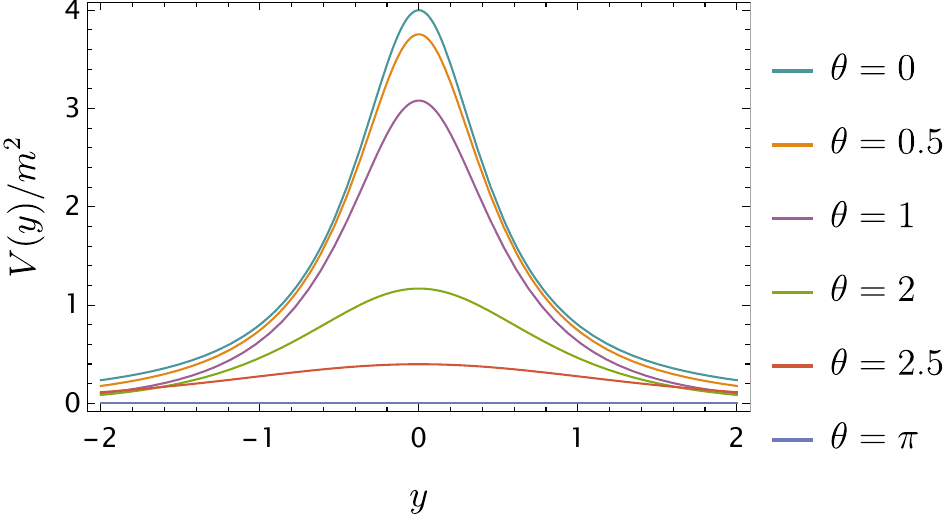}
    \caption{Plot of the potential $V(y)/m^2$ for different values of the black hole horizon's position $\theta$. The value of the peak is $4 \cos^2 (\theta/2)$, so the potential becomes less repulsive as $\theta$ increases.}
    \label{fig:quasinormal1}
\end{figure}

This potential is plotted in Figure \ref{fig:quasinormal1} for different values of $\theta \in [0,\pi]$: as we can see, its shape is always a bell with a peak at $y=0$ and an exponential falloff to $0$ for $y \to \pm \infty$. The appropriate boundary conditions are that the wave should be ingoing near the black hole horizon and ingoing towards the cosmological one: the physical reasoning behind this is that we assume that neither the black hole nor what lies beyond the cosmological horizon are sourcing the wave. Depending on conventions, this either means \mbox{$\psi(y \to \pm \infty) \sim e^{\pm i \omega y}$} or $\psi(y \to \pm \infty) \sim e^{ \mp i \omega y}$. Our choice is that $\psi$ oscillates as $e^{i\omega t}$, so that ingoing waves towards the black hole are $e^{i\omega(t+y)}$, while ingoing waves towards the cosmological horizon are $e^{i\omega(t-y)}$.

If $m = 0$, the solutions are trivial:
\begin{gather}
	\psi(y) = \alpha_1 e^{i\omega y} + \alpha_2 e^{-i\omega y}.
\end{gather}

It is clear that there is no way for $\psi(y)$ to satisfy both boundary conditions. It may seem that, if $\omega$ has an arbitrary positive imaginary part, the $e^{i\omega y}$ piece is exponentially suppressed for $y \to +\infty$, and vice versa. These modes would be stable, as their magnitude would decrease exponentially over time. The problem with this type of wave is clear from a wave packet's (the actual physical object) point of view: a wave packet is not still in space, but travels from one direction to the other as time passes. This means that such a packet does not stand still at $\pm \infty$, and eventually leaves that region: any non-null component, although exponentially suppressed in space, gives a finite contribution \textit{everywhere}. The situation becomes clear once one accounts for time evolution. A wave that leaves the cosmological horizon is actually $\psi(t,y) \sim e^{i\omega(t+y)}$: it was localized at $y = +\infty$ at $t = -\infty$, in such a way that $t+y =$ constant (we are studying massless particles). This means that this component is not negligible at all, and has to be set to $0$ manually. The same reasoning clearly applies to the outgoing wave near the black hole horizon.

In order to find the quasinormal modes of this field when $m^2 > 0$, we can use the WKB approximation. We use the results of \cite{Schutz:1985km}, from which we infer that the frequencies are approximately given by:
\begin{gather}
    \omega_n^2 = V(0) + i \left( n+\frac{1}{2} \right) \sqrt{-2V''(0)} = 4m^2 \cos^2 \left( \frac{\theta}{2} \right) + i \, 4 (2n+1) \, m \ell^{-1} \cos^2 \left( \frac{\theta}{2} \right), \quad n \in \mathbb{N}. \label{eq:dsquasiwave}
\end{gather}

We have restored $\ell \neq 1$ for this result. Given the shape of the potential, it is clear that it is impossible to satisfy the need for an ingoing wave at both horizons if $\Re(\omega^2) > V(0)$ and that the wave's amplitude would be heavily suppressed at $y < 0$ or $y > 0$ unless $\Re(\omega^2) \lesssim V(0)$. We see that all frequencies go to 0 when $\theta \to \pi$, so it is interesting to study this particular case with the parametrization \eqref{eq:dS2}. \\
The Schrödinger equation becomes:
\begin{equation}
\begin{gathered}
    \left( -\frac{d^2}{dy^2} + \frac{m^2 \varepsilon^2}{\cosh^2 (\varepsilon y)} \right) \psi(y) = \omega^2 \psi(y) \\
    \implies  \left( -\frac{d^2}{d\tilde y^2} + \frac{m^2}{\cosh^2 \tilde y} \right) \psi(\tilde y) = \tilde \omega^2 \psi(\tilde y), \label{eq:staticquasinormal}
\end{gathered}
\end{equation}

where we have defined $\tilde y = \varepsilon y, \: \tilde \omega = \omega/\varepsilon$. We recognize a repulsive Pöschl-Teller potential \cite{Cevik:2016mnr}:
\begin{gather}
    V(\tilde y) = -\frac{\lambda(\lambda-1)}{\cosh^2(\tilde y)}, \quad
    \lambda = \frac{1}{2} \left( 1 + \sqrt{1-4m^2} \right).
\end{gather}

We now perform an exact study of this potential. Although it is clear that all the $\omega_n$ are $\mathcal{O}(\varepsilon)$ frequencies (it follows from the $\tilde \omega_n$ being $\mathcal{O}(1)$) and that in the static patch geometry all scalar fields with $\omega = \mathcal{O}(1)$ are basically propagating freely, it is interesting to see the structure of the quasinormal modes that lies beyond the $\varepsilon$ prefactor. We observe, in fact, that a typical frequency will actually be of order $\mathcal{O}(\varepsilon)$ when using the ``wrong'' time coordinate $t$. We recall the form of the metric when $\theta \to \pi$:
\begin{gather}
	ds^2 = -(\varepsilon^2-\delta^2) \, dt^2 + \frac{d\delta^2}{\varepsilon^2-\delta^2}, \qquad \delta \in (-\varepsilon,\varepsilon).
\end{gather}

It is clear that the relevant coordinates in the $\varepsilon \to 0$ limit are $\tilde t = \varepsilon t$ and $\tilde r = \delta/\varepsilon$, with respect to which the metric truly takes the form of a static patch:
\begin{gather}
	ds^2 = -(1-\tilde r^2) \, d\tilde t^2 + \frac{d\tilde r}{1-\tilde r^2}, \qquad \tilde r \in (-1,1).
\end{gather}

The frequency associated to the time coordinate $\tilde t$ is exactly $\tilde \omega = \omega/\varepsilon$, which is of order $\mathcal{O}(1)$.

We will assume $m^2 > 0$ in the following. The most general solution of the differential equation \eqref{eq:staticquasinormal} is:
\begin{gather}
	\psi(\tilde y) = \alpha_P P_{-(1-\sqrt{1-4m^2})/2}^{i\tilde \omega}(\tanh \tilde y) + \alpha_Q Q_{-(1-\sqrt{1-4m^2})/2}^{i\tilde \omega}(\tanh \tilde y).
\end{gather}

$P$ and $Q$ are called the Legendre functions. We are interested in considering the $\tilde y \to \pm \infty \: (\tanh \tilde y \to \pm 1)$ behavior of these functions to impose our boundary conditions and constrain the spectrum.

For $\tanh \tilde y \approx 1 - 2e^{-2\tilde y} \to +1$, we find that the two functions behave in the following way:
\begin{equation}
\begin{aligned}
	P_{-(1-\sqrt{1-4m^2})/2}^{i\tilde \omega}(\tanh \tilde y \to +1) \approx & \: \frac{e^{i \tilde \omega \tilde y}}{\Gamma (1-i \tilde \omega )}, \\
	Q_{-(1-\sqrt{1-4m^2})/2}^{i\tilde \omega}(\tanh \tilde y \to +1) \approx & \: \frac{\left(1+e^{2 \pi \tilde \omega +i \pi  \sqrt{1-4 m^2}}\right) \Gamma (-i \tilde \omega) \, \Gamma (i \tilde \omega -\frac{1}{2} \sqrt{1-4 m^2}+\frac{1}{2})}{2 \left(e^{2 \pi  \tilde \omega }+e^{i \pi  \sqrt{1-4 m^2}}\right) \Gamma(-i \tilde \omega -\frac{1}{2} \sqrt{1-4 m^2}+\frac{1}{2})} e^{-i \tilde \omega \tilde y} \\
	& + \frac{1}{2} \cosh (\pi \tilde \omega ) \Gamma (i \tilde \omega) e^{i \tilde \omega \tilde y}.
\end{aligned}
\end{equation}

For $\tanh \tilde y \approx - 1 + 2e^{2\tilde y} \to -1$, on the other hand, the following asymptotic formulas hold:

\begin{equation}
\begin{aligned}
	P_{-(1-\sqrt{1-4m^2})/2}^{i\tilde \omega}(\tanh \tilde y \to -1) \approx \: & \frac{1}{\tilde \omega \sinh (\pi \tilde \omega)} \Bigg[ \frac{\pi \, e^{i \tilde \omega \tilde y} }{\Gamma (i \tilde \omega ) \, \Gamma (-i \tilde \omega -\frac{1}{2} \sqrt{1-4 m^2}+\frac{1}{2} ) \, \Gamma (- i \tilde \omega +\frac{1}{2}\sqrt{1-4 m^2}+ \frac{1}{2} )} \\
	& + \frac{\cos \left(\frac{1}{2} \pi \sqrt{1-4 m^2}\right) e^{-i \tilde \omega \tilde y}}{\Gamma (-i \tilde \omega )} \Bigg], \\
	Q_{-(1-\sqrt{1-4m^2})/2}^{i\tilde \omega}(\tanh \tilde y \to -1) \approx \: & \frac{\Gamma (-i \tilde \omega ) \, \cos (\frac{1}{2} \pi (\sqrt{1-4 m^2}-2 i \tilde \omega )) \, \Gamma (i \tilde \omega -\frac{1}{2} \sqrt{1-4 m^2}+\frac{1}{2})}{4 \Gamma (-i \tilde \omega -\frac{1}{2} \sqrt{1-4 m^2}+\frac{1}{2})} \\
	& \times \left(\cot \left(\frac{1}{4} \pi  \left(\sqrt{1-4 m^2}+2 i \tilde \omega -1\right)\right)- \frac{\sin \left(\frac{1}{4} \pi \left(\sqrt{1-4 m^2}+2 i \tilde \omega -1\right)\right)}{\sin \left(\frac{1}{4} \pi \left(\sqrt{1-4 m^2}+2 i \tilde \omega +1\right)\right)} \right) \, e^{i \tilde \omega \tilde y} \\
	& -\frac{1}{2} \sin \left(\frac{1}{2} \pi \sqrt{1-4 m^2}\right) \Gamma (i \tilde \omega) \, e^{-i \tilde \omega \tilde y}.
\end{aligned}
\end{equation}

To avoid misleading results, we use the known property of Gamma functions
\begin{gather}
	\Gamma(1 \mp i \tilde \omega) \Gamma(\pm i \tilde \omega) = \pm \frac{\pi}{i \sinh(\pi \tilde \omega)}, \label{eq:Gammafuncprop}
\end{gather}

so that we can rewrite:
\begin{equation}
	\begin{aligned}
		P_{-(1-\sqrt{1-4m^2})/2}^{i\tilde \omega}(\tanh \tilde y \to -1) \approx \: & \frac{i \, \Gamma(1-i \tilde \omega) \, e^{i \tilde \omega \tilde y} }{\tilde \omega \, \Gamma (-i \tilde \omega -\frac{1}{2} \sqrt{1-4 m^2}+\frac{1}{2} ) \, \Gamma (- i \tilde \omega +\frac{1}{2}\sqrt{1-4 m^2}+ \frac{1}{2})} \\
		& - \frac{i \, \Gamma(1+i\tilde\omega) \cos \left(\frac{1}{2} \pi \sqrt{1-4 m^2}\right) e^{-i \tilde \omega \tilde y}}{\pi \tilde \omega}.
	\end{aligned}
\end{equation}

Since conventions are physically irrelevant, we will make the more convenient assumption that $\psi$ oscillates as $e^{-i\tilde\omega \tilde t}$ in this case, so what is ingoing and what is outgoing with respect to $\tilde y$ is switched. For $\tilde y \to +\infty$, then, only the $e^{i\tilde\omega \tilde y}$ wave has to be present: this means that the $e^{-i \tilde\omega \tilde y}$ part of $Q$ has to be absent. Since the Gamma function has no zeroes, we must either impose that $\alpha_Q = 0$ or that $\Gamma(-i \tilde \omega -\frac{1}{2} \sqrt{1-4 m^2}+\frac{1}{2}) = \infty$, that is, the argument is a pole. If we impose that $\alpha_Q = 0$, then $\psi(\tilde y)$ is proportional to $P$, which has both an outgoing and an ingoing component at the black hole horizon. Again, we are forced to choose a value of $\tilde \omega$ such that the outgoing wave is absent by taking a pole of one of the Gamma functions. Since $\Gamma(x) = \infty$ if and only if $x$ is a non-positive integer $-n \leq 0$, this first option gives us two choices that satisfy both boundary conditions:
\begin{equation}
\begin{gathered}
	\tilde \omega_{n \geq 0}^{(\pm)} = -i \ell^{-1} \left( n + \frac{1}{2} \pm \frac{1}{2} \sqrt{1-4m^2 \ell^2} \right). \label{eq:dSexactquasinorm}
\end{gathered}
\end{equation}

We have restored $\ell \neq 1$. The option of not setting $\alpha_Q = 0$ yields the same $\tilde \omega_{n \geq 0}^{(-)}$ reported above: for this specific choice, both boundary conditions are satisfied regardless of the values of $\alpha_{P,Q}$. The absence of frequencies with null or positive imaginary part is both expected and a positive feature of the system. A positive imaginary part, in fact, would imply an instability of the system, as waves would grow exponentially with time, while a null imaginary part would be surprising, as it would mean that there is no dissipation due to the horizons at all. The modes we have obtained, on the other hand, are all stable and decay over time.

The exact result of this specific geometry gives us insight about the validity of \eqref{eq:dsquasiwave}. We first observe that, for $m^2 \leq \ell^{-2}/4$, the $\tilde \omega_n^{(\pm)}$ modes do not propagate in the geometry, but they only decay. The situation changes when $m^2 > \ell^{-2}/4$, as the $\tilde \omega_n^{(\pm)}$ acquire a real part. In the $m \gg \ell^{-1}$ limit, we can keep the leading terms:
\begin{gather}
	\tilde \omega_n^{(\pm)} = \pm m -i\ell^{-1} \left( n + \frac{1}{2} \right) + \mathcal{O}(m^{-1}) \implies \tilde \omega_n^2 = m^2 \mp i (2n + 1) m \ell^{-1} + \mathcal{O}(m^0),
\end{gather}

which is exactly the result that \eqref{eq:dsquasiwave} yields for $V(\tilde y) = m^2 \ell^2/\cosh^2 \tilde y$, up to a sign in front of the imaginary part due to the opposite convention for time oscillations (and the possibility of $\mathrm{Re}(\tilde \omega_n) < 0$). We deduce that the WKB approximation is expected to work for scalars whose Compton wavelength $m^{-1}$ is much smaller than $\ell \sim \ell_P$. One could worry about the $-\ell^{-2}(n+1/2)^2$ contribution we have neglected in $\tilde \omega_n^2$, which would become relevant for $n \gtrsim m \ell$. The point is that $n \sim m \ell$ is arbitrarily large as $m$ increases and is associated to overtones that are irrelevant as they are instantaneously suppressed in this limit, since their decay time is $\tau \sim 1/m$ rather than $\tau \sim \ell$.

Finally, we recognize that the $\tilde \omega_n^{(\pm)}$ can be interpreted as overtones of particles in dS. In the context of inflation, in fact, scalar fields near the reheating surface ($\eta = 0$, where $\eta$ is the conformal time) of the ``flat slicing'' (the one that is used in cosmology) of dS behave like primary fields with scaling dimensions \cite{BaumannJoyce}:
\begin{gather}
	\tilde \Delta = \frac{d}{2} + \frac{1}{2} \sqrt{d^2 - 4m^2 \ell^2_{\rm dS}}, \qquad d - \tilde \Delta =  \frac{d}{2} - \frac{1}{2} \sqrt{d^2 - 4m^2 \ell^2_{\rm dS}}.
\end{gather}

$\tilde \Delta$ is a quantum number of the particle and matches our result with $d = 1$. We also correctly find that the behavior is only oscillatory for ``heavy'' particles, and purely decaying otherwise. The physical explanation is that light particles have associated wavelengths that are much larger than the size of the cosmological horizon and are therefore ``frozen'', so that perturbations can only decay; on the other hand, heavy particles have small associated wavelengths which are not subject to this ``freezing'' phenomenon, hence they are free to oscillate. This relation between $\tilde \Delta$ and $m$ is exactly the one we obtain from \eqref{eq:massdeltarelation} through a naive analytical continuation $\ell_{\rm AdS} \to i \ell_{\rm dS}$, and in fact the two geometries share similar isometry groups which, in turn, give rise to similar scalar representations.

Let us now consider a scalar that is coupled to the effective metric. We have different $f(\rho), \: y(\rho)$:
\begin{equation}
\begin{gathered}
    f(\rho) = \rho^2 - \sin^2\theta, \quad \rho \in [\sin \theta, +\infty], \\
    y(\rho) \equiv -\int_\rho^{+\infty} \frac{d\rho'}{\rho'^2-\sin^2\theta} = \frac{1}{2\sin\theta} \log \left( \frac{\rho-\sin\theta}{\rho+\sin\theta} \right), \\
    \rho = \sin\theta \, \frac{1+e^{2\sin\theta \, y}}{1-e^{2\sin\theta \, y}}.
\end{gathered}
\end{equation}

The resulting Schrödinger equation contains a new potential:
\begin{gather}
    V(y) = m^2 f(\rho(y)) = \frac{m^2 \sin^2\theta}{\sinh^2(\sin\theta \, y)}.
\end{gather}

By using $\tilde y \equiv \sin\theta \, y$ and $\tilde \omega \equiv \omega/\sin\theta$, the equation becomes:
\begin{gather}
		-\psi''(\tilde y) + \frac{m^2}{\sinh^2 \tilde y} \psi(\tilde y) = \tilde \omega^2 \psi(\tilde y). \label{eq:tildeYwAds}
\end{gather}

We have a very important qualitative difference with respect to the previous case: the coordinate $\tilde y(\rho)$ now runs from $-\infty$ to $0$ as $\rho$ runs from $\sin\theta$ to $+\infty$, so it is equivalent to a radial coordinate and not to a ``cartesian'' one. To use the WKB approximation in this context, two operations would be necessary. First, we would have to define $u \equiv -\log(-\tilde y)$, which runs from $-\infty$ to $+\infty$ as $\tilde y$ runs from $-\infty$ to $0^-$. The second derivative would transform into:
\begin{gather}
	- \frac{d^2 \psi}{d\tilde y^2} = - \frac{d^2 u}{d\tilde y^2} \frac{d\psi}{du} - \left( \frac{du}{dy} \right)^2 \frac{d^2 \psi}{du^2} = -e^{2u} \left[ \frac{d^2 \psi}{du^2} - \frac{d\psi}{du} \right].
\end{gather}

By then defining $\psi(u) = e^{u/2} \hat \psi(u)$ and inserting this new function into the differential equation, we would obtain:
\begin{gather}
	-\hat \psi''(u) + \left[ \frac{1}{4} + e^{-2u} \left( \frac{m^2}{\sinh^2(e^{-u})} - \omega^2 \right) \right] \hat\psi(u) = 0.
\end{gather}

These operations yield a Schrödinger equation with a coordinate that correctly ranges from $-\infty$ to $+\infty$, so one could in principle obtain an approximated spectrum through the use of WKB. Instead of doing this, though, we can actually determine the exact frequencies. The most general solution of \eqref{eq:tildeYwAds} is:
\begin{gather}
	\psi(\tilde y) = (\cosh^2 \tilde y)^{i\tilde\omega/2} \left( \sum_\pm \alpha_\pm (\tanh^2 \tilde y)^{(1\pm\sqrt{1+4m^2})/4} {}_2F_1[a_\pm,b_\pm,c_\pm; \tanh^2 \tilde y] \right).
\end{gather}

$\alpha_\pm$ are constants of integration, ${}_2 F_1[a,b,c;z]$ is the hypergeometric function, and:
\begin{gather}
    a_\pm = \frac{1}{4}(1 \pm \sqrt{1+4m^2} -2i\tilde\omega), \quad b_\pm = \frac{1}{4}(3 \pm \sqrt{1+4m^2} -2i\tilde\omega), \quad c_\pm = 1 \pm \frac{1}{2} \sqrt{1+4m^2}.
\end{gather}

\begin{figure}
    \centering
    \includegraphics[width = 0.7 \textwidth]{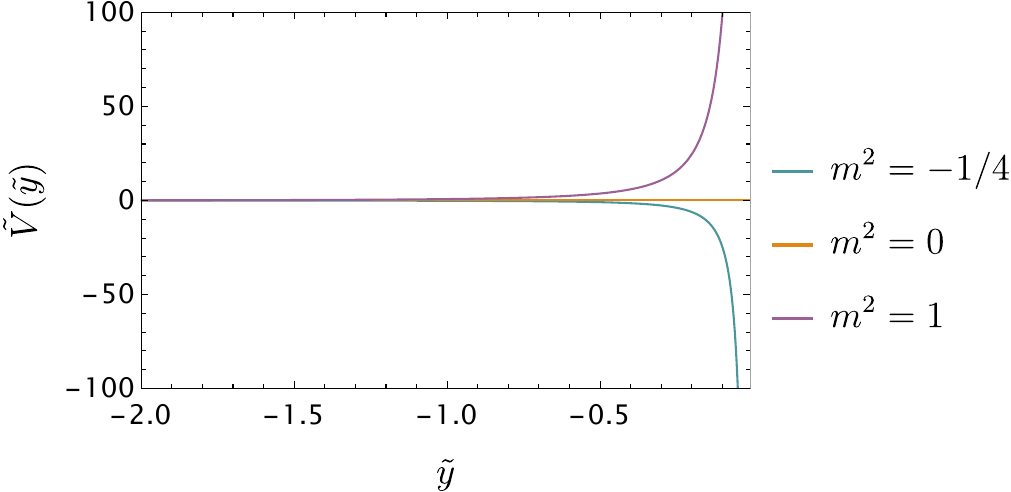}
    \caption{Plot of the potential $\tilde V(\tilde y)$ for different values of the square mass of the scalar $m^2$. For positive $m^2$, the potential becomes infinitely repulsive at the AdS$_2$ boundary, while it becomes infinitely attractive for negative $m^2$. Trivially, there is no potential for massless scalars.}
    \label{fig:quasinormal2}
\end{figure}

Unitarity imposes $m^2 \geq -1/4$. If $m^2 > 0$, the boundary conditions for our problem are $\psi(0) = 0$ and requiring that $\psi(\tilde y \to -\infty)$ is an ingoing wave towards the AdS$_2$ black hole. The requirement for $\psi(0)$ is obvious by looking at the repulsive potential $\tilde V(\tilde y) = m^2/\sinh^2 \tilde y$, which is plotted in Figure \ref{fig:quasinormal2}. From a physical point of view, $\psi(0) = 0$ is equivalent to stating that waves are perfectly reflected at the AdS$_2$ boundary, that is, there is nothing beyond $\tilde y = 0$ to which the wave could be transmitted. Because of this, it makes sense to impose this boundary condition to any kind of wave, regardless of its mass.

Since ${}_2F_1[a,b,c;z\to 0] \to 1$, requiring $\psi(0)=0$ clearly implies $\alpha_- = 0$. The boundary condition at $\tilde y = -\infty$ then yields the spectrum. To this end, we need the asymptotic behavior of the hypergeometric function, using $\tanh^2 \tilde y \approx 1-4e^{2\tilde y}$:
\begin{gather}
	{}_2F_1[a_+,b_+,c_+;1-4e^{2\tilde y}] \propto \frac{\Gamma(1-i\tilde\omega)}{\Gamma(\frac{1}{2} + \frac{1}{2} \sqrt{1+4m^2} - i \tilde\omega)} \, e^{2i\tilde \omega \tilde y} - \frac{\Gamma(1+i\tilde\omega)}{\Gamma(\frac{1}{2} + \frac{1}{2} \sqrt{1+4m^2} + i \tilde\omega)}
\end{gather}

We have not reported the prefactor (which can be absorbed into $\alpha_+$) and we have already rewritten the result using \eqref{eq:Gammafuncprop}. If we also use that, up to irrelevant constants, $(\cosh^2 \tilde y)^{i\tilde\omega/2} = e^{i\tilde\omega \log\cosh \tilde y} \sim e^{-i\tilde\omega \tilde y}$, we finally obtain that:
\begin{gather}
	\psi(\tilde y \to -\infty) \approx \alpha_+ \left( \frac{\Gamma(1-i\tilde\omega)}{\Gamma(\frac{1}{2} + \frac{1}{2} \sqrt{1+4m^2} - i \tilde\omega)} \, e^{i\tilde \omega \tilde y} - \frac{\Gamma(1+i\tilde\omega)}{\Gamma(\frac{1}{2} + \frac{1}{2} \sqrt{1+4m^2} + i \tilde\omega)} e^{-i\tilde \omega \tilde y} \right).
\end{gather}

We now choose the same convention we picked at the beginning, namely that $\psi$ oscillates as $e^{i\omega t}$. Ingoing waves towards the black hole are therefore $e^{i\omega(t+y)}$, so we have to get rid of the $e^{-i\tilde\omega \tilde y}$ wave. Again, this forces $\tilde \omega$ to be a discrete set of values given by: 
\begin{gather}
	\tilde \omega_{n \geq 0} = i \ell^{-1} \left( n + \frac{1}{2} + \frac{1}{2} \sqrt{1+4m^2\ell^2} \right)
\end{gather}

Multiplying these frequencies by $\sin \theta$ finally yields the $\omega_n$. Once again, we only have stable, decaying modes for this scalar field. Similarly to the case studied above, we have found overtones of particles in AdS, with exactly the same $\Delta$ as \eqref{eq:massdeltarelation} when $d = 1$. In this case, it is impossible to have $\mathrm{Re}(\omega_n) \neq 0$, no matter how big $m^2$ is, which implies that field excitations can only decay and never oscillate.

When $m^2 \leq 0$, as already mentioned, imposing $\psi(0) = 0$ is not required by the potential at hand, but is still well physically motivated. For $m^2 = 0$, we have the trivial propagation of free waves with frequency $\omega$:
\begin{gather}
    \psi(y) = \alpha_+ e^{i\omega y} + \alpha_- e^{-i \omega y}.
\end{gather}

$\psi(0) = 0$ implies $\psi(y) \propto \sin(\omega y)$, so we always have the sum of an ingoing and an outgoing wave at $-\infty$. Since there is no way to get rid of the outgoing wave, we deduce that massless particles do not exhibit quasinormal modes.

For $m^2 < 0$, unfortunately, $\psi(0) = 0$ is a condition that is trivially satisfied for all $\alpha_\pm$. The implication is that we now have two independent coefficients, so suppressing the outgoing wave at the horizon translates to a relation between $\alpha_+$ and $\alpha_-$ for any value of $\omega$, rather than to a statement about the frequencies themselves. As a consequence, it would appear that there is no compelling reason to exclude any $\omega$ at all. Interestingly enough, this only happens in the case of particles with imaginary mass. We have to be careful when deducing anything in this case, though, as this potential behaves like $\tilde y^{-2}$ near the origin. This kind of \textit{attractive} potential is known to be problematic in Quantum Mechanics on the real half-line, and therefore requires special care \cite{Gitman_2010}. We expect that a correct treatment of this system will yield the second boundary condition we need, thus greatly restricting the possible modes. We leave this analysis to future work.

To conclude, there is a clear difference with respect to the minimally coupled scalar. In particular, although there is a discrete set of possible modes in both cases, there is no way to have an actual propagation of waves inside the effective geometry. Although \eqref{eq:dsquasiwave} is not exact, we can expect it to be an extremely good approximation for $m \gg \ell^{-1}$, especially when it comes to the dependence on the mass of the particle. The heavy particles are exactly the ones of interest, since they are the ones that propagate in only one scenario. An hypothetical probe that couples to the original metric and not to the combination that yields the effective one, then, would be able to give us information about the most relevant region, namely the one sitting at $r = \pi$ (where $R = +2$), thanks to its distinctive traits. This is clear from its spectrum, which is entirely characterized (in the regime where WKB is a good approximation) by $V(y \approx 0)$, namely by $r \approx \pi$, i.e. the position of a geodesic observer who locally experiences a static patch. Even though we have obtained the spectrum with the assumption that the particle lives on the real axis between $\theta$ and $2\pi-\theta$, this property reassures us that things are going to be similar even along the actual integration contour. Finally, we underline the fact that we do not yet know of an operator in DSSYK that is dual to a probe that couples minimally to gravity. If we were to find one, though, studying its strong coupling behavior on the boundary would allow us to investigate quantum de Sitter physics in a new and extremely interesting way. As we have shown through the calculations in this section, the properties of this probe are expected to be completely different with respect to the random operators on the boundary that are currently at our disposal.

\section{Towards a Bulk Dual of Complex DSSYK}
We have reviewed the relevant material, again filling the several gaps in the literature and expanding on the interpretation of the results. We have also shown how the model has the potential to let us investigate quantum de Sitter physics by considering the appropriate probe. Our objective for the remainder of this section will be to use these instruments to study the semiclassical bulk dual of the complex DSSYK case. We do not review the construction of the chord diagram picture (which is described in \cite{Berkooz:2020uly}) in this case, since it is only a technical involution of the Majorana model that we have already presented, but we take the main result that we will need for granted, i.e. the partition function of the model in the canonical ensemble with fixed mean charge $Q$:
\begin{equation}
\begin{aligned}
    \mathcal{Z}_{\rm cDSSYK}(\beta, Q) & = \frac{2}{\sqrt{2\pi N}} \frac{1}{\sqrt{1-4Q^2}} \exp \left[ NQ \log \left( \frac{1-2Q}{1+2Q} \right) + \frac{N}{2} \log \left( \frac{4}{1-Q^2} \right) \right] \\
    & \int_0^\pi \frac{d\theta}{2\pi} (q^2,e^{\pm 2i\theta};q^2)_\infty \, \exp \left[ - \frac{2\beta \mathcal{J} \cos \theta}{\sqrt{1-q^2}} e^{\lambda/2} (1-4Q^2)^{p/4} \exp \left( \frac{2\lambda Q}{1-4Q^2} \right) \right]. \label{eq:cDSSYKfullQ}
\end{aligned}
\end{equation}

We are using the definitions introduced in the previous chapter and:
\begin{align}
    q^2 = \exp \left( -\frac{4\lambda}{1-4Q^2} \right), \quad \lambda = \frac{p^2}{4N} = \frac{1-4Q^2}{2} \abs{\log q}.
\end{align}

In principle, this partition function is what one obtains by tracing $e^{-\beta H}$ only over the states that are eigenvectors of the charge operator with eigenvalue $Q$. In the path integral formalism, equivalently, this is the result of inserting $\delta \left( 1/N \sum_i \bar\psi_i \psi_i - 1/2 - Q \right)$ inside \eqref{eq:startingCSYK}. The quantum theory we are considering is therefore cDSSYK with a restricted Hilbert space: since the charge is a symmetry that commutes with the Hamiltonian, every subsector of the Hilbert space $\mathcal{H}$ with fixed charge does not mix with the others over time and therefore yields a well-defined theory.

We are interested in matching the partition function of the complex model to a gravitational theory at the semiclassical level. For this purpose, we once again redefine the variance of the couplings in a $Q$-independent way
\begin{gather}
    \langle |J_{i_1 \dots i_{p/2},j_1 \dots j_{p/2}}|^2 \rangle_J = \mathcal{J}^2 \binom{N}{p}^{-2} \to \frac{1}{16\lambda} \mathcal{J}^2 \binom{N}{p}^{-2},
\end{gather}

and we focus on the $\abs{\log q} \to 0 \: (\lambda \to 0)$ expansion of the partition function. We need \eqref{eq:loge2ithapprox} (of which we also keep the first term, since it depends on $Q$) and similarly:
\begin{equation}
\begin{aligned}
    \log (q^2;q^2)_{\infty} & \approx - \frac{\mathrm{Li}_2(q^2)}{2\abs{\log q}} + \frac{1}{2} \log(1-q^2) \\
    & \approx -\frac{1}{2\abs{\log q}} \left( \frac{\pi^2}{6} - \log(1-q^2)\log(q^2) - \mathrm{Li}_2(1-q^2) \right) + \frac{1}{2}\log(2\abs{\log q}) \\
    & \approx -\frac{\pi^2}{24\lambda}(1-4Q^2) + \mathcal{O}(\log \lambda).
\end{aligned}
\end{equation}

We have only kept the $\mathcal{O}(1/\lambda)$ terms that depend on the charge. This allows us to write the partition function at order $\mathcal{O}(1/\lambda)$:
\begin{equation}
\begin{aligned}
    \mathcal{Z}_{\rm cDSSYK}(\beta,Q) & \approx \int_0^\pi d\theta \: (q^2, e^{\pm 2i\theta};q^2)_\infty \, \exp \left( \frac{\beta \mathcal{J} \cos \theta}{2 \abs{\log q}} (1-4Q^2)^{(p-2)/4} \right) \\
    & \propto \int_0^\pi d\cos\theta \: \exp \left( -\frac{\pi^2}{12\abs{\log q}} -\frac{\pi^2}{6\abs{\log q}} + \frac{\pi\theta-\theta^2}{\abs{\log q}} + \frac{\beta \cos\theta}{2\abs{\log q}} \right) \\
    & = \int_0^\pi d\cos\theta \: \exp \left( -\frac{\pi^2}{8\lambda}(1-4Q^2) + \frac{\pi\theta-\theta^2}{2\lambda}(1-4Q^2) + \frac{\beta \cos\theta}{4\lambda}(1-4Q^2) \right). \label{eq:cDSSYKtarget}
\end{aligned}
\end{equation}

We recall that the effective coupling strength that we had found in the previous chapter was:
\begin{gather}
    \mathcal{J}^2_{\rm eff}(Q) = \mathcal{J}^2 (1-4Q^2)^{p/2-1},
\end{gather}

so what we have done in the second line is setting $\mathcal{J}_{\rm eff}(Q)$ to $1$ instead of the usual $\mathcal{J}$. We have dropped the $Q$-dependent prefactor of $\mathcal{Z}_{\rm cDSSYK}$ as it is $\mathcal{O}(\lambda^0)$ and thus only relevant when considering one-loop corrections.

Let us build a gravitational theory that yields the same partition function at the semiclassical level. It is sufficient to focus on reproducing the ``fake'' thermodynamics, i.e. those that arise when one does not remove the negative chord number states in cDSSYK. The reason is that the insertion of a defect at the horizon is then able to complete the matching process at the semiclassical level in this scenario, too. The presence of a $U(1)$ global symmetry on the boundary suggests that a gauge field $A_\mu(x)$ should be introduced. A gauge-invariant modification to \eqref{eq:bulkdualSYK} in two dimensions is:
\begin{equation}
\begin{gathered}
    S_E = -\frac{1}{2} \int d^2x \, \sqrt{g} \, \left( \Phi R + V(\Phi) + W(\Phi) E^{\mu\nu} F_{\mu\nu} -\frac{1}{8\pi} X(\Phi) F^{\mu\nu} F_{\mu\nu} \right) - \int dt \, \sqrt{h} \, \left( \Phi K - G(\Phi) \right). \label{eq:dualconjecture}
\end{gathered}
\end{equation}

We have set $\ell = \ell_P = 1$ and we have chosen CGS normalization for the electromagnetic term when $X(\Phi) = 1$. $V(\Phi), \: W(\Phi)$ and $X(\Phi)$ are functions that we need to determine to perform the match (it is not obvious that $V(\Phi)$ is the same as before), $G(\Phi)$ is an appropriate counterterm. Two dimensions are peculiar, because $F_{\mu\nu} \propto E_{\mu\nu}$ and we are allowed to insert the extra term $E^{\mu\nu} F_{\mu\nu}$ in the action. No new GHY terms are needed, since the electromagnetic terms only contain first derivatives of $A_{\mu}$ so, when extremizing the action with respect to $A_\mu$, we can always integrate by parts up to $\delta A_{\mu}|_{\partial M}$ terms, which are null when taking Dirichlet boundary conditions.

We can say something more regarding the gauge field $A_\mu(x)$. First, we can use the gauge freedom $A_\mu(x) \to A_\mu(x) + \partial_\mu \Lambda(x)$ to fix $A_r(x) = 0$:
\begin{gather}
    \partial_r \Lambda(t,r) = -A_r(t,r) \implies \Lambda(t,r) = -\int^r dr' \: A_r(t,r').
\end{gather}

This leaves us with a residual gauge freedom: $A_0 \to A_0 + \dot f(t)$, with $f(t)$ an arbitrary function that is periodic $(\beta \neq \infty)$ or goes to $0$ at infinity $(\beta = \infty)$, which excludes for example $\dot f(t) = \mathrm{constant}$. The fact that $\dot f(t) =$ constant is not a gauge symmetry implies that the constant term in $A_0(t,r_{\rm bdy})$ is physical, although this is only true at the quantum level, since it is connected to the Wilson line along the Euclidean circle \cite{harlow2019factor}. We can set $A_\mu(x)$ to $0$, i.e. it is the gradient of a function, if and only $\varepsilon^{\mu\nu} \partial_\mu A_\nu(x) = F_{\mu\nu}(x) = 0$, which will generally not be the case. The GKPW dictionary then tells us that a properly rescaled $A_0(x)$ at the boundary will act as a source for a conserved current $J^\mu$, which is clearly the charge $Q$. We deduce that $A_0(x)$ at the boundary is basically a time-dependent chemical potential, which tells us that we should consider semiclassical solutions with the source part of $A_0(t,r)$ set to $0$. The boundary theory we are considering is, in fact, a fixed charge subsector of cDSSYK, which knows nothing of a chemical potential $\mu$. In other terms, we emphasize once more that our partition function is the trace of $e^{-\beta H}$ on a subspace of the total Hilbert space $\mathcal{H}$, not the trace of $e^{-\beta(H - \mu N \hat Q)}$ over the full $\mathcal{H}$ with the right $\mu$ such that the mean charge is $Q$. Computing derivatives of the gravitational partition function with respect to the boundary source piece of $A_0(x)$, which we call $\tilde A_0(\tau)$, should return the two point function of the charge:
\begin{gather}
    \frac{1}{\mathcal{Z}_{\rm grav}[0]} \frac{\delta^2 \mathcal{Z}_{\rm grav}[\tilde A_0(\tau)]}{\delta \tilde A_0(\tau_1) \delta \tilde A_0(\tau_2)} \Bigg|_{\tilde A_0 = 0} = \langle \hat Q(\tau_1) \hat Q(\tau_2) \rangle = Q^2. \label{eq:chargegravmatch}
\end{gather}

We see that $\hat Q$ behaves as a primary of conformal dimension $\Delta = 0$, which correctly implies a dual massless field (recall \eqref{eq:DeltaMassEq}). For a spin-$s$ field, the boundary behavior in AdS/CFT is the following \cite{Sleight:2016hyl}:
\begin{gather}
	A_{a_1 \dots a_s}(z \to 0,x) = z^{d-\Delta-s} \tilde A_{a_1 \dots a_s}(x) + z^{\Delta-s} \hat A_{a_1 \dots a_s}(x). \label{eq:SleightScal}
\end{gather}

$\Delta = 0$ then also implies that $A_0(\tau,\rho \to +\infty) \sim \tilde A_0(\tau) + \rho \hat A_0(\tau)$ as we reach the boundary, so that the source is actually the subleading part of the bulk field $A_0(x)$. This is, in fact, the piece that correctly transforms as $\tilde A_0 \to \lambda^{-1} \tilde A_0$ under $(z, \vec x) \to \lambda (z,\vec x)$, so that $A_0 \to \lambda^{-1} A_0$ (it is a covector) and $\int d\tau \: \tilde A_0 Q$ is invariant.

The $z \to 0$ behavior is easy to prove in the case of interest, i.e. a two-dimensional spin-$1$ gauge field. We can use Poincaré coordinates in Euclidean AdS$_2$, then set $A_z(t,z)$ to $0$ and expand $A_t(t, z \to 0) = z^a A_t(t)$. This way, the equation of motion is:
\begin{equation}
\begin{aligned}
	0 & = D_\mu F^{\mu\nu} = \partial_\mu (z^{-2} g^{\mu\rho} g^{\nu\sigma} F_{\rho\sigma}) = \partial_z (z^2 \partial_z A_t) - \partial_t (z^2 \partial_z A_t) \\
	& = a(a+1) z^a A_t(t) - a z^{a+1} \partial_t A_t(t).
\end{aligned}
\end{equation}

The second term in the second line is subleading for small $z$, so we find two roots $a = 0$ and $a = -1$, as prescribed by \eqref{eq:SleightScal}.

We will not consider the constraints on the dual theory that come from the match of the gauge field's two point function, but it is a clear future work direction. 

Finally, notice that the case under study goes beyond the mere AdS/CFT correspondence, yet matching the partition functions of the two theories allows us to interchange which one we can use to compute functional derivatives anyway. The rest of our insight also lies on the assumption that we can extrapolate the knowledge inherited from AdS/CFT to any other holographic scenario up to small modifications, but this appears to be consistent with our results. It may be that $A_\mu$ also couples to an effective AdS$_2$ geometry, at which point our discussion would be exact with the right adjustments.

We now start our matching procedure by first rescaling the dilaton field, $2\abs{\log q} \Phi \to \Phi$:
\begin{equation}
\begin{aligned}
    S_E = -\frac{1}{2\abs{\log q}} \bigg[ & \frac{1}{2} \int d^2x \, \sqrt{g} \, \left( \Phi R + V(\Phi) + W(\Phi) E^{\mu\nu} F_{\mu\nu} -\frac{1}{8\pi} X(\Phi) F^{\mu\nu} F_{\mu\nu} \right) \\
    & + \int dt \, \sqrt{h} \, \left( \Phi K - G(\Phi) \right) \bigg].
\end{aligned}
\end{equation}

Just like before, the $\lambda \to 0 \: (\abs{\log q} \to 0)$ saddle is manifest this way. All the potentials have been properly rescaled, although we have kept their previous names. This way of writing the dual action agrees with what is said in \cite{Berkooz:2020uly}, namely that $q \to 0$ should yield a strongly coupled theory in the bulk due to the backreaction of the electric flux on spacetime.

We now write the classical equations of motion, starting with:
\begin{gather}
    \frac{\delta S_E}{\delta \Phi} = 0 \implies R+V'(\Phi) + W'(\Phi) E^{\mu\nu} F_{\mu\nu} - \frac{1}{8\pi} X'(\Phi) F^{\mu\nu} F_{\mu\nu} = 0. \label{eq:EOMdual1}
\end{gather}

Next, we vary the action with respect to the EM potential:
\begin{equation}
\begin{aligned}
    \delta S_E & \propto \int d^2x \left( -W(\Phi) \, \varepsilon^{\mu\nu} \partial_\mu \delta A_\nu + \sqrt{g} \frac{1}{4\pi} X(\Phi) F^{\mu\nu} \partial_\mu \delta A_\nu \right) \\
    & = \int d^2x \, \sqrt{g} \: \frac{1}{\sqrt{g}} \partial_\mu \left( W(\Phi) \, \varepsilon^{\mu\nu} - \frac{1}{4\pi} \sqrt{g} X(\Phi) F^{\mu\nu} \right) \delta A_\nu, \\
    \frac{\delta S_E}{\delta A_{\nu}} & = 0 \implies \nabla_\mu (W(\Phi) E^{\mu\nu}) = \frac{1}{4\pi} \nabla_\mu (X(\Phi) F^{\mu\nu}). \label{eq:EOMdual2}
\end{aligned}
\end{equation}

In fact, we recall that, for an antisymmetric tensor:
\begin{gather}
    \nabla_\mu A^{\mu\nu} = \frac{1}{\sqrt{g}} \partial_\mu (\sqrt{g} A^{\mu\nu}). 
\end{gather}

Finally, we need to vary the action with respect to the inverse metric $g^{\mu\nu}$. We do this in Appendix \ref{app:2dDilaton}, but now there is an extra contribution coming from the $\sqrt{g} \, g^{\mu\rho} g^{\nu\sigma} F_{\rho\sigma} F_{\mu\nu}$ kinetic term. Note that the other new term is proportional to $\varepsilon^{\mu\nu} F_{\mu\nu}$, so it does not depend on the metric. The resulting equation of motion is:
\begin{gather}
    \frac{\delta S_E}{\delta g^{\mu\nu}} = 0 \implies \nabla_\mu \nabla_\nu \Phi - g_{\mu\nu} \nabla^2 \Phi + \frac{1}{2} g_{\mu\nu} V(\Phi) + \frac{1}{4\pi} X(\Phi) F_{\mu\rho} {F_{\nu}}^{\rho} - \frac{1}{16\pi} g_{\mu\nu} X(\Phi) F^{\rho\sigma} F_{\rho\sigma} = 0. \label{eq:EOMdual3}
\end{gather}

In Appendix \ref{app:2dDilaton}, we show that $\xi^\mu = E^{\mu\nu} \partial_\nu \Phi$ is a Killing vector using the fact that $\nabla_\mu \nabla_\nu \Phi$ is proportional to $g_{\mu\nu}$, and fortunately this is still the case. To see this, we have to consider the only term that is apparently an issue:
\begin{gather}
    F_{\mu\rho} {F_\nu}^\rho \stackrel{F_{\mu\nu} = \varepsilon_{\mu\nu} f}{=} \varepsilon_{\mu\rho} f \varepsilon_{\nu\lambda} f g^{\rho\lambda} = f^2 (\delta_{\mu\nu}\delta_{\rho\lambda} - \delta_{\mu\lambda} \delta_{\rho\nu} ) g^{\rho\lambda} = f^2 \delta_{\mu\nu} (g^\lambda_\lambda - 1).
\end{gather}

Although, for a generic choice of coordinates, $\delta_{\mu\nu}$ is not proportional to $g_{\mu\nu}$, we do not lose generality in two dimensions if we assume the conformal gauge $g_{\mu\nu} = e^{2\omega(t,r)} \delta_{\mu\nu}$. Since $E^{\mu\nu} \partial_\nu \Phi$ is a Killing vector with this choice of coordinates, then it is one regardless of the way we parametrize the solution. This allows us to make the following choice of coordinates once again, with $\xi^\mu = (1,0)$:
\begin{gather}
    ds^2 = f(r) \, dt^2 + \frac{dr^2}{f(r)}, \quad \Phi(r) = r \implies R(r) = -f''(r).
\end{gather}

\eqref{eq:EOMdual1} reduces to:
\begin{gather}
    -f''(r) + V'(r) + 2 W'(r) F_{tr}(t,r) - \frac{1}{4\pi} X'(r) (F_{tr}(t,r))^2 = 0.
\end{gather}

On the other hand, tracing \eqref{eq:EOMdual3} extracts all the relevant information that it contains:
\begin{gather}
    -\nabla^2\Phi + V(r) + \frac{1}{8\pi} X(r) F^{\rho\sigma}F_{\rho\sigma} = -f'(r) + V(r) + \frac{1}{4\pi} X(r) (F_{tr}(t,r))^2 = 0.
\end{gather}

Finally, \eqref{eq:EOMdual2} becomes:
\begin{gather}
    \varepsilon^{\mu\nu} W'(r) \partial_\mu \Phi - \frac{1}{4\pi} X'(r) F^{\mu\nu} \partial_\mu \Phi - \frac{1}{4\pi} X(r) \partial_\mu F^{\mu\nu} = 0.  
\end{gather}

Taking $\nu = r$ returns:
\begin{gather}
    -\frac{1}{4\pi} X(r) \partial_t F^{tr}(t,r) = 0 \implies F_{tr}(t,r) = F_{tr}(r).
\end{gather}

Taking $\nu = t$, instead, returns:
\begin{equation}
\begin{gathered}
    -W'(r) + \frac{1}{4\pi} X'(r) F_{tr}(r) + \frac{1}{4\pi} X(r) F_{tr}'(r) = 0 = \frac{d}{dr} \left( -W(r) + \frac{1}{4\pi} X(r) F_{tr}(r) \right) \\
    \implies F_{tr}(r) = \frac{\alpha + 4\pi W(r)}{X(r)}. \label{eq:FtrSol}
\end{gathered}
\end{equation}

We observe that, if we derive the trace of \eqref{eq:EOMdual3} using this result, we obtain:
\begin{equation}
\begin{aligned}
    & -f''(r) + V'(r) + \frac{1}{4\pi} X'(r) (F_{tr}(r))^2 + 2 \left( \frac{1}{4\pi} X(r) F'_{tr}(r) \right) F_{tr}(r) \\
    = \: & -f''(r) + V'(r) + 2 W'(r) F_{tr}(r) - \frac{1}{4\pi} X'(r) (F_{tr}(r))^2 = 0.
\end{aligned}
\end{equation}

This is exactly \eqref{eq:EOMdual1}, which is therefore redundant. The trace of \eqref{eq:EOMdual3} finally yields:
\begin{gather}
    f(r) = \int_{\theta}^r dr' \left( V(r') + \frac{(\alpha+4\pi W(r'))^2}{4\pi X(r')} \right). \label{eq:f(r)Sol}
\end{gather}

$\theta$ is the position of the horizon. One immediately obtains that the inverse temperature $\beta$ of this solution is:
\begin{gather}
    \beta = \frac{4\pi}{\left( V(\theta) + \frac{(\alpha+4\pi W(\theta))^2}{4\pi X(\theta)} \right)}.
\end{gather}

To obtain the ``fake'' thermodynamics, we must impose that:
\begin{align}
    V(\theta) + \frac{(\alpha+4\pi W(\theta))^2}{4\pi X(\theta)} = 2 \sin\theta.
\end{align}

We now compute the on-shell action of the classical solution:
\begin{equation}
\begin{aligned}
    S_E & = -\frac{\beta}{4\abs{\log q}} \Bigg[ \int_{\theta}^{r_0} dr \: \left( -r \left( V + \frac{(\alpha + 4\pi W)^2}{4\pi X} \right)'(r) + V(r) + 2 W(r) \frac{\alpha + 4\pi W(r)}{X(r)} - \frac{(\alpha + 4\pi W(r))^2}{4\pi X(r)} \right) \\
    & \hspace{2cm} + r_0 \left( V + \frac{(\alpha + 4\pi W)^2}{4\pi X} \right)'(r_0) - 2 \sqrt{f(r_0)} \, G(r_0) \Bigg] \\
    & = -\frac{\beta}{4\abs{\log q}} \Bigg[ \theta \left( V(\theta) + \frac{(\alpha + 4\pi W(\theta))^2}{4\pi X(\theta)} \right) + 2 \int_{\theta}^{r_0} dr \left( V(r) + \frac{W(r)}{X(r)}(\alpha + 4\pi W(r)) \right) - 2 \sqrt{f(r_0)} \, G(r_0) \Bigg] \\
    & = - \frac{\pi\theta}{\abs{\log q}} - \frac{\beta}{2\abs{\log q}} \Bigg[ \int_{\theta}^{r_0} dr \left( V(r) + \frac{W(r)}{X(r)}(\alpha + 4\pi W(r)) \right) - \sqrt{f(r_0)} \, G(r_0) \Bigg].
\end{aligned}
\end{equation}

We have integrated the first term by parts in the second line. We do not know whether $r_0$ is still exactly $\pi/2 + i\infty$ rather than some new $\Phi_{\rm bdy}$, but it is reasonable to assume that it is still such that a counterterm $G(r_0)$ is needed to subtract the divergent part of the action. Our initial choice for $G(r_0)$ is the following:
\begin{equation}
\begin{aligned}
    G(r_0) & = \frac{\int_{r_c}^{r_0} dr \left( V(r) + \frac{W(r)}{X(r)}(\alpha + 4\pi W(r)) \right)}{\sqrt{\int_{r_c}^{r_0} dr \left( V(r) + \frac{(\alpha+4\pi W(r))^2}{4\pi X(r)} \right)}} \\
    & = \frac{\int_{r_c}^{r_0} dr \left( V(r) + \frac{W(r)}{X(r)}(\alpha + 4\pi W(r)) \right)}{\sqrt{\int_{\theta}^{r_0} dr \left( V(r) + \frac{(\alpha+4\pi W(r))^2}{4\pi X(r)} \right)}} \left[ 1 + \frac{1}{2} \, \frac{\int_{\theta}^{r_c} dr \left( V(r) + \frac{(\alpha+4\pi W(r))^2}{4\pi X(r)} \right)}{\int_{\theta}^{r_0} dr \left( V(r) + \frac{(\alpha+4\pi W(r))^2}{4\pi X(r)} \right)} + \dots \right].
\end{aligned}
\end{equation}

$r_c$ is a fixed value. Again, we make the assumption that we need to renormalize the action in the first place, which is true for Majorana DSSYK, so that the series expansion in the second line makes sense. If we multiply this function by $\sqrt{f(r_0)}$ and we only keep the terms that do not tend to $0$ as $r_0$ tends to $\Phi_{\rm bdy}$, we obtain:
\begin{equation}
\begin{aligned}
    & \int_{r_c}^{r_0} dr \left( V(r) + \frac{W(r)}{X(r)}(\alpha + 4\pi W(r)) \right) \\
    & + \frac{1}{2} \frac{\int_{r_c}^{r_0} dr \left( V(r) + \frac{W(r)}{X(r)}(\alpha + 4\pi W(r)) \right)}{\int_{\theta}^{r_0} dr \left( V(r) + \frac{(\alpha+4\pi W(r))^2}{4\pi X(r)} \right)} \int_{\theta}^{r_c} dr \left( V(r) + \frac{(\alpha+4\pi W(r))^2}{4\pi X(r)} \right).
\end{aligned}
\end{equation}

We now make a useful working assumption: $W(\Phi)$ and $X(\Phi)$ are such that the ratio between the integrals in front of the second term is $1$, which is true provided that:
\begin{gather}
    \int^{r_0} dr \left( V(r) + 4\pi \frac{W^2(r)}{X(r)} \right) \gg \int^{r_0} dr \left( \alpha \, \frac{W(r)}{X(r)} + \frac{\alpha^2}{4\pi X(r)} \right). \label{eq:cSYKworkhyp}
\end{gather}

Taking $r_c = \pi/2$, the resulting renormalized on-shell action is:
\begin{equation}
\begin{aligned}
    S_E & = - \frac{\pi\theta}{\abs{\log q}} - \frac{\beta}{2\abs{\log q}} \int_{\theta}^{\pi/2} dr \: \Bigg[ V(r) + \frac{W(r)}{X(r)}(\alpha + 4\pi W(r)) - \frac{1}{2} \left( V(r) + \frac{(\alpha+4\pi W(r))^2}{4\pi X(r)} \right) \Bigg]. \label{eq:cSYKrenormS}
\end{aligned}
\end{equation}

There are two natural ways to proceed now, which we will investigate in the following. We underline that our aim is to obtain \eqref{eq:cDSSYKtarget} up to the $\theta^2$ term, which will be a consequence of the insertion of the defect at the end. We repeat that the way the defect should be inserted may be different with respect to \cite{Blommaert:2024ydx}, due to the possible inconsistency we have pointed out at the end of Section \ref{sec:DefectOp}, but we do not concern ourselves with this issue in the following.

\subsection{Option 1: \texorpdfstring{$V(\Phi) = 2\sin\Phi$}{V(Phi) = 2 sin(Phi)}}
The first natural option at our disposal is to take the same potential as the Majorana case. In terms of the original dilaton field, the potential is:
\begin{gather}
    V(\Phi) = \frac{\sin(2\abs{\log q}\Phi)}{\abs{\log q}}
\end{gather}

We observe that the potential is affected by the boundary charge $Q$. In order to obtain the correct $\beta = 2\pi/\sin\theta$, this choice implies that $F_{tr}(\theta) = 0$, so that $\alpha = -4\pi W(\theta)$. In this case, the renormalized action \eqref{eq:cSYKrenormS} simplifies to:
\begin{gather}
    S_E = - \frac{\pi\theta}{\abs{\log q}} - \frac{\beta \cos\theta}{2\abs{\log q}} - \frac{\pi\beta}{\abs{\log q}} \int_{\theta}^{\pi/2} dr \left( \frac{W^2(r) - W^2(\theta)}{X(r)} \right). \label{eq:cSYKrenormS11}
\end{gather}

The naive match to \eqref{eq:cDSSYKtarget} implies that:
\begin{gather}
    \int_{\theta}^{\pi/2} dr \left( \frac{W^2(r) - W^2(\theta)}{X(r)} \right) = -\frac{1}{8} \sin\theta.
\end{gather}

To see the implications of this equation, we derive both sides with respect to $\theta$:
\begin{equation}
\begin{gathered}
    -2 W(\theta) W'(\theta) \int_{\theta}^{\pi/2} dr \: \frac{1}{X(r)} = -\frac{1}{8} \cos\theta \\
    \implies W^2(\Phi) = \frac{1}{8} \int^\Phi dx \: \frac{\cos x}{\int_x^{\pi/2} \frac{ds}{X(s)}}.
\end{gathered}
\end{equation}

For every real number, there is a saddle solution where $\Phi$ takes that value. Since $W(\Phi)$ appears in the metric and the electric field, it has to be real for every real $\Phi$, which implies $W^2(\Phi) \geq 0$. The $W^2(\Phi)$ we have found is defined up to an arbitrary integration constant, in particular there exists a constant $C$ such that $W^2(\Phi) \geq 0 \; \forall \Phi \in \mathbb{R}$ if and only if $X(\Phi)$ is such that:
\begin{gather}
    \inf_{\Phi \in \mathbb{R}} \: \int^\Phi dx \: \frac{\cos x}{\int_x^{\pi/2} \frac{ds}{X(s)}} > -\infty.
\end{gather}

This is a really interesting bound. Roughly speaking, it tells us that $\abs{\int_x^{\pi/2} \frac{ds}{X(s)}}$ should not have $0$ as its infimum (with the exception of all $x$ such that $\cos x = 0$) and preferably grow with $x$ or, at most, behave like a constant. This way, in fact, $W^2(\Phi)$ is either suppressed or oscillates for big $\Phi$ and there exists a constant $C$ that makes it always non-negative.

This result tells us that if we rescale $X \to \gamma X$ with $\gamma$ a constant, then $W \to \sqrt{\gamma} \, W$: since these functions are coupled in the action to $F^2$ and $F$ respectively, the action is left invariant by then taking $F \to F/\sqrt{\gamma}$, coherently with \eqref{eq:FtrSol} (we also impose that $\alpha \to \sqrt{\gamma} \, \alpha$). If we take $X \propto 1/Q^2$, for example, which is a common normalization for gauge fields when there is no charged matter in the theory, we have that $F \propto Q$. On the other hand, these rescalings do not change \eqref{eq:f(r)Sol}. For simplicity, we will consider charge-independent scalings of the functions below.

Imagine taking $X(\Phi) = V(\Phi) = 2\sin\Phi$, by analogy with some abelian dilaton models studied in the literature \cite{Frolov_2006}. In this case, we obtain:
\begin{gather}
    W^2(\Phi) = \frac{1}{2} \int^\Phi dx \: \frac{\cos x}{\log \left( \frac{1+\cos x}{1- \cos x} \right)}.
\end{gather}

The integrand is a non-negative, periodic and even function, which is regular for $\cos x = 0$ (its value is $1/2$ there). This means that it diverges to $+ \infty$ as $\Phi$ tends to $+\infty$, but it also diverges to $-\infty$ as $\Phi$ tends to $-\infty$. Such a model violates our bound and therefore cannot be dual to cDSSYK.

A first example of an interesting model that does not violate this bound is $X(\Phi) = 1$, i.e. the gauge field's kinetic term not coupling to the dilaton. In this case, we have:
\begin{gather}
    W^2(\Phi) = \frac{1}{8} \int^\Phi dx \: \frac{\cos x}{\frac{\pi}{2}-x} = \frac{1}{8} \mathrm{Si}\left( \Phi - \frac{\pi}{2} \right) + C. \label{eq:WPhi1}
\end{gather}

\begin{figure}
    \centering
    \includegraphics[width = 0.47 \textwidth]{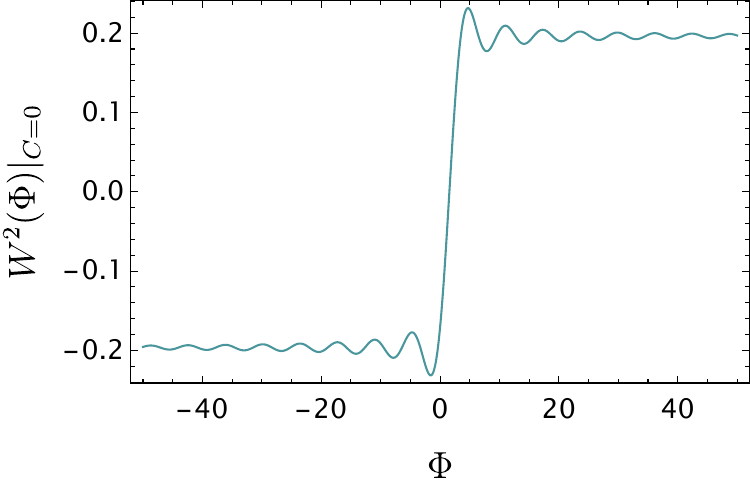} \quad
    \includegraphics[width = 0.47 \textwidth]{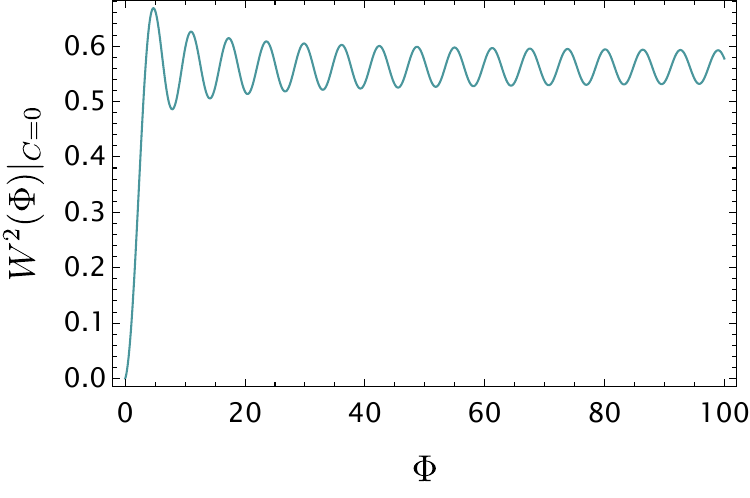}
    \caption{Left: Plot of \eqref{eq:WPhi1} for $C = 0$. Right: Plot of \eqref{eq:WPhi2} for $C=0$. The convergence of the second function to its asymptotic limit is extremely slow due to the slow growth of the logarithm at the denominator of the integrand. In practice, we observe the oscillations of the sine (integral of the cosine), whose amplitude tends to $0$ as $\Phi$ tends to $+\infty$. Qualitatively speaking, the behavior of this $W^2(\Phi)$ is similar to the sine integral result, with the main difference being that in that case the oscillations settle down much more rapidly.}
    \label{fig:WPhi}
\end{figure}

$\mathrm{Si}(x)$ is the sine integral function, whose maximum is achieved for $x = \pi$ and whose minimum is achieved for $x = -\pi$: $\mathrm{Si}(\pi) = - \mathrm{Si}(-\pi) \approx 1.85194$. Therefore, $W(\Phi)$ is real for any $C \geq \mathrm{Si}(\pi)/8$. The behavior of the function is shown in Figure \ref{fig:WPhi}.

We observe that this model also satisfies the working hypothesis \eqref{eq:cSYKworkhyp}. In particular, $\mathrm{Si}(i \ell) = i \,\mathrm{Shi}(\ell) = i \int_0^\ell dx \: \sinh x/x$, so $V(\pi/2+i\ell) = 2 \cosh \ell$ dominates over every other term. Using L'Hôpital's rule, in fact, one can easily show that $\mathrm{Shi}(x) \sim \cosh x/x$ for $\abs{x} \gg 1$, i.e. that:
\begin{gather}
    \lim_{x\to \pm \infty} \frac{x \, \mathrm{Shi}(x)}{\cosh x} = 1.
\end{gather}

We underline that we do not know whether $\Phi_{\rm bdy}$ is now $\pi/2 + i\infty$, but any other value with an infinite imaginary part satisfies the working hypothesis.

A second example of a relevant model that satisfies this bound is $X(\Phi) = \Phi$, i.e. the gauge field's kinetic term coupling to the dilaton in the same way as the Ricci scalar. In this case, we have:
\begin{gather}
    W^2(\Phi) = \frac{1}{8} \int^\Phi dx \: \frac{\cos x}{\log \left( \frac{\pi}{2\abs{x}} \right)} = \frac{1}{8} \int_0^\Phi dx \: \frac{\cos x}{\log \left( \frac{\pi}{2\abs{x}} \right)} + C. \label{eq:WPhi2}
\end{gather}

The integrand is an even function and tends to $\pi/2$ when $\abs{x}$ tends to the potentially singular value $\pi/2$, so we can focus on $\Phi > 0$. The function satisfies our bound if and only if:
\begin{gather}
    \lim_{\Phi \to +\infty} \abs{\int_0^\Phi dx \frac{\cos x}{\log \left( \frac{\pi}{2\abs{x}} \right)}} \neq +\infty.
\end{gather}

We now show that this is indeed the case. We can consider the asymptotic behavior of the integral:
\begin{equation}
\begin{aligned}
    \int_{N\pi}^{+\infty} dx \: \frac{\cos x}{-\log x} = - \sum_{n=0}^{+\infty} \int_{(N+n)\pi}^{(N+n+1)\pi} dx \: \frac{\cos x}{\log x} = - \sum_{n=0}^{+\infty} \int_{0}^{\pi} dx \: \frac{(-1)^{N+n} \cos x}{\log (N\pi + n\pi + x)}.
\end{aligned}
\end{equation}

This series is absolutely convergent:
\begin{equation}
\begin{aligned}
    \sum_{n=0}^{+\infty} \abs{\int_0^\pi dx \: \frac{\cos x}{\log (N\pi + n\pi + x)}} & \approx \sum_{n=0}^{+\infty} \frac{1}{\log((N+n)\pi)} \abs{\int_0^\pi dx \: \cos x \left( 1 - \frac{x}{(N+n)\pi \, \log((N+n)\pi)} \right)} \\
    & \approx \sum_{n=0}^{+\infty} \frac{2}{(N+n)\pi \, \log^2((N+n)\pi)}.
\end{aligned}
\end{equation}

Convergence follows, for example, from the integral test or the Cauchy condensation test. The behavior of the function is shown in Figure \ref{fig:WPhi}. It is then sufficient to take $8C \geq \sup_{\Phi \geq 0} \int_0^\Phi \cos x / \log(\pi/(2x))$ to obtain $W^2(\Phi) \geq 0$. This model also satisfies \eqref{eq:cSYKworkhyp}. In fact, by taking $\Phi = \pi/2 + i \ell$, we have that:
\begin{gather}
    W^2(\Phi) \sim -\frac{1}{8} \int_0^\ell dx \: \frac{\sinh x}{\log \left(1 + \frac{2i x}{\pi} \right)} \sim \frac{\cosh \ell}{\log \ell}.
\end{gather}

The second equivalence is up to constants and can be proven through L'Hôpital's rule. Once again, the term that dominates is $V(\Phi)$, which makes our computations consistent.

We further observe that, if \eqref{eq:dualconjecture} is a UV-complete theory of quantum gravity, we expect all our functions to have no poles on the real axis. For example, imagine taking $X(\Phi) = 1/\sin\Phi$. From a purely semiclassical point of view, this function is perfectly fine at least for any $\theta \neq k \pi$ saddle. If we now consider quantum fluctuations of the dilaton field, $\Phi = \Phi_{\rm cl} + \delta \Phi$, with $\Phi_{\rm cl} = r$ and $\delta \Phi$ its fluctuations, this theory appears to be an EFT with an UV cutoff scale that is at most $\sim h(\theta) \, \ell^{-1}$, with $h(\theta)$ some regular function:
\begin{equation}
    \frac{1}{\sin\Phi} = \frac{1}{\sin \Phi_{\rm cl}} \left[ 1 - \frac{\Phi_{\rm cl}}{\tan{\Phi_{\rm cl}}} \left( \frac{\delta \Phi}{\Phi_{\rm cl}} \right) + \Phi^2_{\rm cl} \left( \frac{1}{2} + \cot^2 \Phi_{\rm cl} \right) \left( \frac{\delta \Phi}{\Phi_{\rm cl}} \right)^2 - \frac{\Phi^3_{\rm cl} (5+6 \cot^2 \Phi_{\rm cl})}{6 \tan \Phi_{\rm cl}} \left( \frac{\delta \Phi}{\Phi_{\rm cl}} \right)^3 + \dots \right]. \label{eq:cSYKeft}
    \vspace{0.01 cm}
\end{equation}

\begin{figure}
    \centering
    \includegraphics[width = 0.7 \textwidth]{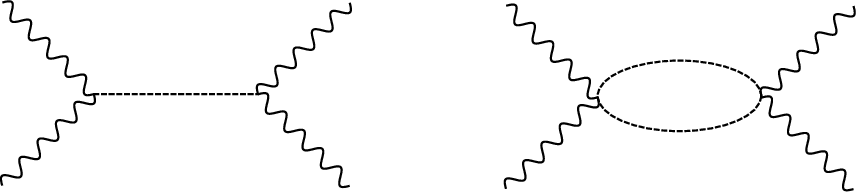}
    \caption{Contributions to the $\gamma+\gamma \to \gamma+\gamma$ scattering built with the vertices originating from the second and third term in \eqref{eq:cSYKeft}, respectively. The wavy lines are photons, while the dashed lines are dilatons.}
    \label{fig:gammascatt}
\end{figure}

This is clearly not a UV-complete theory and it stops working for $\mathcal{O}(1)$ quantum fluctuations. One can find an upper bound on the cutoff of the theory by considering the contribution of these terms to the $\gamma + \gamma \to \gamma + \gamma$ scattering in Figure \ref{fig:gammascatt}, and imposing that the loop diagram (which has an extra $(\mathrm{energy})^2$ dependence due to the $\int d^2k$ loop integral, since the dilaton propagator has no kinetic part) is subleading with respect to the tree level one. In general, a function that has poles is the reciprocal of a function that has zeroes for certain values of $\Phi$, so there is a limit on the magnitude of quantum fluctuations that we can consider before the theory breaks down: a UV-complete theory cannot admit this possibility.

In any case, we need to slightly modify the counterterm $G(r_0)$ in order to actually obtain \eqref{eq:cDSSYKtarget}. Given our choice of $W^2(\Phi)$ in terms of $X(\Phi)$, we can plug it back into the last term of \eqref{eq:cSYKrenormS11}:
\begin{gather}
    \int_\theta^{\pi/2} dr \: \frac{1}{8 X(r)} \int^r_\theta dx \: \frac{\cos x}{\int_x^{\pi/2} \frac{ds}{X(s)}}. 
\end{gather}

We define the second integral as $F(r)$ and we choose the primitive of $1/X(r)$ to be $H(r) = \int_{\pi/2}^r \, ds/X(s)$, so that an integration by parts returns:
\begin{gather}
    \frac{1}{8} \left[ H \left( \frac{\pi}{2} \right) F \left( \frac{\pi}{2} \right) - H(\theta)F(\theta) - \int_\theta^{\pi/2} dr \: H(r) \, \frac{\cos r}{(-H(r))} \right] = \frac{1}{8} - \frac{1}{8} \sin\theta.
\end{gather}

We find an extra $-\pi\beta/(8\abs{\log q})$ term in the action, which is easily removed by taking:
\begin{gather}
    G(r_0) \to G(r_0) + \frac{\pi}{4} \frac{1}{\sqrt{\int_{\pi/2}^{r_0} dr \left( V(r) + \frac{(\alpha+ 4\pi W(r))^2}{4\pi X(r)} \right)}}.
\end{gather}

We have severely restricted the space of possible dual theories, but there is still room for a lot of different options. On the other hand, it is interesting to observe that two physically motivated choices for $X(\Phi)$ are allowed by our calculations. For any choice of $W(\Phi)$ and $X(\Phi)$ that is compatible with our bounds, matching the two partition functions is finalized through the same steps (up to a resolution of their possible inconsistency) that we have described in Section \ref{sec:DefectOp}, which we do not repeat here.

To visualize which geometries arise from this discussion, we compute the Ricci scalar:
\begin{gather}
    R(r) = -2\cos r + \frac{4\pi X'(r) (W(r)-W(\theta))^2}{X^2(r)} - \frac{8\pi W'(r) (W(r)-W(\theta))}{X(r)}. \label{eq:newCurv}
\end{gather}

\begin{figure}
    \centering
    \includegraphics[width = 0.47 \textwidth]{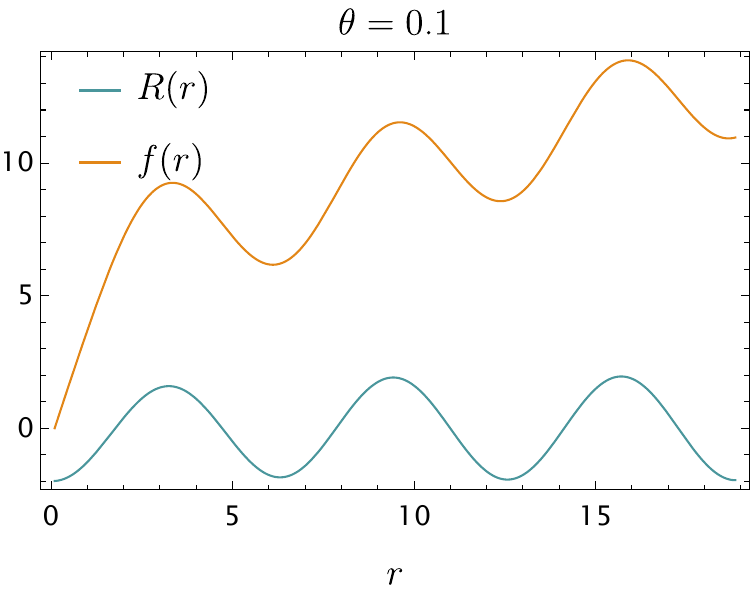} \quad
    \includegraphics[width = 0.47 \textwidth]{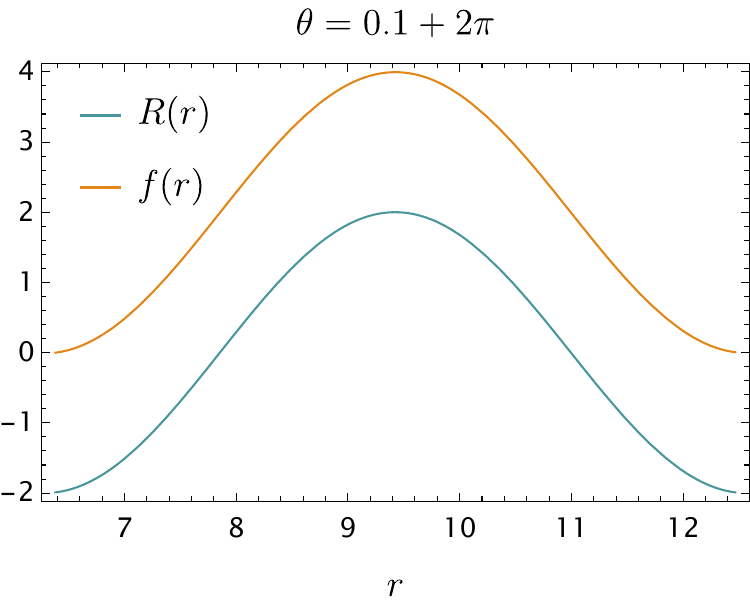}
    \includegraphics[width = 0.47 \textwidth]{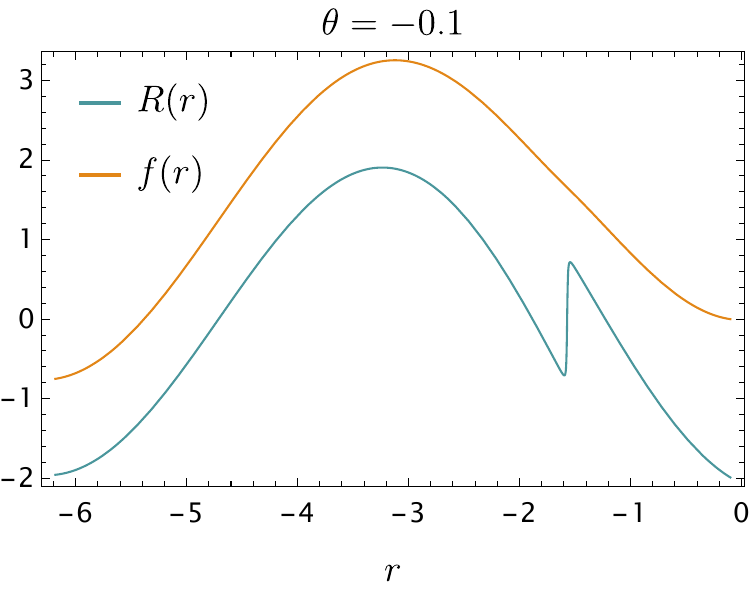} \quad
    \includegraphics[width = 0.47 \textwidth]{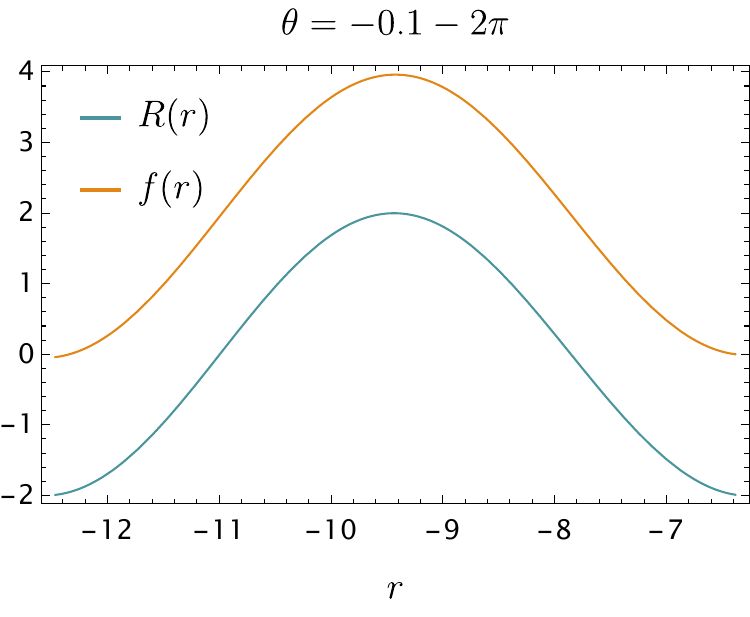}
    \caption{Plot of $R(r)$ and $f(r)$ for the first few saddles in the case of $X(\Phi) = 1$ and $8C = \mathrm{Si}(\pi)+10^{-5}$ in \eqref{eq:WPhi1}. We see that only the first positive saddle and the first negative saddle are different from the Majorana dual. The $\theta = 0.1$ saddle does not have a cosmological horizon at $\bar\theta = 2\pi-\theta$, but rather $f(r)$ grows indefinitely as $r$ increases. The $\theta = -0.1$ saddle has a cosmological horizon at $|\bar r| < |2\pi + \theta|$ and exhibits a sudden decrease in its curvature for $r=-\pi/2$ due to $W'(\Phi) = (W^2)'(\Phi)/(2W(\Phi))$ becoming relevant in the region where $W(\Phi) \sim 0$. This behavior smoothens as $C$ increases.}
    \label{fig:geometry1}
\end{figure}

\begin{figure}
    \centering
    \includegraphics[width = 0.47 \textwidth]{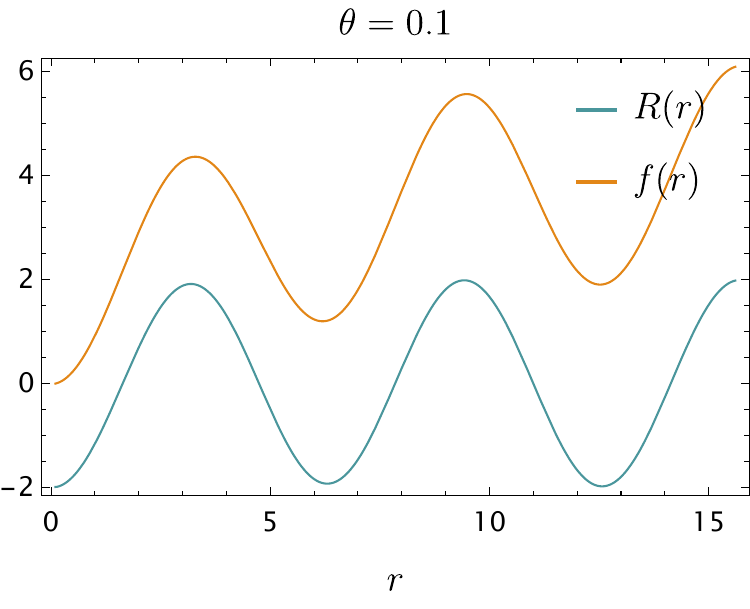} \quad
    \includegraphics[width = 0.47 \textwidth]{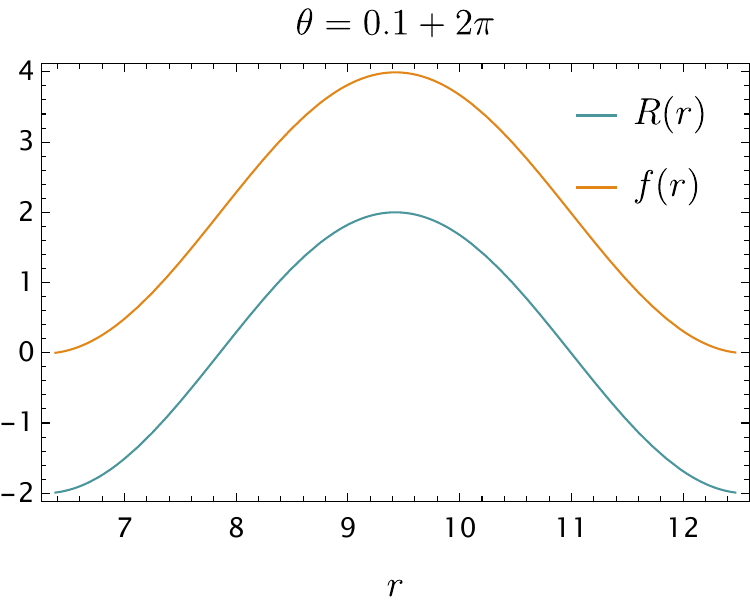} \\
    \includegraphics[width = 0.47 \textwidth]{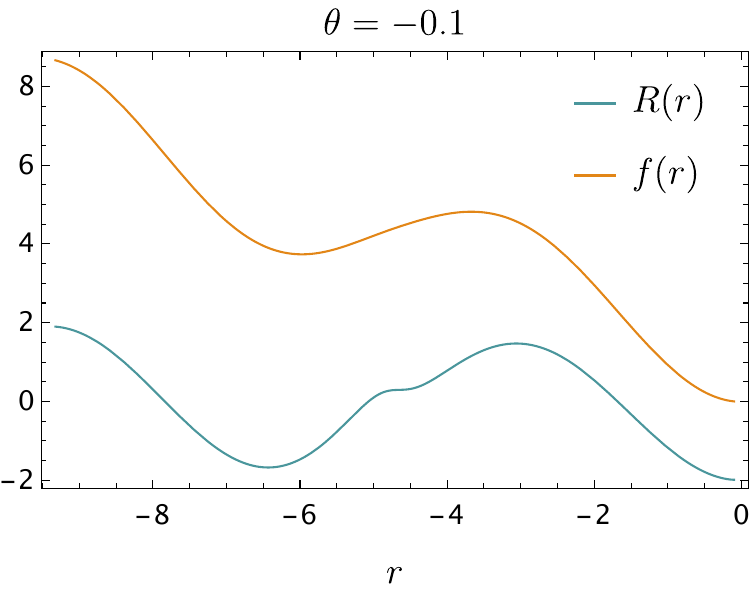} \quad
    \includegraphics[width = 0.47 \textwidth]{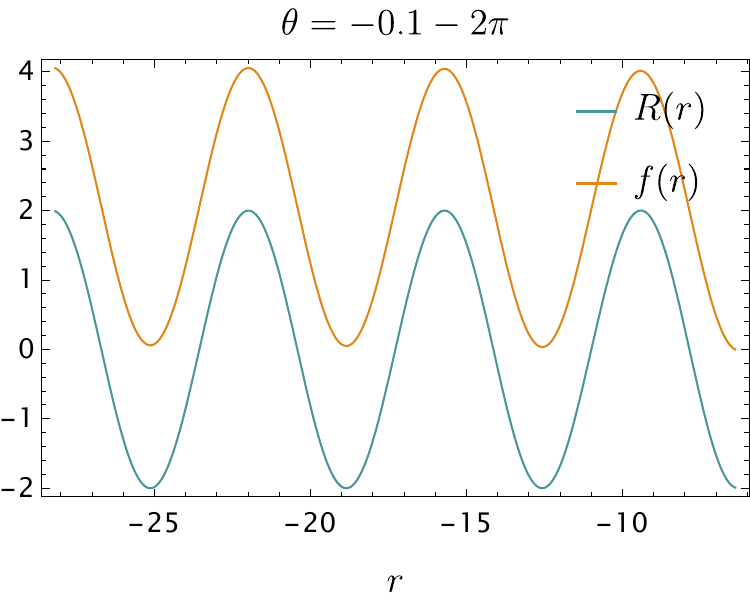}
    \caption{Plot of $R(r)$ and $f(r)$ for the first few saddles in the case of $X(\Phi) = \Phi$ and $C = 0.68$ (which is slightly above the maximum of the right plot in Figure \ref{fig:WPhi}) in \eqref{eq:WPhi2}. We see that only the first positive saddle and the first negative saddle are visibly different from the Majorana dual. The $\theta = 0.1$ saddle does not have a cosmological horizon at $\bar\theta = 2\pi-\theta$, but rather $f(r)$ grows indefinitely as $r$ increases. The $\theta = -0.1$ saddle exhibits a sudden plateau in its curvature for $r \approx -4.5 \sim -5$ due to $W'(\Phi)$ becoming relevant in the region where $W(\Phi) \sim 0$. This behavior smoothens as $C$ increases. The slight difference of the $\theta = -0.1 - 2\pi$ saddle with respect to $f_{\rm old}(r) = 2\cos\theta-2\cos r$ is responsible for $f(r)$ never reaching $0$, but rather growing indefinitely as $r$ decreases. Both here and in the $\theta = 0.1$ case, the asymptotic behavior of \eqref{eq:f(r)Sol} is $f(r) \propto \log |r|$, although the coefficient in front is visibly smaller for the negative saddle since $(W(-\infty)-W(-0.1-2\pi))^2 < (W(+\infty)-W(0.1))^2$.}
    \label{fig:geometry2}
\end{figure}

There is a new, interesting feature: although we still have an infinite amount of saddles with a black hole or a cosmological horizon, we do not have a periodic repetition of identical geometries because $f(r)$ is not periodic anymore. If $f(\theta) = 0$, it is not true anymore that $f(2\pi-\theta)=0$, so a naive black hole solution with a real $r$ coordinate does not end on a cosmological horizon and viceversa. Still, for big $\abs{\theta}$, only the first term in \eqref{eq:f(r)Sol} and \eqref{eq:newCurv} is expected to dominate whenever $W(\Phi)$ has a limit for $\Phi \to \pm \infty$ (which is true in both of our examples), so only the first few saddles actually exhibit different characteristics. We plot $R(r)$ and $f(r)$ for different saddles and for both examples in Figures \ref{fig:geometry1} and \ref{fig:geometry2}. The lesson we learn is that this model also offers the possibility of investigating quantum de Sitter physics through holography.

It may seem worrying that the $Q = 0$ limit of these geometries does not simplify to the Majorana geometries. We observe, however, that the $Q=0$ cDSSYK model we are considering is different from the $\mu = 0$ full model, as it is a restricted theory where only $Q = 0$ eigenvectors appear, as opposed to the entire Hilbert space: this is a great qualitative difference with respect to the Majorana DSSYK case, which may ultimately be responsible for this effect.

\subsection{Option 2: \texorpdfstring{$f(r) = 2 \cos\theta - 2\cos r$}{f(r) = 2cos(theta) - 2cos(r)}}
The second natural option at our disposal is to take the same geometry as the Majorana model, which implies:
\begin{gather}
    V(\Phi) + \frac{(\alpha + 4\pi W(\Phi))^2}{4\pi X(\Phi)} = 2\sin\Phi. \label{eq:option2}
\end{gather}

This option gives the right inverse temperature $\beta$ automatically. There exists a choice of $\alpha$ (i.e. of a boundary condition for $F_{tr}$) that makes this statement true for some choices of the three functions. This is clearly a less interesting scenario, where saddles exhibit identical geometries and the usual $R(r)$ and $f(r)$. Nevertheless, studying it shows a different type of bounds on the dual theory that a semiclassical match is able to yield. In this case, the renormalized action \eqref{eq:cSYKrenormS} simplifies to:
\begin{gather}
    S_E = - \frac{\pi\theta}{\abs{\log q}} - \frac{\beta\cos\theta}{2\abs{\log q}} + \frac{\beta}{2\abs{\log q}} \int_{\theta}^{\pi/2} dr \left( \frac{\alpha W(r)}{X(r)} + \frac{\alpha^2}{4\pi X(r)} \right).
\end{gather}

The naive match to cDSSYK implies that:
\begin{gather}
    \int_{\theta}^{\pi/2} dr \left( \frac{\alpha W(r)}{X(r)} + \frac{\alpha^2}{4\pi X(r)} \right) = \frac{\pi}{4} \sin\theta.
\end{gather}

Again, we derive both sides with respect to $\theta$:
\begin{gather}
    \frac{\alpha W(\theta)}{X(\theta)} + \frac{\alpha^2}{4\pi X(\theta)} = -\frac{\pi}{4} \cos\theta \implies W(\Phi) = -\frac{\pi}{4\alpha} \cos\Phi \, X(\Phi) - \frac{\alpha}{4\pi}.
\end{gather}

The condition that defines $\alpha$ is $\theta$-independent, so the derivative was much more straightforward here. If we substitute this equality into the renormalized action, we find an extra $-\pi\beta/(8\abs{\log q})$ term that can be removed through the same exact modification to $G(r_0)$ as the one in the previous subsection. This seems to suggest that the final choice of $G(r_0)$ is always the same, regardless of which starting option one picks for the unknown functions. In particular, one should take precisely this shifted $G(r_0)$ when determining the unique dual theory.

Plugging this result back into \eqref{eq:option2}, we also find:
\begin{gather}
    V(\Phi) = 2\sin\Phi - \frac{\pi^3}{4\alpha^2} \cos^2 \Phi \, X(\Phi).
\end{gather}

Operatively speaking, after having picked any real $X(\Phi)$ with no real poles, then any $V(\Phi)$ and $W(\Phi)$ such that there exsists a unique constant $\alpha$ that satisfies the equations above will be allowed by this match. The $\alpha$ that makes these equalities possible will then be the one that defines which $F_{tr}$ we have to pick in the semiclassical geometry. The final result for the electric field is always going to be:
\begin{gather}
    F_{tr}(r) = -\frac{\pi^2}{\alpha} \cos r,
\end{gather}

so that $\alpha$ is fundamentally related to the boundary value $F_{tr}(r_0)$. In this case, the effect of $Q$ is only seen in the rescaling of the dilaton through $\abs{\log q}$ and in possibly defining $\alpha$. Again, the match is completed through the same steps that we have described in Section \ref{sec:DefectOp}, up to a resolution of their possible inconsistency.

In principle, one could consider also more exotic options, but we believe these two to be the only plausible ones. There are several ways to proceed in order to better localize the unique dual theory: we present them in the Conclusions and we leave them to future work.

\newpage
\chapter{Conclusions}
In this thesis, we have followed two very distinct paths and we have shown how the holographic principle connects them on a profound level. On one hand, we have considered a simple $(0+1)$ quantum mechanical system, the SYK model, and we have studied its properties. We have considered its large $p$ limit, its IR limit, its double-scaled limit, and its charged version. In doing so, we have learned a great deal about this model in itself. On the other hand, we have focused on gravity. We have shown how the gravitational path integral allows us to determine the thermodynamic properties of spacetimes and we have then focused our attention on a specific class of two-dimensional models, the dilaton-gravity models. By studying the most famous one, i.e. JT gravity, we have seen its connection to (near-)extremal black holes and to the IR sector of the SYK model. This duality is a very important result, which ultimately inspired everything that was done in the last three chapters. The issues of this match are that it is not a UV-complete duality and that it fully lives in AdS$_2$ on the bulk side. For this reason, we have considered more recent works in the literature, which directed their attention towards the DSSYK model and its holographic dual. The DSSYK model is a simplification of the usual one with deep connections to combinatorics and representation theory. It also has features that heavily suggest the presence of a de Sitter region inside its dual gravitational theory.

In Chapter 8, then, we have reviewed how the duality between the DSSYK model and sine dilaton gravity has been estabilished. This duality solves the issues associated to the connection between the SYK model and JT gravity, namely both are UV-complete theories and there exists a de Sitter region on the gravitational side. After having expanded on several missing details, we have made a necessary comparison between this result and other previous claims in the literature regarding the properties of the bulk dual. Since the final objective of studying the DSSYK model is using it to probe quantum de Sitter physics, this discussion has naturally led ourselves to investigate the properties of scalar probes that couple minimally to gravity and to compare them to those of the probes at our disposal, namely scalar fields that couple to an effective AdS$_2$ metric. Through the study of their respective quasinormal modes, we have been able to see that their spectra exhibit completely different features. Unfortunately, though, we currently do not know of any boundary operator that is dual to a probe of the first kind. It would be interesting to study $m^2 < 0$ scalar particles in the AdS$_2$ black hole geometry, with the objective of determining their quasinormal modes after properly accounting for the issues associated to the $m^2 (\sinh \tilde y)^{-2}$ potential on the real half-line.

In the Appendix, we have shown our proofs of several known facts in the literature and we have presented useful notions that tie naturally to what has been discussed throughout the thesis.

As for the cDSSYK model, finally, we have worked on it from two different points of view. First, in Chapter 7, we have built its partition function and rewritten it in terms of a single field $g(\tau)$, which accounts for deviations of the fermionic two point function from the free theory. This simplified partition function is a new result in itself, which we have then used to determine what equations are satisfied by $g(\tau)$ in the $\lambda \to 0$ limit and how they depend on the chemical potential $\mu$. Through our approach, we have been able to determine $g(\tau)$'s shape, up to two coefficients that are fixed by complicated relations. We have studied their dependence on $\mu$ numerically and, in the $\beta\abs{\mu} \ll 1$ and $\beta\abs{\mu} \gg 1$ limits, also analytically. Extremely interesting features have emerged, namely a notion of critical temperature $T_{\rm crit}$ and a discontinuous behavior of $g(\tau)$ with respect to $\mu$. On this purely boundary side, one could further investigate the nature of these discontinuities, namely their physical origin and how to resolve the ambiguity in the choice of the coefficients among the infinitely many possibilities. A way of investigating these discontinuities would be to compute the $\lambda \to 0$ ensemble-averaged partition function $\langle \mathcal{Z} \rangle_J$, and study the associated thermodynamics. By then performing a one-loop expansion of the partition function we have built, we would simultaneously be able to determine the $\mathcal{O}(\lambda)$ corrections to $g(\tau)$ and to the thermodynamic quantities.

Secondly, in Chapter 8, we have focused on its currently unknown gravitational dual. Starting from the known partition function of a fixed charge subsector, we have extracted the information on the dual theory that a semiclassical match is able to give us. We have seen that we have a great amount of freedom, in the form of three functions appearing in the action, which cannot be narrowed down to a single possibility through this coarse-grained analysis alone. However, several theories have been excluded since they do not satisfy the semiclassical bounds we have determined. We conjecture that \eqref{eq:dualconjecture} is indeed the dual theory to cDSSYK for a unique choice of the three functions $V(\Phi), \: W(\Phi), \: X(\Phi)$. This model, in fact, is the minimal modification to an already estabilished duality that implements the necessary $U(1)$ gauge symmetry.

In this case, there are at least three ways to proceed, which would further constrain the space of possible dual theories. 
\begin{itemize}
    \item First, one could match semiclassical observables on both sides of the duality, for example two point functions of random operators and of the charge. In particular, the gauge field $A_\mu$ could be a probe capable of experiencing the de Sitter region of the semiclassical spacetime. To this end, it may be interesting to redo all our computations from scratch through the use of the conformal gauge:
    \begin{gather}
        ds^2 = e^{2\omega(t,r)}(dt^2+dr^2).
    \end{gather}
    
   \item Secondly, the possible inconsistency we have pointed out at the end of section \ref{sec:DefectOp} should be addressed through an explicit and consistent computation. This check may reassure us that their final result is still correct, or that the defect should be introduced in a different way. This is a crucial step in finalizing the match between the two theories, so it is important to make sure that it works correctly.
    
    \item Thirdly, one could go beyond the semiclassical expansion of the partition functions and consider $\mathcal{O}(\abs{\log q}^0)$ corrections to both theories. At the one-loop order, the partition function of cDSSYK \eqref{eq:cDSSYKfullQ} becomes, up to constant terms:
    \begin{equation}
    \begin{aligned}
        \mathcal{Z}_{\rm cDSSYK} \approx \: & \exp \left[ NQ \log \left( \frac{1-2Q}{1+2Q} \right) - \frac{N+1}{2} \log(1-4Q^2) \right] \\
        & \int_0^\pi d\cos\theta \: \exp \bigg[ -\frac{\pi^2}{6\abs{\log q}} + \frac{\pi\theta-\theta^2}{\abs{\log q}} - \frac{\pi^2}{12\abs{\log q}} - \frac{1}{2} \log(2\abs{\log q}) \\
        & + \frac{\beta \cos\theta}{2\abs{\log q} \sqrt{1-\abs{\log q}}} \left( 1+\frac{\lambda}{2} \right) \left( 1 + \frac{2\lambda Q}{1-4Q^2} \right) \bigg] \\
        \approx \: & \exp \left[ NQ \log \left( \frac{1-2Q}{1+2Q} \right) - \frac{N}{2} \log(1-4Q^2) \right] \\
        & \int_0^\pi d\cos\theta \: \exp \left[ -\frac{\pi^2}{4\abs{\log q}} + \frac{\pi\theta-\theta^2}{\abs{\log q}} + \frac{\beta \cos\theta}{2\abs{\log q}} \left( 1+ \abs{\log q} \left( 1 - (Q-1/2)^2 \right) \right) \right].
    \end{aligned}
    \end{equation}

    On the gravitational side, one would instead have to compute:
    \begin{equation}
    \begin{gathered}
        \mathcal{Z}_{\rm grav} \approx e^{-S_E[g_{\rm cl},\Phi_{\rm cl},A_{\rm cl}]} \int \mathcal{D}h_{\mu\nu} \, \mathcal{D}\delta \Phi \, \mathcal{D} \delta A_\mu \exp \left[ -\frac{1}{2} \int d^2x \, d^2y \: \delta \psi_i(x) \frac{\delta^2 S_E[g_{\rm cl},\Phi_{\rm cl},A_{\rm cl}]}{\delta \psi_i(x) \delta \psi_j(y)} \delta \psi_j(y) \right], \\
        g_{\mu\nu} = g_{\mu\nu, \rm cl} + h_{\mu\nu}, \quad \Phi = \Phi_{\rm cl} + \delta \Phi, \quad A_\mu = A_{\mu, \rm cl} + \delta A_\mu,
    \end{gathered}
    \end{equation}
    with $\delta \psi_i$ a shorthand for the perturbation of the classical configuration of any field in the theory. This type of one-loop expansion for gravity is also a possible explanation for the minus sign in front of the cosmological saddles, which was already observed in the Majorana model.
    
    \item Finally, one could consider the partition function of cDSSYK in the presence of a chemical potential rather than a fixed value of the charge. This is also discussed in \cite{Berkooz:2020uly}. This study could provide a way to connect our analysis of $g(\tau)$ in Chapter 7 to an holographic setup, where bulk fields dual to the Dirac fermions may emerge. Functional derivatives with respect to their source piece would then yield the Green function $G(\tau)$, thus estabilishing an interesting link between our computations and gravity.
\end{itemize}

It is clear that there is still a lot left to be said about the DSSYK model and the possibilities it provides to further our knowledge of quantum gravitational theories and of de Sitter in particular, but all the topics and mathematical connections we have discussed in this thesis should motivate us to keep working along the designated directions.

\newpage
\appendix

\chapter{Anti-de Sitter Spacetime} \label{app:AdS}
In this appendix, we review several properties of AdS$_d$, from its definition to some examples of coordinate choices. Our reference for this presentation is \cite{natsuume2016adscft}.

AdS$_d$ can be defined as an hypersurface in $\mathbb{R}^{d+1}$ with metric $g_{\mu\nu} = \mathrm{diag}(+1,\dots,+1,-1,-1)$:
\begin{equation}
\begin{gathered}
    X_1^2 + X_2^2 + \dots + X_{d-1}^2 - X_d^2 - X_{d+1}^2 = -\ell_{\rm AdS}^2, \\
    ds^2 = dX_1^2 + dX_2^2 + \dots + dX_{d-1}^2 - dX_d^2 - dX_{d+1}^2. \label{eq:AdSdef}
\end{gathered}
\end{equation}

We notice that this spacetime is maximally symmetric, as it possesses $d(d+1)/2$ Killing vectors: this stems from the fact that transformations belonging to $SO(d-1,2)$ and acting on $\mathbb{R}^{d+1}$ map the spacetime into itself while leaving the metric invariant and are thus isometries. In fact, the generators of this group are exactly $d(d+1)/2$. 

Another characterization of AdS$_d$ is that it is the solution to Einstein's equations with a null stress-energy tensor and a negative cosmological constant $\Lambda$:
\begin{gather}
    R_{\mu\nu} - \frac{1}{2}R g_{\mu\nu} + \Lambda g_{\mu\nu} = 0. \label{eq:EinEq}
\end{gather}

We recall that the sign in front of $\Lambda$ depends on the chosen convention and that we are using a negative signature for the timelike coordinate. It is interesting to notice that if we take the trace of the above equation we obtain:
\begin{gather}
    R- \frac{d}{2} R + d\Lambda = 0 \implies \Lambda = \frac{d-2}{2d} R,
\end{gather}

which in the case of $d = 2$ can only be satisfied if $\Lambda = 0$. In this case, then, our AdS spacetime should solve Einstein's equations without a cosmological constant.

For a maximally symmetric spacetime:
\begin{equation}
\begin{gathered}
    R_{ABCD} = -\frac{1}{\ell^2_{\rm AdS}} (g_{AC} g_{BD} - g_{AD} g_{BC}), \\
    R_{MN} = -\frac{d-1}{\ell^2_{\rm AdS}} g_{MN}, \quad R = -\frac{d(d-1)}{\ell^2_{\rm AdS}}
\end{gathered}
\end{equation}

The indices above run from $1$ to $d$. One can explicitly verify that these formulas hold for our metric after expressing $X_{d+1}$ in \eqref{eq:AdSdef} in terms of the other $d$ coordinates and substituting it into $ds^2$. Finally, inserting them into Einstein's equations yields the relation:
\begin{gather}
    \Lambda = -\frac{(d-1)(d-2)}{2\ell^2_{\rm AdS}},
\end{gather}

thus proving that this spacetime is indeed a solution of \eqref{eq:EinEq} for all $d \geq 2$. It is now useful to consider the cases $d = 2$ and $d > 2$ separately.

\section{\texorpdfstring{$d = 2$}{d = 2}}
In this case, the global coordinate choice is the following:
\begin{equation}
\begin{gathered}
    X_1 = \ell_{\rm AdS} \sinh \rho, \quad X_2 = \ell_{\rm AdS} \cosh \rho \sin t, \quad X_3 = \ell_{\rm AdS} \cosh \rho \cos t, \\
    ds^2 = \ell_{\rm AdS}^2 (-\cosh^2 \hspace{-0.04 in} \rho \, dt^2 + d\rho^2).
\end{gathered}
\end{equation}

\begin{figure}
    \centering
    \includegraphics[width = 0.5 \textwidth]{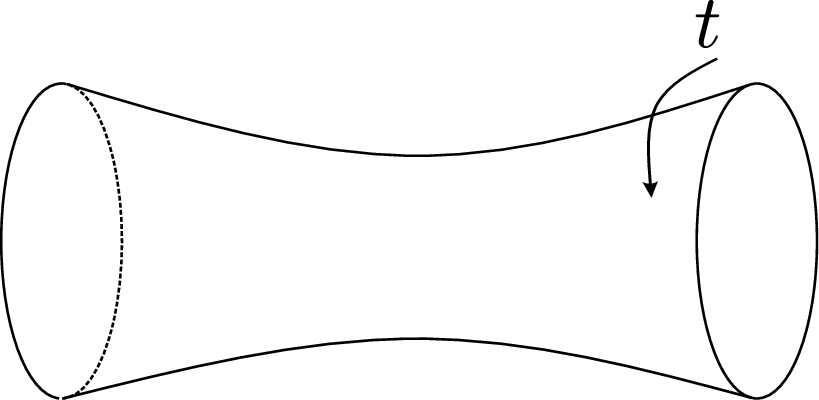}
    \caption{The embedding of AdS$_2$ in $\mathbb{R}^{1,2}$. The timelike direction $t$ is periodic, so we consider the covering space of the manifold instead.}
    \label{fig:AdS2}
\end{figure}

The spacelike coordinate $\rho$ spans the entire real axis, while the timelike coordinate $t$ has periodicity $2\pi$: this allows for closed timelike curves (one being $\rho = 0$ with a varying $t$), which are obviously in conflict with the request of causality. A solution is to ``unwrap'' the time coordinate and to consider the covering space of AdS$_2$, which is a natural operation given the metric we have just obtained. This can be done by cutting open the hyperboloid in Figure \ref{fig:AdS2} horizontally, flattening the spacetime and gluing an infinite amount of copies together along these cuts, with the final result being the covering. The possibility of $\rho$ being also negative is a property of AdS$_2$ only, and the reason why there are two disconnected timelike boundaries ($\rho = \pm \infty$) instead of only one: what will be $S^{d-2}$ in the next section reduces here to two points.

We can recover the global coordinates used in Section \ref{sec:NearHorizon} by switching to $\tan \theta = \sinh \rho$ and then taking $\nu = t, \, \sigma = \theta + \pi/2 \in [0,\pi]$, so that the metric becomes:
\begin{gather}
    ds^2 = \ell_{\rm AdS}^2 \frac{-d\nu^2 + d\sigma^2}{\sin^2 \sigma}.
\end{gather}

Another choice of global coordinates that connects us to \eqref{eq:metric} instead is:
\begin{equation}
\begin{gathered}
    r \equiv \ell_{\rm AdS} \sinh \rho, \quad \tau \equiv \ell_{\rm AdS} \, t, \\
    ds^2 = -\left( 1 + \frac{r^2}{\ell_{\rm AdS}^2} \right) d\tau^2 + \frac{1}{1 + \frac{r^2}{\ell_{\rm AdS}^2} } \, dr^2.
\end{gathered}
\end{equation}

The Poincaré patch is instead found by making the following coordinate choice:
\begin{equation}
\begin{gathered}
    X_1 = \frac{\ell_{\rm AdS}}{2z} \left( -t^2 + z^2 - 1 \right), \quad X_2 = \ell_{\rm AdS} \frac{t}{z}, \quad X_3 = \frac{\ell_{\rm AdS}}{2z} \left( -t^2 + z^2 + 1 \right), \\
    ds^2 = \ell_{\rm AdS}^2 \frac{-dt^2 + dz^2}{z^2}, \quad z > 0, \, t \in \mathbb{R}.
\end{gathered}
\end{equation}

\section{\texorpdfstring{$d > 2$}{d > 2}}
In this case, the global coordinate choice is the following:
\begin{equation}
\begin{gathered}
    X_{1 \leq i \leq d-1} = \ell_{\rm AdS} \sinh \rho \, \omega_i, \quad X_d = \ell_{\rm AdS} \cosh \rho \sin t, \quad X_{d+1} = \ell_{\rm AdS} \cosh \rho \cos t, \\
    ds^2 = \ell_{\rm AdS}^2 (-\cosh^2 \hspace{-0.04in} \rho \, dt^2 + d\rho^2 + \sinh^2 \hspace{-0.04in} \rho \, d\Omega_{d-2}^2).
\end{gathered}
\end{equation}

The index $i$ runs from $1$ to $d-1$ and the $\omega_i$ are the coordinates of a unit versor in $\mathbb{R}^{d-1}$. Again, we consider the covering space. As already mentioned in the previous section, $\rho$ is now a non-negative radial coordinate, so that we only have one boundary located at $\rho = +\infty$. This is evident by choosing conformal coordinates, once again defining $\tan \theta = \sinh \rho, \, \theta \in [0, \pi/2]$, so that we have:
\begin{gather}
    ds^2 = \frac{\ell_{\rm AdS}^2}{\cos^2 \theta} (-dt^2 + d\theta^2 + \sin^2 \theta \, d\Omega_{d-2}^2).
\end{gather}

Indeed, the metric is only singular in $\theta = \pi/2$. The picture of AdS$_{d>2}$ that emerges here is that of a full cylinder of radius $\pi/2$, whose boundary is exactly its border at $\theta = \pi/2$.

We can connect this geometry to \eqref{eq:metric} again by performing the same coordinate redefinition as before:
\begin{equation}
\begin{gathered}
    r \equiv \ell_{\rm AdS} \sinh \rho, \quad \tau \equiv \ell_{\rm AdS} \, t, \\
    ds^2 = -\left( 1 + \frac{r^2}{\ell_{\rm AdS}^2} \right) d\tau^2 + \frac{1}{1 + \frac{r^2}{\ell_{\rm AdS}^2} } \, dr^2 + r^2 d\Omega_{d-2}^2.
\end{gathered}
\end{equation}

Finally, we can also generalize the Poincaré coordinates in the following way, where $i$ runs from $1$ to $d-2$:
\begin{equation}
\begin{gathered}
    X_i = \ell_{\rm AdS} \frac{x_i}{z}, \quad X_{d-1} = \frac{\ell_{\rm AdS}}{2z} \left( \sum_i x_i^2 - t^2 + z^2 -1 \right), \quad X_d = \ell_{\rm AdS} \frac{t}{z}, \quad X_{d+1} = \frac{\ell_{\rm AdS}}{2z}\left( \sum_i x_i^2 - t^2 + z^2 + 1 \right), \\
    ds^2 = \ell_{\rm AdS}^2 \frac{-dt^2 + d\vec x^2 + dz^2}{z^2}, \quad z > 0, \, t \in \mathbb{R}, \, \vec x \in \mathbb{R}^{d-2}.
\end{gathered}
\end{equation}

\newpage
\chapter{A Corollary of Stokes' Theorem} \label{app:Stokes}
In this appendix, we show our original proof of \eqref{eq:Stokes}:
\begin{gather}
    \int_M d^dx \, \partial_\mu (\sqrt{-g} \, v^\mu) = \int_{\partial M} d^{d-1}y \, \sqrt{|h|} \, \varepsilon n_\mu v^\mu. \label{eq:Stokes2}
\end{gather}

In particular, we will focus on the case that interests us, namely timelike boundaries and $\varepsilon = 1$. To this end, we first define a $(d-1)$ form:
\begin{equation}
\begin{gathered}
    \alpha = \frac{1}{(d-1)!} \alpha_{\nu_1 \dots \nu_{d-1}} \, dx^{\nu_1}\wedge\dots\wedge dx^{\nu_{d-1}},\\
    \alpha_{\nu_1 \dots \nu_{d-1}} = \sqrt{-g} \, \varepsilon_{\lambda \nu_1 \dots \nu_{d-1}} v^\lambda, \quad v^\mu = -\frac{1}{(d-1)!} \frac{1}{\sqrt{-g}} \varepsilon^{\mu \nu_1 \dots \nu_{d-1}} \alpha_{\nu_1 \dots \nu_{d-1}}.
\end{gathered}
\end{equation}

The minus sign in $v^\mu$ is due to the signature of the metric. $\alpha$ is indeed a $(d-1)$ form, as its components transform correctly under a change of coordinates as a consequence of these two tensors being invariant in form:
\begin{gather}
    E^{\mu_0 \dots \mu_{d-1}} \equiv \frac{1}{\sqrt{-g}} \varepsilon^{\mu_0 \dots \mu_{d-1}}, \quad E_{\mu_0 \dots \mu_{d-1}} \equiv \sqrt{-g} \, \varepsilon_{\mu_0 \dots \mu_{d-1}}.
\end{gather}

These two tensors are connected by lowering and raising indices with the (inverse) metric, while we choose the following convention for the antisymmetric Levi-Civita symbols:
\begin{gather}
    \varepsilon_{0 \dots d-1} = 1, \quad \varepsilon^{0 \dots d-1} = -1.
\end{gather}

Let us now consider the l.h.s. of \eqref{eq:Stokes2}:
\begin{gather}
    \int_M dx^0 \wedge \dots \wedge dx^{d-1} \, \frac{-1}{(d-1)!} \varepsilon^{\mu \nu_1 \dots \nu_{d-1}} \partial_\mu (\alpha_{\nu_1 \dots \nu_{d-1}}) = \int_M dx^\mu \wedge \dots \wedge dx^{\nu_{d-1}} \frac{1}{(d-1)!} \partial_\mu (\alpha_{\nu_1 \dots \nu_{d-1}}).
\end{gather}

We recognize that we have found the integral on $M$ of the exterior derivative $d\alpha$, so we can use Stokes' theorem:
\begin{gather}
    \int_M d\alpha = \int_{\partial M} \alpha = \frac{1}{(d-1)!} \int_{\partial M} dx^{\nu_1}\wedge \dots \wedge dx^{\nu_{d-1}} \, \sqrt{-g} \, \varepsilon_{\mu \nu_1 \dots \nu_{d-1}} v^\mu.
\end{gather}

If we parametrize the boundary with $\mathbf{y} \in A \subset \mathbb{R}^{d-1}$, we can perform a pullback of the embedding map $x^\mu(\mathbf{y})$:
\begin{gather}
    \frac{1}{(d-1)!} \int_{\partial M} dy^{a_1}\wedge \dots \wedge dy^{a_{d-1}} \sqrt{-g} \, \varepsilon_{\mu \nu_1 \dots \nu_{d-1}} \frac{\partial x^{\nu_1}}{\partial y^{a_1}} \dots \frac{\partial x^{\nu_{d-1}}}{\partial y^{a_{d-1}}} v^\mu.
\end{gather}

For timelike boundaries, we also have to be careful of the minus signs in $\partial M$. In this case, it is true that
\begin{gather}
    dy^{a_1} \wedge ... \wedge dy^{a_{d-1}} = - \varepsilon^{a_1 \dots a_{d-1}} dy^0 \wedge ... \wedge dy^{d-2},
\end{gather}

so that the l.h.s. can be rewritten as:
\begin{equation}
\begin{gathered}
    \int_{\partial M} dy^0 \wedge \dots \wedge dy^{d-2} \sqrt{-g} \, K_\mu v^\mu = (-1)^d \int_{\partial M} d^{d-1}y \, \sqrt{-g} K_\mu v^\mu, \\
    N_\mu \equiv (-1)^d K_\mu = \frac{(-1)^{d-1}}{(d-1)!} \varepsilon_{\mu \nu_1 \dots \nu_{d-1}} \varepsilon^{a_1 \dots a_{d-1}} \frac{\partial x^{\nu_1}}{\partial y^{a_1}} \dots \frac{\partial x^{\nu_{d-1}}}{\partial y^{a_{d-1}}}.
\end{gathered}
\end{equation}

The $(-1)^d$ sign comes from the following observation, that is actually part of the proof of Stokes' theorem itself. A $d$-dimensional manifold with a border can be locally described with a chart whose domain is $\mathbb{R}^d_+ = (x^1,x^2,\dots,x^d), \: x^{i < d} \in \mathbb{R}, \: x^d \geq 0$, that is, the upper semi-region of $\mathbb{R}^d$, so that the border is exactly $x^d = 0$. Consider taking a $(d-1)$-form, that is described in this chart by:
\begin{equation}
\begin{gathered}
    \omega = \sum_{j=1}^{d} f_j \, dx^1 \wedge \dots \wedge \hat{dx^j} \wedge \dots \wedge dx^d, \\
    d\omega = \sum_{j=1}^{d} (-1)^{j-1} \partial_j f_j \, dx^1 \wedge \dots \wedge dx^j \wedge \dots \wedge dx^d,
\end{gathered}
\end{equation}

where $\hat{dx^j}$ denotes the absence of the covector. When integrating $d\omega$ on $\mathbb{R}^d_+$ and $\omega$ on its border $\partial \mathbb{R}^d_+$, only the $j=d$ term gives a non-null contribution for both. In particular, it is easy to see that we obtain:
\begin{equation}
\begin{gathered}
    \int_{\partial \mathbb{R}^d_+} \omega = \int_{\partial \mathbb{R}^d_+} dx^1 \wedge \dots \wedge dx^{d-1} \: f_d(x^1, \dots, x^{d-1},0), \\
    \int_{\mathbb{R}^d_+} d\omega = (-1)^d \int_{\mathbb{R}^{d-1}} dx^1 \dots dx^{d-1} \: f_d(x^1, \dots, x^{d-1},0).
\end{gathered}
\end{equation}

The crucial observation is the following: Stokes' theorem assumes a positive orientation of both $M$ and $\partial M$, but the orientation induced by $\mathbb{R}^d_+$ on its $(d-1)$-dimensional border is not always positive. In order to fix this, an extra sign of $(-1)^d$ is needed when switching from the wedge product to the usual integration measure.

We will now focus on $N_\mu$. First of all, we want to prove that it is orthogonal to the boundary, hence that it is proportional to $n_\mu$. To do so, we consider its scalar product with an arbitrary vector lying on $\partial M$:
\begin{gather}
    u^\mu = c^a \frac{\partial x^\mu}{\partial y^a},
\end{gather}

where $c^a$ are its coefficients with respect to the chosen basis. By linearity, it suffices to consider the scalar product with each $\partial x^\mu/\partial y^a$, with $a$ fixed:
\begin{gather}
    N_\mu \frac{\partial x^\mu}{\partial y^a} = \frac{(-1)^{d-1}}{(d-1)!} \varepsilon_{\mu \nu_1 \dots \nu_{d-1}} \varepsilon^{a_1 \dots a_{d-1}} \frac{\partial x^{\nu_1}}{\partial y^{a_1}} \dots \frac{\partial x^{\nu_{d-1}}}{\partial y^{a_{d-1}}} \frac{\partial x^\mu}{\partial y^a}. \label{eq:Nmuprod}
\end{gather}

Since the $a_i$ range from $0$ to $(d-2)$ and we have $d$ of them, there is necessarily a $x^{\nu_j}$ that is derived with respect to the same $y^{a_i}$ as $x^\mu$, hence we can permute them and obtain the same result. Because of the presence of $\varepsilon_{\mu \nu_1 \dots \nu_{d-1}}$, though, this operation flips the overall sign. We can thus conclude that \eqref{eq:Nmuprod} is always 0 and that indeed $N_\mu \propto n_\mu$.

To conclude our proof, then, we only need to show that:
\begin{gather}
    \sqrt{-g} \, N_\mu = \sqrt{|h|} \, \varepsilon n_\mu \iff
    \begin{cases} 
    \mathrm{sgn}(N^\mu n_\mu) = \mathrm{sgn}(\varepsilon) \\
    N_\mu N^\mu = \frac{|h|}{|g|} n_\mu n^\mu = \pm \frac{|h|}{|g|}
    \end{cases}.
\end{gather}

For a timelike boundary, we recall that $n^\mu$ is spacelike and $n^\mu n_\mu = 1$.

We start by considering $|h|$. Derivatives with latin indices are done with respect to $y^a$, those with greek indices are done with respect to $x^\mu$. We know that:
\begin{gather}
    h_{ab} = \partial_a x^\mu \partial_b x^\nu g_{\mu \nu} \implies h_{ab} = \left( \frac{\partial x}{\partial y} \right)^T g_{\mu \nu} \left( \frac{\partial x}{\partial y} \right).
\end{gather}

We can use Cauchy-Binet's formula by applying it to the following matrices:
\begin{gather}
    A_{i \nu} = \partial_i x^\mu g_{\mu \nu}, \quad B_{\mu j} = \partial_j x^\mu \implies h_{ab} = A_{a\nu} B_{\nu b}.
\end{gather}

This lets us link the determinant of $h_{ab}$ to that of $g_{\mu\nu}$. Let $S = \{ S_1, \dots, S_{d-1} \}$ be an ordered choice of $d-1$ indices from $\{ 0, \dots, d-1 \}$: there are $d$ of them in total. By defining $A_S$ as the square sub-matrix obtained from $A$ by taking the columns whose indices belong to $S$ (and similarly for $B_S$ and the rows of $B$), the formula states that:
\begin{equation}
\begin{aligned}
    h & = \sum_S \det(A_S) \det(B_S) \\
    & = \sum_S \left( -\varepsilon^{a_1 \dots a_{d-1}} \partial_{a_1} x^{\mu_1} g_{\mu_1 S_1} \dots \partial_{a_{d-1}} x^{\mu_{d-1}} g_{\mu_{d-1} S_{d-1}} \right) \left( -\varepsilon^{b_1 \dots b_{d-1}} \partial_{b_1} x^{S_1} \dots \partial_{b_{d-1}} x^{S_{d-1}} \right) \\
    & = \sum_S \varepsilon^{a_1 \dots a_{d-1}} \varepsilon^{b_1 \dots b_{d-1}} \partial_{a_1} x^{\mu_1} g_{\mu_1 S_1} \partial_{b_1} x^{S_1} \dots \partial_{a_{d-1}} x^{\mu_{d-1}} g_{\mu_{d-1} S_{d-1}} \partial_{b_{d-1}} x^{S_{d-1}}. 
\end{aligned}
\end{equation}

We now turn to the computation of $N_\mu N^\mu$:
\begin{gather}
    N_\mu N^\mu = \frac{1}{[(d-1)!]^2} \varepsilon_{\mu \nu_1 \dots \nu_{d-1}} \varepsilon^\mu_{\rho_1 \dots \rho_{d-1}} \varepsilon^{a_1 \dots a_{d-1}} \varepsilon^{b_1 \dots b_{d-1}} \partial_{a_1} x^{\nu_1} \partial_{b_1} x^{\rho_1} \dots \partial_{a_{d-1}} x^{\nu_{d-1}} \partial_{b_{d-1}} x^{\rho_{d-1}}.
\end{gather}

We can simplify the product of the first two $\varepsilon$ tensors \cite{carroll2004spacetime}:
\begin{equation}
\begin{gathered}
    \varepsilon_{\mu \nu_1 \dots \nu_{d-1}} \varepsilon^\mu_{\rho_1 \dots \rho_{d-1}} = \frac{1}{|g|} g^{\mu\rho_0} E_{\mu \nu_1 \dots \nu_{d-1}} E_{\rho_0 \rho_1 \dots \rho_{d-1}} = - \frac{1}{|g|} g^{\mu\rho_0} g_{\mu\beta_0} g_{\nu_1 \beta_1} \dots g_{\nu_{d-1} \beta_{d-1}} \delta^{\beta_0 \beta_1 \dots \beta_{d-1}}_{\rho_0 \rho_1 \dots \rho_{d-1}}, \\
    \delta^{\beta_0 \beta_1 \dots \beta_{d-1}}_{\rho_0 \rho_1 \dots \rho_{d-1}} = \det 
    \begin{bmatrix}
        \delta^{\beta_0}_{\rho_0} & \delta^{\beta_0}_{\rho_1} & \dots & \delta^{\beta_0}_{\rho_{d-1}} \\
        \delta^{\beta_1}_{\rho_0} & \ddots &  & \vdots \\
        \vdots &  & \ddots & \vdots \\
        \delta^{\beta_{d-1}}_{\rho_0} & \dots & \dots & \delta^{\beta_{d-1}}_{\rho_{d-1}}
    \end{bmatrix} = \varepsilon_{\rho_0 \rho_1 \dots \rho_{d-1}} \sum_k \mathrm{sgn}(k) \, \delta^{\beta_0}_{k(0)} \delta^{\beta_1}_{k(1)} \dots \delta^{\beta_{d-1}}_{k(d-1)}.
\end{gathered}
\end{equation}

In the second line, we have factored out the permutation of the indices $\{ 0, \dots, d-1 \}$ due to $\{ \rho_0, \dots, \rho_{d-1} \}$. We now have:
\begin{gather}
    \hspace{-1.0 cm} N_\mu N^\mu = -\frac{1}{[(d-1)!]^2} \frac{1}{|g|} \sum_k \mathrm{sgn}(k) \, \varepsilon_{k(0) \rho_1 \dots \rho_{d-1}} \varepsilon^{a_1 \dots a_{d-1}} \varepsilon^{b_1 \dots b_{d-1}} g_{\nu_1 k(1)} \partial_{a_1} x^{\nu_1} \dots g_{\nu_{d-1} k(d-1)} \partial_{a_{d-1}} x^{\nu_{d-1}} \partial_{b_1} x^{\rho_1} \dots \partial_{b_{d-1}} x^{\rho_{d-1}}.
\end{gather}

The presence of $\varepsilon^{b_1 \dots b_{d-1}}$, allows us to reorder each of the $(d-1)!$ possible sequences of $\rho_i$ to match the order of the $k(i)$ by also permuting the $b_i$. We always end up with a plus sign, so we can use \mbox{$\varepsilon_{k(0) k(1) \dots k(d-1)} = \mathrm{sgn}(k)$} and obtain:
\begin{gather}
    N_\mu N^\mu = -\frac{1}{(d-1)!} \frac{1}{|g|} \sum_k \varepsilon^{a_1 \dots a_{d-1}} \varepsilon^{b_1 \dots b_{d-1}} g_{\nu_1 k(1)} \partial_{a_1} x^{\nu_1} \dots g_{\nu_{d-1} k(d-1)}\partial_{a_{d-1}} x^{\nu_{d-1}} \partial_{b_1} x^{k(1)} \dots \partial_{b_{d-1}} x^{k(d-1)}.
\end{gather}

At this point, we notice that all the permutations that have the same $k(0)$ give the same contribution thanks to the presence of both $\varepsilon^{a_1 \dots a_{d-1}}$ and $\varepsilon^{b_1 \dots b_{d-1}}$ and that there are $(d-1)!$ of them. We can thus sum on the $d$ possible choices of $k(0)$ (that does not appear in the contractions) and assume the $k(i \geq 1)$ to be ordered. This is precisely the sum over the sets $S$ defined earlier, so that we can finally conclude:
\begin{equation}
\begin{aligned}
    N_\mu N^\mu & = -\frac{1}{|g|} \sum_S \varepsilon^{a_1 \dots a_{d-1}} \varepsilon^{b_1 \dots b_{d-1}} \partial_{a_1} x^{\mu_1} g_{\mu_1 S_1} \dots \partial_{a_{d-1}} x^{\mu_{d-1}} g_{\mu_{d-1} S_{d-1}} \partial_{b_1} x^{S_1} \dots \partial_{b_{d-1}} x^{S_{d-1}} \\
    & = -\frac{h}{|g|} = \frac{|h|}{|g|}.
\end{aligned}
\end{equation}

This result is compatible with $n_\mu n^\mu = +1$, as it should. We have still left to prove that the orientation of $N^\mu$ is the right one, i.e. that $N^\mu n_\mu > 0$, with $n_\mu$ the ``outgoing'' versor. In order to check this, let us write this vector locally by using the coordinates that map a neighborhood of a point of $M$ near the boundary to $\mathbb{R}^d_+$, as we have seen earlier. With this choice, $\partial M$ is parametrized by $\mathbf{y} = (x^0, \dots, x^{d-2})$ and the partial derivatives in $N_\mu$ are trivial:
\begin{gather}
    N_{\mu} = \frac{(-1)^{d-1}}{(d-1)!} \varepsilon_{\mu \nu_1 \dots \nu_{d-1}} \varepsilon^{a_1 \dots a_{d-1}} \delta^{\nu_1}_{a_1} \dots \delta^{\nu_{d-1}}_{a_{d-1}} = \frac{(-1)^{d-1}}{(d-1)!} \varepsilon_{\mu \nu_1 \dots \nu_{d-1}} \varepsilon^{\nu_1 \dots \nu_{d-1}}.
\end{gather}

Clearly, only $N_{d-1}$ is non-null. By keeping a consistent choice of convention for the Levi-Civita symbols, we can write $\varepsilon_{d-1 \, \nu_1 \dots \nu_{d-1}} = (-1)^{d-1} \varepsilon_{\nu_1 \dots \nu_{d-1}}$. Using the known contraction $\varepsilon_{\nu_1 \dots \nu_{d-1}} \varepsilon^{\nu_1 \dots \nu_{d-1}} = -(d-1)!$, we are left with:
\begin{gather}
    N_\mu = -\delta_{\mu,d-1} \implies N_{d-1} < 0.
\end{gather}

Since our manifold is locally $\mathbb{R}^d_+$, the fact that $N_{d-1}$ is negative means that our vector is indeed ``outgoing'' and that $N_\mu n^\mu > 0$. As is clear from \eqref{eq:Stokes2}, it is the vector with the lowered index that should be orthogonal to the manifold.

In the case of spacelike boundaries, one can proceed in the same way as above while correctly accounting for the different signs appearing in several passages. With these computations, then, \eqref{eq:Stokes} has been shown to be true.

\newpage
\chapter{Schwarzian Derivative} \label{app:Schwarz}
Let us recall the definition of the Schwarzian derivative:
\begin{gather}
    S(\phi(\tau),\tau) \equiv \{ \phi, \tau \} \equiv \left( \frac{\phi''(\tau)}{\phi'(\tau)} \right)' - \frac{1}{2} \left( \frac{\phi''(\tau)}{\phi'(\tau)} \right)^2 = \frac{\phi'''(\tau)}{\phi'(\tau)} - \frac{3}{2} \left( \frac{\phi''(\tau)}{\phi'(\tau)} \right)^2.
\end{gather}

We start by looking for which functions have null Schwarzian derivative:
\begin{equation}
\begin{gathered}
    \left( \frac{\phi''(\tau)}{\phi'(\tau)} \right)' = \frac{1}{2} \left( \frac{\phi''(\tau)}{\phi'(\tau)} \right)^2 \implies \frac{\phi''(\tau)}{\phi'(\tau)} = -\frac{2}{\tau+A} \\
    \implies \phi(\tau) = -\frac{B}{\tau + A} + C = \frac{\frac{C}{\sqrt{B}} \tau + \frac{AC-B}{\sqrt{B}}}{\frac{1}{\sqrt{B}}\tau + \frac{A}{\sqrt{B}}} = \frac{\alpha \tau + \beta}{\gamma \tau + \delta}, \quad \alpha\delta - \beta\gamma = 1.
\end{gathered}
\end{equation}

All and only Möbius transformations are then those with null Schwarzian derivative. Another property of this operator is a chain rule when considering the Schwarzian derivative of a composition:
\begin{gather}
    S(\psi \circ \phi, \tau) = S(\psi, \phi(\tau)) \, \phi'(\tau)^2 + S(\phi,\tau). \label{eq:SchComp}
\end{gather}

One can prove this identity by substituting the derivatives inside the definition. It follows that, when $\psi$ is a Möbius transformation, the Schwarzian derivative does not vary.

A nice interpretation of the Schwarzian is the following. Given $f$ a conformal (holomorphic with non-zero derivative) mapping in a neighborhood of $z_0 \in \mathbb{C}$, there exists a unique Möbius transformation $M$ such that $M$ and $f$ share the same derivatives of order $0,1,2$ in $z_0$: this is made possible by fixing the three free parameters of $M$ appropriately. Because of this, it is true that 
\begin{gather}
    \label{eq:mobf} (M^{-1} \circ f) (z) = z_0 + (z-z_0) + \frac{1}{6}a (z-z_0)^3 + \mathcal{O}((z-z_0)^4).
\end{gather}

Equivalently, this composition deviates from the identity near $z_0$ only starting from the third order term. We show our proof of this statement and of the value of $a$ in particular. First, we write:
\begin{equation}
\begin{gathered}
    M^{-1}(z) = \frac{Az+B}{Cz+\frac{1+BC}{A}}, \\
    f(z) = f(z_0) + f'(z_0)(z-z_0) + \frac{1}{2}f''(z_0)(z-z_0)^2 + \frac{1}{6}f'''(z_0)(z-z_0)^3 + \mathcal{O}((z-z_0)^4), \\
    (M^{-1} \circ f)(z) = \sum_{j = 0}^3 h_j(A,B,C,f^{(i \leq j)}(z_0)) (z-z_0)^j + \mathcal{O}((z-z_0)^4),
\end{gathered}
\end{equation}

where $h_j$ are functions that we do not report here and that depend on the parameters of the Möbius transformation and on the derivatives of $f$ evaluated in $z_0$. By imposing that $h_{0 \leq j \leq 2}$ match \eqref{eq:mobf}, we find only two non-trivial solutions ($A \neq 0$) that are linked by a $A,B,C \longleftrightarrow -A,-B,-C$ $\mathbb{Z}_2$ symmetry (they are, in fact, the same transformation) and that fix the value of $a$:
\begin{equation}
\begin{gathered}
    A = \frac{2f'(z_0)+f''(z_0)z_0}{2f'(z_0)^{3/2}}, \quad B = \frac{2 f'(z_0)^2 z_0 - f(z_0) f''(z_0) z_0 - 2f(z_0)f'(z_0)}{2f'(z_0)^{3/2}}, \quad C = \frac{f''(z_0)}{2f'(z_0)^{3/2}} \\
    \implies a = \frac{f'''(z_0)}{f'(z_0)} - \frac{3}{2} \left( \frac{f''(z_0)}{f'(z_0)} \right)^2 = S(f,z_0).
\end{gathered}
\end{equation}

We underline the fact that we are assuming $f$ to be a conformal map, so $f'(z_0) \neq 0$. This computation suggests a possible interpretation of the Schwarzian derivative: it is a measure of how much a conformal map locally deviates from a Möbius transformation.

We conclude this appendix by proving that the Schwarzian derivative is the unique lowest order in derivatives Lagrangian that is invariant under $SL(2,\mathbb{R})$, up to total derivatives. Consider a Möbius transformation:
\begin{gather}
    f(x) \to \frac{a f(x) + b}{c f(x) + d}, \quad ad - bc = 1.
\end{gather}

The invariance of the operator we are looking for under translations $(f(x) \to f(x) + b)$ implies that it can only be a function of derivatives, $f^{(n \geq 1)}(x)$. The invariance under rescalings $(f(x) \to a f(x))$ tells us that it has to be a function of ratios, $f^{(n)}(x)/f^{(m)}(x)$.

The first object that satisfies these requirements is $f''(x)/f'(x)$, but it is the derivative of $\log f'(x)$. The next objects that the operator we are building can depend on are:
\begin{gather}
    \left( \frac{f''(x)}{f'(x)} \right)^2, \quad \frac{f'''(x)}{f'(x)}, \quad \left( \frac{f'''(x)}{f''(x)} \right)^2.
\end{gather}

We have also excluded $f'''(x)/f''(x)$, since it is the derivative of $\log f''(x)$. The most general object we can build, up to a global constant, is:
\begin{gather}
    S(f(x),x) = \frac{f'''(x)}{f'(x)} + r \left( \frac{f''(x)}{f'(x)} \right)^2 + s \left( \frac{f'''(x)}{f''(x)} \right)^2, \quad r, \: s \in \mathbb{R}.
\end{gather}

We are restricting ourselves to use at most the third derivative of $f(x)$ and the lowest possible power of each ratio. We now have to impose invariance under $f(x) \to 1/f(x)$, which causes the derivatives to transform in the following way:
\begin{gather}
    f'(x) \to -\frac{f'(x)}{f(x)^2}, \quad f''(x) \to -\frac{f''(x)}{f(x)^2} + 2 \, \frac{f'(x)^2}{f(x)^3}, \quad f'''(x) \to -\frac{f'''(x)}{f(x)^2} + 6 \, \frac{f'(x)f''(x)}{f(x)^3} - 6 \, \frac{f'(x)^3}{f(x)^4}.
\end{gather}

As a consequence, the operator we are building becomes:
\begin{equation}
\begin{aligned}
    S(f(x),x) = \: & \frac{f'''(x)}{f'(x)} - 6 \, \frac{f''(x)}{f(x)} + 6 \left( \frac{f'(x)}{f(x)} \right)^2 + r \left[ \left( \frac{f''(x)}{f'(x)} \right)^2 + 4 \left( \frac{f'(x)}{f(x)} \right)^2 - 4 \, \frac{f''(x)}{f(x)} \right] \\
    & + s \left( \frac{f'''(x)}{f''(x)} \right)^2 \left( \frac{1 - 6\, \frac{f'(x) f''(x)}{f(x) f'''(x)} + 6 \, \frac{f'(x)^3}{f(x)^2 f'''(x)} }{1-2 \, \frac{f'(x)^2}{f(x) f''(x)}} \right)^2.
\end{aligned}
\end{equation}

It is invariant if and only if $r = -3/2$ and $s = 0$, which is exactly the definition of the Schwarzian derivative.

\newpage
\chapter{Continuous \texorpdfstring{$q$-Hermite}{q-Hermite} Polynomials} \label{app:qHermite}
In this appendix, we report several properties of the continuous $q$-Hermite polynomials \cite{koekoek1996askeyscheme,koekoek2010,Gasper_Rahman_2004, Szabłowski+2013+679+708}.
\begin{itemize}
    \item They are defined as:
    \begin{gather}
        H_n(x|q) \equiv \sum_{k=0}^n \frac{(q;q)_n}{(q;q)_k (q;q)_{n-k}} e^{i(n-2k)\theta} \equiv \sum_{k=0}^n \binom{n}{k}_q e^{i(n-2k)\theta}, \quad x = \cos \theta.
    \end{gather}
    
    A $q$-binomial coefficient appears in the above formula:
    \begin{equation}
    	\begin{gathered}
    		\binom{n}{m}_q \equiv \frac{(q;q)_n}{(q;q)_m (q;q)_{n-m}}, \\
    		\binom{n}{n_1, \dots, n_k}_q \equiv \frac{(q;q)_n}{\prod_{j=1}^k (q;q)_{n_j}}, \quad \sum_{j=1}^k n_j = n.
    	\end{gathered}
    \end{equation}
    
    These objects reduce to the usual ratios of factorials when $q = 1-\varepsilon, \: \varepsilon \to 0$.
    
    The $q$-Hermite polynomials are real functions: if one takes their complex conjugate, in fact, the result is equivalent to renaming $k \to n-k$ in the sum, which obviously leaves the function invariant. They are actually a specific basic hypergeometric series, which in general is defined as:
    \begin{gather}
        \prescript{}{r}{\Phi}_s \left[ {a_1, \dots, a_r \atop b_1, \dots, b_s}; q, z \right] \equiv \sum_{k=0}^\infty \frac{(a_1, \dots, a_r; q)_k}{(b_1, \dots, b_s, q; q)_k} \left[ (-1)^k q^{\binom{k}{2}}\right]^{1+s-r} z^k.
    \end{gather}
    We have used the shorthand notation $(a_1, a_2, \dots a_n; q)_k \equiv (a_1; q)_k (a_2; q)_k \dots (a_n;q)_k$. If one of the upper parameters belongs to $q^{\mathbb{Z}_{\leq 0}}$, the series reduces to a finite sum. Assuming $0 < |q| < 1$, the series converges absolutely for any $z$ if $r \leq s$ and for $|z| < 1$ if $r = s+1$ (unless, again, it is actually a finite sum): this follows trivially from the root test. In terms of these series, we can write:
    \begin{gather}
        H_n(x|q) = e^{in\theta} \prescript{}{2}{\Phi}_0 \left[ {q^{-n}, 0 \atop -}; q, q^n e^{-2i\theta} \right].
    \end{gather}
    
    \item Their generating function is:
    \begin{gather}
        \frac{1}{(t e^{i\theta}, te^{-i\theta};q)_\infty} = \sum_{n=0}^\infty H_n(x|q) \frac{t^n}{(q;q)_n}.
    \end{gather}

    \item They satisfy the following recursion relation:
    \begin{gather}
        2x H_n(x|q) = H_{n+1}(x|q) + (1-q^n) H_{n-1}(x|q), \qquad H_{-1}(x|q) = 0, \quad H_0(x|q) = 1.
    \end{gather}

    For $q = 1$, this is solved by $H_n(x|1) = (2x)^n$.

    \item There is a connection between polynomials associated to $p \neq q$:
    \begin{equation}
    \begin{gathered}
        H_n(x|p) = \sum_{m=0}^{\floor{n/2}} c_{n,m}(p,q) H_{n-2m}(x|q), \\
        c_{n,m}(p,q) = \sum_{j=0}^m (-1)^j p^{m-j} q^{j(j+1)/2} \binom{n-2m+j}{j}_q \left( \binom{n}{m-j}_p - p^{n-2m+2j+1} \binom{n}{m-j-1}_p \right). \label{eq:qHermConn}
    \end{gathered}
    \end{equation}
    
    \item They are orthogonal, in the sense that the following relation holds:
    \begin{gather}
        \int_0^\pi d\theta \: H_n(\cos \theta|q) \, H_m(\cos\theta|q) |(e^{2i\theta};q)_\infty|^2 = 2\pi \frac{(q;q)_n}{(q;q)_\infty} \delta_{nm}. \label{eq:qHermOrt}
    \end{gather}

    \item They also satisfy the following orthogonality relation:
    \begin{gather}
        \sum_{n=0}^\infty H_n(\cos\theta|q) H_n(\cos\phi|q) \frac{t^n}{(q;q)_n} = \frac{(t^2;q)_\infty}{(te^{i(\theta+\phi)}, te^{i(\theta-\phi)}, te^{-i(\theta-\phi)}, te^{-i(\theta+\phi)}; q)_\infty}. \label{eq:qHermOrtSum}
    \end{gather}
\end{itemize}

We would like to consider the $t \to 1$ limit of the last identity. We can rewrite the right hand side as:
\begin{gather}
    \frac{(1-t^2)}{|1-te^{i(\theta+\phi)}|^2 |1-te^{i(\theta-\phi)}|^2} \frac{(qt^2;q)_\infty}{|(qte^{i(\theta+\phi)};q)_\infty|^2 |(qte^{i(\theta-\phi)};q)_\infty|^2}.
\end{gather}

If $\theta \pm \phi \neq 2 k \pi$, $k \in \mathbb{Z}$, this quantity goes to $0$ as $t \to 1$, otherwise it blows up. We observe that, in the range $\theta, \: \phi \in [0,\pi]$, the only achievable equality is $\theta-\phi = 0$: even if we were to account for the two degenerate cases $\theta = \phi = 0, \: \pi$ (such that $\theta + \phi = 0, \: 2\pi$ respectively), they would give a null contribution in any double integral over $\theta, \: \phi$ and can therefore be neglected (they would not even be possible for an open interval). This means that the $t \to 1$ limit of the expression is proportional to $\delta(\theta - \phi)$, with some prefactor. We can determine the prefactor by rewriting $t = 1 - \varepsilon, \: \varepsilon \to 0$ and focusing on $\theta \to \phi$, so that the expression simplifies to:
\begin{gather}
    \lim_{\varepsilon \to 0} \frac{2\varepsilon}{|-i(\theta - \phi) + \varepsilon|^2} \frac{(q;q)_\infty}{|(e^{ 2i\theta};q)_\infty|^2 |(q;q)_\infty|^2} = 2\pi \delta(\theta - \phi) \frac{1}{|(e^{ 2i\theta};q)_\infty|^2 (q;q)_\infty}. 
\end{gather}

We recall that we had defined the components of the normalized eigenfunctions of the DSSYK's transition matrix according to \eqref{eq:EigenThL}:
\begin{gather}
    \hat \psi_l(\cos \theta|q) = N(\theta,q) \frac{H_l(\cos\theta|q)}{\sqrt{(q;q)_l}} = \hat \psi_0(\cos\theta|q) \frac{H_l(\cos\theta|q)}{\sqrt{(q;q)_l}}.
\end{gather}

We can fix the normalization factor by writing:
\begin{gather}
    \sum_{n=0}^\infty \hat \psi_n(\cos\theta |q) \hat \psi_m(\cos\theta |q) = \frac{2\pi N^2(\theta,q)}{|(e^{2i\theta};q)_\infty|^2 (q;q)_\infty} \delta(\theta-\phi),
\end{gather}

so that the factor in front of the delta equals $1$ if we choose
\begin{gather}
    N(\theta,q) = \hat \psi_0(\cos\theta|q) = \frac{\sqrt{(q;q)_\infty} |(e^{2i\theta};q)_\infty|}{\sqrt{2\pi}}.
\end{gather}

This choice also guarantees that these functions are orthonormal (using \eqref{eq:qHermOrt}):
\begin{gather}
    \int_0^{\pi} d\theta \: \hat \psi_n(\cos\theta|q) \hat \psi_m(\cos\theta|q) = \delta_{nm}.
\end{gather}

\newpage
\chapter{Partition Function of DSSYK} \label{app:ZDSSYK}
In this appendix, we compute the partition function of the DSSYK model and evaluate several limits. Since the energy spectrum of this theory is truncated, there are no issues associated to negative temperatures. In this appendix, though, we will only consider $\beta > 0$, since extending these results to $\beta < 0$ (and $\beta \to -\infty$ in particular) is trivial.

We start from \eqref{eq:qHermConn} in the case of $p = 1$:
\begin{equation}
\begin{gathered}
    x^n = \frac{1}{2^n} \sum_{m=0}^{\floor{n/2}} c_{n,m}(q) H_{n-2m}(x|q), \\
    c_{n,m}(q) = \sum_{j=0}^m (-1)^j q^{j(j+1)/2} \binom{n-2m+j}{j}_q \frac{n-2m+2j+1}{n+1} \binom{n+1}{m-j}.
\end{gathered}
\end{equation}

We can compute the moments $m_k$ for even $k$ by using the above formula and the orthogonality relation \eqref{eq:qHermOrt} (recall that $H_0(\cos\theta|q) = 1$) in \eqref{eq:DSSYKmoment}:
\begin{equation}
\begin{gathered}
    m_k = \int_0^\pi \frac{d\theta}{2\pi} \: (q,e^{\pm 2i\theta}; q)_\infty \frac{1}{(1-q)^{k/2}} \sum_{m=0}^{k/2} c_{k,m} H_{k-2m}(\cos\theta|q) = \frac{c_{k,k/2}}{(1-q)^{k/2}}.
\end{gathered}
\end{equation}

Due to the orthogonality relation, only $m = k/2$ gives a non-null contribution. The averaged partition function is therefore:
\begin{gather}
    \langle \mathcal{Z} \rangle_J(\beta) = \sum_{k = 0}^{\infty} \frac{\beta^{2k}}{(2k)!} \frac{c_{2k,k}}{(1-q)^k} = \sum_{k=0}^\infty \frac{\beta^{2k}}{(1-q)^k (2k)!} \sum_{j=0}^k (-1)^j q^{j(j+1)/2} \frac{2j+1}{2k+1} \binom{2k+1}{k-j}. \label{eq:appZ1}
\end{gather}

We will now show that the partition function can be rewritten in terms of modified Bessel functions of the first kind $I_\nu(z)$ (this relation is not proven in \cite{Berkooz:2018jqr}):
\begin{equation}
\begin{gathered}
    \langle \mathcal{Z} \rangle_J(\beta) = \frac{\sqrt{1-q}}{\beta} \sum_{j=0}^\infty (-1)^j q^{j(j+1)/2} (2j+1) I_{2j+1} \left( \frac{2\beta}{\sqrt{1-q}} \right), \\
    I_\nu(z) \equiv \left( \frac{z}{2} \right)^\nu \sum_{\ell=0}^\infty \left( \frac{z}{2} \right)^{2\ell} \frac{1}{\ell! \: \Gamma(\nu+\ell+1)}. \label{eq:appZ2}
\end{gathered}
\end{equation}

To do this, we expand \eqref{eq:appZ2} and show that it can be recast into \eqref{eq:appZ1}:

\begin{equation}
\begin{aligned}
    \langle \mathcal{Z} \rangle_J(\beta) & \stackrel{?}{=} \frac{\sqrt{1-q}}{\beta} \sum_{j=0}^\infty (-1)^j q^{j(j+1)/2} (2j+1) \left( \frac{\beta}{\sqrt{1-q}} \right)^{2j+1} \sum_{\ell=0}^\infty \left( \frac{\beta}{\sqrt{1-q}} \right)^{2\ell} \frac{1}{\ell! (2j+\ell+1)!} \\
    & = \sum_{j=0}^\infty \sum_{\ell=0}^\infty (-1)^j q^{j(j+1)/2} \frac{2j+1}{\ell! (2j+\ell+1)!} \frac{\beta^{2(j+\ell)}}{(1-q)^{j+\ell}} \\
    & \stackrel{k = j+\ell}{=} \sum_{k=0}^\infty \sum_{j=0}^k \frac{\beta^{2k}}{(1-q)^k} (-1)^j q^{j(j+1)/2} \frac{2j+1}{(k-j)! (k+j+1)!}.
\end{aligned}
\end{equation}

This proves our statement. We can already consider \eqref{eq:appZ1} to be a high temperature ($\beta \to 0$) expansion of the partition function, although we will say more about this later. We can look for the low temperature behavior ($\beta \to +\infty$), using the asymptotic form of the modified Bessel functions \cite{NIST}:
\begin{gather}
    I_\nu(z \to \infty) = \frac{e^z}{\sqrt{2\pi z}} \sum_{k=0}^\infty (-1)^k \frac{(4\nu^2-1)(4\nu^2-9)\dots(4\nu^2-(2k-1)^2)}{(8z)^k k!}. \label{eq:BesselApprox}
\end{gather}

First of all, the argument of the Bessel functions is ``big'' if and only if $\beta \gg \sqrt{1-q}$. We would like to only keep the first term in the sum: this is possible only if $\nu^2 \ll z$. The $q^{j(j+1)/2} \sim e^{-j^2 \lambda}$ factor in \eqref{eq:appZ2} seems to tell us that only the first $\sim \lambda^{-1/2}$ terms contribute, so that the maximum order that we have to worry about is $\nu \sim \lambda^{-1/2}$. We disagree with this reasoning appearing in \cite{Berkooz:2018jqr}. The reason is that the final result \eqref{eq:DSSYKZapproach1} contains $\mathcal{O}(e^{-1/\lambda})$ terms, so all the terms that contribute at least at order $\mathcal{O}(e^{-1/\lambda})$ should be considered relevant to be consistent. As a consequence, we actually have to consider $\nu \lesssim \lambda^{-1}$ and require that $z \gg \lambda^{-2}$.

We can make the desired approximation, then, only if:
\begin{gather}
    \beta \gg \max\left( \sqrt{1-q}, \frac{\sqrt{1-q}}{\lambda^2} \right).
\end{gather}

For small $\lambda$, this means that we require $\beta \gg \lambda^{-3/2}$, instead of the claimed $\lambda^{-1/2}$. For small temperatures, we obtain:
\begin{gather}
    \langle \mathcal{Z} \rangle_J (\beta \to +\infty) = \frac{(1-q)^{3/4}}{\sqrt{4\pi}} \left( \sum_{j=0}^\infty (-1)^j q^{j(j+1)/2} (2j+1) \right) \frac{e^{2\beta/\sqrt{1-q}}}{\beta^{3/2}}.
\end{gather}

The prefactor is only a function of $\lambda$, while we have completely extracted the dependence on $\beta$. We are actually able to compute the series when $\lambda \to 0$, as we now show. We start by observing that:
\begin{equation}
\begin{gathered}
    \sum_{j=0}^\infty (-1)^j q^{j(j+1)/2}(j+1) = \sum_{j=1}^\infty (-1)^j q^{(j-1)j/2} (-j) = \sum_{j=-\infty}^{-1} (-1)^j q^{j(j+1)/2} j \\
    \implies \sum_{j=0}^\infty (-1)^j q^{j(j+1)/2} (2j+1) = \sum_{j=-\infty}^\infty (-1)^j q^{j(j+1)/2} j.
\end{gathered}
\end{equation}

We can link this sum to the Jacobi theta functions \cite{NIST} by introducing their arguments:
\begin{gather}
    z = e^{2\pi i \nu}, \quad q = e^{2\pi i \tau}.
\end{gather}

In our model, $\tau = i\lambda/(2\pi)$. Keeping in mind that we are taking $\nu = \tau/2$, we can write:
\begin{gather}
    f(\lambda) \equiv \sum_{j=-\infty}^\infty (-1)^j q^{j(j+1)/2} j = \sum_{j=-\infty}^\infty (-1)^j q^{j^2/2} z^j j = \frac{1}{2\pi i} \partial_\nu \sum_{j=-\infty}^\infty (-1)^j q^{j^2/2} z^j \Big|_{\nu = \tau/2} = \frac{1}{2\pi i} \partial_\nu \theta_{01}(\nu,\tau)|_{\nu = \tau/2}.
\end{gather}

It is best to transform $\theta_{01}$ into $\theta_{10}$:
\begin{equation}
\begin{aligned}
     f(\lambda) & = \frac{1}{2\pi i} \partial_\nu \left[ (-i\tau)^{-1/2} e^{-\pi i \nu^2/\tau} \theta_{10}\left( \frac{\nu}{\tau}, -\frac{1}{\tau} \right) \right] \Bigg|_{\nu = \tau/2} \\
     & = \frac{1}{2\pi i} \partial_\nu \Bigg[ 2 (-i\tau)^{-1/2} e^{-\pi i \nu^2/\tau} e^{-\pi i/(4\tau)} \cos \left( \frac{\pi \nu}{\tau} \right) \\ 
     & \hspace{1.6 cm} \times \prod_{m=1}^\infty \left( 1 - e^{-2\pi i m/\tau} \right) \left( 1 + 2\cos \left( \frac{2 \pi \nu}{\tau} \right) e^{-2\pi i m/\tau} + e^{-4 \pi i m/\tau} \right) \Bigg] \Bigg|_{\nu = \tau/2}
\end{aligned}
\end{equation}

\begin{equation}
\begin{aligned}
     \implies f(\lambda) & = \frac{i}{\tau} (-i\tau)^{-1/2} e^{-\pi i \tau/4} e^{-\pi i/(4\tau)} \prod_{m=1}^\infty \left( 1 - e^{-2\pi i m/\tau} \right)^3 \\
     & = \frac{2\pi}{\lambda} \left( \frac{\lambda}{2\pi} \right)^{-1/2} e^{\lambda/8} e^{-\pi^2/(2\lambda)} \prod_{m=1}^{\infty} \left( 1-e^{-4\pi^2 m/\lambda} \right)^3.
\end{aligned}
\end{equation}

Note that the only derivative we have to perform is that of $\cos(\pi\nu/\tau)$, since all the other terms vanish when evaluating them at $\nu = \tau/2$ because of this cosine. This result is valid for all $\lambda$ and it is easy to evaluate when $\lambda \to 0$:
\begin{gather}
    f(\lambda \to 0) = \left( \frac{2\pi}{\lambda} \right)^{3/2} e^{-\pi^2/(2\lambda)} \implies \langle \mathcal{Z} \rangle_J(\beta \gg \lambda^{-3/2})|_{\lambda \to 0} = \frac{\sqrt{2} \pi}{\beta^{3/2} \lambda^{3/4}} \exp \left( \frac{2\beta}{\sqrt{\lambda}} - \frac{\pi^2}{2\lambda} \right). \label{eq:DSSYKZapproach1}
\end{gather}

An alternative and perhaps more useful approach to compute $\langle \mathcal{Z} \rangle_J (\beta)$ is to rewrite and then approximate the distribution of the energy eigenstates:
\begin{gather}
    \Psi(\theta,q) = \frac{(q,e^{\pm 2i\theta};q)_\infty}{2\pi}. \label{eq:DSSYKdistr}
\end{gather}

We start by defining $\bar q = e^{i \pi \tau}$ and considering different Jacobi theta functions:
\begin{equation}
\begin{gathered}
    \theta_1(\theta,\tau) = -\theta_{11}(\theta,\tau) = 2 \bar q^{1/4} \sin(\pi \theta) \prod_{m=1}^\infty (1-\bar q^{2m}) (1-2\cos(2\pi\theta)\bar q^{2m}+\bar q^{4m}), \\
    \theta_1 \left( \frac{\theta}{\pi}, \frac{i\lambda}{2\pi} \right) = 2 q^{1/8} \sin\theta \: (q,e^{\pm 2i\theta}q;q)_\infty = \frac{2q^{1/8} \sin \theta}{(1-e^{2i\theta})(1-e^{-2i\theta})} (q,e^{\pm 2i\theta};q)_\infty = \frac{q^{1/8}}{2\sin\theta} (q,e^{\pm 2i\theta};q)_\infty \\
    \implies \Psi(\theta,q) = \frac{\sin\theta}{\pi q^{1/8}} \: \theta_1 \left( \frac{\theta}{\pi}, \frac{i\lambda}{2\pi} \right). \label{eq:civitavecchia}
\end{gathered}
\end{equation}

We can use the modular transformation of the function:
\begin{equation}
\begin{gathered}
    \theta_1(\theta,\tau) = i (-i\tau)^{-1/2} \exp \left(- i \frac{\pi}{\tau} \theta^2 \right) \theta_1 \left( \frac{\theta}{\tau}, -\frac{1}{\tau}\right) \\
    \implies \theta_1 \left( \frac{\theta}{\pi}, \frac{i\lambda}{2\pi} \right) = i \sqrt{\frac{2\pi}{\lambda}} e^{-2\theta^2/\lambda} \: \theta_1 \left( -\frac{2i\theta}{\lambda}, \frac{2 i \pi}{\lambda} \right).
\end{gathered}
\end{equation}

This expression has a clear $\lambda \to 0$ limit:
\begin{align}
    \nonumber \theta_1 \left( -\frac{2i\theta}{\lambda}, \frac{2 i \pi}{\lambda} \right) & = 2 e^{-\pi^2/(2\lambda)} \sin \left( -\frac{2 i \pi\theta}{\lambda} \right) \prod_{m=1}^\infty (1-e^{-4m\pi^2/\lambda}) \left( 1- 2 \cos \left( -\frac{4i\pi\theta}{\lambda} \right) e^{-4m\pi^2/\lambda} + e^{-8m\pi^2/\lambda} \right) \\
    & \approx -2i e^{-\pi^2/(2\lambda)} \sinh \left( \frac{2\pi\theta}{\lambda} \right) \prod_{m=1}^\infty \left( 1- 2 \cosh \left( \frac{4\pi\theta}{\lambda} \right) e^{-4m\pi^2/\lambda} \right) \\
    \nonumber & \approx -2i e^{-\pi^2/(2\lambda)} \sinh \left( \frac{2\pi\theta}{\lambda} \right) \left( 1- 2 \cosh \left( \frac{4\pi\theta}{\lambda} \right) e^{-4\pi^2/\lambda} \right).
\end{align}

Note that in the third line we have kept the $m=1$ term of the infinite product because in this case (and only this one) the hyperbolic cosine counterbalances the exponential suppression when $\theta \sim \pi$.

We insert this expression back into the original theta function:
\begin{equation}
\begin{aligned}
    \theta_1 \left( \frac{\theta}{\pi}, \frac{i\lambda}{2\pi} \right) & \approx 2 \sqrt{\frac{2\pi}{\lambda}} e^{-2 \left( \theta - \frac{\pi}{2}\right)^2/\lambda - 2\pi\theta/\lambda} \sinh \left( \frac{2\pi\theta}{\lambda} \right) \left( 1- 2 \cosh \left( \frac{4\pi\theta}{\lambda} \right) e^{-4\pi^2/\lambda} \right) \\
    & = 4 \sqrt{\frac{2\pi}{\lambda}} e^{-2\pi^2/\lambda - 2 \left( \theta - \frac{\pi}{2}\right)^2/\lambda} \sinh \left( \frac{2\pi\theta}{\lambda} \right) \left( \sinh \left( \frac{2\pi(\pi-\theta)}{\lambda} \right) - e^{-2\pi(\pi+3\theta)/\lambda} \right). \label{eq:sumthetaDSSYK}
\end{aligned}
\end{equation}

In the $\lambda \to 0$ limit, since $\theta \geq 0$, we can neglect $e^{-2\pi(\pi+3\theta)/\lambda}$. This finally yields:
\begin{gather}
    \Psi(\theta,q) \approx 4 \sqrt{\frac{2}{\pi\lambda}} e^{-2\pi^2/\lambda} e^{-2 \left( \theta - \frac{\pi}{2}\right)^2/\lambda} \sin (\theta) \sinh \left( \frac{2\pi\theta}{\lambda} \right) \sinh \left( \frac{2\pi(\pi-\theta)}{\lambda} \right). \label{eq:ApproxDistrDSSYK}
\end{gather}

\begin{figure}
    \centering
    \includegraphics[width = 0.45 \textwidth]{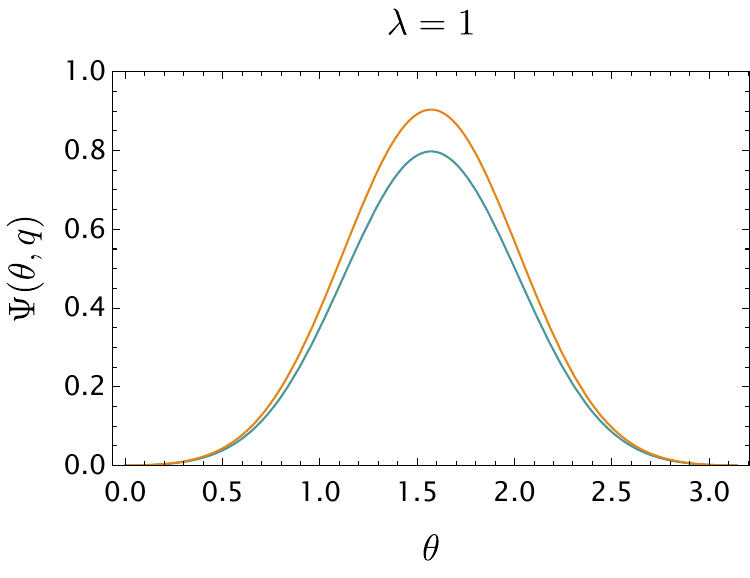} \quad
    \includegraphics[width = 0.45 \textwidth]{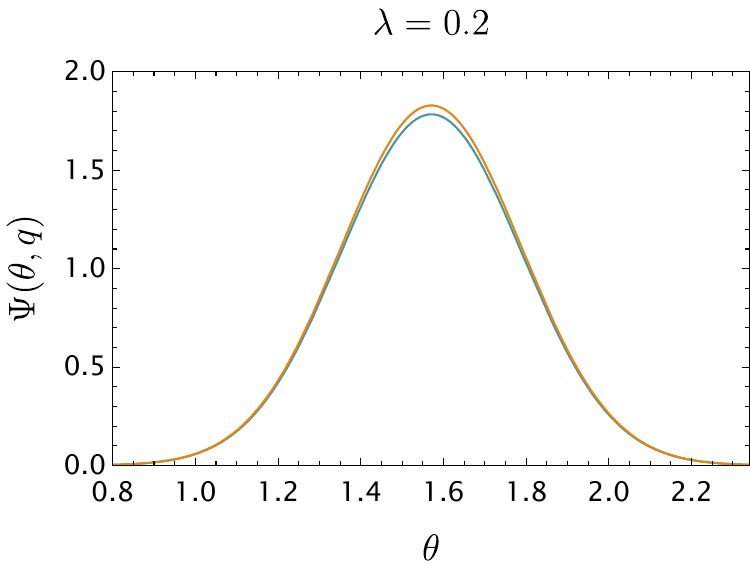} \\
    \includegraphics[width = 0.45 \textwidth]{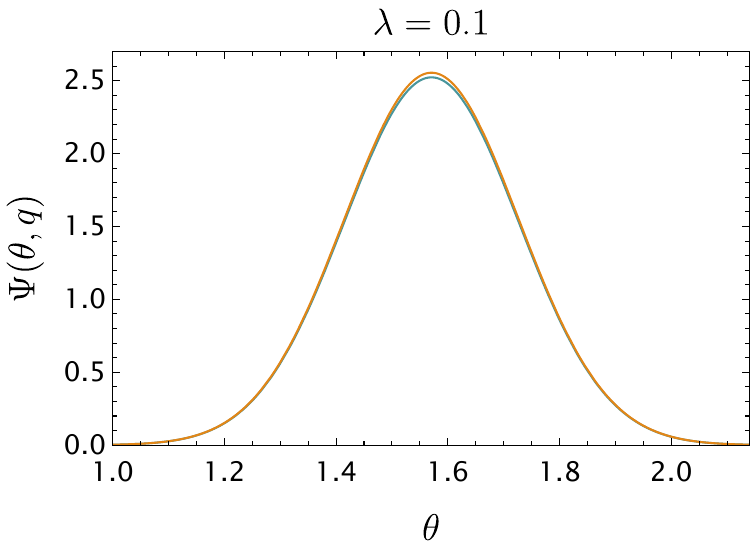} \quad
    \includegraphics[width = 0.45 \textwidth]{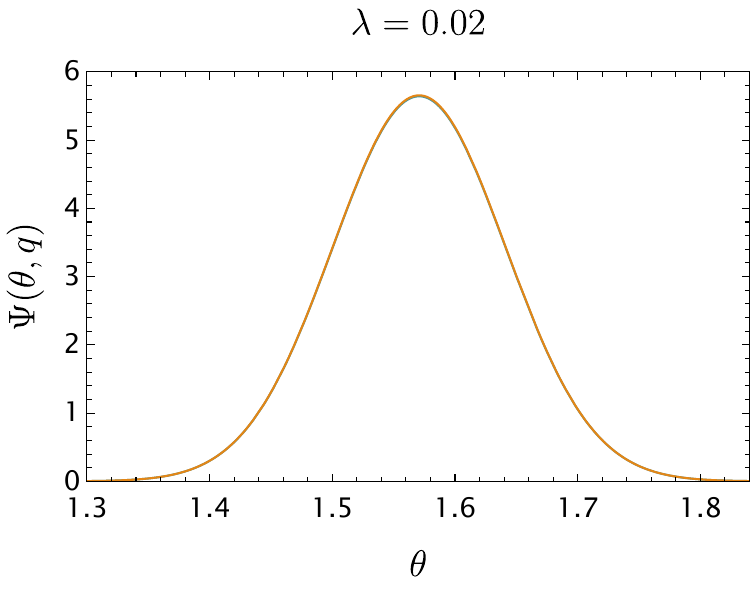}
    \caption{Comparison between \eqref{eq:DSSYKdistr} (orange line) and \eqref{eq:ApproxDistrDSSYK} (light blue line) for different values of $\lambda$. $(q,e^{\pm 2i\theta};q)_\infty$ has been approximated by $(q,e^{\pm 2i\theta};q)_{k(\lambda)}$, with $k(\lambda)$ chosen in such a way that the error was below machine precision. In particular, $k(\lambda)$ increases as $\lambda$ decreases. We can see that the approximation always lies below the exact distribution, but the two curves are practically indistinguishable at $\lambda = 0.02$. Roughly speaking, this approximation is perfect for $\lambda \lesssim 10^{-2}$, but already works extremely well for $\lambda \lesssim 10^{-1}$.}
    \label{fig:DSSYKdistrVS}
\end{figure}

This function is symmetric under $\theta \to \pi-\theta \: (E \to -E)$ and vanishes at the edges $\theta = 0, \: \pi$. This approximation is contrasted with the exact distribution \eqref{eq:DSSYKdistr} in Figure \ref{fig:DSSYKdistrVS}.

If we now send $\lambda \to 0$ while keeping $\theta \: (E)$ fixed, i.e. we perform a pointwise limit of the distribution, the product of the two hyperbolic sines is approximately independent of $\theta$, since the positive exponentials dominate and their product is $\exp(2\pi^2/\lambda)/4$. As a consequence, only the region near $\theta \sim \pi/2$ is relevant and the distribution has a Gaussian behavior there, hence we obtain:
\begin{gather}
    \Psi(\theta,q) \propto e^{-E^2/2}, \quad E \approx -\frac{2}{\sqrt{\lambda}} \left( \theta - \frac{\pi}{2} \right).
\end{gather}

From the point of view of the $\theta$ variable (to which $\Psi(\theta,q)$ is associated), the Gaussian distribution becomes increasingly narrow with variance $\sigma = \sqrt{\lambda}/2$. From \eqref{eq:DSSYKdistr}, we observe that the distribution is actually symmetric under $\theta \to \pi-\theta$ and vanishes at the edges for all $\lambda$. It is also easy to find the $\lambda \to +\infty$ limit:
\begin{gather}
    \Psi(\theta,q \to 0) = \lim_{q\to 0} \frac{1}{2\pi} \prod_{i=0}^{\infty} (1-q^{i+1})(1-e^{2i\theta}q^i)(1-e^{-2i\theta}q^i) = \frac{1}{2\pi}(1-e^{2i\theta})(1-e^{-2i\theta}) = \frac{2}{\pi} \sin^2 \theta.
\end{gather}

\begin{figure}
    \centering
    \includegraphics[width = 0.42 \textwidth]{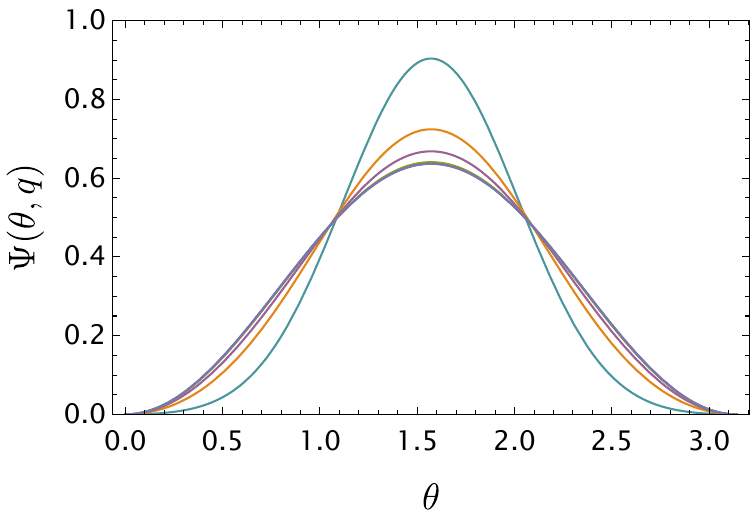} \quad
    \includegraphics[width = 0.55 \textwidth]{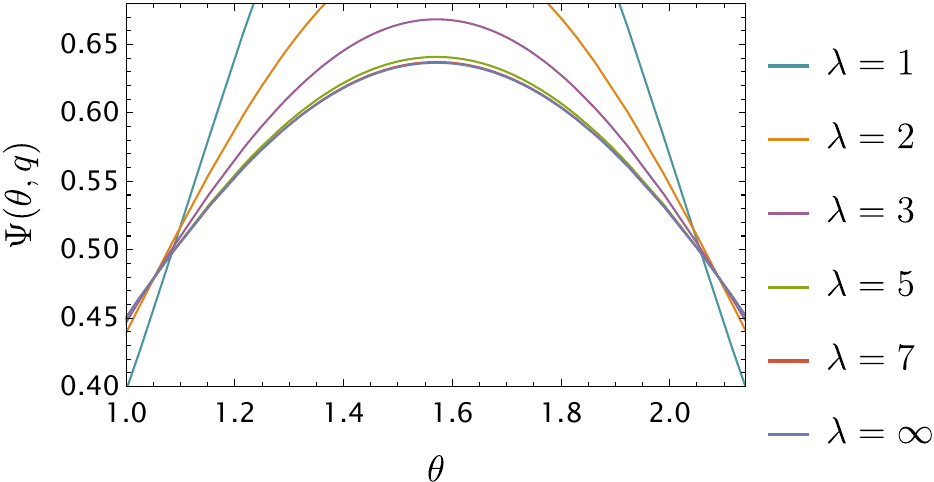}
    \caption{Behavior of \eqref{eq:DSSYKdistr} for $\lambda \geq 1$, with a magnification of the peak. $(q,e^{\pm 2i\theta};q)_\infty$ has been approximated by $(q,e^{\pm 2i\theta};q)_{1000}$, whose error is always below machine precision. We can see that we can describe the curves with $\lambda \gtrsim 5$ through the limit distribution $\Psi(\theta,0)$, with the error committed being absolutely negligible for $\lambda \gtrsim 10$ (as an order of magnitude).}
    \label{fig:DSSYKdistr}
\end{figure}

We have plotted the exact distribution for different values of $\lambda \geq 1$ in Figure \ref{fig:DSSYKdistr} to show the convergence to this limit.

Another interesting limit is obtained by varying the angle that we consider as we send $\lambda \to 0$: we can consider being near one of the edges by taking, for example, $\phi = \pi - \theta \propto \lambda, \: \phi/\lambda$ fixed. In this regime, the positive exponential of the first hyperbolic sine in \eqref{eq:ApproxDistrDSSYK} dominates, so we are left with:
\begin{gather}
    \Psi(\phi,q) \approx 2 \sqrt{\frac{2}{\pi\lambda}} e^{-\pi^2/(2\lambda) - 2\phi^2/\lambda} \sin(\phi) \sinh \left( \frac{2\pi\phi}{\lambda} \right). \label{eq:smallphiDSSYKdistr}
\end{gather}

In principle, we could also neglect the quadratic term in the exponential as it is $\mathcal{O}(\lambda)$. The edge we have chosen here is that of the lower part of the energy spectrum, with:
\begin{gather}
    E-E_{\rm min} \approx \frac{\phi^2}{\sqrt{\lambda}}.
\end{gather}

The approximated distribution \eqref{eq:ApproxDistrDSSYK} can be used to compute any relevant observable of the system. In particular, we can use this distribution to compute the partition function again in the $\lambda \to 0$ limit at high and low temperatures. We perform all the computations explicitly and correct some typos in \cite{Berkooz:2018qkz}, explaining in detail how the assumptions on the temperature are used. We also merge the ``low'' and ``very low'' temperature scenarios presented in the paper into a single one, since the partition function varies smoothly from one regime to the other and can be computed without actually needing the specific approximations associated to each regime. We are going to use:
\begin{gather}
    \langle \mathcal{Z} \rangle_J (\beta) = \int_0^\pi d\phi \: e^{2\beta \cos\phi/\sqrt{\lambda}} \: \Psi(\phi,q).
\end{gather}

\begin{itemize}
    \item $\beta \ll \lambda^{-1/2}$ (high temperatures): At high temperatures, we are able to investigate the entire energy spectrum of the theory, so the biggest contribution to the partition function comes from the Gaussian region around $E = 0, \: |\phi - \pi/2| \ll 1$. We can approximate the hyperbolic sines in \eqref{eq:ApproxDistrDSSYK} and write:
    \begin{gather}
        \langle \mathcal{Z} \rangle_J (\beta \ll \lambda^{-1/2}) \approx \sqrt{\frac{2}{\pi\lambda}} \int_0^\pi d\phi \: e^{2\beta\cos\phi/\sqrt{\lambda} - 2 \left( \frac{\pi}{2} - \phi \right)^2/\lambda} \: \sin\phi.
    \end{gather}
    The Gaussian allows $\phi$ to deviate from $\pi/2$ by a distance of order $\sqrt{\lambda}$:
    \begin{gather}
        x \equiv \frac{\pi}{2}-\phi = \mathcal{O}\left(\sqrt{\lambda}\right),
    \end{gather}
    so we can approximate $\cos \phi \approx x$ and complete the square at the exponential. We obtain:
    \begin{equation}
    \begin{aligned}
        \langle \mathcal{Z} \rangle_J (\beta \ll \lambda^{-1/2}) & \approx \sqrt{\frac{2}{\pi\lambda}} \left[ \int_{-\frac{\pi}{2}}^{\frac{\pi}{2}} dx \: e^{-2\left( x - \frac{\beta\sqrt{\lambda}}{2} \right)^2/\lambda} \: \cos x \right] e^{\beta^2/2} \\
        & \approx  \sqrt{\frac{2}{\pi\lambda}} \left[ \int_{-\infty}^{+\infty} dx \: e^{-2x^2/\lambda} \right] e^{\beta^2/2} = e^{\beta^2/2}.
    \end{aligned}
    \end{equation}

    In the hypothesis of $\beta \sqrt{\lambda} \ll 1$, we can assume that $\cos x \approx 1$ in the relevant region of integration, that is, an interval of width $\sim \sqrt{\lambda}$ centered around $x = \beta\sqrt{\lambda}/2$: this is true up to $\mathcal{O}\left( \sqrt{\lambda} \right)$ corrections to the partition function coming from $\sqrt{2/(\pi\lambda)} \cos x \approx \sqrt{2/(\pi\lambda)} \, (1 - x^2/2)$. At the same time, extending the region of integration to the entire real axis is responsible for an error of order $e^{-1/\lambda}$ only.

    \item $\beta \gg \lambda^{-1/2}$ (low temperatures): As opposed to the previous situation, here we are only able to investigate the eigenstates at low energy, so only the region with $\phi \ll 1$ is relevant. If we assume that we can expand $\cos \phi \approx 1 - \phi^2/2$, the thermal exponential tells us that our maximum relevant $\phi$ should be of order $\lambda^{1/4}\beta^{-1/2}$ (up to shifts of the center of the distribution, which will be present but will not disrupt this analysis), which is indeed much smaller than $1$ since $\beta \gg \lambda^{-1/2}$ here: to be more precise, this hypothesis tells us (again, up to shifts of the center of the distribution) that $\phi \ll \sqrt{\lambda}$. If we also linearize the sine in \eqref{eq:smallphiDSSYKdistr}, we obtain:
    \begin{gather}
        \langle \mathcal{Z} \rangle_J (\beta \gg \lambda^{-1/2}) \approx 2 \sqrt{\frac{2}{\pi\lambda}} e^{-\pi^2/(2\lambda)} \int_0^\pi d\phi \: \phi \, e^{2\beta/\sqrt{\lambda} - \beta\phi^2/\sqrt{\lambda}} \, e^{- 2\phi^2/\lambda} \, \sinh \left( \frac{2\pi\phi}{\lambda} \right). \label{eq:ApproxDistr2DSSYK}
    \end{gather}
    
    The expansion of the cosine and the sine is consistent, since it only requires that $\phi \ll 1$. The prefactor of $\phi^2$ in the exponential is made of two pieces, but we can neglect the term proportional to $\lambda^{-1}$ (which would instead allow us to have $\phi_{\rm max} \sim \sqrt{\lambda}$) since $\beta/\sqrt{\lambda} \gg \lambda^{-1}$.

    At this point, the reference considers two different regimes. This distinction is unnecessary, yet we will still make it due to the insight it gives. If $\lambda^{-1/2} \ll \beta \ll \lambda^{-3/2}$, then $\phi_{\rm max} \gg \lambda$ (but still $\phi_{\rm max} \ll \sqrt{\lambda}$, up to shifts) and we can approximate the hyperbolic sine with an exponential:
    
    \begin{equation}
    \begin{aligned}
        \langle \mathcal{Z} \rangle_J (\lambda^{-1/2} \ll \beta \ll \lambda^{-3/2}) & \approx \sqrt{\frac{2}{\pi\lambda}} e^{-\pi^2/(2\lambda) + 2\beta/\sqrt{\lambda}} \int_0^\pi d\phi \: \phi \, e^{- \beta\phi^2/\sqrt{\lambda} + 2\pi\phi/\lambda}, \\
        \int_0^\pi d\phi \: \phi \, e^{- \beta\phi^2/\sqrt{\lambda} + 2\pi\phi/\lambda} & \approx \int_0^{+\infty} d\phi \left( \phi - \frac{\pi}{\beta \sqrt{\lambda}} + \frac{\pi}{\beta \sqrt{\lambda}} \right) \exp \left(-\frac{\beta}{\sqrt{\lambda}} \left( \phi - \frac{\pi}{\beta \sqrt{\lambda}} \right)^2 + \frac{\pi^2}{\beta \lambda^{3/2}} \right) \\
        & = e^{\pi^2/(\beta\lambda^{3/2})} \int_{-\frac{\pi}{\beta \sqrt{\lambda}}}^{+\infty} d\phi \: \left[ \frac{\pi}{\beta \sqrt{\lambda}} e^{-\beta \phi^2/\sqrt{\lambda}} + \phi \, e^{-\beta \phi^2/\sqrt{\lambda}} \right] \\ 
        & = \frac{\pi^{3/2} e^{\pi^2/(\beta\lambda^{3/2})}}{2\beta^{3/2}\lambda^{1/4}} \left( 1 - \mathrm{erf} \left( -\frac{\pi}{\sqrt{\beta}\lambda^{3/4}} \right) \right) +\frac{\sqrt{\lambda}}{2\beta} \\
        & \approx \frac{\pi^{3/2} e^{\pi^2/(\beta\lambda^{3/2})}}{\beta^{3/2}\lambda^{1/4}}.
    \end{aligned}
    \end{equation}
    Extending the region of integration to the positive semiaxis is responsible for a $\mathcal{O}\left(e^{-\beta/\sqrt{\lambda}}\right)$ error. In the last line, we have only kept the dominant contribution: since $\beta \ll \lambda^{-3/2}$, the second term is irrelevant when compared to the blowing-up exponential and we can use $\mathrm{erf}(-\infty) = -1$. We can interpret this result as the majority of the contribution coming from a region of width $\sim \lambda^{1/4}\beta^{-1/2}$ near $\phi = \pi/(\beta\sqrt{\lambda})$ which, although being a big shift with respect to this width (so that actually $\phi_{\rm max} \sim \beta^{-1}\lambda^{-1/2}$), is still such that the expansion of the cosine and the sine is a good approximation (it is not true that $\lambda \ll \phi_{\rm max} \ll \sqrt{\lambda}$, as claimed earlier, but it is still true that $\lambda \ll \phi_{\rm max} \ll 1$, which is what we need for our computation to be consistent). Plugging this result back into the partition function, we obtain the following result in this regime:
    \begin{gather}
        \langle \mathcal{Z} \rangle_J (\lambda^{-1/2} \ll \beta \ll \lambda^{-3/2}) \approx \frac{\sqrt{2} \pi}{\beta^{3/2} \lambda^{3/4}} \exp \left( \frac{2\beta}{\sqrt{\lambda}} - \frac{\pi^2}{2\lambda} + \frac{\pi^2}{\beta \lambda^{3/2}} \right).
    \end{gather}

    A different approximation emerges when $\beta \gg \lambda^{-3/2}$: in this case, $\phi_{\rm max} \sim \lambda^{1/4} \beta^{-1/2} \ll \lambda$ (no shifts occur) and we can take a linear approximation of the hyperbolic sine, so that we are left with:
    \begin{equation}
    \begin{aligned}
        \langle \mathcal{Z} \rangle_J (\beta \gg \lambda^{-3/2}) & \approx \frac{4}{\lambda} \sqrt{\frac{2\pi}{\lambda}} e^{-\pi^2/(2\lambda) + 2\beta/\sqrt{\lambda}} \int_0^\pi d\phi \: \phi^2 \, e^{-\beta\phi^2/\sqrt{\lambda}} \\
        & \approx \frac{\sqrt{2} \pi}{\beta^{3/2} \lambda^{3/4}} \exp \left( \frac{2\beta}{\sqrt{\lambda}} - \frac{\pi^2}{2\lambda} \right).
    \end{aligned}
    \end{equation}

    Again, we have extended the region of integration to the entire positive semiaxis. Notice that this result exactly matches both \eqref{eq:DSSYKZapproach1} and the one we have obtained for $\lambda^{-1/2} \ll \beta \ll \lambda^{-3/2}$, up to a missing factor of $e^{\pi^2/(\beta\lambda^{3/2})}$, which is suppressed in this regime anyway. As a consequence, we can merge these two partial results into a single one, which can be considered valid for all $\beta \gg \lambda^{-1/2}$:
    \begin{gather}
        \langle \mathcal{Z} \rangle_J (\beta \gg \lambda^{-1/2}) \approx \frac{\sqrt{2} \pi}{\beta^{3/2} \lambda^{3/4}} \exp \left( \frac{2\beta}{\sqrt{\lambda}} - \frac{\pi^2}{2\lambda} + \frac{\pi^2}{\beta \lambda^{3/2}} \right).
    \end{gather}

    The observation we make is that the integral in \eqref{eq:ApproxDistr2DSSYK} can be computed directly by integrating $\phi$ in $[0,+\infty]$, thus obtaining exactly the above result. This result supports our claim that $\beta \gg \lambda^{-3/2}$ was needed to derive \eqref{eq:DSSYKZapproach1}, instead of $\beta \gg \lambda^{-1/2}$ (as is stated in \cite{Berkooz:2018jqr}). Otherwise, there would be a mismatch with the partition function we have just derived by a term that would not be negligible for $\lambda^{-1/2} \ll \beta \ll \lambda^{-3/2}$. More specifically, for $\beta \gtrsim \lambda^{-1/2}$, the exponent would be wrong at the relevant order $\mathcal{O}(\lambda^{-1})$, which is exactly the order of the terms that \cite{Berkooz:2018jqr} failed to consider appropriately in their computation when only requiring $\beta \gg \lambda^{-1/2}$.
\end{itemize}

\newpage
\chapter{KMS Relation in the cSYK Model} \label{app:KMSSYK}
In this appendix, we show our proof of the fact that the fermionic two point function of the SYK model satisfies the KMS relation, provided that fermions are time-evolved with $H$:
\begin{gather}
    G(\tau+\beta) = -e^{-\beta\mu} G(\tau).
\end{gather}

The main idea is to adapt \eqref{eq:AntisymProof} to Dirac fermions in a grandcanonical ensemble. The $U(1)$ charge can be rescaled to be the number operator:
\begin{gather}
    \mathcal{N} = \sum_{i=1}^N \psi_i^\dagger \psi_i - \frac{N}{2}, \quad [\mathcal{N},H] = 0.
\end{gather}

First, we prove a preliminary result:
\begin{equation}
\begin{gathered}
    [\mathcal{N}, \psi_i^\dagger] = [\psi_i^\dagger \psi_i, \psi_i^\dagger] = \psi_i^\dagger \psi_i \psi_i^\dagger = \psi_i^\dagger \{ \psi_i, \psi_i^\dagger \} = \psi_i^\dagger, \\
    -[\mathcal{N}, \psi_i^\dagger]^{\dagger} = [\mathcal{N}, \psi_i] = -\psi_i.
\end{gathered}
\end{equation}

We have used that $\psi^\dagger_i$ anticommutes with $\psi_{j \neq i}, \: \psi^\dagger_{j \neq i}$, and also that:
\begin{gather}
    \{ \psi_i, \psi_i^\dagger \} = 1, \quad \{ \psi_i^\dagger, \psi_i^\dagger \} = 0.
\end{gather}

We now define a sort of time evolution induced by the number operator:
\begin{gather}
    \psi_i^\dagger(\mu) \equiv e^{-\beta \mu \mathcal{N}} \psi_i^\dagger(0) e^{+\beta \mu \mathcal{N}}.
\end{gather}

By taking its derivative with respect to $\mu$, we can find $\psi_i^\dagger(\mu)$ and $\psi_i(\mu)$:
\begin{gather}
    \psi_i^\dagger(\mu)' = -\beta e^{-\beta \mu \mathcal{N}} [\mathcal{N}, \psi_i^\dagger(0)] e^{+\beta \mu \mathcal{N}} = -\beta \psi_i^\dagger(\mu) \implies \psi_i^\dagger(\mu) = e^{-\beta\mu} \psi_i^\dagger, \: \psi_i(\mu) = e^{+\beta\mu} \psi_i.
\end{gather}

In the grandcanonical ensemble, $G_{ij}(\tau)$ is:
\begin{gather}
    G_{ij}(\tau) = \tr[e^{-\beta (H-\mu \mathcal{N})} \mathrm{T}[\psi_i(\tau) \psi^\dagger_j(0)]]/\tr[e^{-\beta (H-\mu \mathcal{N})}].
\end{gather}

The result is extended from $\tau \in [-\beta/2, \beta/2)$ to the whole real axis in the following way, which is also the final step of our proof. If $\tau > 0$, we have:
\begin{equation}
\begin{aligned}
    G_{ij}(\tau) & \propto \tr[e^{-\beta(H-\mu\mathcal{N})} \psi_i(\tau) \psi_j^\dagger(0)] \\
    & = \tr[e^{-\beta(H-\mu\mathcal{N})} e^{\beta H} e^{-\beta\mu\mathcal{N}} \psi^\dagger_j(0) e^{+\beta\mu\mathcal{N}} e^{-\beta H} \psi_i(\tau)] \\
    & = e^{-\beta\mu} \tr[e^{-\beta(H-\mu\mathcal{N})} \psi^\dagger_j(\beta) \psi_i(\tau)] \\
    & = -e^{-\beta\mu} \tr[e^{-\beta(H-\mu\mathcal{N})} \mathrm{T}[\psi_i(\tau) \psi^\dagger_j(\beta)]] \\ 
    \implies G_{ij}(\tau) & = -e^{-\beta\mu} G_{ij}(\tau-\beta).
\end{aligned}
\end{equation}

If $\tau < 0$, similarly:
\begin{equation}
\begin{aligned}
    G_{ij}(\tau) & \propto -\tr[e^{-\beta(H-\mu\mathcal{N})} \psi_j^\dagger(0) \psi_i(\tau)] \\
    & = -\tr[e^{-\beta(H-\mu\mathcal{N})} e^{\beta H} e^{-\beta\mu\mathcal{N}} \psi_i(\tau) e^{+\beta\mu\mathcal{N}} e^{-\beta H} \psi^\dagger_j(0)] \\
    & = -e^{+\beta\mu} \tr[e^{-\beta(H-\mu\mathcal{N})} \psi_i(\tau+\beta) \psi^\dagger_j(0)] \\
    & = -e^{+\beta\mu} \tr[e^{-\beta(H-\mu\mathcal{N})} \mathrm{T}[\psi_i(\tau+\beta) \psi^\dagger_j(0)]] \\
    \implies G_{ij}(\tau) & = -e^{+\beta\mu} G_{ij}(\tau+\beta).
\end{aligned}
\end{equation}

With these computations, we have shown that the cSYK model satisfies the KMS relation. A slight modification to these computations is also able to prove that, if we instead time-evolve fermions with $H-\mu \mathcal{N}$, the two point function satisfies:
\begin{gather}
	G(\tau+\beta) = -G(\tau).
\end{gather}

This is a trivial consequence of the KMS relation proven above:
\begin{equation}
\begin{aligned}
	G_{H-\mu\mathcal{N}}(\tau) & = \tr[e^{-\beta (H-\mu \mathcal{N})} \mathrm{T}[e^{\tau(H-\mu\mathcal{N})}\psi_i(0) e^{-\tau(H-\mu\mathcal{N})} \psi^\dagger_j(0)]]/\tr[e^{-\beta (H-\mu \mathcal{N})}] \\
	& = \tr[e^{-\beta (H-\mu \mathcal{N})} \mathrm{T}[e^{\tau H} e^{\mu\tau} \psi_i(0) e^{-\tau H} \psi^\dagger_j(0)]]/\tr[e^{-\beta (H-\mu \mathcal{N})}] \\
	& = e^{\mu\tau} G_H(\tau) \\
	\implies G_{H-\mu\mathcal{N}}(\tau+\beta) & = e^{\mu(\tau+\beta)} G_H(\tau+\beta) = - e^{\mu\tau} G_H(\tau) = -G_{H-\mu\mathcal{N}}(\tau).
\end{aligned}
\end{equation}

\newpage
\chapter{\texorpdfstring{$\langle Q \rangle (\mu)$}{Q(mu)} in the cSYK Model} \label{app:chargeSYK}

In this appendix, we prove that:
\begin{gather}
	\langle Q \rangle_{\rm free}(\mu) \equiv \langle Q \rangle (\mu) = \frac{1}{2} \tanh \left( \frac{\beta\mu}{2} \right).
\end{gather}

The first way to obtain this result is to start from the free two point function in the grandcanonical ensemble:
\begin{gather}
	G_{ij}(\tau) = \tr[e^{-\beta(H-\mu N Q)} \mathrm{T}[\psi_i(\tau) \psi^\dagger_j(0)]]/\tr[e^{-\beta(H-\mu N Q)}].
\end{gather}

In the free case, $H = 0$. The trace of the partition function factorizes into the product of the traces over the individual fermions, whose basis is simply $\{ |0 \rangle_i, \: |1 \rangle_i \}_{1 \leq i \leq N}$:
\begin{gather}
	\tr[e^{\beta \mu (\sum_i \psi_i^\dagger \psi_i - N/2)}] = e^{-\beta\mu N/2} \left( 1 + e^{\beta\mu} \right)^N = \left (e^{\beta\mu/2} + e^{-\beta\mu/2} \right)^N.
\end{gather}

We turn to the numerator. We know that $G_{ij}(\tau) \propto \delta_{ij}$, so we fix $i = j$ and we do not sum over it. We are evolving fermions with $H - \mu N Q = -\mu \, (\sum_k \psi^\dagger_k \psi_k - N/2)$, so $\psi_i(\tau) = e^{\mu\tau} \psi_i$ (this is shown in Appendix \ref{app:KMSSYK}). This trace also factorizes into a product where only the $i$-th fermion is affected by the time ordered product. For $0 < \tau < \beta/2$, we have:
\begin{equation}
	\begin{aligned}
		\tr[e^{\beta \mu (\sum_k \psi_k^\dagger \psi_k - N/2)} \psi_i(\tau) \psi^\dagger_i(0)] & = e^{\mu\tau - \beta \mu N/2} \left( 1 + e^{\beta\mu} \right)^{N-1} \tr[e^{\beta\mu \psi_i^\dagger \psi_i} \psi_i \psi^\dagger_i] \\
		& = e^{\mu\tau} \left( e^{\beta\mu/2} + e^{-\beta\mu/2} \right)^{N-1} e^{-\beta\mu/2} (1+0).
	\end{aligned}
\end{equation}

For $-\beta/2 < \tau < 0$, instead, we have:
\begin{equation}
	\begin{aligned}
		-\tr[e^{\beta \mu (\sum_k \psi_k^\dagger \psi_k - N/2)} \psi^\dagger_i(0) \psi_i(\tau)] & = -e^{\mu\tau - \beta \mu N/2} \left( 1 + e^{\beta\mu} \right)^{N-1} \tr[e^{\beta\mu \psi_i^\dagger \psi_i} \psi^\dagger_i \psi_i] \\
		& = -e^{\mu\tau} \left( e^{\beta\mu/2} + e^{-\beta\mu/2} \right)^{N-1} e^{-\beta\mu/2} \left(0 + e^{\beta\mu} \right).
	\end{aligned}
\end{equation}

Putting these two results together, we obtain $G_{\rm free}(\tau)$:
\begin{equation}
	\begin{aligned}
		G_{\rm free}(\tau) & = \left( \frac{e^{-\beta\mu/2}}{e^{\beta\mu/2} + e^{-\beta\mu/2}} \, \theta(\tau) - \frac{e^{\beta\mu/2}}{e^{\beta\mu/2} + e^{-\beta\mu/2}} \, \theta(-\tau) \right) e^{\mu\tau} \\
		& = \left( \frac{1}{2} \mathrm{sgn}(\tau) - \frac{1}{2} \tanh \left( \frac{\beta\mu}{2} \right) \right) e^{\mu\tau}.
	\end{aligned}
\end{equation}

This result assumes $-\beta < \tau < \beta$ (this is the range for the difference between the times of the two fermions), but to obtain its extension to $\tau \in \mathbb{R}$ it is once again sufficient to use the sine function (as was done in \eqref{eq:circlemap}). By comparing this result to \eqref{eq:AtauFreeMu}, we immediately find:
\begin{gather}
	\langle Q \rangle(\mu) = \frac{1}{2} \tanh \left( \frac{\beta\mu}{2} \right).
\end{gather}

A second, faster but less instructive way to obtain the same result is to simply require the Green function to satisfy the KMS relation:
\begin{equation}
\begin{gathered}
	G_{\rm free}(0 < \tau < \beta) = \left( \frac{1}{2} - \langle Q \rangle \right) e^{\mu\tau} = -G_{\rm free}(-\beta < \tau-\beta < 0) = -\left( -\frac{1}{2} - \langle Q \rangle \right) e^{\mu(\tau-\beta)} \\
	\implies \langle Q \rangle(\mu) = \frac{1}{2} \tanh \left( \frac{\beta\mu}{2} \right).
\end{gathered}
\end{equation}

The two procedures correctly yield the same relation, as expected.

\newpage
\chapter{Dilaton-Gravity Solutions} \label{app:2dDilaton}
In this appendix, we show our determination of the solution of the classical equations of motion of the most general Euclidean dilaton-gravity model, which is described (up to a prefactor) by:
\begin{gather}
	S = \int d^2x \, \sqrt{g} \left[ \Phi R + \frac{\lambda}{4\Phi} \partial_\mu \Phi \, \partial^\mu \Phi + V(\Phi) \right].
\end{gather}

Most importantly, with this appendix, we will show that the metric \eqref{eq:notansatzgmn} is the most general solution and not merely an ansatz. We are not reporting the boundary terms, but we know that it is their presence that allows us to proceed correctly. We recall that we can always perform a Weyl transformation that sets $\lambda = 0$, so we will assume that this is the case without loss of generality. Varying the action with respect to the dilaton yields the first equation:
\begin{gather}
	R + V'(\Phi) = 0.
\end{gather}

By varying it with respect to the metric, using \eqref{eq:MetricVariation} and
\begin{gather}
	\delta \sqrt{g} = -\frac{1}{2} \sqrt{g} \, g_{\mu\nu} \delta g^{\mu\nu},
\end{gather}

we obtain:
\begin{gather}
	\delta S = \int d^2x \left[ \Phi \partial_\mu (\sqrt{g} \, v^\mu) -\frac{1}{2} \sqrt{g} \, g_{\mu\nu} V(\Phi) \, \delta g^{\mu\nu} \right], \quad v_\mu = \nabla^\nu (\delta g_{\mu\nu}) - g^{\nu\rho} \nabla_\mu (\delta g_{\nu\rho}).
\end{gather}

We have used that, in two dimensions, $R_{\mu\nu} = Rg_{\mu\nu}/2$. Up to a boundary term that is canceled by the $\propto \Phi K$ GHY term\footnote{We first write $\Phi \partial_\mu (\sqrt{g} \, v^\mu) = \partial_\mu (\Phi \sqrt{g} \, v^\mu) - \sqrt{g} \, v_\mu \partial^\mu \Phi$, then we notice that the first term is canceled by $\Phi K$.}, we can integrate the first term by parts \textit{twice} and obtain
\begin{gather}
	\delta S = \int d^2x \, \sqrt{g} \left[ -\nabla_\mu \nabla_\nu \Phi + g_{\mu\nu} \nabla^2 \Phi - \frac{1}{2} g_{\mu\nu} V(\Phi) \right] \delta g^{\mu\nu},
\end{gather}

from which we read the second equation of motion:
\begin{gather}
	\nabla_\mu \nabla_\nu \Phi - g_{\mu\nu} \nabla^2 \Phi + \frac{1}{2} g_{\mu\nu} V(\Phi) = 0. \label{eq:appdilatonmotion}
\end{gather}

Integrating the covariant derivative by parts was possible because $\delta g_{\mu\nu} = 0$ on the boundary. We have also used that $g^{\nu\rho} \delta g_{\nu\rho} = -g_{\nu\rho} \delta g^{\nu\rho}$ and that $\nabla^\mu \nabla^\nu \Phi \, \delta g_{\mu\nu} = - \nabla_\mu \nabla_\nu \Phi \, \delta g^{\mu\nu}$.

This second equation of motion implies the existence of a Killing vector field:
\begin{gather}
	\xi^\mu = \frac{1}{\sqrt{g}} \varepsilon^{\mu\nu} \nabla_\nu \Phi = E^{\mu\nu} \partial_\nu \Phi.
\end{gather}

The use of $E^{\mu\nu}$ rather than $\varepsilon^{\mu\nu}$ is necessary, since $\xi^\mu$ would not transform as a vector under a change of coordinates otherwise. Let us verify explicitly that it satisfies the Killing equation:
\begin{equation}
\begin{aligned}
	0 \stackrel{?}{=} \nabla_\mu \xi_\nu + \nabla_\nu \xi_\mu & = \nabla_\mu (g_{\nu\rho} E^{\rho\lambda} \partial_\lambda \Phi) + (\nu \leftrightarrow \mu) = g_{\nu\rho} E^{\rho\lambda} \nabla_\mu \nabla_\lambda \Phi + g_{\nu\rho} \partial_\lambda \Phi \nabla_\mu E^{\rho\lambda} + (\nu \leftrightarrow \mu).
\end{aligned}
\end{equation}

First, we observe that $\nabla_\mu E^{\rho\lambda} = 0$. The most obvious way to see this is to consider the locally inertial system with $g_{\mu\nu}(x^\mu + \delta x^\mu) = \delta_{\mu\nu} + \mathcal{O}((\delta x^\mu)^2)$, where all covariant derivatives reduce to partial derivatives and $E^{\rho\lambda} = \varepsilon^{\rho\lambda}$. Secondly, we observe that the second equation of motion implies that $\nabla_\mu \nabla_\nu \Phi \propto g_{\mu\nu}$, therefore:
\begin{gather}
	g_{\nu\rho} E^{\rho\lambda} \nabla_\mu \nabla_\lambda \Phi + (\nu \leftrightarrow \mu) \propto \varepsilon^{\rho \lambda} (g_{\nu\rho} g_{\mu\lambda} + g_{\mu\rho} g_{\nu\lambda}) = 0.
\end{gather}

This proves that $\nabla_\mu \xi_\nu + \nabla_\nu \xi_\mu = 0$, and therefore it is a Killing vector field. We now choose a coordinate system $(\tilde t, \tilde r)$ where $\xi^\mu = (1,0)$. The Killing equation implies that the metric is independent of $\tilde t$:
\begin{equation}
\begin{aligned}
	0 & = \partial_\mu \xi_\nu - \Gamma^{\lambda}_{\mu\nu} \xi_\lambda + (\nu \leftrightarrow \mu) \\
	& = \partial_\mu (g_{\nu\alpha} \xi^\alpha) + \partial_\nu (g_{\mu\alpha} \xi^\alpha) - \xi^\alpha (\partial_\mu g_{\nu\alpha} + \partial_\nu g_{\mu\alpha} - \partial_\alpha g_{\mu\nu}) \\
	& = \xi^\alpha \partial_\alpha g_{\mu\nu} = \partial_0 g_{\mu\nu}.
\end{aligned}
\end{equation}

With this choice of coordinates, the most general metric is therefore:
\begin{gather}
	ds^2 = A(\tilde r) \, d\tilde t^2 + 2C(\tilde r) \, d\tilde t \, d\tilde r + B(\tilde r) \, d\tilde r^2.
\end{gather}

We first redefine the time coordinate:
\begin{gather}
	t = \tilde t + \int^{\tilde r} dx \: \frac{C(x)}{A(x)} \implies dt = d\tilde t + \frac{C(\tilde r)}{A(\tilde r)} \, d\tilde r.
\end{gather}

This gives us a diagonal metric:
\begin{gather}
	ds^2 = A(\tilde r) \, dt^2 + \frac{A(\tilde r)B(\tilde r) - C^2(\tilde r)}{A(\tilde r)} \, d\tilde r^2.
\end{gather}

The positivity of $g_{\tilde r \tilde r}$ is due to the positivity of the determinant of the metric. We finally define a new coordinate $r$:
\begin{gather}
	r = \int^{\tilde r} dx \, \sqrt{A(x)B(x)- C^2(x)} \implies ds^2 = f(r) \, dt^2 + \frac{dr^2}{f(r)}.
\end{gather}

We have renamed $A(r) \to f(r)$. The Killing vector also fixes $\Phi(r,t)$:
\begin{equation}
\begin{aligned}
	0 & = \xi^1 = -\partial_0 \Phi(r,t) && \implies \Phi(r,t) = \Phi(r), \\
	1 & = \xi^0 = \partial_1 \Phi(r) && \implies \Phi(r) = r + c.
\end{aligned}
\end{equation}

Up to a shift of the radial coordinate, we can take $c = 0$ and use $\Phi$ and $r$ interchangeably. \eqref{eq:appdilatonmotion} has still some information left that one can find by taking its trace:
\begin{gather}
	V(\Phi) = V(r) = \nabla^2 \Phi = f'(r) \implies f(r) = \int^r_{r_h} d\Phi \: V(\Phi),
\end{gather}

where $r_h$ is an horizon, namely $f(r_h) = 0$.

The first equation is now redundant and actually carries less information about $f(r)$:
\begin{gather}
	R + V'(\Phi) = 0 \implies f''(r) = V'(\Phi) \implies f(r) = \int^r_{r_h} d\Phi \: V(\Phi) + \alpha (r-r_h),
\end{gather}

so it is unable to fix $\alpha = 0$, as we have just proven through the other equation. Interestingly, $\alpha \neq 0$ would be equivalent to a constant shift of the potential: $V(\Phi) \to V(\Phi) + \alpha$.

A Wick rotation back to the Lorentzian signature concludes the proof. Basically, the existence of a Killing vector field is crucial to the possibility of rewriting every metric in this very useful diagonal form. 

We could have also inserted this form of the metric directly inside the action, although this would not have told us anything about the fact that a Killing vector field is always present. This would have yielded the equations of motion more easily. Starting from the action
\begin{gather}
    S = \int dt \, dr \left[ -f''(r) \, \Phi + V(\Phi) \right],
\end{gather}

one immediately finds:

\begin{equation}
\begin{aligned}
    \frac{\delta S}{\delta f} = 0 & \implies \Phi'' = 0, \\
    \frac{\delta S}{\delta \Phi} = 0 & \implies f''(r) = V'(\Phi).
\end{aligned}
\end{equation}

Coherently with what we have found above, these equations are solved by $\Phi = r$ (up to translations and rescalings of $r$) and $f(r) = \int_{r_h}^r d\Phi \: V(\Phi)$ (up to translations of $V(\Phi)$). It is clear that, although much simpler, this way of obtaining the desired metric lacks the completeness of the above procedure.

\newpage
\chapter{Worldline Formalism} \label{app:worldline}
In this appendix, we show our proof of the following identity for heavy scalars in the semiclassical limit:
\begin{gather}
	\langle \phi(x_2) \phi(x_1) \rangle \equiv G_F(x_2-x_1) \approx \sum_g e^{-m L_g}. \label{eq:appworldline}
\end{gather}

$\phi$ is a real scalar field of mass $m$ that is minimally coupled to the Euclidean metric $g_{\mu\nu}$ of a $D$-dimensional manifold, while the $L_g$ are the proper lengths of all the possible geodesics that connect $x_1$ to $x_2$. We know that the Euclidean propagator is the inverse of the quadratic kernel in the action:
\begin{gather}
	(-\nabla_\mu \nabla^\mu + m^2) G_F(x_2-x_1) = \frac{1}{\sqrt{g}} \delta^{(D)}(x_2 - x_1) \implies G_F(x_2-x_1) = \langle x_2 | [-\nabla_\mu \nabla^\mu + m^2]^{-1} | x_1 \rangle.
\end{gather}

The $|x_i \rangle$ are the usual position eigenvectors that one can use to write functions belonging to $L^2(\mathbb{R}^D)$, on which operators such as $-\nabla_\mu \nabla^\mu + m^2$ act. We now introduce a Schwinger parameter $T$:
\begin{gather}
	G_F(x_2-x_1) = \int_0^{+\infty} dT \: \langle x_2 | e^{-T(-\nabla_\mu \nabla^\mu + m^2)} | x_1 \rangle = \int_0^{+\infty} dT \: e^{-m^2 T} \, \langle x_2 | e^{-T(-\nabla_\mu \nabla^\mu)} | x_1 \rangle.
\end{gather}

We can identify $e^{-T(-\nabla_\mu \nabla^\mu)}$ as an Euclidean time evolution operator with Hamiltonian $-\nabla_\mu \nabla^\mu$, so that the term that appears in the integral is the propagator of a quantum particle from $x_1$ to $x_2$. In usual non-relativistic Quantum Mechanics in flat space, it is known that such an object can be rewritten in terms of a path integral:
\begin{gather}
	\langle x_2 | e^{-T (-\nabla^2/(2m))} |x_1 \rangle = \int_{X(0) = x_1}^{X(T) = x_2} \mathcal{D}X \: \exp \left( -\int_0^T d\tau \: \frac{1}{2}m \dot X^2 \right).
\end{gather}

The situation here is analogous, provided we take $m = 1/2$ and we write a covariant Lagrangian that reduces to the simple kinetic energy in the local inertial frame (with respect to \textit{all} the $D$ coordinates of the manifold). There is only one result that is compatible with these basic requirements, which is:
\begin{gather}
	\langle x_2| e^{-T(-\nabla_\mu \nabla^\mu)} |x_1 \rangle = \int_{X(0) = x_1}^{X(T) = x_2} \mathcal{D}X \: \exp \left( -\int_0^T d\tau \: \frac{1}{4} g_{\mu\nu}(X) \dot X^\mu \dot X^\nu \right), \quad \dot X^\mu \equiv \frac{dX^\mu}{d\tau}.
\end{gather}

There are now two complementary ways to obtain \eqref{eq:appworldline}.

The first method starts with approximating the path integral with its saddle solutions:
\begin{gather}
	\langle x_2| e^{-T(-\nabla_\mu \nabla^\mu)} |x_1 \rangle \approx \sum_g \exp \left( -\int_0^T d\tau \: \frac{1}{4} g_{\mu\nu} \dot X_g^\mu \dot X_g^\nu \right), \label{eq:appworldsaddle}
\end{gather}

where the classical equations of motion for the $X^\mu_g$ are precisely the geodesic equations: 
\begin{gather}
	\ddot X^\mu_g + \Gamma^\mu_{\nu\rho} \dot X_g^\nu \dot X_g^\rho = 0, \quad X^\mu_g(0) = x_1, \: X^\mu_g(T) = x_2.
\end{gather}

The fact that the geodesic equations hold with respect to $\tau$ means that it is an affine parameter, proportional to the proper time $s$ of the massive particle:
\begin{gather}
	\tau = a s + b \implies T - 0 = a (L_g - 0) \implies \tau = \frac{T}{L_g} s,
\end{gather}

with $L_g$ the geodesic length of the trajectory. This result gives a physical interpretation to the Schwinger parameter $T$: it is (proportional to) the proper time of the quanta excited by the field $\phi$, whose trajectories are responsible for the correlations between different points of the spacetime. We can plug this result back into \eqref{eq:appworldsaddle} to obtain:
\begin{gather}
	\langle x_2| e^{-T(-\nabla_\mu \nabla^\mu)} |x_1 \rangle \approx \sum_g \exp \left( - \frac{T}{4} \frac{L_g^2}{T^2} \right),
\end{gather}

so that the Feynman propagator simplifies to:
\begin{gather}
	G_F(x_2-x_1) \approx \sum_g \int_0^{+\infty} dT \: e^{-m^2 T - L_g^2/(4T)} = \sum_g \int_0^{+\infty} \frac{dT}{m} \: e^{-m \left( T + L_g^2/(4T) \right)}.
\end{gather}

The rescaled time $T \to T/m$ has the correct dimension $[T] = -1$. For heavy scalars (which roughly translates to $m \gg 1/L_g$), we can use the saddle point approximation with respect to $T$ to obtain that the main contribution to the integral comes from the region near $T_* = L_g/2$, so that we finally obtain the desired result:
\begin{gather}
	G_F(x_2-x_1) \approx \sum_g e^{-m L_g}.
\end{gather}

The second method to obtain this result is slightly more convoluted, but it yields an intermediate formula for heavy scalars that does not assume the semiclassical limit. We would like to commute the path integral over trajectories with the integral over $T$ in the propagator:
\begin{gather}
	G_F(x_2 - x_1) = \int_0^{+\infty} dT \: e^{-m^2 T} \int_{X(0) = x_1}^{X(T) = x_2} \mathcal{D}X \: \exp \left( -\int_0^T d\tau \: \frac{1}{4} g_{\mu\nu}(X) \dot X^\mu \dot X^\nu \right).
\end{gather}

\begin{figure}
	\centering
	\includegraphics[width = 0.4 \textwidth]{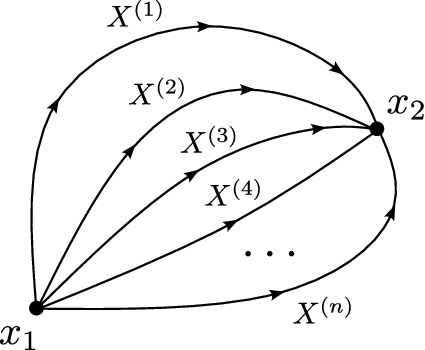}
	\caption{A path integral is a sum over all the possible trajectories that connect two points in a given time $T$. If we rescale the time coordinate properly, the original set of paths is mapped one-to-one to the set of those that contribute to the path integral for a different time $T'$.}
	\label{fig:worldintegral}
\end{figure}

To do this, we make the following observation. A path integral is essentially a sum over different trajectories $X^{(i)}(s)$ that connect $x_1$ to $x_2$, as shown in Figure \ref{fig:worldintegral}. For a fixed time $T$, then, we can characterize it through the continuous set of trajectories $\{ X^{(i)}(\tau) \, | \, X^{(i)}(0) = x_1, \: X^{(i)}(T) = x_2 \}_{i \in \mathbb{R}}$ that contribute to it. If we now consider a different time $T'$, we can easily see that there is an injection from the first set to the second one by simply rescaling the time coordinate: $X(\tau) \to X(\tau T'/T), \: 0 \leq \tau \leq T$. The inverse rescaling is an injection from the second set to the first one, from which we infer that this transformation gives us a bijection between the two sets, and that the set of paths for every $T$ is completely determined by considering the one associated to an arbitrary time $T_0$.

This observation tells us that switching the two integrals means that, given a fixed trajectory belonging to the set of some time $T_0$, we are integrating over all possible rescalings of the speed of the particle. If we switch the integrals and we take the geodesic length $L$ as the reference time for a fixed trajectory, to which we can trace back all the others through $\tau \to \tau T/L$ (the new $\tau$ ranges from $0$ to $L$), we obtain:
\begin{equation}
	\begin{aligned}
		G_F(x_2 - x_1) & = \int_{X(0) = x_1}^{X(L) = x_2} \mathcal{D}X \int_0^{+\infty} dT \: \exp \left( -m^2 T -\frac{L}{T} \int_0^L d\tau \: \frac{1}{4} g_{\mu\nu}(X) \dot X^\mu \dot X^\nu \right) \\
		& \stackrel{T \to T/m}{=} \int_{X(0) = x_1}^{X(L) = x_2} \mathcal{D}X \int_0^{+\infty} \frac{dT}{m} \exp \left[ -m \left( T + \frac{L^2}{4T} \right) \right].
	\end{aligned}
\end{equation}

In the second line, we have used that $d\tau = ds$. For heavy scalars, we can once again use the saddle point approximation with respect to $T$: just like before, the main contribution is due to the region near $T_* = L/2$. This line of reasoning then gives us exactly \eqref{eq:worldstart}, of which one can then take the semiclassical limit:
\begin{gather}
	G_F(x_2-x_1) \approx \int_{X(0) = x_1}^{X(L) = x_2} \mathcal{D}X \exp \left( -m \int_{X(s)} ds \right) \approx \sum_g e^{-m L_g}.
\end{gather}

The worldline formalism is an interesting tool, which allows us to trade a formulation of QFT based on fields with one based on particles. As for this thesis, its utility is the possibility of easily computing approximated two point functions in curved spacetimes in terms of geodesics, as is done in \cite{Blommaert:2024ydx}.

\newpage
\printbibliography

\end{document}